\documentclass[12pt,twoside,singlespace]{mitthesis}

\usepackage{lgrind}
\usepackage{cmap}
\usepackage[T1]{fontenc}
\def\all{all}
\ifx\files\all  \else 

\usepackage[pdfencoding=auto, psdextra]{hyperref}
\usepackage{amsmath}
\usepackage{amssymb}
\usepackage{cleveref}
\usepackage{relsize}
\usepackage{siunitx}

\usepackage{comment}

\usepackage{adjustbox}
\usepackage{subcaption}
\usepackage{xspace}
\usepackage{cases}
\usepackage{isotope}
\usepackage{pdfpages}
\usepackage{indentfirst}
 \usepackage{feynmp-auto}
\usepackage[style=phys, biblabel=brackets, eprint=true, backref=true, url=true]{biblatex}
\usepackage{multirow}

\newcommand{\nue}{\ensuremath{\nu_{e}}\xspace}
\newcommand{\numu}{\ensuremath{\nu_{\mu}}\xspace}
\newcommand{\nutau}{\ensuremath{\nu_{\tau}}\xspace}
\newcommand{\nuebar}{\ensuremath{\bar{\nu}_e}\xspace}
\newcommand{\numubar}{\ensuremath{\bar{\nu}_\mu}\xspace}
\newcommand{\nutaubar}{\ensuremath{\bar{\nu}_\tau}\xspace}

\newcommand{\dCP}{\delta_\mathrm{CP}}
\newcommand{\eVq}{\ensuremath{\text{eV}^2}\xspace}
\newcommand{\keV}{\ensuremath{\text{keV}}\xspace}
\newcommand{\MeV}{\ensuremath{\text{MeV}}\xspace}
\newcommand{\GeV}{\ensuremath{\text{GeV}}\xspace}
\newcommand{\TeV}{\ensuremath{\text{TeV}}\xspace}
\newcommand{\Dmq}{\ensuremath{\Delta m^2}\xspace}
\newcommand{\Dmqfo}{\ensuremath{\Delta m_{41}^2}\xspace}
\newcommand{\Dmqfiveo}{\ensuremath{\Delta m_{51}^2}\xspace}
\newcommand{\sinsqtth}{\sin^2 2 \theta}
\newcommand{\sinsqtthtf}{\ensuremath{\sin^2 2 \theta_{24}}\xspace}

\newcommand{\Uefsq}{\ensuremath{|U_{e4}|^2}\xspace}
\newcommand{\Umufsq}{\ensuremath{|U_{\mu4}|^2}\xspace}
\newcommand{\Utaufsq}{\ensuremath{|U_{\tau4}|^2}\xspace}
\newcommand{\Uef}{\ensuremath{|U_{e4}|}\xspace}
\newcommand{\Umuf}{\ensuremath{|U_{\mu4}|}\xspace}

\newcommand{\Uefive}{\ensuremath{|U_{e5}|}\xspace}
\newcommand{\Umufive}{\ensuremath{|U_{\mu5}|}\xspace}

\DeclareSIUnit\nucleon{nucleon}

\begin{document}

\title{Through Iron \& Ice: Searching for Sterile Neutrinos at the IceCube Neutrino Observatory}

\author{Alejandro Diaz}
\prevdegrees{B.A., University of Chicago (2016)}
\department{Department of Physics}

\degree{Doctor of Philosophy}

\degreemonth{February}
\degreeyear{2023}
\thesisdate{September 30, 2022}


\supervisor{Janet M. Conrad}{Professor of Physics}

\chairman{Lindley Winslow}{Associate Department Head of Physics}

\maketitle



\cleardoublepage
\setcounter{savepage}{\thepage}
\begin{abstractpage}
%
%
%
Despite the rapid progression in our understanding of neutrinos over the last half century, much is left unknown about their properties.
This leaves neutrinos as the most promising portal for Beyond Standard Model (BSM) physics, and neutrinos have already provided fruitful surprises.

A number of neutrino experiments in the last three decades have observed anomalous oscillation signals consistent with a mass-squared splitting of $\Dmq \sim \SI{1}{\eV\squared}$, motivating the existence and search for sterile neutrinos.
On the other hand, other experiments have failed to see such a signal.

In this thesis, we present two analyses.
The first is an update to the sterile neutrino global fits with the inclusion of recent experimental data.
We find that the 3+1 model provides a better fit to the global data set compared to the null, with an improvement of $\Delta \chi^2 = 51$ with the addition of only 3 degrees of freedom, corresponding to $6.6\sigma$.
While a substantial improvement, we also find a irreconcilable tension between the data sets of $5.1\sigma$, calculated using the parameter goodness-of-fit test.
This motivates the exploration of expanded models: a 3+2 model, and a 3+1+Decay model.
In the 3+2 model, we find negligible improvement to the fit, and an even worse tension of $5.5\sigma$.
In the more exotic 3+1+Decay model, we find the tension reduced to $3.6\sigma$.
While a substantial improvement compared to the 3+1 model with the introduction of only one additional parameter, the tension is still too large to assuage concerns. 

The second analysis is the results of an expanded IceCube sterile neutrino search.
A previous sterile neutrino search found no evidence for sterile neutrinos, finding a p-value of 8\%.
Of the three sterile mixing angles, $\theta_{14}, \theta_{24}$, and $\theta_{34}$, only $\theta_{24}$ was fitted for, as $\theta_{14}$ was negligible and $\theta_{34} = 0$ was considered a conservative assumption.
We present results of an analysis where we include $\theta_{34}$ to the fitted model.
Both a frequentist and Bayesian analysis were conducted, with fits done in terms of the mass-squared splitting \Dmqfo and the mixing matrix parameters \Umufsq and \Utaufsq.
The frequentist analysis finds a best fit at $\Dmqfo = \SI{5.0}{\eV\squared}$, $\Umufsq = 0.04$, and $\Utaufsq = 0.006$, with a p-value of 5.2\% assuming Wilks' Theorem with 3 degrees of freedom.
Pseudoexperiments are indicating a smaller p-value 2.7\%.
The Bayesian analysis finds a similar best fit point at $\Dmqfo = \SI{5.0}{\eV\squared}$, $\Umufsq = 0.02$, and $\Utaufsq = 0.006$, with a Bayes factor indicating a ``Very Strong'' preference for this sterile hypothesis over the null hypothesis.

\end{abstractpage}


\cleardoublepage

\section*{Acknowledgments}

It's impossible to properly acknowledge and thank everyone who has had an impact on me over the last six years, but I will try.

First, I have to thank my advisor, Professor Janet Conrad.
An omnipotent force, Janet has guided me through the maze of neutrino physics with an uncanny intuition towards the profound and interesting.
The scientist I am today would not have existed without Janet's hard work and dedication to my success.
Thank you, Janet, for having reached out to me after I submitted my application to MIT.

Along with Janet, any success of mine must be shared with the whole of the Conrad research group.
The work in this thesis is truly the outcome of a collaborative effort amongst this formidable group of up-and-coming scientists.

To the postdocs, Carlos Arg\"{u}elles, Daniel Winklehner, Taritree Wongjirad, Adrien Hourlier, David Vannerom, Austin Schneider, and John Hardin: while only a few years stand between myself and them, their knowledge and experience feels decades ahead of mine.
Beyond the physics, they have taught me what the life of an academic entails.
In particular, I'd like to thank (now Professor) Carlos Arg\"{u}elles.
His first task as a postdoc with Janet was to talk to me as a prospective student; and I'm grateful, both professionally and personally, to have known him through my entire grad school career.

To Janet's grad students that I got to know well, Gabriel Collin, Spencer Axani, Jarrett Moon, Marjon Moulai, Lauren Yates, Loyd Waits, Joe Smolsky, and Nick Kamp:
my relationship with each of you has been lopsided, having gained more from you than you did from me.
Either in teaching me all the research know-how, or having the shared experience of stumbling through neutrino physics, I'm indebted to each of you.

To the younger grad students, Darcy Newmark and Philip Weigel: unfortunately, worldwide circumstances took away our chance to learn from each other.
I only have one piece of wisdom: Don't work so hard, you're fine.

Outside of my research group, I've been blessed with a multitude of people that have ridden through MIT alongside me.
To Field, Joe, Efrain, Cedric, Afro, Nick, Sangbaek, Francesco, and Dani: You know I hate to do things alone, and that includes struggling. Thank you for being there with me while we studied and cried, played board and video games, and explored bits of the world outside of Cambridge.
To my roommate, Michael Calzadilla: thank you for the late-night company and the frequent trips for ice cream; I'm sorry for inflicting my social needy-ness onto you.
To the Astro and LIGO boys and girls, Ben, Chris, Kaley, David, and Nick: thank you for soaking up the sun with me at Provincetown and Spectacle Island, and for taking spontaneous trips to Walden Pond and Iceland. 
Outside of MIT, I'd like to thank Chris Barnes and Adam Lister for keeping me sane in Fermilab.
And to Alejandro Buendia, thank you for your support and comfort throughout the pain of the last year.

I'm lucky to have kept in touch with my closest friends from UChicago, and fortunate that many lived near where my research took me.
I'm thankful to have spent Thanksgivings with Max, Tres, Brian, and Hunter, and celebrating a New Year with Jenni and Sal at Medieval Times.
As I'm currently sitting on a plane to Los Angeles for the bachelor party, I'd like to wish Jenni and Sal a happy life together.
I'd also like to thank Raul Zalda\~{n}a-Calles, for having to deal with me in my first years at MIT from a distance.

To those in my hometown of Miami, Lazaro Rodriguez, Brandon Castro, and Carlos Morales: thank you for maintaining our friendships, despite the months and years that pass between our hangouts.

I'd like to thank my family, my dad and sister (the real doctor in the family), for having supported and encouraged me throughout my entire academic career.
I also need to thank my extended family in Colombia, for providing invaluable support especially over the last year.

Above and beyond, I have to thank my mother, Angela Maria.
I adore her, and any success in my life should be attributed to her love, care, and unreasonable pride for me.
I miss her dearly, \textit{Te quiero}.


\pagestyle{plain}

\tableofcontents

\begin{fmffile}{diagram}

    \chapter{Neutrino Oscillations}

\section{Theory}

Let us consider neutrino oscillation in the case of $N$-neutrino mixing.
In this discussion, we will denote the neutrino mass states with Latin subscripts (e.g. $\nu_{i}$, $\nu_{j}$), and the neutrino flavor states with Greek subscripts (e.g. $\nu_{\alpha}$, $\nu_{\beta}$), unless otherwise stated. 
The neutrino mass states are related to the flavor states by the $N \times N$ matrix
\begin{equation}
    \begin{pmatrix}
        \nu_\alpha \\
        \nu_\beta \\
        \vdots
    \end{pmatrix}
    = 
    \begin{pmatrix}
        U_{\alpha 1} & U_{\alpha 2} & \dots \\
        U_{\beta 1} & U_{\beta 2} & \\
        \vdots & & \ddots
    \end{pmatrix}
    \begin{pmatrix}
        \nu_1 \\
        \nu_2 \\
        \vdots
    \end{pmatrix}.
\end{equation}
For now, we simply quote here the $N$-neutrino oscillation formula. 
\begin{equation}
    \begin{split}
        P(\nu_\alpha \to \nu_\beta )= \delta_{\alpha \beta} 
        &- 4 \sum_{i<j} \Re(U_{\alpha i}^{*} U_{\beta i}U_{\alpha j} U_{\beta j}^{*}) \sin^{2}\left( 1.27 \Delta m_{ji}^2 [\si{\eV}] \frac{L [\si{\kilo\meter}]}{E [\si{\GeV}]} \right) \\
        &-2 \sum_{i<j} \Im(U_{\alpha i}^{*} U_{\beta i}U_{\alpha j} U_{\beta j}^{*}) \sin \left(2.54 \Delta m_{ji}^2 [\si{\eV}] \frac{L [\si{\kilo\meter}]}{E [\si{\GeV}]} \right),
    \end{split}
    \label{eq:finaloscequation}
\end{equation}
where the mass-squared splitting $\Delta m_{ji}^2 = m_{j}^{2} - m_{i}^{2}$.
The notation ``$P(\nu_\alpha \to \nu_\beta)$'' is understood as the probability that a neutrino of original flavor $\nu_\alpha$ is later measured as $\nu_\beta$.
In the case that $\alpha \neq \beta$, ``$P(\nu_\alpha \to \nu_\beta)$'' is referred to as an \textit{appearance} probability, and an experiment that makes this kind of measurement is referred to as an appearance experiment.
When $\alpha = \beta$, ``$P(\nu_\alpha \to \nu_\alpha)$'' is referred to as a \textit{disappearance} probability, and an experiment that makes this measurement is called a disappearance experiment.

In \Cref{eq:finaloscequation}, the mass-squared splitting $\Delta m_{ji}^2$ is in units of \si{\eV\squared}, the neutrino energy $E$ in \si{\GeV}, and the distance $L$ in kilometers. 
This is the standard in the neutrino community, and we will use these units unless otherwise stated.
A complete derivation of \Cref{eq:finaloscequation} is provided in \Cref{sec:NeutrinoOscillationsDerivation}. 

Let's quickly note some CP-related properties of \Cref{eq:finaloscequation}. 
To get the CP conjugated oscillation equation \(\hat{C}\hat{P} P(\nu_\alpha \to \nu_\beta ) = P(\bar{\nu}_\alpha \to \bar{\nu}_\beta )\), we would simply replace each mixing matrix parameter with its complex conjugate \(U\to U^{*}\). This results in flipping the sign of the second line of \Cref{eq:finaloscequation}. 
Therefore, if the mixing matrix contains complex terms, \( P(\nu_\alpha \to \nu_\beta )  \neq P(\bar{\nu}_\alpha \to \bar{\nu}_\beta )\) and CP-symmetry is violated in the neutrino sector. 
An exception occurs when we consider \( \nu_\beta = \nu_\alpha \) (disappearance). 
In that case, the term \(U_{\alpha i}^{*} U_{\beta i}U_{\alpha j} U_{\beta j}^{*}\) becomes \(|U_{\alpha i}|^{2} |U_{\alpha j}|^{2}\), which is entirely real. Therefore \(\Im(U_{\alpha i}^{*} U_{\beta i}U_{\alpha j} U_{\beta j}^{*}) = 0\) and \Cref{eq:finaloscequation} does not change with the transformation \(U\to U^{*}\). CP-violation in the lepton sector is thus not observable in disappearance experiments. 

\section{Two Neutrinos}
\label{sec:twonu}

As an example, it is useful to first consider the case where we have only two neutrinos mixing. We'll consider the weak eigenstates \(\nu_{e}\) \& \(\nu_{\mu}\), and the two neutrino mass eigenstates \(\nu_{1}\) \& \(\nu_{2}\). 

We write our mixing relationship as
\begin{equation}
    \begin{pmatrix}
        \nu_e \\
        \nu_\mu
    \end{pmatrix}
    = 
    \begin{pmatrix}
        U_{e 1} & U_{e 2} \\
        U_{\mu 1} & U_{\mu 2}
    \end{pmatrix}
    \begin{pmatrix}
        \nu_1 \\
        \nu_2 
    \end{pmatrix}.
\end{equation}

As we'll come to see, the mixing matrix is frequently written as a rotation matix, with the matrix elements witten in terms of some mixing ``angle.'' In the two-neutrino case this is 
\begin{equation}
    \begin{pmatrix}
        \nu_e \\
        \nu_\mu
    \end{pmatrix}
    = 
    \begin{pmatrix}
        \cos\theta & \sin\theta \\
        -\sin\theta & \cos\theta
    \end{pmatrix}
    \begin{pmatrix}
        \nu_1 \\
        \nu_2 
    \end{pmatrix},
    \label{eq:2neutrinowithmixing}
\end{equation}
where the mixing is parametarized by the single angle $\theta$ (the remaining degrees of freedom for the $2\times2$ unitary matrix can be absorbed into the definition of the neutrino states).

Reading off \Cref{eq:finaloscequation} and using some trigonometric identities, we end up with the oscillation equations

\begin{align}
    P(\nue \to \numu) &= \sin^2(2\theta) \sin^2\left(1.27 \frac{\Delta m^{2} L}{E}\right) \label{eq:2n_appearance}\\
    &= P(\nuebar \to \numubar)\\
    &= P(\numu \to \nue)\\
    &= P(\numubar \to \nuebar)\\
    P(\nue \to \nue) &= 1 - \sin^2(2\theta) \sin^2\left(1.27 \frac{\Delta m^{2} L}{E}\right)\\
    &= P(\nuebar \to \nuebar)\\
    &= P(\numu \to \numu)\\
    &= P(\numubar \to \numubar)\label{eq:2n_disappearance}.
\end{align}

\begin{figure}
    \centering
    \includegraphics[width=0.9\textwidth]{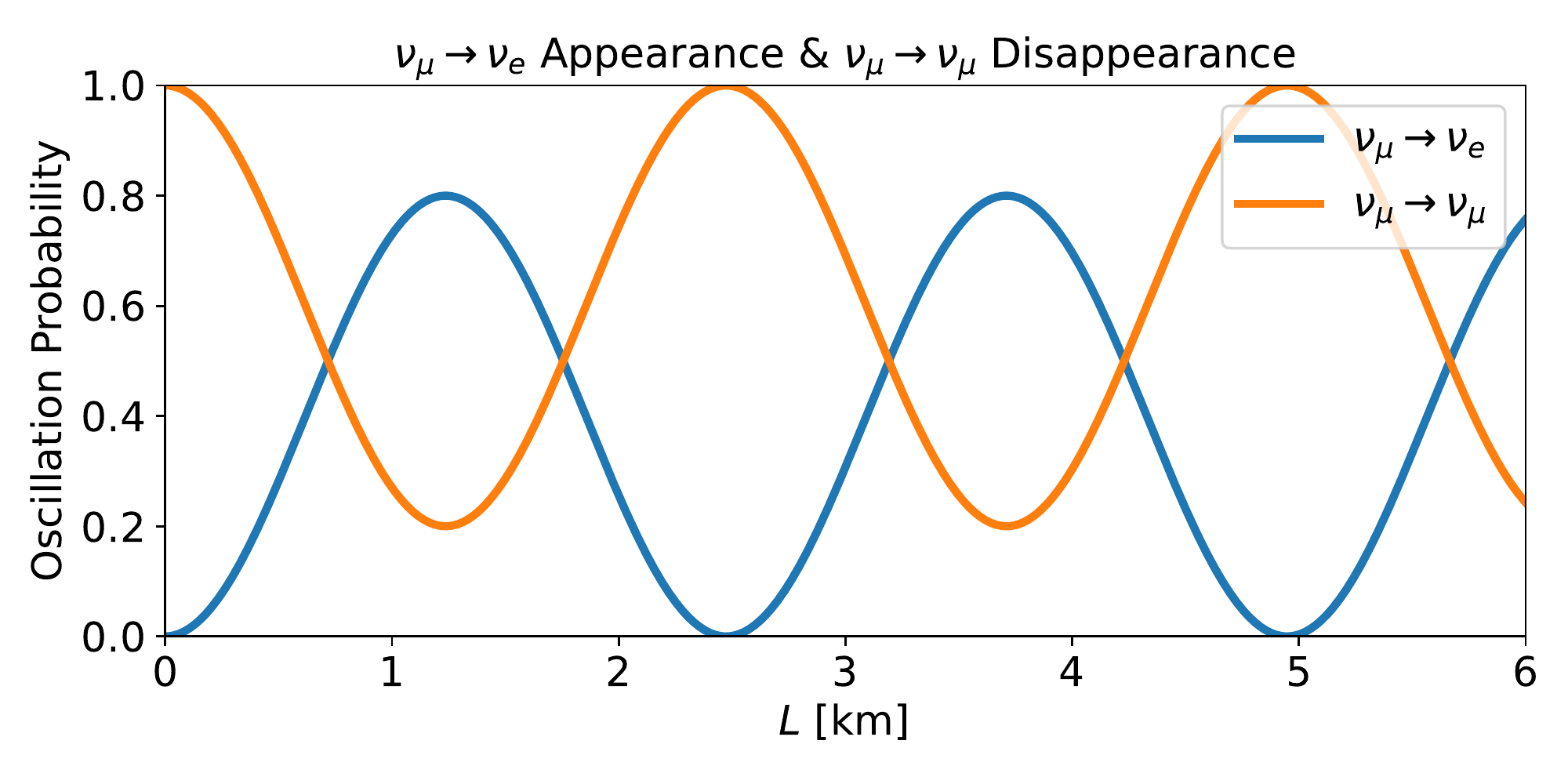}
    \caption{Appearance probability $P(\numu \to \nue)$ and disappearance probability $P(\numu \to \numu)$, as a function of baseline $L$, for a $\numu$ beam of \SI{1}{\giga\electronvolt} and oscillation parameters $\sin^2(2\theta) = 0.8$ and $\Delta m^{2} = \SI{1}{\electronvolt\squared}$.}
    \label{fig:osc_prob}
\end{figure}

Suppose that we have a
\SI{1}{\GeV} \numu beam produced at some source, and we measure the flavor composition some distance $L$ away. 
For the mixing parameters $\sin^2(2\theta) = 0.8$ and $\Delta m^{2} = \SI{1}{\eV\squared}$, we would have an oscillation probability as a function of distance $L$ as shown in \Cref{fig:osc_prob}.
In the figure, the oscillation amplitude is determined by $\sin^2(2\theta)$ and the frequency by $\Delta m^{2}$. 

It is convenient for us to define the \textit{oscillation length}
\begin{equation}
    \label{eq:osclength}
    L_o = \frac{\pi E}{1.27 \Dmq},
\end{equation}
which is the propagation distance over which a complete oscillation takes place. 
In \Cref{fig:osc_prob}, this would be about 2.5 km. 

The oscillation length also dictates the $\Dmq$ an experiment is sensitive to given an $L$ and $E$; or, alternatively, what $L$ and $E$ to choose given a known $\Dmq$. 
In the example of \Cref{fig:osc_prob}, where we assume to know $\Dmq$ and have a fixed $E$, we would like to place our detector at $L_o/2$ where the oscillation is at its maximum. 

In practice, if a detector is placed $L \gg L_o$, then uncertainties in $L$ and $E$ (due to production and detection uncertainties) will cause the oscillation curves to average out, such that $\sin^2\left(1.27 \frac{\Delta m^{2} L}{E}\right) \to 1/2$, and 
\begin{align}
    P(\nu_\alpha \to \nu_\beta) &= \frac{1}{2} \sin^2(2\theta)  \quad \beta \neq \alpha\\
    P(\nu_\alpha \to \nu_\alpha) &= 1 - \frac{1}{2} \sin^2(2\theta). 
\end{align}
In this case, the experiment will have no sensitivity to $\Dmq$, only the mixing angle $\theta$.

If, on the other hand, the detector is placed such that $L \ll L_o$, then the neutrinos will have propagated for too little distance (i.e. time) to have observably oscillated. Therefore, there is no sensitivity to any oscillation parameters. 

While nature is known to have more than two neutrinos, the two-neutrino model is often a valid approximation. 
For $N$ neutrinos, there are oscillation lengths corresponding to each pair of $\Dmq$
\begin{equation}
    L_{o_{ij}} = \frac{\pi E}{1.27 \Dmq_{ij}}.
\end{equation}
If there exists a $\Dmq$ (or a set of degenerate $\Dmq$s) that is much larger than the remaining $\Dmq$s, then the corresponding oscillation length $L_o^{*}$ would be much shorter than the remaining oscillation lengths $\bar L_o^{*}$s.
If the detector is placed such that $L \sim L_{o}^{*} \ll \bar L_o^{*}$, then the detector would be sensitive only to the one larger $\Dmq$, approximating two-neutrino oscillations.

\section{Three Neutrinos}
\label{sec:3neutrinos}

In the Standard Model (SM), there are three neutrinos, and therefore a $3\times3$ mixing matrix, called the Pontecorvo-Maki-Nakagawa-Sakata (PMNS) matrix. 

\begin{equation}
    \begin{pmatrix}
        \nu_e \\
        \nu_\mu \\
        \nu_\tau
    \end{pmatrix}
    = 
    \begin{pmatrix}
        U_{e 1} & U_{e 2} & U_{e 3} \\
        U_{\mu 1} & U_{\mu 2} & U_{\mu 3}\\
        U_{\tau 1} & U_{\tau 2} & U_{\tau 3}
    \end{pmatrix}
    \begin{pmatrix}
        \nu_1 \\
        \nu_2 \\
        \nu_3
    \end{pmatrix}.
\end{equation}

We will refer to Ref.~\cite{Giunti:2007ry} for the details of three neutrino oscillations. 
For now, we will only note that three neutrino oscillations, like two neutrino oscillations, is typically written in terms of unitary rotations. 
In this convention, the PMNS matrix is written as 
\begin{equation}
    U_{\textrm{PMNS}} = R^{23}(\theta_{23}) R^{13}(\theta_{13}, \delta) R^{12}(\theta_{12}),
\end{equation}
or, 
\begin{equation}
    \begin{pmatrix}
        U_{e 1} & U_{e 2} & U_{e 3} \\
        U_{\mu 1} & U_{\mu 2} & U_{\mu 3}\\
        U_{\tau 1} & U_{\tau 2} & U_{\tau 3}
    \end{pmatrix}
    \\
    = \begin{pmatrix}
        1 & 0 & 0 \\
        0 & c_{23} & s_{23} \\
        0 & -s_{23} & c_{23}
    \end{pmatrix}
    \begin{pmatrix}
        c_{13} & 0 & s_{13} e^{-i\delta} \\
        0 & 1 & 0 \\
        -s_{13} e^{i\delta} & 0 & c_{13}
    \end{pmatrix}
    \begin{pmatrix}
        c_{12} & s_{12} & 0 \\
        -s_{12} & c_{12} & 0 \\
        0 & 0 & 1
    \end{pmatrix},
\end{equation}
where $s_{ij}$ and $c_{ij}$ is shorthand for $\sin\theta_{ij}$ and $\cos\theta_{ij}$ respectively.

In this model, there are 6 independent parameters, $\Delta m_{21}^{2}, \Delta m_{31}^{2}, \theta_{12}, \theta_{13},  \theta_{23}$, and $\delta$. 

\subsection{Best Fit}
\label{sec:3nubf}

Combined fits of the three neutrino model parameters are periodically conducted by the NuFit organization. The results of their most recent fit \cite{Esteban:2020cvm} are printed in \Cref{table:3nubf}.

\begin{table}\centering
    \begin{footnotesize}
        \begin{tabular}{l|cc|cc}

            & \multicolumn{2}{c|}{Normal Ordering (best fit)}
            & \multicolumn{2}{c}{Inverted Ordering ($\Delta\chi^2=2.7$)}
            \\
            \cline{2-5}
            & bfp $\pm 1\sigma$ & $3\sigma$ range
            & bfp $\pm 1\sigma$ & $3\sigma$ range
            \\
            \cline{1-5}
            \rule{0pt}{4mm}\ignorespaces
            $\sin^2\theta_{12}$
            & $0.304_{-0.012}^{+0.013}$ & $0.269 \to 0.343$
            & $0.304_{-0.012}^{+0.013}$ & $0.269 \to 0.343$
            \\[1mm]
            $\theta_{12}/^\circ$
            & $33.44_{-0.75}^{+0.78}$ & $31.27 \to 35.86$
            & $33.45_{-0.75}^{+0.78}$ & $31.27 \to 35.87$
            \\[3mm]
            $\sin^2\theta_{23}$
            & $0.570_{-0.024}^{+0.018}$ & $0.407 \to 0.618$
            & $0.575_{-0.021}^{+0.017}$ & $0.411 \to 0.621$
            \\[1mm]
            $\theta_{23}/^\circ$
            & $49.0_{-1.4}^{+1.1}$ & $39.6 \to 51.8$
            & $49.3_{-1.2}^{+1.0}$ & $39.9 \to 52.0$
            \\[3mm]
            $\sin^2\theta_{13}$
            & $0.02221_{-0.00062}^{+0.00068}$ & $0.02034 \to 0.02430$
            & $0.02240_{-0.00062}^{+0.00062}$ & $0.02053 \to 0.02436$
            \\[1mm]
            $\theta_{13}/^\circ$
            & $8.57_{-0.12}^{+0.13}$ & $8.20 \to 8.97$
            & $8.61_{-0.12}^{+0.12}$ & $8.24 \to 8.98$
            \\[3mm]
            $\dCP/^\circ$
            & $195_{-25}^{+51}$ & $107 \to 403$
            & $286_{-32}^{+27}$ & $192 \to 360$
            \\[3mm]
            $\dfrac{\Dmq_{21}}{10^{-5}~\eVq}$
            & $7.42_{-0.20}^{+0.21}$ & $6.82 \to 8.04$
            & $7.42_{-0.20}^{+0.21}$ & $6.82 \to 8.04$
            \\[3mm]
            $\dfrac{\Dmq_{3\ell}}{10^{-3}~\eVq}$
            & $+2.514_{-0.027}^{+0.028}$ & $+2.431 \to +2.598$
            & $-2.497_{-0.028}^{+0.028}$ & $-2.583 \to -2.412$
            \\[2mm]

        \end{tabular}
    \end{footnotesize}
    \caption{Best fit three-neutrino oscillation parameters as fitted by the NuFit group \cite{Esteban:2020cvm}. The first column gives the best fit values assuming normal ordering (i.e. $\Delta m_{31}^{2}>0$), while the second column gives the best fit values assuming inverted ordering (i.e. $\Delta m_{31}^{2}<0$).}
    \label{table:3nubf}
\end{table}

With these parameters, NuFit finds the $3\sigma$ range of the mixing parameters to be 
\begin{equation}
    \label{eq:umatrix}
    \begin{aligned}
        |U|_{3\sigma}=
        \begin{pmatrix}
            0.801 \to 0.845 &\qquad
            0.513 \to 0.579 &\qquad
            0.143 \to 0.156
            \\
            0.233 \to 0.507 &\qquad
            0.461 \to 0.694 &\qquad
            0.631 \to 0.778
            \\
            0.261 \to 0.526 &\qquad
            0.471 \to 0.701 &\qquad
            0.611 \to 0.761
        \end{pmatrix}.
    \end{aligned}
\end{equation}

\begin{figure}
    \centering
    \includegraphics[width=0.46\textwidth]{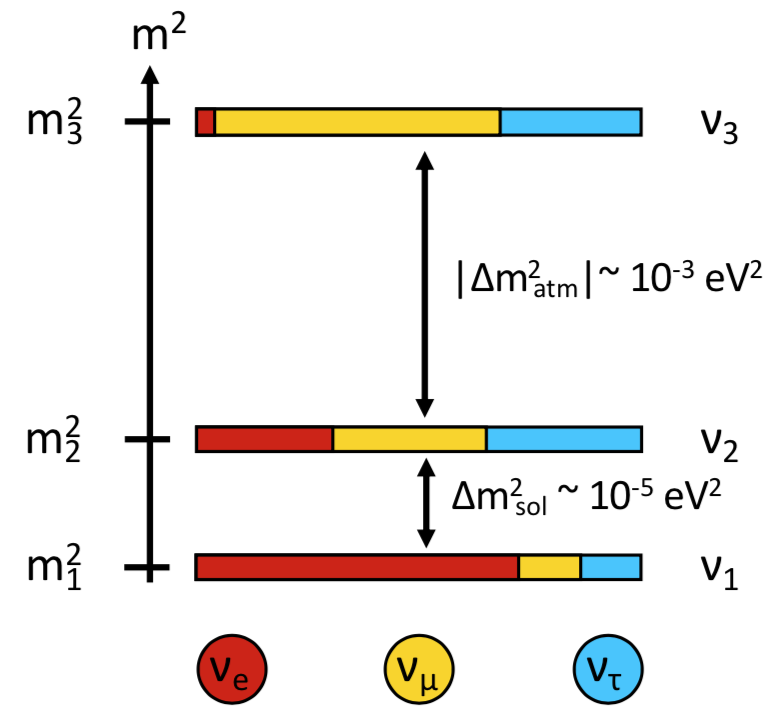}
    \caption{A visualization of the mass-squared splittings and mixing of the three neutrinos in the SM. Each horizontal bar corresponds to a neutrino mass state $\nu_i$, where their spacing illustrates the mass squared differences (normal ordering, $\Delta m_{31}^{2}>0$, is assumed). The colors within each bar represents the content of the flavor eigenstates $\nu_\alpha$ within each mass state. Figure taken from Ref.~\cite{Moulai:2021zey}.}
    \label{fig:3nu}
\end{figure}

A visualization of the mass-squared splittings and the mixing elements are shown in \Cref{fig:3nu}.

\section{Neutrino Oscillation in Matter}
\label{sec:matteroscillation}

In the preceding sections, we have described neutrinos oscillating specifically in a \textit{vacuum}.
In reality, neutrinos will propagate through matter, interacting with the particles it traverses. 
While all three SM neutrino types will experience neutral-current (NC) interactions, only electron neutrinos will experience charged-current (CC) interactions since matter contains free electrons but no free muons or taus.
This will alter how the neutrinos will oscillate compared with how neutrinos oscillate through vacuum. 

The following description follows Ref.~\cite{Giunti:2007ry}.
In the flavor basis, the neutrino evolution equation can be written as 
\begin{equation}
    i \frac{d}{dx}\Psi_\alpha = \mathcal{H}_{F} \Psi_\alpha,
\end{equation}
where time $t$ usually found in the Schr{\"o}dinger equation is replaced by $x$ (due to the approximation that $x=t$, as done in \Cref{sec:NeutrinoOscillationsDerivation}), $\Psi_\alpha$ is a column vector describing a neutrino state initially produced in the $\alpha$ state, and $\mathcal{H}_{F}$ is the effective Hamiltonian in the flavor basis. 
In matter, $\mathcal{H}_{F}$ is described by
\begin{equation}
    \mathcal{H}_{F}=\frac{1}{2E}(U \mathbb{M}^{2} U^{\dagger} + \mathbb{A}),
    \label{eq:HF}
\end{equation}
where, $U$ is the neutrino mixing matrix.
For three neutrinos,
\begin{equation}
    \Psi_\alpha =     \begin{pmatrix}
        \psi_{\alpha e} \\
        \psi_{\alpha \mu} \\
        \psi_{\alpha \tau}
    \end{pmatrix},
    \quad
    \mathbb{M}^{2} =
    \begin{pmatrix}
        0 & 0 & 0 \\
        0 & \Delta m_{21}^{2} & 0 \\
        0 & 0 & \Delta m_{31}^{2}
    \end{pmatrix}, 
    \quad
    \mathbb{A} =
    \begin{pmatrix}
        A_{\textrm{CC}} & 0 & 0 \\
        0 & 0 & 0 \\
        0 & 0 & 0
    \end{pmatrix},
\end{equation}
with 
\begin{equation}
    A_\textrm{CC} \equiv 2 E V_\textrm{CC} = 2 \sqrt{2} E G_{F} N_e,
\end{equation}
where $G_F$ is the Fermi constant and $N_e$ is the density of electrons in the propagation medium. 
NC interactions are ignored since all neutrino types would undergo NC interactions in matter equally; the NC terms in $ \mathbb{A}$ can therefore be removed by a common phase. 

After simplifying our problem to two neutrinos, and applying a phase shift,
\begin{equation}
    \psi_{\alpha \beta} (x)
    \to 
    \psi_{\alpha \beta} (x)
    e^{
        -i \Dmq x/4E -\frac{i}{2}
        \int_{0}^{x}V_{\textrm{CC}}(x')dx'
    }.
\end{equation}
the evolution equation can be written as 
\begin{equation}
    i\frac{d}{dx}
    \begin{pmatrix}
        \psi_{\alpha e} \\
        \psi_{\alpha \mu}
    \end{pmatrix}
    = \frac{1}{4E}
    \begin{pmatrix}
        -\Delta m^2 \cos 2\theta + A_{\textrm{CC}} & \Delta m^2 \sin 2\theta  \\
        \Delta m^2 \sin 2\theta & \Delta m^2 \cos 2\theta - A_{\textrm{CC}}
    \end{pmatrix}
    \begin{pmatrix}
        \psi_{\alpha e} \\
        \psi_{\alpha \mu}
    \end{pmatrix},
\end{equation}
where $\theta$ is the two-neutrino mixing angle, as in \Cref{eq:2neutrinowithmixing}.
We can diagonalize this matrix, giving us the effective Hamiltonian matrix in the mass basis when in matter of constant density,
\begin{equation}
    U_{\textrm{M}}^{T}\mathcal{H}_{\textrm{F}} U_{\textrm{M}} = \mathcal{H}_{\textrm{M}},
\end{equation}
where
\begin{equation}
    \mathcal{H}_{\textrm{M}} = \frac{1}{4E}\textrm{diag}(-\Delta m_\textrm{M}^{2}, \Delta m_\textrm{M}^{2})
\end{equation}
is the effective Hamiltonian in the mass basis. The mixing matrix $U_\textrm{M}$ is given by 
\begin{equation}
    U_\textrm{M} = 
    \begin{pmatrix}
        \cos \theta_\textrm{M} & \sin \theta_\textrm{M} \\
        -\sin \theta_\textrm{M}  & \cos \theta_\textrm{M} 
    \end{pmatrix}.
\end{equation}
The new parameters $\Delta m_{\textrm{M}}^{2}$ and $\theta_\textrm{M}$ are given by 
\begin{equation}
    \Delta m_{\textrm{M}}^{2}=
    \sqrt{(\Delta m^{2} \cos 2\theta - A_{\textrm{CC}})^{2} + (\Delta m^{2} \sin 2\theta)^{2}}.
    \label{eq:229}
\end{equation}
and 
\begin{equation}
    \tan 2 \theta_\textrm{M} = 
    \frac{\tan 2 \theta}{ 1 - \frac{A_\textrm{CC}}{\Delta m^{2} \cos 2 \theta } },
\end{equation}
or 
\begin{align}
    \cos 2 \theta_\textrm{M} & = \frac{\Delta m^{2} \cos 2 \theta - A_\textrm{CC} }{\Delta m_\textrm{M}^{2}} &= \frac{\Delta m^{2} \cos 2 \theta - A_\textrm{CC} }{\sqrt{(\Delta m^{2} \cos 2\theta - A_{\textrm{CC}})^{2} + (\Delta m^{2} \sin 2\theta)^{2}}}  \\
    \sin 2 \theta_\textrm{M} & = \frac{\Delta m^{2} \sin 2 \theta}{\Delta m_\textrm{M}^{2}} &= \frac{\Delta m^{2} \sin 2 \theta}{\sqrt{(\Delta m^{2} \cos 2\theta - A_{\textrm{CC}})^{2} + (\Delta m^{2} \sin 2\theta)^{2}}}.
    \label{eq:sin2thmatter}
\end{align}
In this scenario, where the matter density is constant, we find that the oscillation parameters $\Delta m^{2}$ and $\theta$ pick up an effective value $\Delta m^{2}_\textrm{M}$ and $\theta_\textrm{M}$. 
They would simply replace the parameters as seen in  \Crefrange{eq:2n_appearance}{eq:2n_disappearance}.

An interesting phenomena can be seen in \Cref{eq:sin2thmatter}. If we set
\begin{equation}
    A_\textrm{CC}^{\textrm{R}}=\Delta m^{2} \cos 2 \theta,
\end{equation}
which is equivalent to setting the electron density to
\begin{equation}
    N_e^{\textrm{R}}= 
    \frac{ \Delta m^{2} \cos 2 \theta} {2 \sqrt{2} E G_\textrm{F}},
    \label{eq:234}
\end{equation}
then $\sin \theta_\textrm{M}$ is maximised to a value of $1$, i.e. we see complete disappearance of the produced flavor eigenstate. This is regardless of the vacuum value of $\theta$.
The phenomena of matter oscillation was first described in \cite{PhysRevD.17.2369,Mikheyev:1985zog,Mikheev:1986wj}.

While a treatment of neutrino oscillation through \textit{changing} matter density is beyond the scope of this thesis, a complete treatment can be found in \cite{Giunti:2007ry,Caldwell:2001pc}.

For the experiments used in our global fits describe in \Cref{ch:globalfits}, the neutrino energies are too low, the baselines too short, and the medium too low density to make matter effects observable. Therefore, we simply assume vacuum oscillations for those experiments. 
Matter oscillations will only become relevant when we discuss IceCube in \Cref{ch:IceCube,ch:MEOWSplusth34,ch:ICresults}.

    \chapter{Anomalous Results \& Sterile Neutrinos}
\label{ch:anomalousresults}

In this chapter, we will first introduce a number of experiments and observations that have motivated the search for sterile neutrinos. 
These experiments can typically be categorized into three types: accelerator-source neutrinos, reactor-source neutrinos, and radioactive-source neutrinos. 
We will then introduce a handful of sterile neutrino models and their phenomenology.

\section{Accelerator Source Neutrinos}
\label{sec:acceleratorsourceneutrinos}

\subsection{LSND}
\label{sec:anomaliesLSND}

The earliest experiment that suggested the existence of sterile neutrinos was the Liquid Scintillator Neutrino Detector (LSND) experiment \cite{LSND:2001aii}, which ran 1993-1998 at Los Alamos National Laboratory (LANL).

The purpose of the experiment was to observe $\numubar$ of energy $20-52.8~\si{\MeV}$ oscillating into $\nuebar$ over a \SI{30}{\m} baseline. 
Referring to \Cref{eq:osclength}, this gave LSND sensitivity to an oscillation with $\Dmq \sim \SI{1}{\eV\squared}$, while being insensitive to the two SM mass squared splittings given in \Cref{sec:3nubf}.

The decay-at-rest (DAR) neutrino source was created by impinging a \SI{\sim 1}{\mA} beam of \SI{798}{\MeV} protons on a target, producing mainly pions. 
The negatively charged $\pi^{-}$'s are mostly absorbed.
On the other hand, the positively charged $\pi^{+}$'s are likely to decay as $\pi^{+}\to \mu^{+} \numu$.
The $\mu^{+}$'s then decay at rest as $\mu^{+}\to e^{+} \nue \numubar$.
The fact that the $\mu^{+}$'s decay at rest means that the $\numubar$ energy distribution is well understood with an end point at \SI{52.8}{\MeV}.

The detector was a cylindrical tank filled with 167 metric tons of mineral oil acting as a liquid scintillator.
The event of interest, a $\nuebar + p \to e^{+} + n$ interaction, would produce two correlated signals. 
First, the outgoing positron produces scintillation light, while the outgoing neutron later captures on a free proton and emits a \SI{2.2}{\MeV} photon.

\begin{figure}
    \centering
    \begin{subfigure}{0.45\textwidth}
        \includegraphics[width=\textwidth]{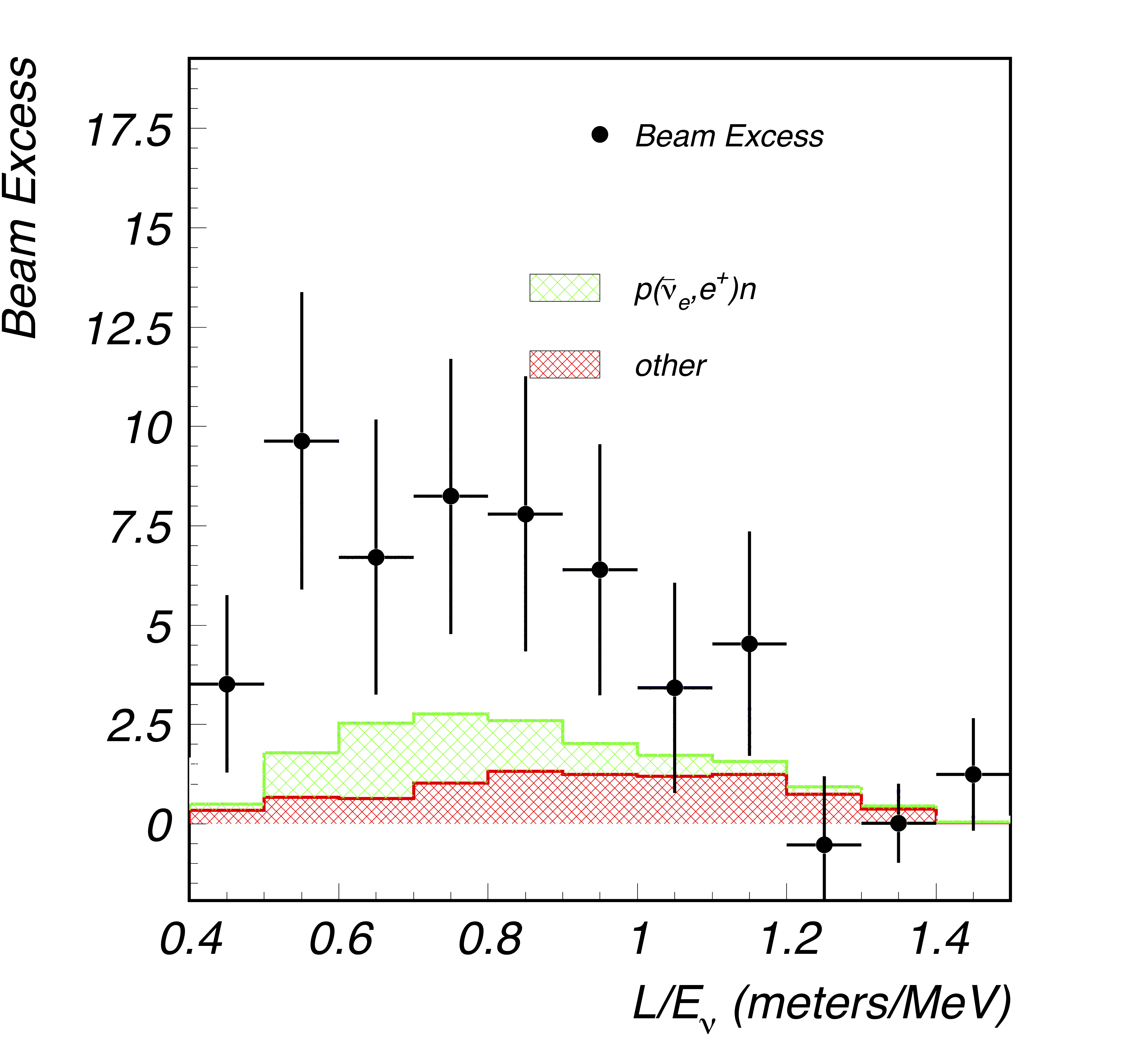}
        \caption{}
        \label{fig:LSNDexcess}
    \end{subfigure}
    \begin{subfigure}{0.45\textwidth}
        \includegraphics[width=\textwidth]{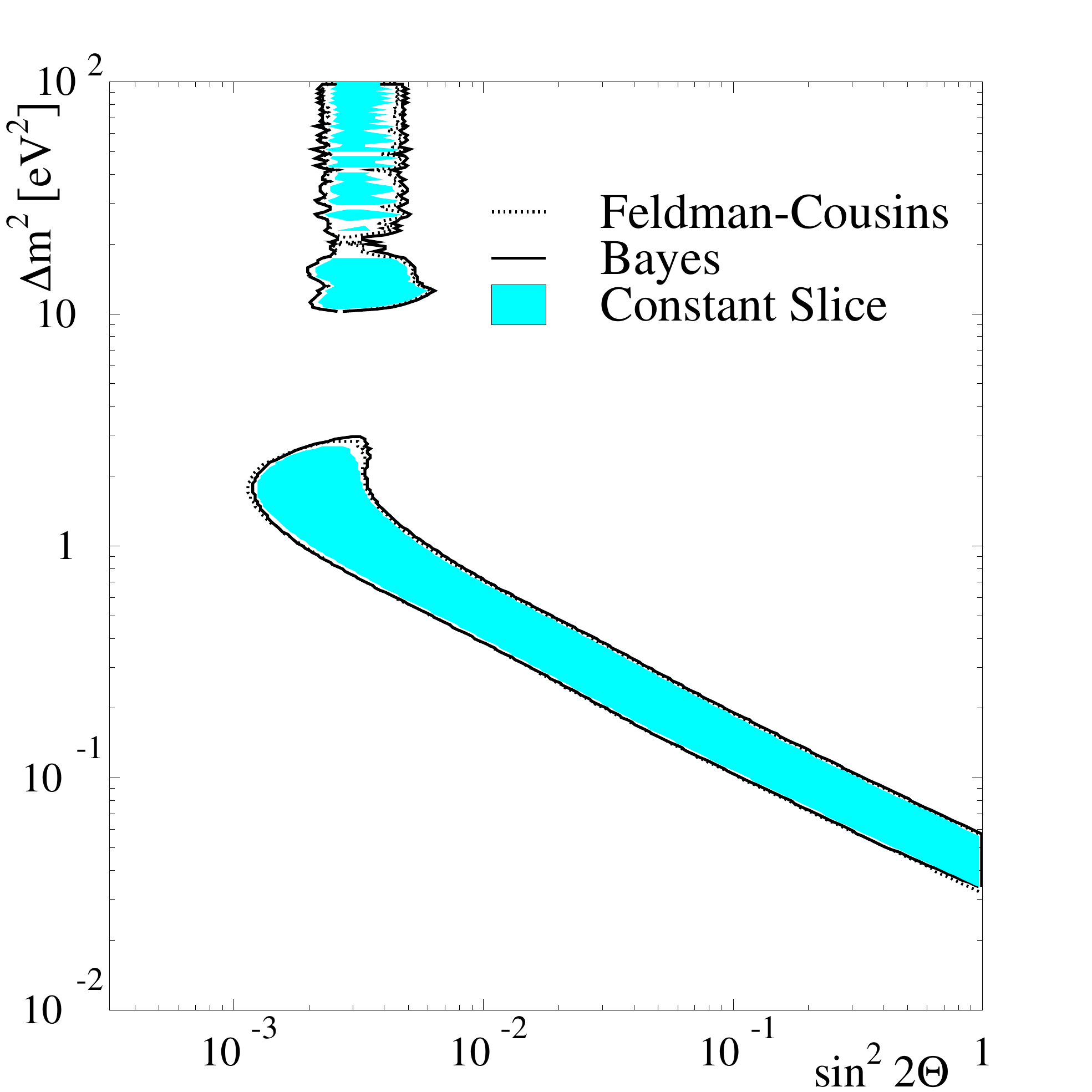}
        \caption{}
        \label{fig:LSNDbestfit}
    \end{subfigure}
    \caption{(a) The beam excess event distribution observed by LSND. The colored histograms are the expected beam-on backgrounds, while beam-off backgrounds have been subtracted. The figure is a modified figure from Ref.~\cite{LSND:2001aii}. (b) The 90\% confidence level of the LSND observation when fitted to a two-neutrino oscillation model. Figure taken from Ref.~\cite{LSND:2001aii}.}
    \label{fig:LSNDresults}
\end{figure}

LSND observed an excess of $87.9 \pm 22.4 \pm 6.0$ \nuebar events above the expected backgrounds with no oscillations.
This excess is shown \Cref{fig:LSNDexcess}. 
If this event distribution is modeled as two-neutrino oscillations, the best fit parameters would predict an excess of 89.5 events, agreeing very well with the observed data. 
The favored oscillation parameters are shown in \Cref{fig:LSNDbestfit}. 
The plot shows a best fit regions with $\Dmq>10^{-2}\ \eVq$, a $\Dmq$ larger than the SM $\Dmq$'s discussed in \Cref{sec:3nubf}.
Ultimately, in the neutrino oscillation picture, the observed LSND data hints towards a mass splitting $\Dmq$ inconsistent with those in the three neutrino SM picture: $\Delta m^{2}_{21}$ and $\Delta m^{2}_{31}$. 
To reiterate, any oscillations observed by LSND would not be due to $\Delta m^{2}_{21}$ or $\Delta m^{2}_{31}$, since the corresponding oscillation lengths would be too long for LSND to observe.

\subsection{MiniBooNE}
\label{sec:minibooneintro}

The MiniBooNE experiment was another accelerator neutrino experiment conducted to further study the LSND anomaly \cite{MiniBooNE:2020pnu}. MiniBooNE is located at Fermilab, having collected data 2002--2019.

Unlike LSND, MiniBooNE used a decay-in-flight (DIF) neutrino beam. 
An \SI{8}{\GeV} proton beam from Fermilab's Booster Neutrino Beam (BNB) was impinged on a beryllium target, where the resulting mesons then travel down a decay pipe and decay in flight to produce $\numu$'s or $\numubar$'s. 
These neutrinos then travel \num{\sim 500} meters before reaching the MiniBooNE detector. 
A magnetic focusing horn placed around the target allowed the experiment to selectively focus positive $\pi^{+}/K^{+}$ mesons or negative $\pi^{-}/K^{-}$ mesons, letting the experiment run in either neutrino or antineutrino mode. 
The $\numu$ flux peaked at around \SI{600}{\MeV}, while the $\numubar$ flux peaked at around \SI{400}{\MeV}. 
This gave the MiniBooNE experiment a $L/E\sim1$, similar to LSND and thus giving MiniBooNE sensitivity to the same $\Dmq$ parameter space.
Further information on the MiniBooNE detector can be found in \Cref{chapter:miniboone}.

\begin{figure}
    \centering
    \begin{subfigure}{0.45\textwidth}
        \includegraphics[width=\textwidth]{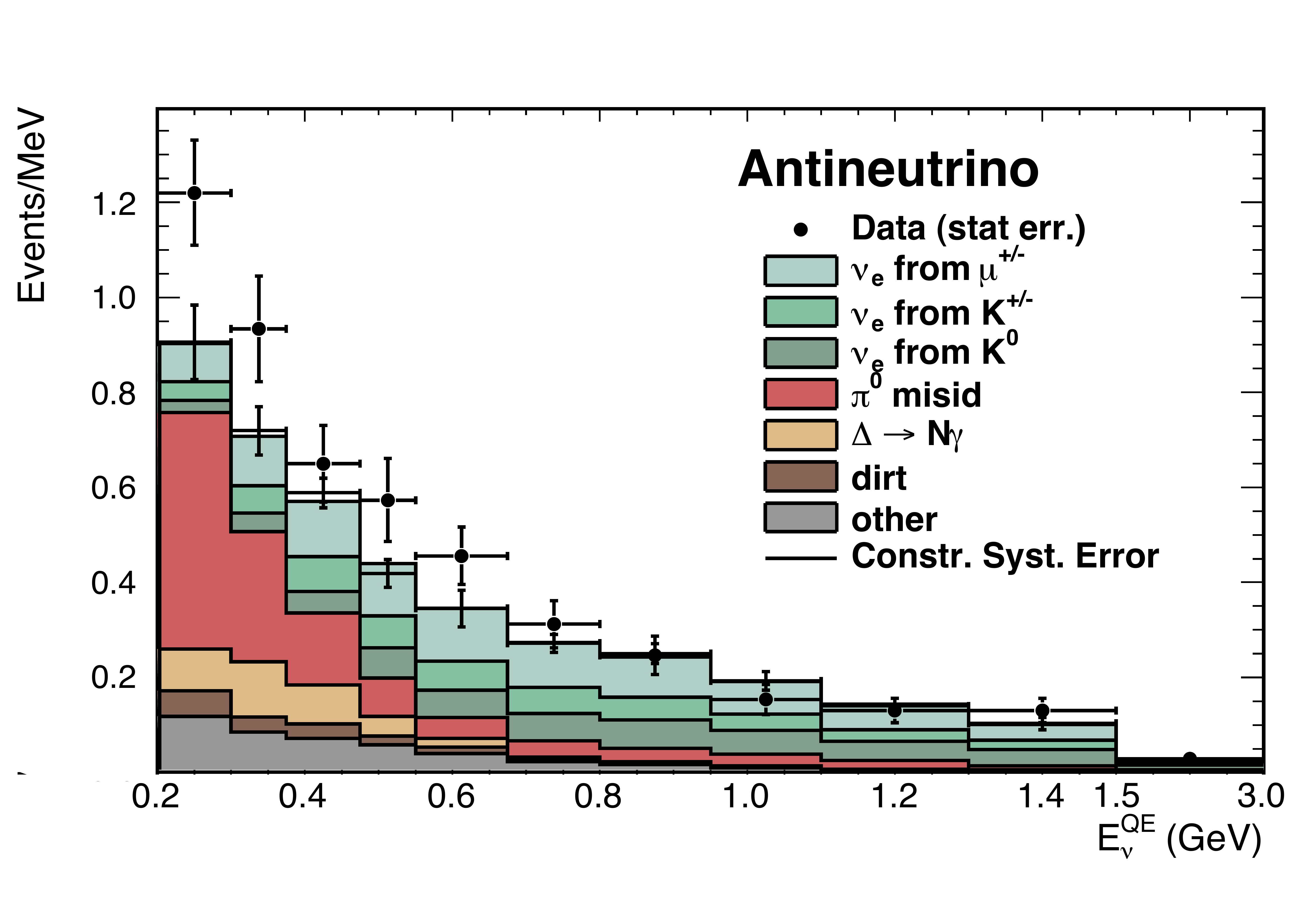}
        \caption{}
        \label{fig:MBantineutrinoexcess}
    \end{subfigure}
    \begin{subfigure}{0.45\textwidth}
        \includegraphics[width=\textwidth]{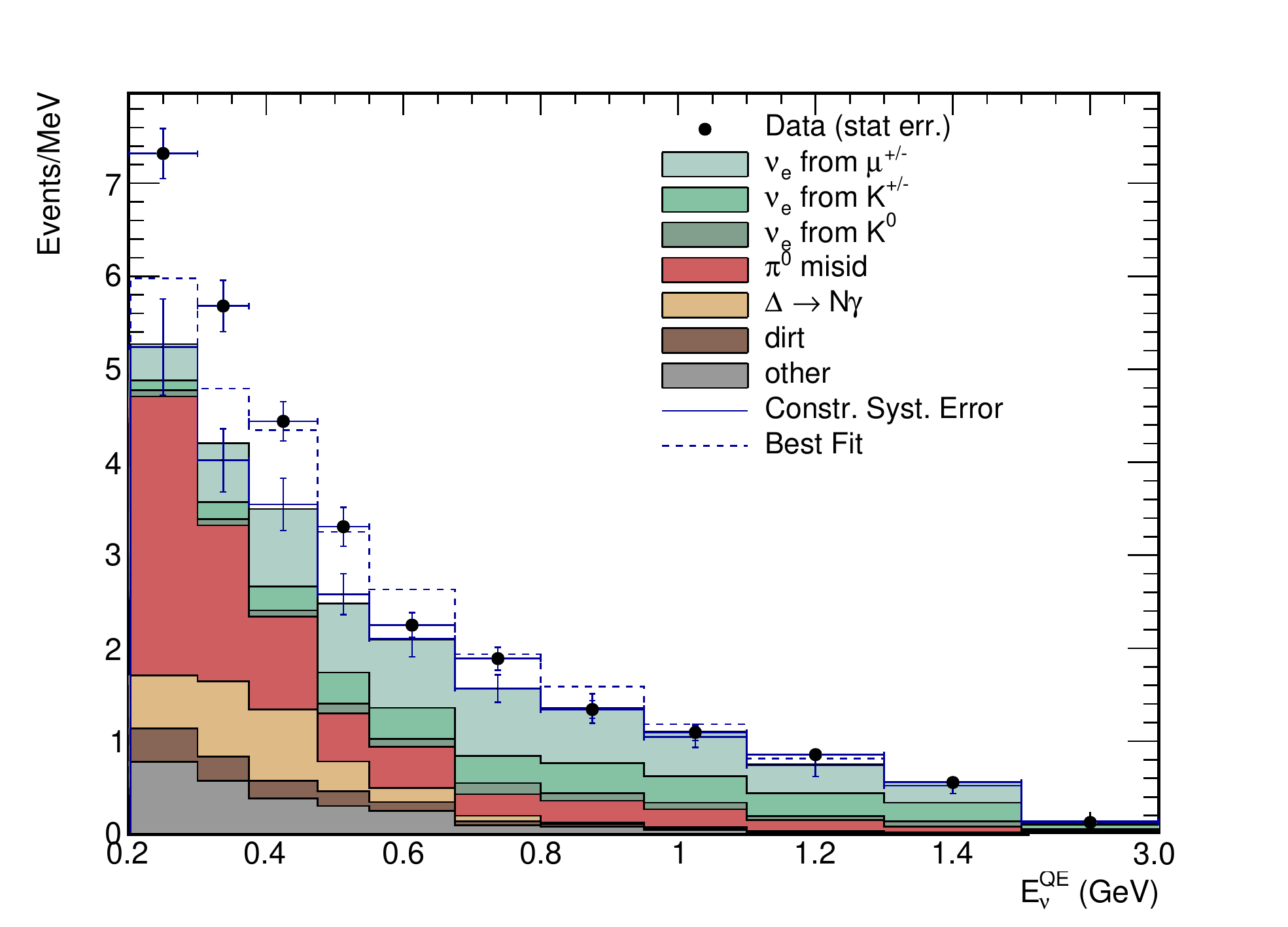}
        \caption{}
        \label{fig:MBneutrinoexcess}
    \end{subfigure}
    \caption{(a) The event distribution of \nuebar-like events observed by MiniBooNE in antineutrino mode. The data are given by the black crosses, while the colored histograms are the expected background events. Figure taken from Ref.~\cite{MiniBooNE:2013uba}. (b) The event distribution of \nue-like events observed in neutrino mode. Figure taken from Ref.~\cite{MiniBooNE:2020pnu}.}
    \label{fig:MBexcess}
\end{figure}

In its 17 years of running, MiniBooNE observed an excess above expectation in both neutrino and antineutrino modes \cite{MiniBooNE:2013uba,MiniBooNE:2020pnu}. 
In neutrino mode, the excess was $560.6 \pm 119.6$ events, while the excess in antineutrino mode was $78.4 \pm 28.5$. 
Combined, this is a $4.8\sigma$ observed anomaly, corresponding to a p-value of $p=\num{1.59e-6}$. 
The excess is plotted for antineutrino mode in \Cref{fig:MBantineutrinoexcess}, and for neutrino mode in \Cref{fig:MBneutrinoexcess}.

\begin{figure}
    \centering
    \includegraphics[width=0.4\textwidth]{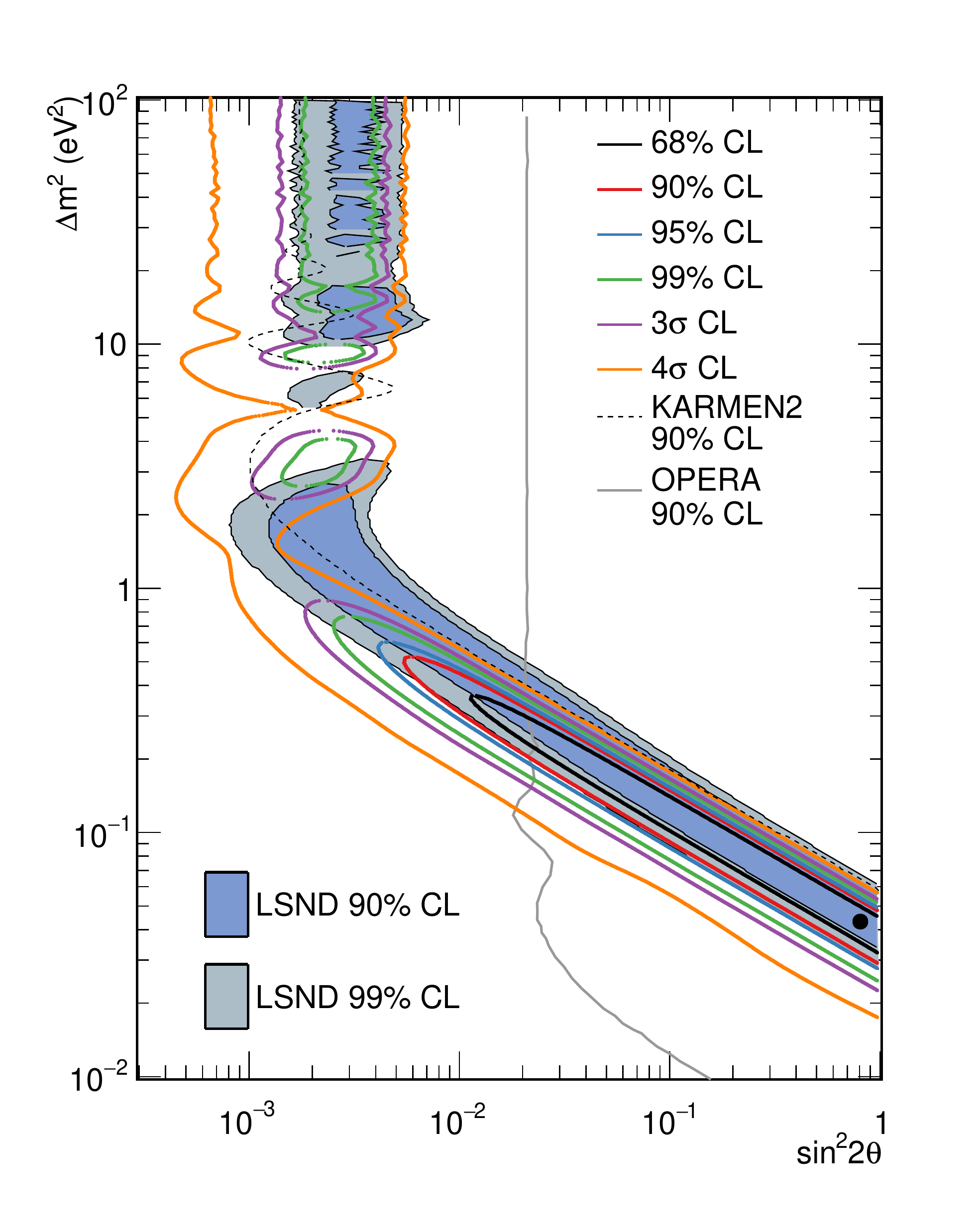}
    \caption{The hollow contours show the preferred parameter space for a two neutrino oscillation model given the MiniBooNE neutrino-antineutrino combined data. Also shown are the LSND best fit contours. KARMEN2 and OPERA are other experiments which did not see an anomalous signal, but will not be discussed in this section.}
    \label{fig:MBcontours}
\end{figure}

When the data are fitted to a two neutrino model, the preferred sterile parameters are shown in \Cref{fig:MBcontours}. 
Using the best fit point as the hypothesis, the p-value of the data increases dramatically to $p=0.123$.
Like LSND, the best fit parameters are found to be at a $\Dmq$ larger than the SM neutrinos. Furthermore, the MiniBooNE preferred parameters have considerable overlap with LSND's. 

\section{Sterile Neutrino?}

Before going over the other types of experiments that have seen anomalous data, let's first briefly introduce the focus of this thesis, \textit{sterile neutrinos}.

The SM already predicts three neutrinos, with two corresponding independent mass-squared splittings, $\Delta m_{21}^{2}$ and $\Delta m_{31}^{2}$. 
This model is very well established with overwhelming data supporting it. 
However, as seen in \Cref{sec:acceleratorsourceneutrinos}, LSND and MiniBooNE have observed an excess of neutrino events above the SM expectation.
If attributed to neutrino oscillations, then remarkable agreement is found between the data and model.
Further, LSND and MiniBooNE would predict neutrino oscillation parameters compatible with each other.

\begin{figure}
    \centering
    \includegraphics[width=0.5\textwidth]{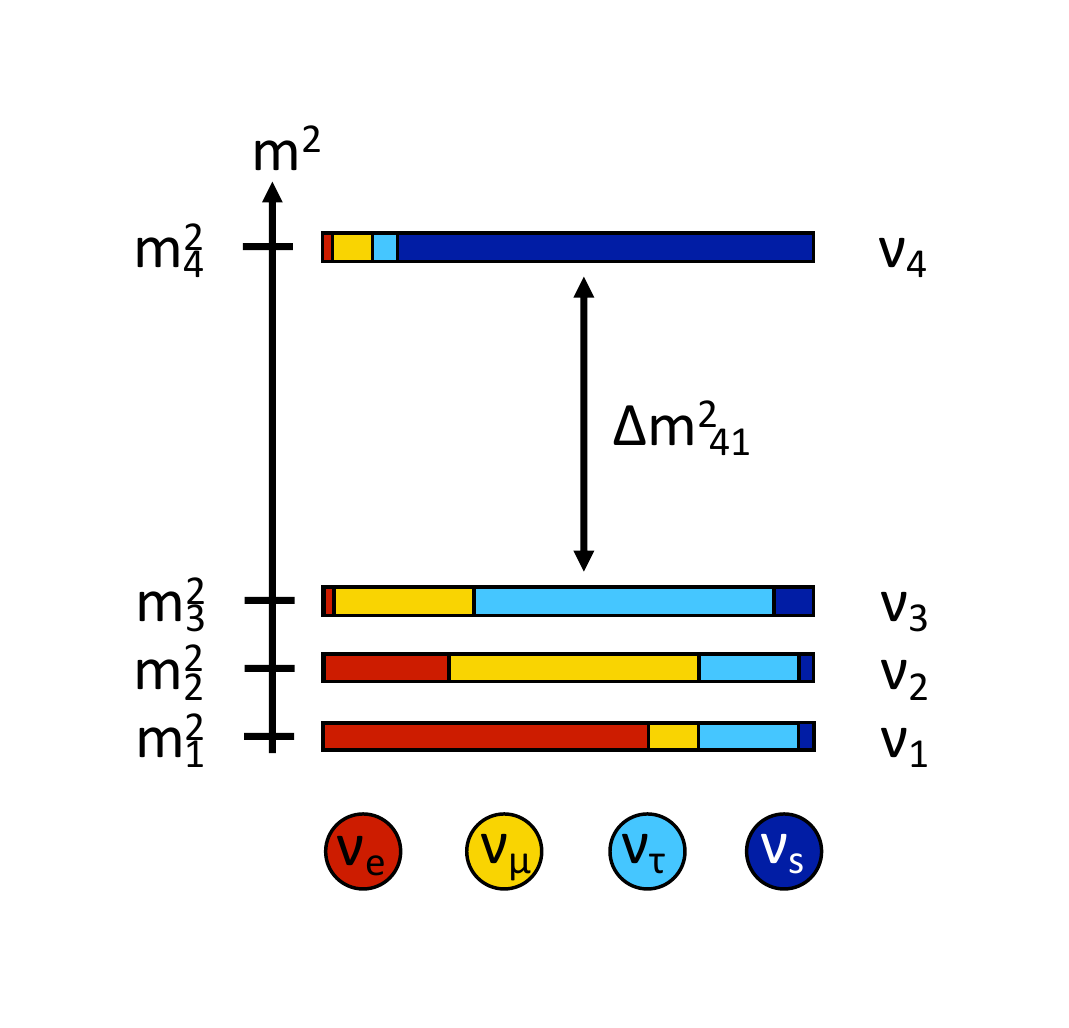}
    \caption{A visualization of the mass squared splittings and mixing, with the introduction of a fourth neutrino mass state $\nu_{4}$ and weak state $\nu_{s}$. Figure taken from Ref.~\cite{Moulai:2021zey}.}
    \label{fig:4nu}
\end{figure}

As \Cref{fig:LSNDresults} and \Cref{fig:MBcontours} show, the $\Delta m^{2}$ of such an oscillation would be too large to be compatible with the two established mass splittings, $\Delta m_{21}^{2} \approx 7.4 \pm 0.2\ \eVq$ and $|\Delta m_{31}^{2}| \approx 2.5 \pm 0.03\ \eVq$. 
Therefore, interpreting the LSND and MiniBooNE results as neutrino oscillations would require the introduction of a \textit{third} independent mass-squared splitting, and a \textit{fourth} neutrino mass state, $\nu_4$.
Building off of \Cref{fig:3nu}, the addition of a fourth mass state $\nu_4$ with mass-squared splitting $\Delta m_{41}^{2} \sim 1\ \eVq$ can be represented as in \Cref{fig:4nu}. 

However, LEP data has shown that the $Z$-boson only decays to \textit{three} neutrino types with mass $m_\nu < m_{Z}/2 \approx \SI{45.6}{\GeV}$ \cite{ALEPH:2005ab}. 
Therefore, in order to have an additional mass and weak neutrino state to contribute to oscillations, we require that the new weak eigenstate $\nu_s$ does not interact weakly, i.e. it is a \textit{sterile} neutrino. 

Because we are considering a third mass-squared splitting $\Delta m_{41}^2$ that is much larger than $\Delta m_{21}^{2}$ and $|\Delta m_{31}^{2}|$, we are justified in treating our observations as simply following two-neutrino oscillations, as described in \Cref{sec:twonu} and as will be further demonstrated in \Cref{sec:3plus1}.
This approximation, in the context of sterile neutrino oscillations, is referred to as the Short-Baseline (SBL) approximation.
Further, experiments that are sensitive to this $\Delta m_{41}^2$ are referred to as SBL experiments.\footnote{``Short-Baseline'' is a bit of a misnomer. ``Short Baseline'' refers to relatively small values of $L/E$, not just $L$.} 

\section{Reactor Neutrinos}
\label{sec:RAA}

\begin{figure}
    \centering
    \includegraphics[width=0.95\textwidth]{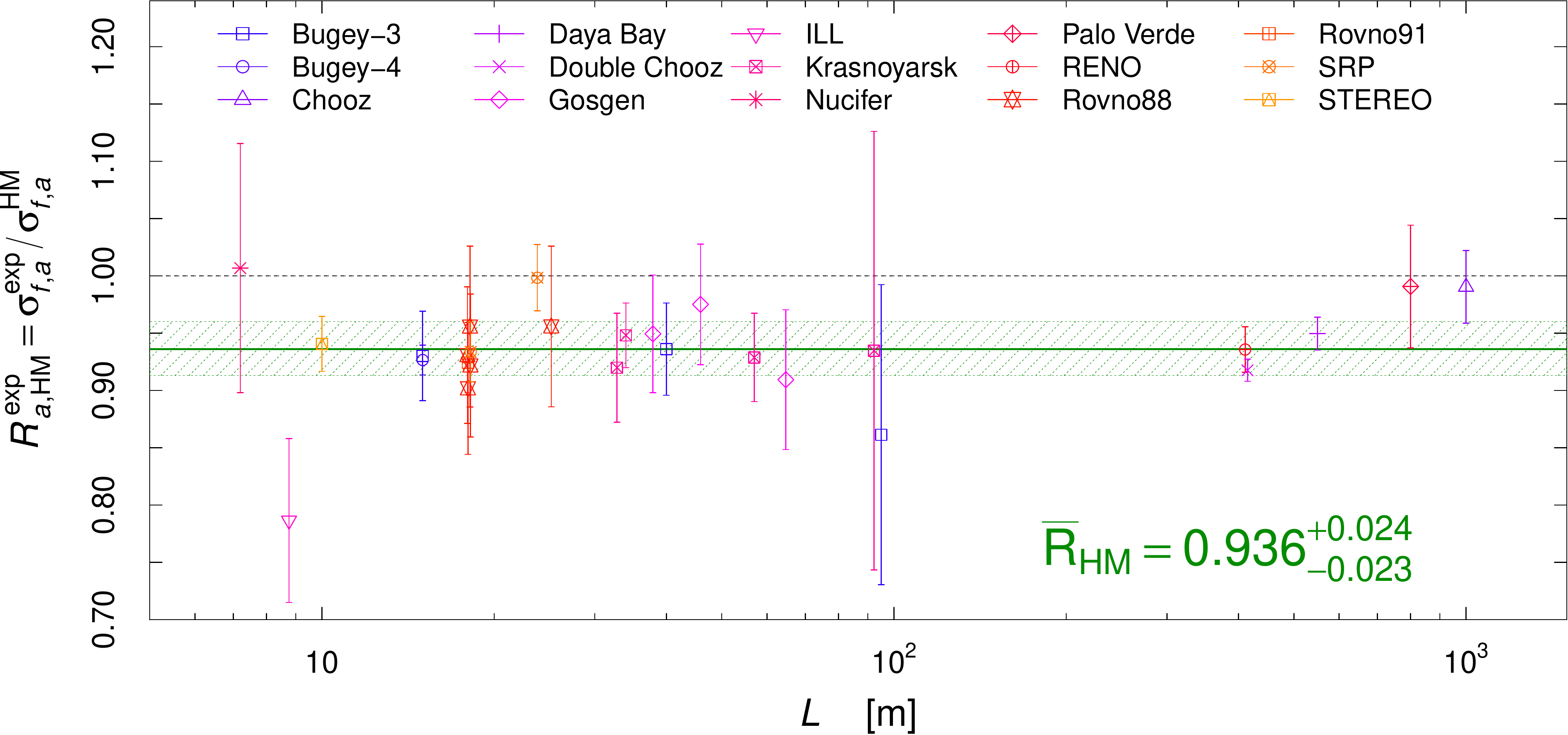}
    \caption{Ratios of the observed inverse beta decay (IBD) yields over the expectation from the HM model. The plot shows that the deficit is baseline independent, and universal amongst various reactors and detectors. Figure taken from Ref.~\cite{Giunti:2021kab}.}
    \label{fig:RAA}
\end{figure}

Nuclear reactors are a good source of $\mathcal{O}(1)\ \MeV\ \nuebar$'s.
Up until the start of the last decade, good agreement was found between the expected flux of reactor $\nuebar$'s versus the observed $\nuebar$ event rates.
However, in 2011, a reevaluation of the expected reactor $\nuebar$ flux \cite{PhysRevC.84.024617,Mueller:2011nm} resulted in the observed event rate to now have a \num{\sim 6}\% deficit compared to the models \cite{Mention:2011rk}.
This reevaluated flux model is commonly referred to as the ``Huber-Mueller'' (HM) flux, while the observed deficit of reactor $\nuebar$'s is referred to as the Reactor Antineutrino Anomaly (RAA). The RAA is an established phenomena observed over a range of reactors, detectors, and baselines, as shown in \Cref{fig:RAA}.

\begin{figure}
    \centering
    \includegraphics[width=0.95\textwidth]{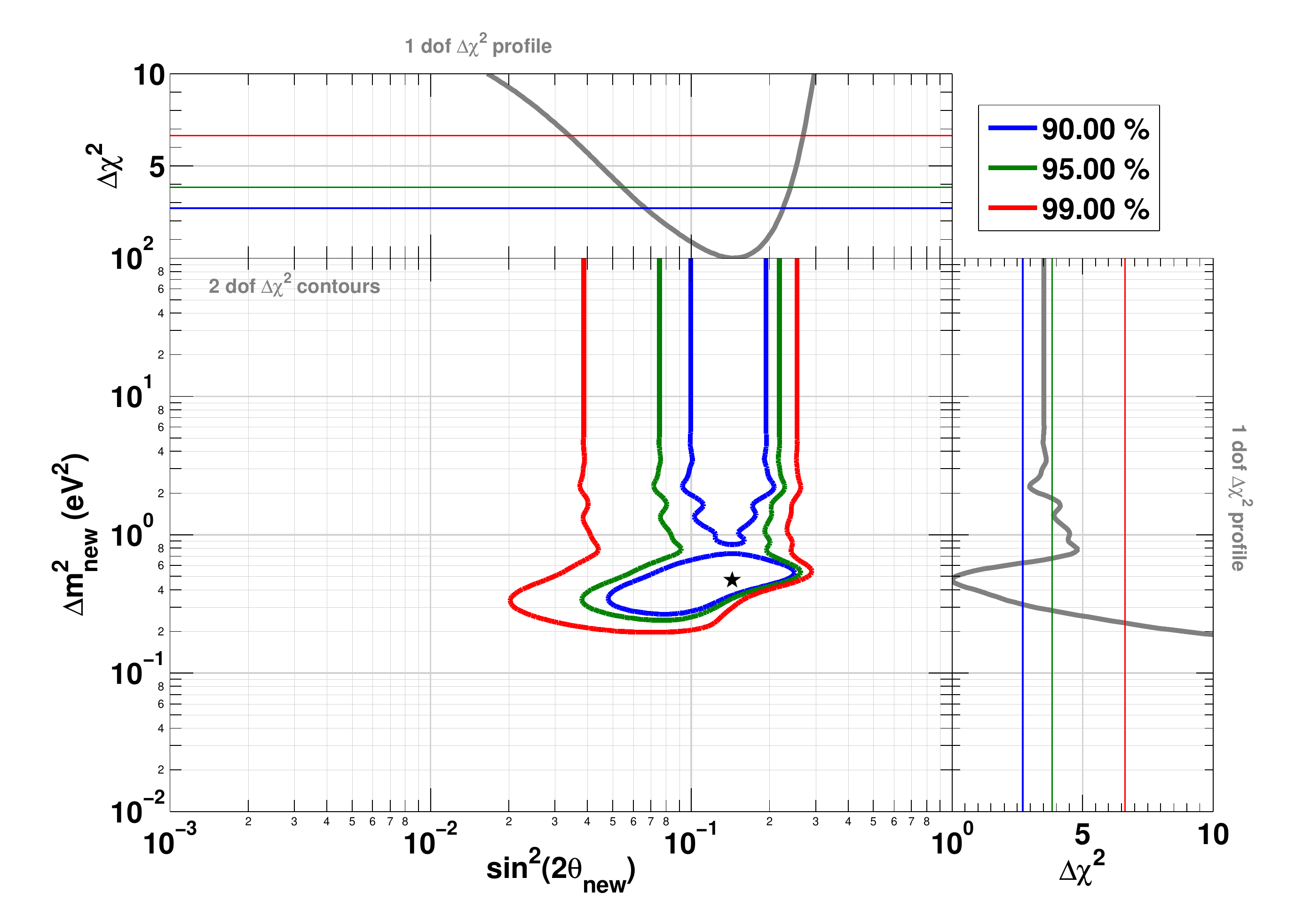}
    \caption{The plot shows the best fit oscillation parameters when the RAA is modeled as neutrino oscillations. Figure taken from Ref.~\cite{Abazajian:2012ys}.}
    \label{fig:RAAfit}
\end{figure}

Like the SBL accelerator results described in \Cref{sec:acceleratorsourceneutrinos}, the observed deficit of reactor $\nuebar$'s versus expectation can be fit to a neutrino oscillation model. 
\Cref{fig:RAAfit} shows the best fit oscillation region when this is done. 
Again, as seen in \Cref{sec:acceleratorsourceneutrinos}, a $\Delta m^{2}>0.1\ \eVq$ can explain the anomalous observations.  

Before continuing, we must note that the RAA is a deficit compared to nuclear reactor models.
Therefore, there does exist the possibility that it's the models that are overestimating the $\nuebar$ flux, as opposed to a real disappearance of $\nuebar$.
In fact, the HM model is known to be incorrect since a spectral distortion is found at \SI{\sim 5}{\MeV} in observed prompt energy \cite{RENO:2015ksa,DayaBay:2015lja,DoubleChooz:2015mfm}, referred to as the ``5 MeV bump,'' which cannot be explained by neutrino oscillations. 
Further, as more reactor models have been published, some strengthen the RAA \cite{Hayen:2019eop}, while others weaken it \cite{Estienne:2019ujo}. 
The conclusion to draw is that the reactor $\nuebar$ flux is difficult to predict, and these models are not reliable enough to let the RAA be convincing proof of neutrino oscillations. 
In \Cref{sec:nuedis}, we discuss modern reactor experiments that work around this limitation.

\section{Gallium Anomalies}
\label{sec:gallium}

GALLEX\cite{Kaether:2010ag} and SAGE\cite{SAGE:2009eeu} were two solar neutrino experiments that used \isotope[71]{Ga} detectors to observe solar $\nue$'s through the process 
\begin{equation}
    \nue + \isotope[71]{Ga} \to \isotope[71]{Ge} + e^{-}.
\end{equation}
The produced \isotope[71]{Ge} would then be collected and counted to calculate the $\nue$ flux. 

Both experiments used intense radioactive electron-capture \nue sources to calibrate their detectors. GALLEX ran two measurements of \isotope[51]{Cr}, while SAGE ran once with \isotope[51]{Cr} and again with \isotope[37]{Ar}. These isotopes would decay like
\begin{align}
    e^{-} + \isotope[51]{Cr} &\to \isotope[51]{V} + \nue \\ 
    e^{-} + \isotope[37]{Ar} &\to \isotope[37]{Cl} + \nue,
\end{align}
producing mono-energetic $\nue$ lines to use as calibration.

Combined, the observed ratio of observed \isotope[71]{Ge} production to expectation was $R=0.87 \pm 0.05$. 
Like the $\nuebar$'s from reactors, it appeared as if the $\nue$'s from the sources were disappearing before interacting with the detector. 

More recently, the BEST experiment \cite{Barinov:2021asz,Barinov:2022wfh} ran to probe this ``gallium anomaly.''
A \isotope[51]{Cr} source was placed within two concentric containers filled with \isotope[71]{Ga}. 
Again, a deficit of \isotope[71]{Ge} production was observed with an observed over expected ratio of $R_\textrm{In} = 0.79 \pm 0.05$ in the inner shell, and $R_\textrm{Out} = 0.77 \pm 0.05$ in the outer shell.

\begin{figure}
    \centering
    \begin{subfigure}{0.45\textwidth}
        \includegraphics[width=\textwidth]{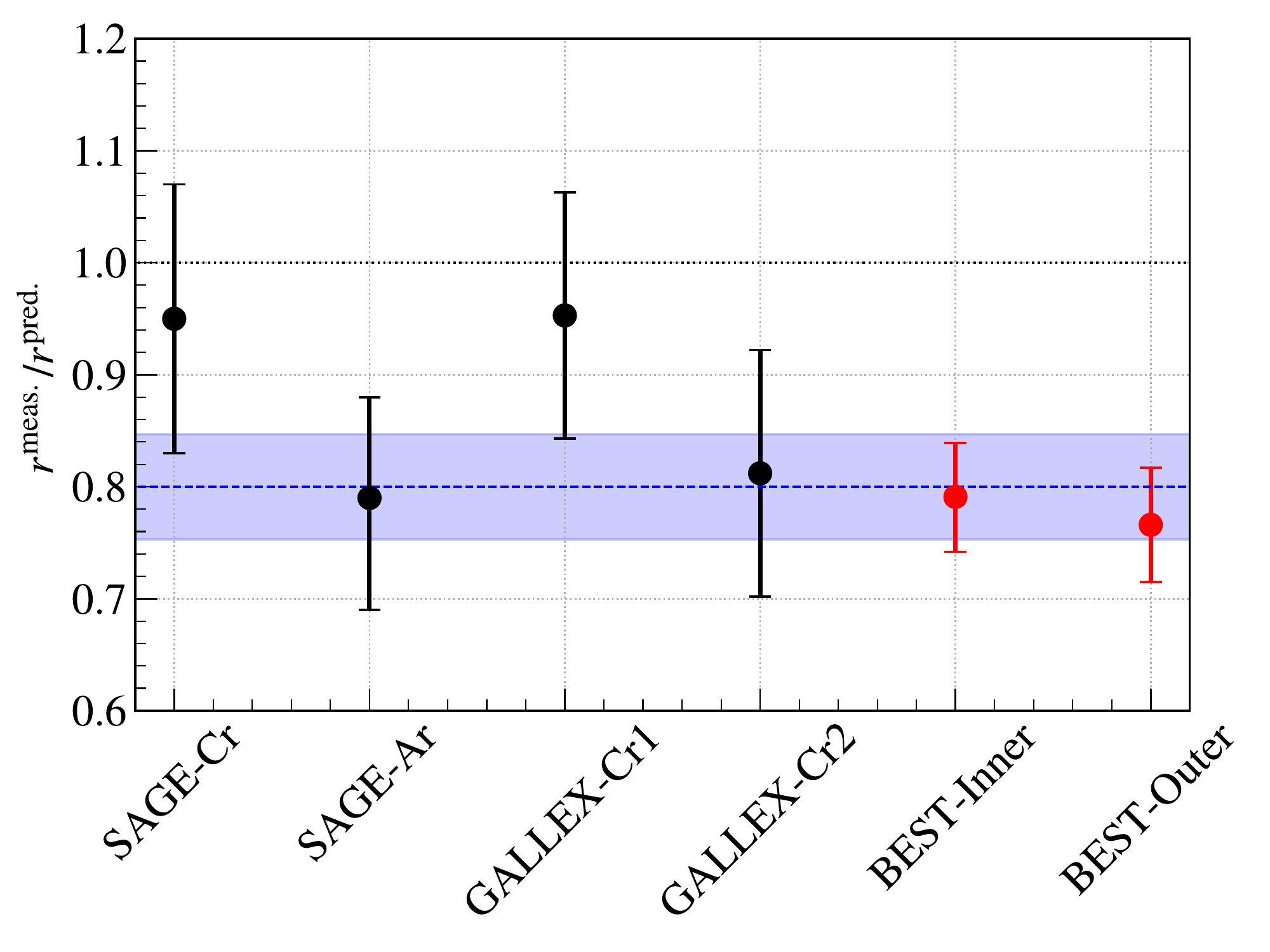}
        \caption{}
        \label{fig:galliumdata}
    \end{subfigure}
    \begin{subfigure}{0.45\textwidth}
        \includegraphics[width=\textwidth]{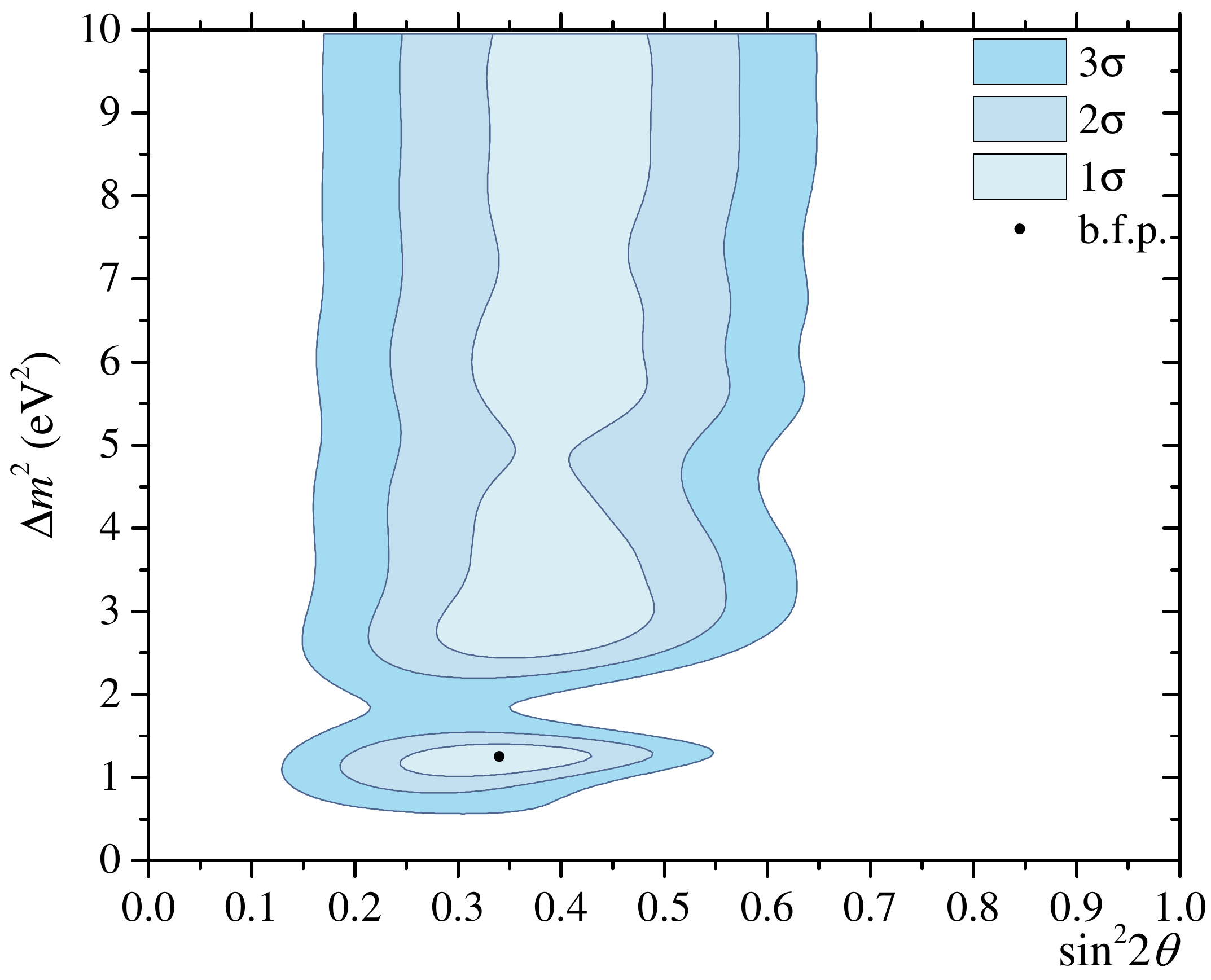}
        \caption{}
        \label{fig:galliumbestfit}
    \end{subfigure}
    \caption{(a) The observed \isotope[71]{Ge} production rate over expectation for the various runs for SAGE, GALLEX, and BEST. Figure taken from Ref.~\cite{Barinov:2022wfh}. 
    (b) Allowed parameter regions of the combined SAGE, GALLEX, and BEST data when modeled as two neutrino oscillations. Figure taken from Ref.~\cite{Barinov:2021asz}.}
    \label{fig:Galliumresults}
\end{figure}

The ratios of observed \isotope[71]{Ge} production over expectations are shown in \Cref{fig:galliumdata}. 
\Cref{fig:galliumbestfit} shows the best fit regions of the oscillation parameters if the data were fitted to neutrino oscillations.
Like the previous anomalies, these results fit well within a neutrino oscillation picture, with a $\Dmq > 1\ \eVq$. 

\section{Sterile Neutrino Models}
\label{sec:sterilemodels}

Now that we have summarized the experimental observations that have motivated the search for sterile neutrinos, let us now introduce some phenomenological models of sterile neutrinos. 
These models will be the basis of the global fit results which will be presented in \Cref{ch:globalfits}.

\subsection{\texorpdfstring{3+1}{3+1} Neutrinos}
\label{sec:3plus1}

Suppose that there exists an additional neutrino on top of the three SM neutrinos. In this scenario, we simply expand the mixing matrix to

\begin{equation}
    \begin{pmatrix}
        \nu_e \\
        \nu_\mu \\
        \nu_\tau \\
        \nu_s
    \end{pmatrix}
    = 
    \begin{pmatrix}
        U_{e 1} & U_{e 2} & U_{e 3} & U_{e 4}\\
        U_{\mu 1} & U_{\mu 2} & U_{\mu 3} & U_{\mu 4}\\
        U_{\tau 1} & U_{\tau 2} & U_{\tau 3 }& U_{\tau 4} \\
        U_{s 1} & U_{s 2} & U_{s 3} & U_{s 4}
    \end{pmatrix}
    \begin{pmatrix}
        \nu_1 \\
        \nu_2 \\
        \nu_3 \\
        \nu_4
    \end{pmatrix}.
\end{equation}

Per usual, the mixing matrix can be written out as a series of unitary rotations
\begin{equation}
    U= R^{34}(\theta_{34}) R^{24}(\theta_{24}, \delta_{24}) R^{14}(\theta_{14}, \delta_{14}) R^{23}(\theta_{23}) R^{13}(\theta_{13}, \delta_{13}) R^{12}(\theta_{12}).
\end{equation}
This increases the number of free parameters to 12, introducing $\Delta m_{41}^{2}, \theta_{14}, \theta_{24}, \theta_{34}, \delta_{14},$ and $\delta_{24}$ on top of the parameters in the three neutrino model. 

Now, let us assume that this additional neutrino state $\nu_4$ has a mass much larger than the other neutrinos such that $\Delta m_{41}^2 \gg \Delta m_{31}^2 > \Delta m_{21}^2 \approx 0$.
Further, let us also suppose that our experiments are set up such that $L/E \ll \Delta m_{21}^2 < \Delta m_{31}^2$. 
We'll refer to these assumptions as the Short-Baseline (SBL) approximation, as mentioned earlier.
Under these assumptions, we take \Cref{eq:finaloscequation} and set $\Delta m_{31}^2 = \Delta m_{21}^2 =0$, and $\Delta m_{41}^2 = \Delta m_{42}^2 = \Delta m_{43}^2$, giving

\begin{equation}
    \begin{split}
        P(\nu_\alpha \to \nu_\beta )= \delta_{\alpha \beta} 
        &- 4 \sum_{i<4} \Re(U_{\alpha i}^{*} U_{\beta i}U_{\alpha 4} U_{\beta 4}^{*}) \sin^{2}\left( 1.27 \Delta m_{41}^2 [\si{\electronvolt}] \frac{L [\si{\kilo\meter}]}{E [\si{\giga\electronvolt}]} \right) \\
        &-2 \sum_{i<4} \Im(U_{\alpha i}^{*} U_{\beta i}U_{\alpha 4} U_{\beta 4}^{*}) \sin \left(2.54 \Delta m_{41}^2 [\si{\electronvolt}] \frac{L [\si{\kilo\meter}]}{E [\si{\giga\electronvolt}]} \right) \\
        = \delta_{\alpha \beta} 
        &- 4 \sin^{2}\left( 1.27 \Delta m_{41}^2 [\si{\electronvolt}] \frac{L [\si{\kilo\meter}]}{E [\si{\giga\electronvolt}]} \right)  \Re(\sum_{i<4} U_{\alpha i}^{*} U_{\beta i}U_{\alpha 4} U_{\beta 4}^{*})  \\
        &-2  \sin \left(2.54 \Delta m_{41}^2 [\si{\electronvolt}] \frac{L [\si{\kilo\meter}]}{E [\si{\giga\electronvolt}]} \right)  \Im(\sum_{i<4} U_{\alpha i}^{*} U_{\beta i}U_{\alpha 4} U_{\beta 4}^{*}). 
        \label{eq:3p1_0}
    \end{split}
\end{equation}

We can rewrite $\sum_{i<4} U_{\alpha i}^{*} U_{\beta i}U_{\alpha 4} U_{\beta 4}^{*}$, utilizing the unitarity of the mixing matrix
\begin{equation}
    \sum_{i<4} U_{\alpha i}^{*} U_{\beta i}U_{\alpha 4} U_{\beta 4}^{*} = U_{\alpha 4} U_{\beta 4}^{*} \sum_{i<4} U_{\alpha i}^{*} U_{\beta i} = U_{\alpha 4} U_{\beta 4}^{*} (\delta_{\alpha\beta} - U_{\alpha 4}^{*} U_{\beta 4}).
    \label{eq:2.7}
\end{equation}
Note that the above equation is real regardless if $\alpha = \beta$ or $\alpha \neq \beta$; therefore, we can drop the imaginary term in \Cref{eq:3p1_0} to give us
\begin{equation}
    P(\nu_\alpha \to \nu_\beta )
    = \delta_{\alpha \beta} -
    4 U_{\alpha 4} U_{\beta 4}^{*} (\delta_{\alpha\beta} - U_{\alpha 4}^{*} U_{\beta 4}) \sin^{2}\left( 1.27 \Delta m_{41}^2 [\si{\electronvolt}] \frac{L [\si{\kilo\meter}]}{E [\si{\giga\electronvolt}]} \right). 
\end{equation}

For the specific case of appearance ($\alpha \neq \beta$), this gives
\begin{equation}
    P(\nu_\alpha \to \nu_\beta )
    =  
    4 |U_{\alpha 4}|^{2} |U_{\beta 4}|^{2} \sin^{2}\left( 1.27 \Delta m_{41}^2 [\si{\electronvolt}] \frac{L [\si{\kilo\meter}]}{E [\si{\giga\electronvolt}]} \right)    
    \quad
    (\alpha\neq\beta),
    \label{eq:3p1_appearance}
\end{equation}
while for disappearance ($\alpha = \beta$) we get
\begin{equation}
    P(\nu_\alpha \to \nu_\alpha )
    = 1 -
    4 |U_{\alpha 4}|^{2} (1 - |U_{\alpha 4}|^{2}) \sin^{2}\left( 1.27 \Delta m_{41}^2 [\si{\electronvolt}] \frac{L [\si{\kilo\meter}]}{E [\si{\giga\electronvolt}]} \right).
    \label{eq:3p1_disappearance}
\end{equation}

Notice the similarities between \Cref{eq:3p1_appearance,eq:3p1_disappearance} and \Crefrange{eq:2n_appearance}{eq:2n_disappearance}. The analogies become clearer when we use effective mixing angles
\begin{equation}
    \sin^2 2\theta_{\alpha\beta} \equiv 4|U_{\alpha 4}|^{2}|U_{\beta 4}|^{2}
    \quad
    (\alpha\neq\beta),
    \quad
    \sin^2 2\theta_{\alpha\alpha} \equiv 4 |U_{\alpha 4}|^{2} (1 - |U_{\alpha 4}|^{2}) 
\end{equation}
so that \Cref{eq:3p1_appearance} and \Cref{eq:3p1_disappearance} can be written as 
\begin{align}
    P(\nu_\alpha \to \nu_\beta )
    & =  
    \sin^2 2\theta_{\alpha\beta} \sin^{2}\left( 1.27 \Delta m_{41}^2 [\si{\electronvolt}] \frac{L [\si{\kilo\meter}]}{E [\si{\giga\electronvolt}]} \right)    
    \quad
    (\alpha\neq\beta)\\
    P(\nu_\alpha \to \nu_\alpha )
    &= 1 - \sin^2 2\theta_{\alpha\alpha} \sin^{2}\left( 1.27 \Delta m_{41}^2 [\si{\electronvolt}] \frac{L [\si{\kilo\meter}]}{E [\si{\giga\electronvolt}]} \right).
\end{align}

We therefore see the similarities between a 3+1 neutrino model under the SBL approximation and a simple two-neutrino model. 

In order to keep track of the different mixing parameter conventions, \Cref{table:params} provides the relations between these different conventions.

\begin{table*}[tbp] \centering
    \begin{adjustbox}{max width=\textwidth}
        \begin{tabular}{lllll}\hline
            $\sin^2 2 \theta_{ee}$ &=& 
            $\sin^2 2 \theta_{14}$ &=& 
            $ 4 (1-|U_{e4}|^2)|U_{e4}|^2$\\
            $\sin^2 2 \theta_{\mu\mu}$ &=& 
            $4 \cos^2 \theta_{14} \sin^2 \theta_{24} (1 - \cos^2 \theta_{14} \sin^2 \theta_{24})$ &=& 
            $4 (1-|U_{\mu4}|^2)|U_{\mu4}|^2$ \\
            $\sin^2 2 \theta_{\tau\tau}$ &=& 
            $4 \cos^2 \theta_{14} \cos^2 \theta_{24} \sin^2 \theta_{34}(1 - \cos^2 \theta_{14} \cos^2 \theta_{24} \sin^2 \theta_{34})$ &=& 
            $4 (1-|U_{\tau4}|^2)|U_{\tau4}|^2$ \\
            $\sin^2 2 \theta_{\mu e}$ &=& 
            $\sin^2 2 \theta_{14} \sin^2 \theta_{24}$ &=& 
            $4|U_{\mu 4}|^2 |U_{e 4}|^2$ \\
            $\sin^2 2 \theta_{e \tau}$ &=& 
            $\sin^2 2 \theta_{14} \cos^2 \theta_{24} \sin^2 \theta_{34}  $&= &
            $4|U_{e 4}|^2 |U_{\tau 4}|^2$\\
            $\sin^2 2 \theta_{\mu \tau}$ &=& 
            $\sin^2 2 \theta_{24} \cos^4 \theta_{14} \sin^2 \theta_{34} $&= &
            $4|U_{\mu 4}|^2 |U_{\tau 4}|^2$\\ \hline 
        \end{tabular}
    \end{adjustbox}
    \caption{3+1 neutrino model mixing parameters with the SBL approximation. \label{table:params}}
\end{table*}

\subsection{\texorpdfstring{3+2}{3+2} Model}

An obvious extension to the 3+1 would be to simply add more neutrinos.
In our global fits, we consider a 3+2 model.
If we continue to use the SBL approximation, where $\Delta m_{51}^2 > \Delta m_{41}^2 \gg \Delta m_{31}^2 > \Delta m_{21}^2 \approx 0$ and $L/E \ll \Delta m_{21}^2 < \Delta m_{31}^2$, then the general oscillation equation \Cref{eq:finaloscequation} can be written as 

\begin{equation}
    \begin{split}
        P(\nu_\alpha \to \nu_\beta )= 
        \delta_{\alpha \beta} 
        &-4 \sum_{i<j} \Re(U_{\alpha i}^{*} U_{\beta i}U_{\alpha j} U_{\beta j}^{*}) \sin^{2}\left( 1.27 \Delta m_{ji}^2  L/E \right) \\
        &-2 \sum_{i<j} \Im(U_{\alpha i}^{*} U_{\beta i}U_{\alpha j} U_{\beta j}^{*}) \sin \left(2.54 \Delta m_{ji}^2  L/E \right)\\
        =\delta_{\alpha \beta} 
        &- 4 \Re(U_{\alpha 4}^{*} U_{\beta 4}U_{\alpha 5} U_{\beta 5}^{*}) \sin^{2}\left( 1.27 \Delta m_{54}^2  L/E \right) \\
        &-2 \Im(U_{\alpha 4}^{*} U_{\beta4i}U_{\alpha 5} U_{\beta 5}^{*}) \sin \left(2.54 \Delta m_{54}^2  L/E \right)\\
        &-4 \sum_{i<4} \Re(U_{\alpha i}^{*} U_{\beta i}U_{\alpha 4} U_{\beta 4}^{*}) \sin^{2}\left( 1.27 \Delta m_{4i}^2  L/E \right) \\
        &-2 \sum_{i<4} \Im(U_{\alpha i}^{*} U_{\beta i}U_{\alpha 4} U_{\beta 4}^{*}) \sin \left(2.54 \Delta m_{4i}^2  L/E \right)\\
        &-4 \sum_{i<4} \Re(U_{\alpha i}^{*} U_{\beta i}U_{\alpha 5} U_{\beta 5}^{*}) \sin^{2}\left( 1.27 \Delta m_{5i}^2  L/E \right) \\
        &-2 \sum_{i<4} \Im(U_{\alpha i}^{*} U_{\beta i}U_{\alpha 5} U_{\beta 5}^{*}) \sin \left(2.54 \Delta m_{5i}^2  L/E \right)\\
        =\delta_{\alpha \beta} 
        &- 4 \Re(U_{\alpha 4}^{*} U_{\beta 4}U_{\alpha 5} U_{\beta 5}^{*}) \sin^{2}\left( 1.27 \Delta m_{54}^2  L/E \right) \\
        &-2 \Im(U_{\alpha 4}^{*} U_{\beta4i}U_{\alpha 5} U_{\beta 5}^{*}) \sin \left(2.54 \Delta m_{54}^2  L/E \right)\\
        &-  4(\Re (\delta_{\alpha \beta}U_{\alpha 4} U_{\beta 4}^{*} -  U_{\alpha 5}^{*} U_{\beta 5}U_{\alpha 4} U_{\beta 4}^{*}) - |U_{\alpha 4} U_{\beta 4}^{*}|^2) \sin^{2}\left( 1.27 \Delta m_{41}^2  L/E \right) \\
        &- 2\Im( \delta_{\alpha \beta}U_{\alpha 4} U_{\beta 4}^{*}  - U_{\alpha 5}^{*} U_{\beta 5}U_{\alpha 4} U_{\beta 4}^{*}) \sin \left(2.54 \Delta m_{41}^2  L/E \right)\\
        &- 4 (\Re (\delta_{\alpha \beta}U_{\alpha 5} U_{\beta 5}^{*}  - U_{\alpha 4}^{*} U_{\beta 4}U_{\alpha 5} U_{\beta 5}^{*} )- |U_{\alpha 5}^{*} U_{\beta 5}|^2 )  \sin^{2}\left( 1.27 \Delta m_{51}^2  L/E \right) \\
        &-  2\Im( \delta_{\alpha \beta} U_{\alpha 5} U_{\beta 5}^{*} - U_{\alpha 4}^{*} U_{\beta 4} U_{\alpha 5} U_{\beta 5}^{*}) \sin \left(2.54 \Delta m_{51}^2  L/E \right),\\
    \end{split}
\end{equation}
where the last step uses \Cref{eq:2.7}.

For an appearance experiment,   
\begin{equation}
    \begin{split}
        P(\nu_\alpha \to \nu_\beta ) =&- 4 |U_{\alpha 4}^{*} U_{\beta 4}U_{\alpha 5} U_{\beta 5}^{*}| \cos(\phi_{\alpha \beta}) \sin^{2}\left( 1.27 \Delta m_{54}^2  L/E \right) \\
                                      &+2 |U_{\alpha 4}^{*} U_{\beta4}U_{\alpha 5} U_{\beta 5}^{*}| \sin(\phi_{\alpha \beta}) \sin \left(2.54 \Delta m_{54}^2  L/E \right)\\
                                      &+ 4(|U_{\alpha 5}^{*} U_{\beta 5}U_{\alpha 4} U_{\beta 4}^{*}| \cos(\phi_{\alpha \beta}) + |U_{\alpha 4} U_{\beta 4}^{*}|^2) \sin^{2}\left( 1.27 \Delta m_{41}^2  L/E \right) \\
                                      &+ 2|U_{\alpha 5}^{*} U_{\beta 5}U_{\alpha 4} U_{\beta 4}^{*}| \sin(\phi_{\alpha \beta}) \sin \left(2.54 \Delta m_{41}^2  L/E \right)\\
                                      &+4 (|U_{\alpha 4}^{*} U_{\beta 4}U_{\alpha 5} U_{\beta 5}^{*}|\cos(\phi_{\alpha \beta})+ |U_{\alpha 5}^{*} U_{\beta 5}|^2 )  \sin^{2}\left( 1.27 \Delta m_{51}^2  L/E \right) \\
                                      &- 2|U_{\alpha 4}^{*} U_{\beta 4} U_{\alpha 5} U_{\beta 5}^{*}| \sin(\phi_{\alpha \beta}) \sin \left(2.54 \Delta m_{51}^2  L/E \right)\\
    \end{split}
\end{equation} 
where
\begin{equation}
    \phi_{\alpha \beta} = \mathrm{arg}(U_{\alpha 4}^{*} U_{\beta4i}U_{\alpha 5} U_{\beta 5}^{*}).
\end{equation}

For a disappearance experiment, 
\begin{equation}
    \begin{split}
        P(\nu_\alpha \to \nu_\alpha ) =1
        &- 4 |U_{\alpha 4}|^2 |U_{\alpha 5}|^2 \sin^{2}\left( 1.27 \Delta m_{54}^2  L/E \right) \\
        &-  4( 1 - |U_{\alpha 4}|^2 -  |U_{\alpha 5}|^2 )|U_{\alpha 4}|^2 \sin^{2}\left( 1.27 \Delta m_{41}^2  L/E \right) \\
        &- 4 ( 1 - |U_{\alpha 4}|^2  - |U_{\alpha 5}|^2 )|U_{\alpha 5}|^2  \sin^{2}\left( 1.27 \Delta m_{51}^2  L/E \right). \\
    \end{split}
\end{equation} 

Similar to the case when we reached three neutrinos in \Cref{sec:3neutrinos}, we now have a CP-violating phase $\phi_{\alpha \beta}$ appearing in the appearance equation. In this scenario, switching from $\nu$ to $\bar\nu$ flips $\phi_{\alpha \beta}\to -\phi_{\alpha \beta}$, leading to different oscillation equations for $P(\nu_\alpha \to \nu_\beta )$ versus $P(\bar\nu_\alpha \to \bar\nu_\beta)$.

We choose to stop at 3+2 since since any further sterile neutrinos lead to too many parameters that could be fit for, and the experimental data are too limited to be used to fit to so many parameters.

\subsection{\texorpdfstring{3+1+Decay}{3+1+Decay} Model}
\label{sec:3plus1plusdecay}

We now consider a more exotic model of sterile neutrinos: an unstable sterile neutrino. 

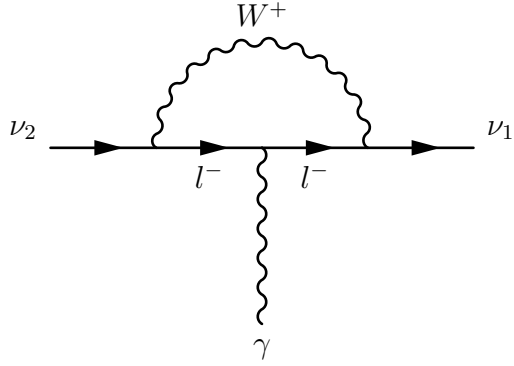
\begin{figure}
    \vspace{1in}
    \centering
    \begin{fmfgraph*}(200,100)
        \fmfleft{pi1,i1}
        \fmfright{po1,o1}
        \fmf{fermion}{i1,v1}
        \fmf{fermion, label=$l^-$}{v1,v2}
        \fmf{fermion, label=$l^-$}{v2,v3}
        \fmf{fermion}{v3,o1}
        \fmf{boson,left,label=$W^+$,tension=0}{v1,v3}
        \fmffreeze
        \fmf{phantom}{pi1,v4,po1}
        \fmf{boson}{v2,v4}
        \fmfv{label=$\gamma$}{v4}
        \fmflabel{$\nu_2$}{i1}
        \fmflabel{$\nu_1$}{o1}
    \end{fmfgraph*}
    \caption{An example of a Standard Model $\nu_2\to\nu_1 + \gamma$ decay.}
    \label{fig:nudecay}
\end{figure}

Strictly speaking, the Standard Model neutrinos are already unstable. \Cref{fig:nudecay} shows an example of a $\nu_2 \to \nu_1 + \gamma$ decay. 

Lifetimes for $\nu_2 \to \nu_1 + \gamma$ and $\nu_2 \to \nu_1 + \gamma + \gamma$ decays are printed below.
\begin{align}
    \nu_2 &\to \nu_1 + \gamma &\tau \approx \mathcal{O}(10^{36})\ (m_2 / \si{\eV})^{-5}\ \textrm{years \cite{Pal:1981rm}}\\
    \nu_2 &\to \nu_1 + \gamma +\gamma &\tau \approx \mathcal{O}(10^{66})\ (m_2 / \si{\eV})^{-9}\ \textrm{years \cite{Nieves:1982bq}}
\end{align}
These lifetimes are well beyond the age of the universe, so for practical purposes the SM neutrinos are treated as stable particles. 

However, models can be introduced where neutrinos are allowed to decay through some new interaction, e.g. $\nu_i \to \nu_j + \phi$ \cite{Kim:1990km}.

Beyond Standard Model (BSM) decays of SM neutrinos have been used in the past to explain $\nu_\mu$ disappearance in atmospheric data \cite{Barger:1998xk}. And a decaying sterile neutrino model has been used to explain the LSND anomaly \cite{Palomares-Ruiz:2005zbh}.

For our studies, we consider sterile neutrino decays as described in Ref.~\cite{Moss:2017pur} and shown in \Cref{fig:nudecayfromMarjonspaper}. Here, the mostly-sterile mass state $\nu_4$ is allowed to decay into some scalar $\phi$, with the Lagrangian \cite{Kim:1990km,Lindner:2001fx}
\begin{equation}
    \mathcal{L_\mathrm{int}}= \frac{g_{4j}^s}{2} \bar{\nu_j} \nu_4 \phi + i \frac{g_{4j}^p}{2} \bar{\nu_j}\gamma_{5}\nu_4,
\end{equation}
where $g_{4j}^s$ is the scalar coupling between $\nu_4$ and $\nu_j$, and $g_{4j}^p$ is the pseudoscalar coupling. In the  limit where $m_4\gg m_j$, both the helicity-preserving and helicity-violating decay rates are given as 
\begin{equation}
    \Gamma(\nu_4\to \nu_j) =  \Gamma(\nu_4\to \bar{\nu_j})
    =\frac{({g^s_{4j}}^2 + {g^p_{4j}}^2 ) m_4^2}{32 \pi E_4}.
    \label{eq:decayrate}
\end{equation}

\begin{figure}
    \vspace{1cm}
    \centering
    \begin{subfigure}{0.45\textwidth}
        \centering
        \begin{fmfgraph*}(100,100)
            \fmfstraight
            \fmfleft{i1}
            \fmfright{o1,o2}
            \fmf{fermion}{i1,v1,o1}
            \fmfblob{30}{v1}
            \fmf{dashes}{v1,o2}
            \fmflabel{$\nu_i$}{i1}
            \fmflabel{$\nu_j$}{o1}
            \fmflabel{$\phi$}{o2}
        \end{fmfgraph*}
        \vspace*{5mm}
        \caption{The \textit{visible} decay $\nu_i \to \nu_j + \phi$}
        \label{fig:nuvisibledecay}
    \end{subfigure}
    \begin{subfigure}{0.45\textwidth}
        \centering
        \begin{fmfgraph*}(100,100)
            \fmfstraight
            \fmfleft{i1}
            \fmfright{o1,o2}
            \fmf{fermion}{i1,v1}
            \fmfblob{30}{v1}
            \fmf{dashes}{o1,v1,o2}
            \fmflabel{$\nu_i$}{i1}
            \fmflabel{$\psi$}{o1}
            \fmflabel{$\phi$}{o2}
        \end{fmfgraph*}
        \vspace*{5mm}
        \caption{The \textit{invisible} decay $\nu_i \to \psi + \phi$}
        \label{fig:nuinvisibledecay}
    \end{subfigure}
    \caption{}
    \label{fig:nudecayfromMarjonspaper}
\end{figure}

In Ref.~\cite{Moss:2017pur}, both decays in \Cref{fig:nudecayfromMarjonspaper} were considered in the context of the IceCube experiment. 
The $\nu_4\to \nu_j + \phi $ decay in \Cref{fig:nuvisibledecay} is referred to as a \textit{visible} decay, since $\nu_j$ is taken to be an active neutrino that can be detected, in principle. 
The $\nu_4\to \psi + \phi$ decay shown in \Cref{fig:nuinvisibledecay}, on the other hand, is referred to as an \textit{invisible} decay, since $\psi$ is considered to be some fermion that cannot be detected through conventional means. 
For our work, we only consider the invisible decay shown in \Cref{fig:nuinvisibledecay}. 
We also assume that the coupling is either purely scalar or pseudo-scalar.

To model the invisible decay, we use the non-Hermitian Hamiltonian
\begin{equation}
    H = H_0 - i\frac{1}{2} \Gamma(E),
\end{equation}
where $\Gamma$ is a diagonal matrix with $\Gamma_{ii} = \Gamma_i(E)$ and $\Gamma_i(E)$ is the  decay rate for the $i^{\textrm{th}}$ neutrino given by \textit{twice} \Cref{eq:decayrate}.
The factor of two comes from the fact that \Cref{eq:decayrate} gives the decay width for only the helicity-preserving or helicity-violating decay.
The total decay width would be the sum of the two. 
In our model, we assume that the only non-zero term is $\Gamma_{4}$, so that only the fourth mass state $\nu_4$ decays.

The neutrino vacuum Hamiltonian $H_0$ in the ultra-relativistic limit can be written as $H_{0_{ii}} = \frac{\Delta m^2_i}{2E}$, where $\Delta m^2_i = m_i^2 - m_1^2$. 
Further, we again take the SBL approximation and assume that $\Delta m_{31}^2 = \Delta m_{21}^2 = 0$. Finally, the $\Gamma(E)$ described above is given in the lab frame. 
When we later wish to compare the decay coupling between different experiments, it's more convenient to deal with the \textit{rest-frame} coupling. 
Therefore, we make the shift $\Gamma(E) \to \Gamma/\gamma = \Gamma m_4/E$, where $\gamma$ is the lorentz factor. 
Together, this gives
\begin{equation}
    H = \frac{1}{2E}
    \begin{pmatrix}
        0 & 0 & 0 & 0\\
        0 & 0 & 0 & 0\\
        0 & 0 & 0 & 0\\
        0 & 0 & 0 & \Dmq - i m_4 \Gamma
    \end{pmatrix}.
\end{equation}

The oscillation probabilities can be found by evaluating
\begin{equation}
    P(\nu_\beta \to \nu_\alpha) = \langle \nu_\beta | U \exp{[-i H L]} U^\dagger | \nu_\alpha \rangle.
\end{equation}

We simply write the solution below for appearance 
\begin{equation}
    P(\nu_\alpha \to \nu_\beta) = 4 |U_{\alpha 4}|^2 |U_{\beta 4}|^2   \left(\sinh^2 \left(\frac{\Gamma  L m_4}{4 E}\right) +\sin^2 \left(\frac{\Dmq L}{4 E}\right)\right) e^{-\frac{\Gamma  L m_4}{2 E}} \quad (\alpha \neq \beta)
\end{equation}
and disappearance
\begin{multline}
    P(\nu_\alpha \to \nu_\alpha) = 2 \left(1-|U_{\alpha 4}|^2\right) |U_{\alpha 4}|^2 \cos \left(\frac{\Dmq L}{2 E}\right) e^{-\frac{\Gamma  L m_4}{2 E}}\\
    +|U_{\alpha 4}|^4 e^{-\frac{\Gamma  L m_4}{E}}+\left(1-|U_{\alpha 4}|^2\right)^2.
\end{multline}

Written in terms of the 3+1 effective angles
\begin{equation}
    \sin^2 2\theta_{\alpha\beta} \equiv 4|U_{\alpha 4}|^{2}|U_{\beta 4}|^{2}
    \quad
    (\alpha\neq\beta),
    \quad
    \sin^2 2\theta_{\alpha\alpha} \equiv 4 |U_{\alpha 4}|^{2} (1 - |U_{\alpha 4}|^{2}),
\end{equation}
the decay oscillation equations can be rewritten as 
\begin{equation}
    P(\nu_\alpha \to \nu_\beta) = \sin^2 2\theta_{\alpha\beta} \left(\sinh^2 \left(\frac{\Gamma  L m_4}{4 E}\right) +\sin^2 \left(\frac{\Dmq L}{4 E}\right)\right) e^{-\frac{\Gamma  L m_4}{2 E}} \quad (\alpha \neq \beta)
\end{equation}
and 
\begin{multline}
    P(\nu_\alpha \to \nu_\alpha) = \frac{1}{2} \sin^2 2\theta_{\alpha\alpha}  \cos \left(\frac{\Dmq L}{2 E}\right) e^{-\frac{\Gamma  L m_4}{2 E}}+ \sin^{4}\theta_{\alpha\alpha} e^{-\frac{\Gamma  L m_4}{E}}+ \cos^{4}\theta_{\alpha\alpha}.
\end{multline}

    \chapter{MiniBooNE}
\label{chapter:miniboone}

Below we present the MiniBooNE publication \cite{MiniBooNE:2018esg} on which the author had the lead contribution. The Letter provides a concise description of the MiniBooNE detector and analysis.
The corresponding Supplemental Material for this work is also provided in \Cref{ch:miniboonesupp}.

The most recent MiniBooNE results are given in \Cref{sec:currentresults} for completeness. 

\clearpage
\phantomsection
\addcontentsline{toc}{section}{\textit{Publication: Significant Excess of Electronlike Events in the MiniBooNE Short-Baseline Neutrino Experiment}}
\includepdf[pages={-}]{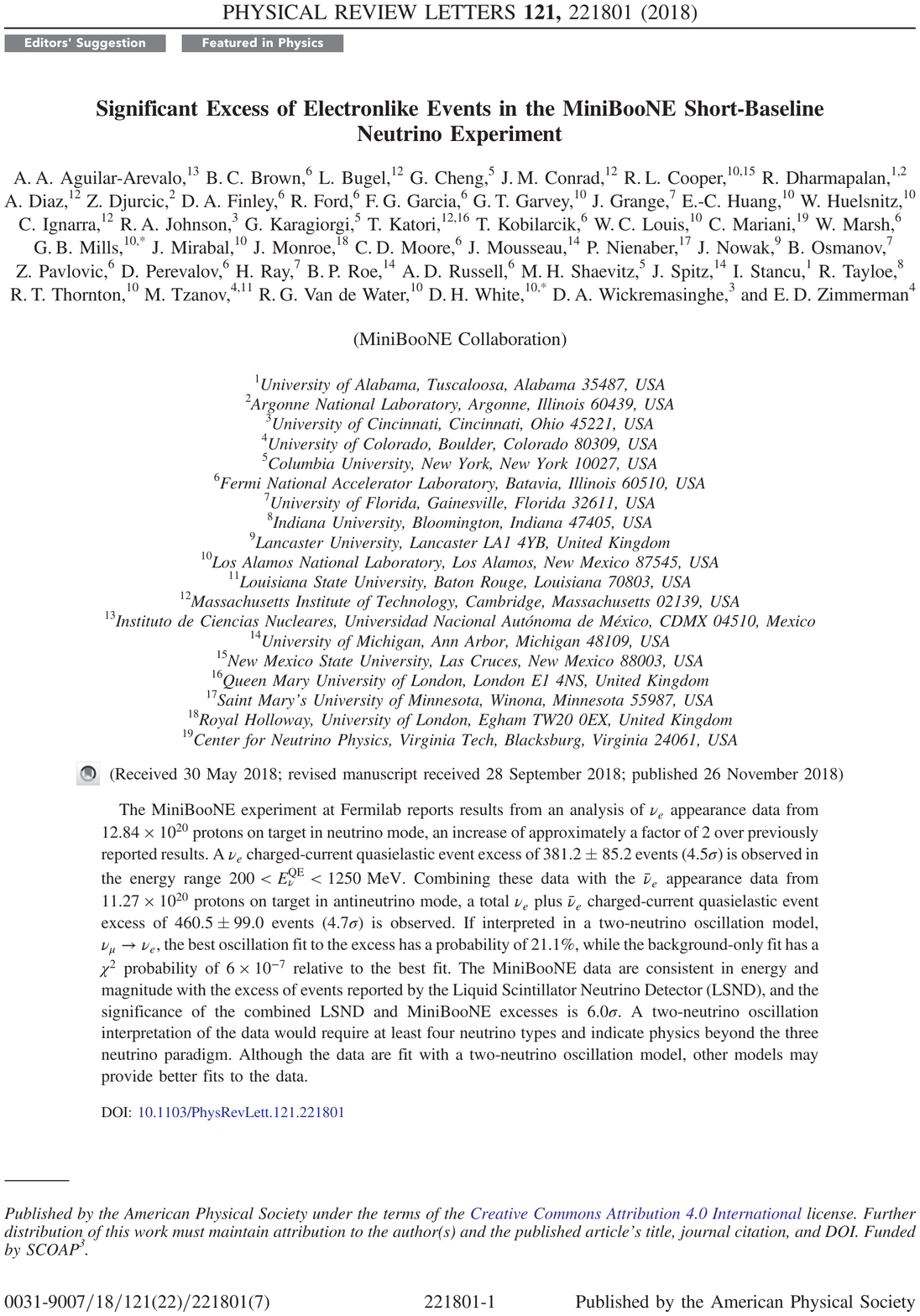}

\section{Current Results}
\label{sec:currentresults}

Since the publication above, MiniBooNE collected its final batch of data and published results in Ref \cite{MiniBooNE:2020pnu}. 
Compared to the publication above, the data sample increased from $12.84 \times 10^{20}$ protons-on-target to $18.75 \times 10^{20}$.
The primary results were already presented in \Cref{sec:minibooneintro}, but we reiterate them here. 

In neutrino mode, 2870 events were observed, with an expectation of $2309.4 \pm 48.1 \textrm{(stat.)} \pm 109.5 \textrm{(syst.)}$, giving an excess of $560.6\pm119.6 (4.7\sigma)$. 
In antineutrino mode, 478 events were observed with an expectation of $400.6 \pm 28.5$, giving an excess of $77.4 \pm 28.5$. 
Combined, this gives a total excess of $638.0 \pm 52.1 \textrm{(stat.)} \pm 122.2 \textrm{(syst.)}$, with a significance of $4.8\sigma$, corresponding to a p-value of $p=1.59\times10^{-6}$.
The excess is plotted in \Cref{fig:MBexcess}.
If fitted to a 3+1 model, the p-value jumps to $p=0.123$.

    \chapter{Global Data Fits to Sterile Neutrino Models}
\label{ch:globalfits}

In \Cref{ch:anomalousresults}, we introduced a few experimental observations that have pointed toward the existence of sterile neutrinos, as well as introduced some sterile neutrino models. 
In this chapter we will test these sterile neutrino models against the global collection of SBL neutrino oscillation data.

For this study, we consider $P(\nue \to \nue)$, $P(\numu \to \numu)$, and $P(\numu \to \nue)$ CC neutrino and antineutrino oscillation channels. 
Assuming the 3+1 model, \Cref{eq:3p1_appearance,eq:3p1_disappearance} show that these oscillation channels are respectively sensitive to \Uefsq, \Umufsq, and $\Uef\Umuf$; 
the equations governing the different oscillation channels are not independent.
Therefore, not only can preferred values of the mixing parameters be determined from global fits, the internal consistency of the models can also be tested. 
In fact, as we will see in \Cref{sec:globalfitresults}, the minimal 3+1 sterile neutrino model suffers from internal inconsistencies that motivate the consideration of more complex models.

In this chapter we summarize the experiments that go into our fits, as well as the limits they have placed on their own. A table of the experiments in our fits is shown in \Cref{table:experiments}. We then review the methodology of our fits, and end with the results. 

\begin{table}
\centering
\resizebox{\textwidth}{!}{
\begin{tabular}{l || c | c | c}
& $\nu_\mu \to \nu_e$ & $\nu_\mu \to \nu_\mu$ & $\nu_e \to \nu_e$ \\
\hline
\hline
\multirow{4}{6em}{Neutrino} & MiniBooNE (BNB)  & SciBooNE/MiniBooNE & KARMEN/LSND Cross Section\\
& MiniBooNE(NuMI) & CCFR & Gallium  \\
& NOMAD & CDHS & BEST\\
& & MINOS & \\
\hline
\multirow{6}{6em}{Antineutrino} 
& LSND  & SciBooNE/MiniBooNE & Bugey \\
& KARMEN & CCFR & NEOS \\
& MiniBooNE (BNB) & MINOS & DANSS  \\
& & & PROSPECT \\
& & & STEREO \\ 
& & & Neutrino-4
\end{tabular}
}
\caption{The collection of experiments that go into the global fits in this thesis. }
\label{table:experiments}
\end{table}

The contents of this chapter can be seen as an update of the work in Ref.~\cite{Diaz:2019fwt}. 

\section{Experiments}
\label{sec:experiments}

The experiments included in our fits fall into one of three groups: \nue appearance ($P(\numu \to \nue)$ \& $P(\numubar \to \nuebar)$), \nue disappearance ($P(\nue \to \nue)$ \& $P(\nuebar \to \nuebar)$), and \numu disappearance ($P(\numu \to \numu)$ \& $P(\numubar \to \numubar)$).

In this section, we give a short summary of each experience and their individual findings.
For each experiment, we provide the 3+1 confidence regions released by the collaborations, if available. 
In \Cref{fig:3+1appearancefits,fig:3+1galliumdisappearance,fig:3+1reactor,fig:numufits}, we also provide the 3+1 confidence regions we recover in our implementation of these datasets for our global fits. 
Later we will look at what the combined data tells us. 

\subsection{\texorpdfstring{$P(\numu \to \nue)$ \& $P(\numubar \to \nuebar)$}{P(\nu \mu \rightarrow \nu \mu) \& P(\=\nu \mu \ \rightarrow \=\nu \mu)}}

\begin{figure}
    \centering
    \begin{subfigure}{0.47\linewidth}
        \centering
        \includegraphics[width=\linewidth]{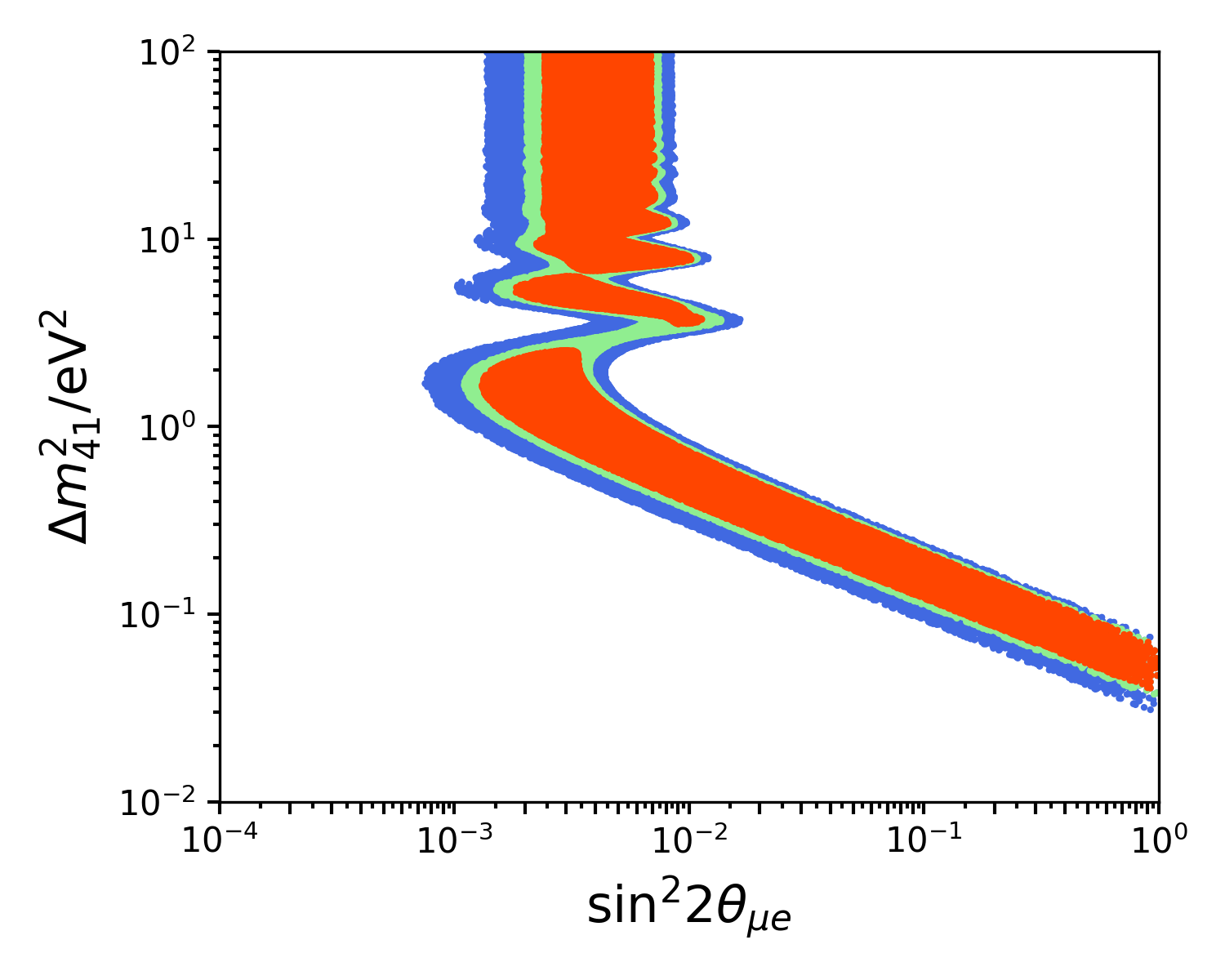}
        \caption{LSND}
        \label{fig:LSNDfit}
    \end{subfigure}
    \hfill
    \begin{subfigure}{0.47\linewidth}
        \centering
        \includegraphics[width=\linewidth]{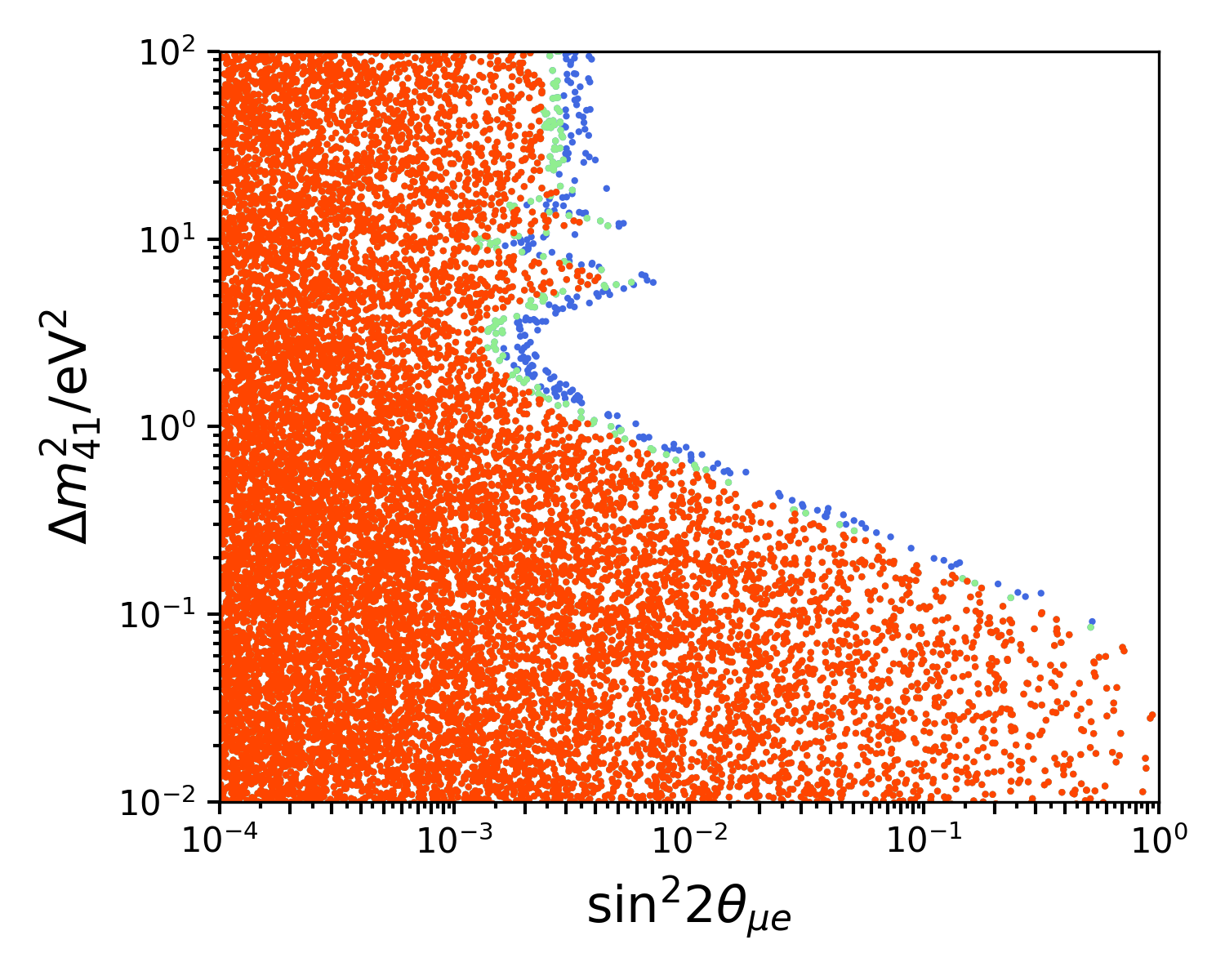}
        \caption{KARMEN}
        \label{fig:KARMENfit}
    \end{subfigure}

    \begin{subfigure}{0.47\linewidth}
        \centering
        \includegraphics[width=\linewidth]{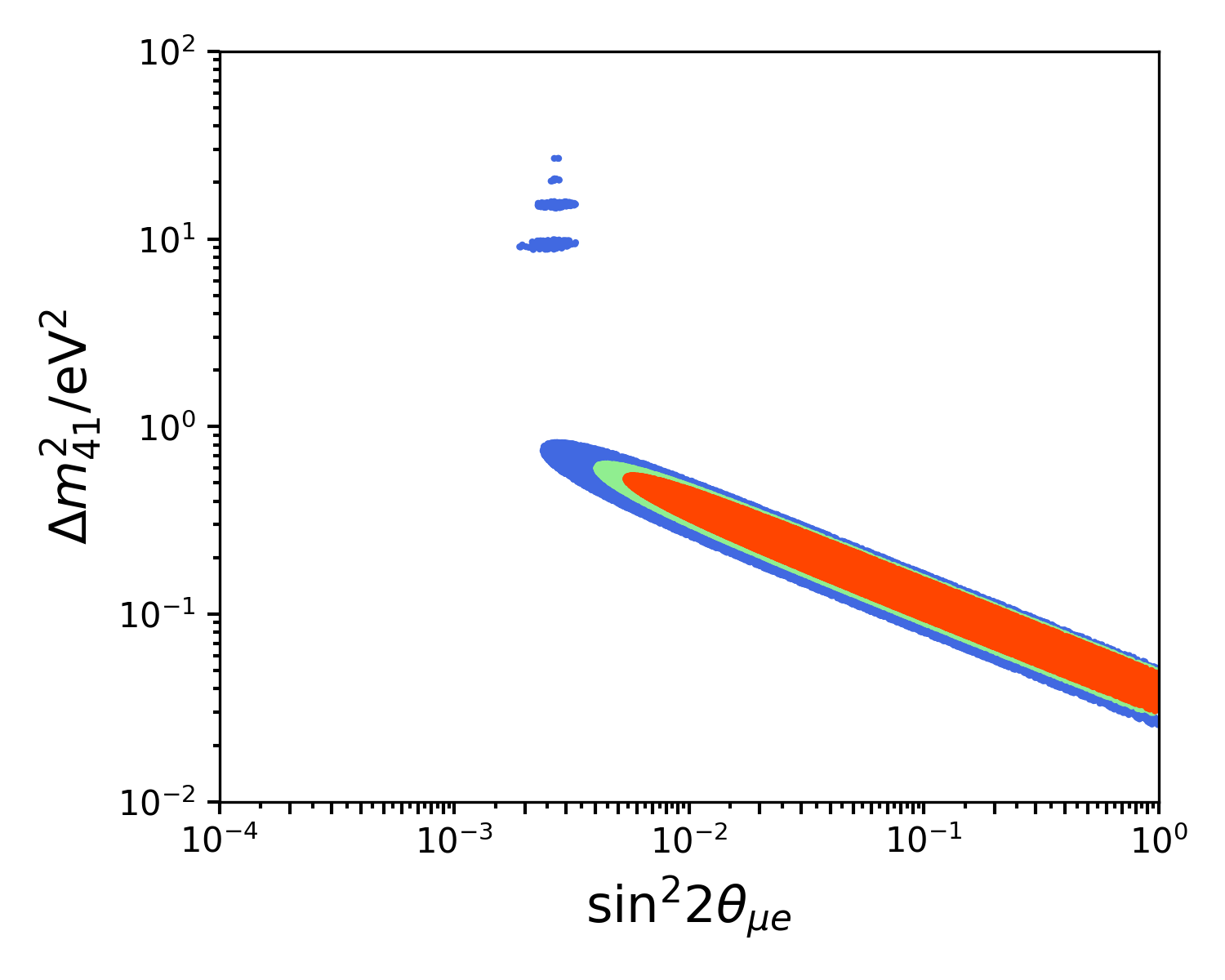}
        \caption{MiniBooNE (BNB)}
        \label{fig:MBfit}
    \end{subfigure}
    \hfill
    \begin{subfigure}{0.47\linewidth}
        \centering
        \includegraphics[width=\linewidth]{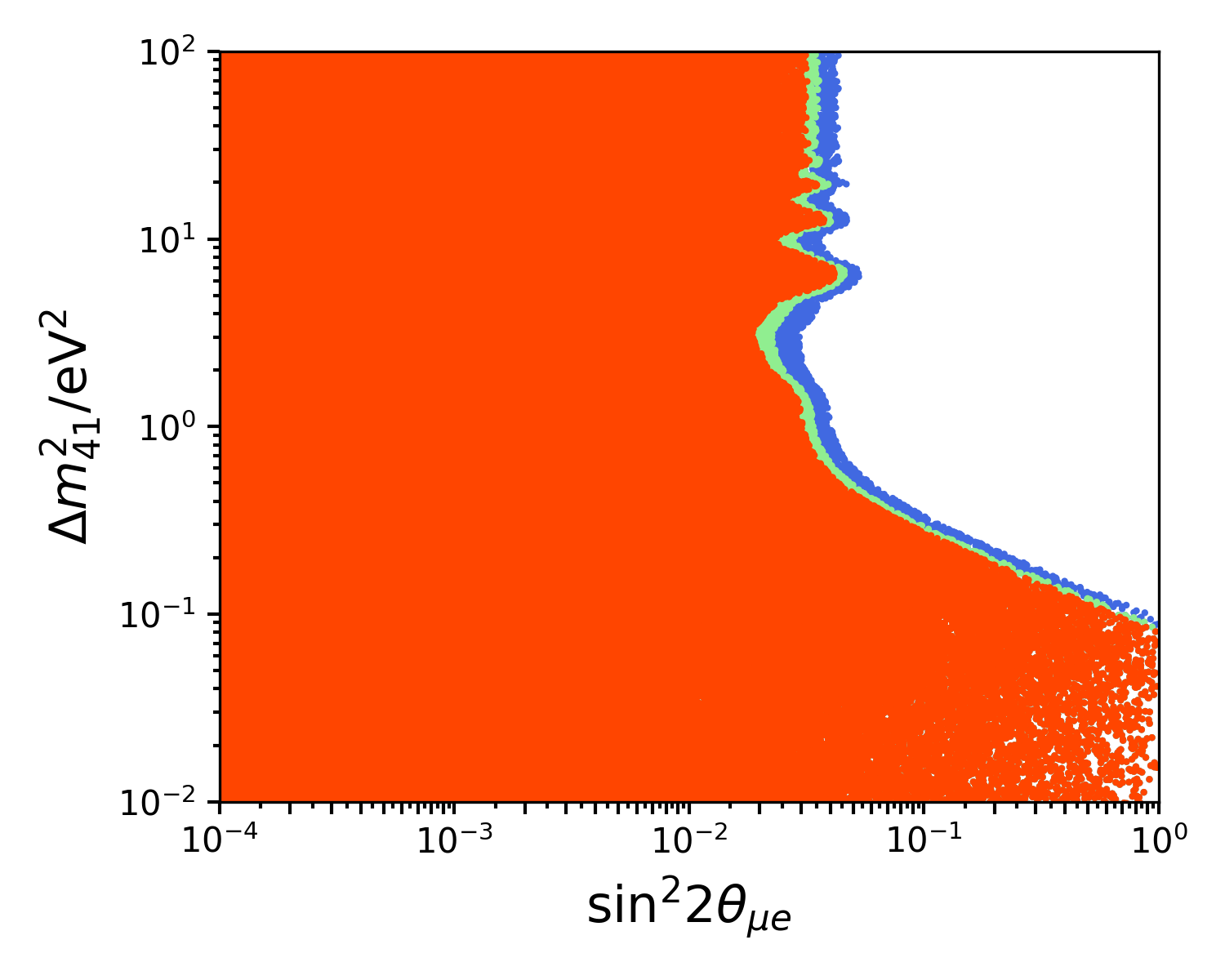}
        \caption{MiniBooNE (NuMI)}
        \label{fig:MB-NMfit}
    \end{subfigure}

    \begin{subfigure}{0.47\linewidth}
        \centering
        \includegraphics[width=\linewidth]{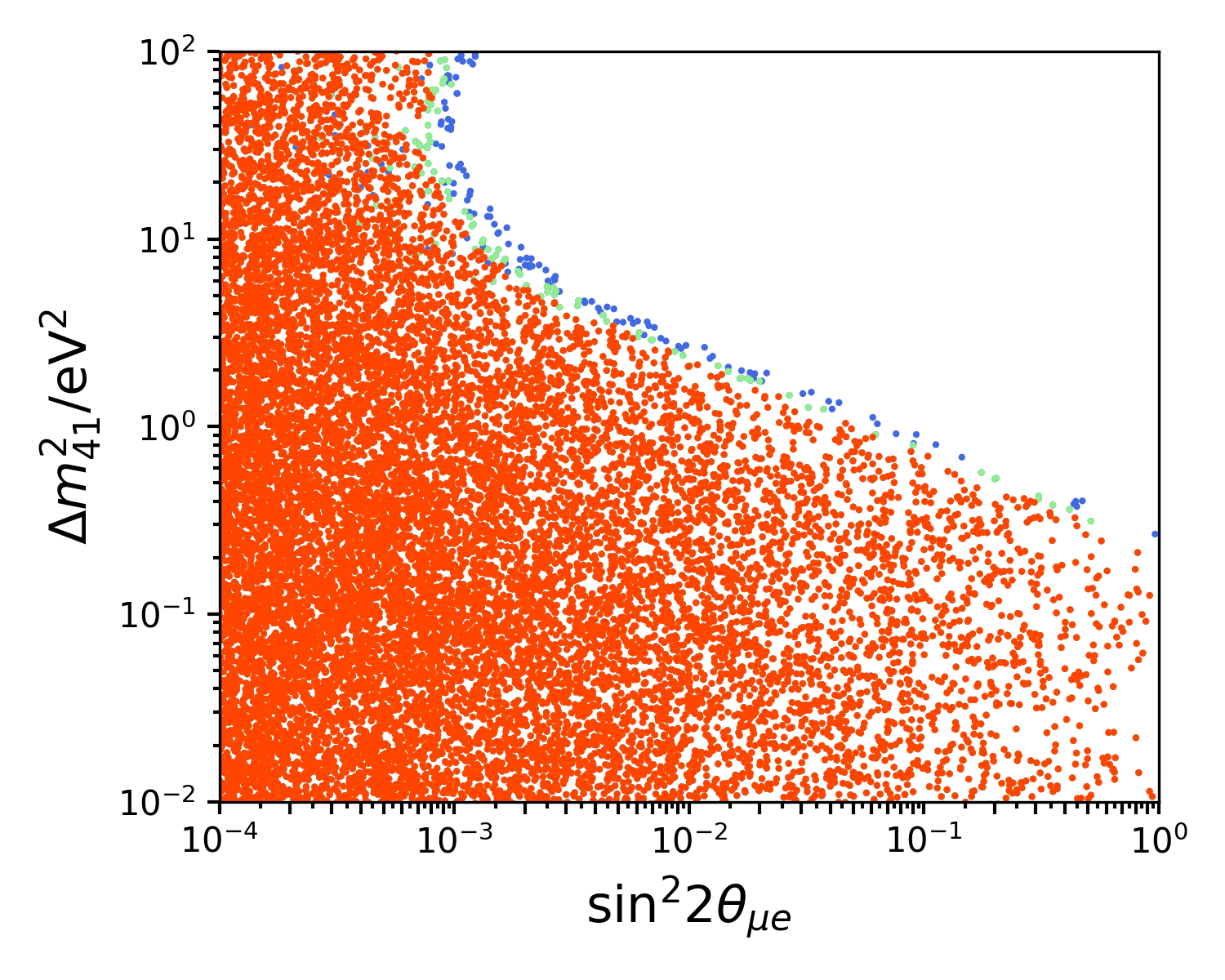}
        \caption{NOMAD}
        \label{fig:NOMADfit}
    \end{subfigure}
    \caption{3+1 fits to the $\numu \to \nue$ \& $\numubar \to \nuebar$ appearance oscillation data used in our global fits. The 90\%, 95\%, and 99\% confidence regions correspond to the red, green, and blue points respectively.}
    \label{fig:3+1appearancefits}
\end{figure}

\begin{description}
    \item[LSND \cite{LSND:2001aii}] \hfill \\
        The \textbf{L}iquid \textbf{S}cintillator \textbf{N}eutrino \textbf{D}etector (LSND) experiment ran 1993--1998 at the Los Alamos Neutron Science Center (LANSCE), searching for $\numubar \to \nuebar$ oscillations. 
        As reviewed in \Cref{sec:anomaliesLSND}, the \numubar beam was created by impinging a \SI{798}{\MeV} proton beam on a target and allowing the subsequent $\mu^+$'s to decay at rest into \numubar's. 
        The neutrinos would then propagate \SI{30}{\meter} towards a cylindrical tank \SI{8.3}{\meter} long by \SI{5.7}{\meter} in diameter filled with 167 metric tons of liquid scintillator. 
        The oscillated \nuebar would then inverse beta decay like $\nuebar + p \to e^+ + n$, producing a signal from the positron followed by a coincident $2.2\ \MeV\ \gamma$ when the neutron captures. With this required coincident signal, LSND would select events with positron energies in the range $20<E_e<60\ \MeV$.

        LSND observed an excess of $87.9 \pm 22.4 \pm 6.0$ events above background, corresponding to an oscillation probability of $(0.264 \pm 0.067 \pm 0.04)\%$.
        The 90\% confidence region is shown in \Cref{fig:lsndcontour}. 

        For our fits, we use the data shown in \Cref{fig:lsndrgtgamma10}.
        Note that these data are a cleaner subset of the total LSND data.
        Events with $R_\gamma > 10$ are chosen, where $R_\gamma$ is defined as the likelihood ratio that the neutron-captured $\gamma$ observed is correlated to the initial signal versus accidental.
        Here, the event excess is $32.2 \pm 9.4 \pm 2.3$. 
        More details on the $R_\gamma$ selection can be found in Ref.~\cite{LSND:2001aii}.

        The 3+1 results of our implementation of LSND is shown in \Cref{fig:LSNDfit}.

        \begin{figure}
            \centering
            \begin{subfigure}{0.45\textwidth}
                \includegraphics[width=\textwidth]{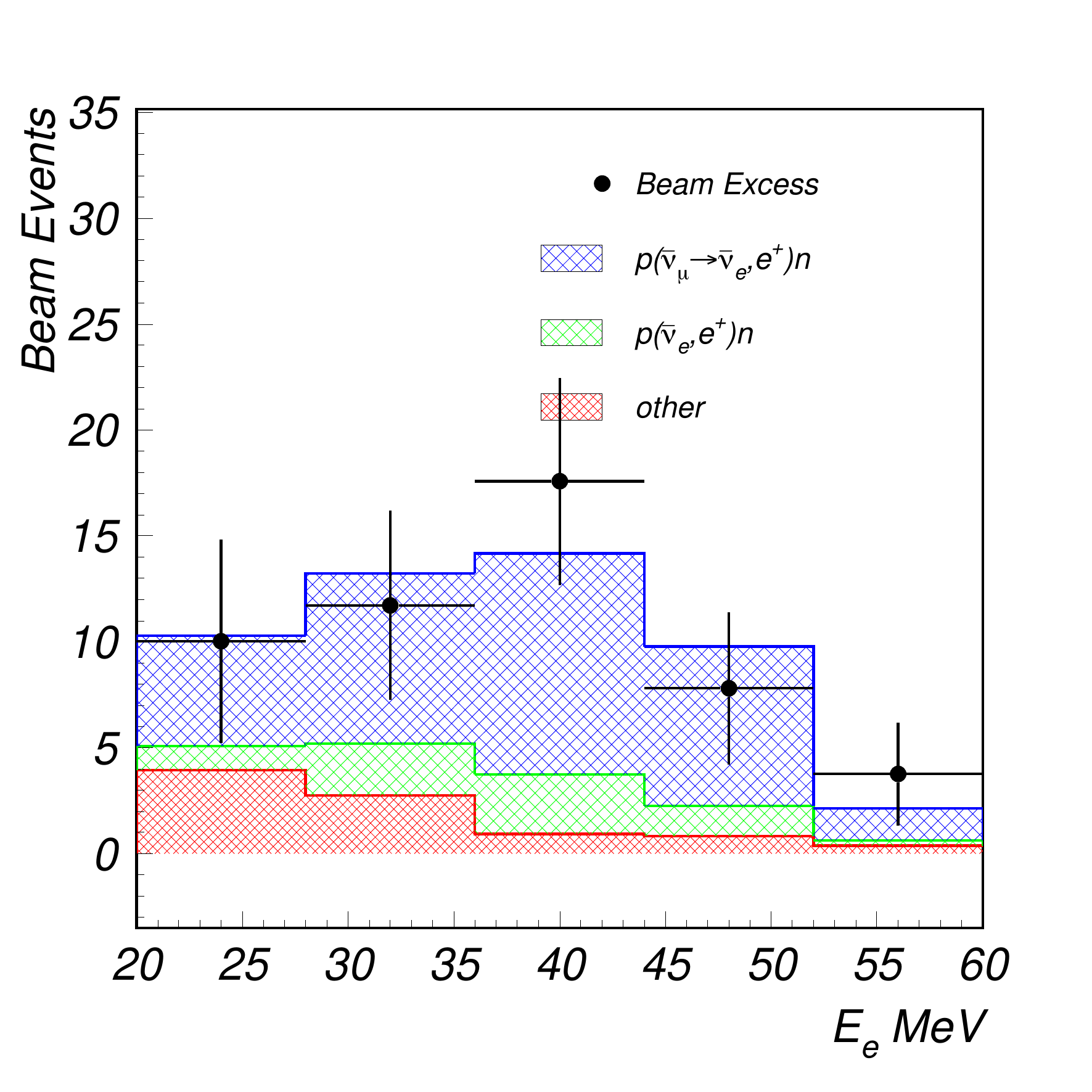}
                \caption{}
                \label{fig:lsndrgtgamma10}
            \end{subfigure}
            \begin{subfigure}{0.45\textwidth}
                \includegraphics[width=\textwidth]{images/AnomalousResultsandSterileNeutrinos/Fig26_cl90.pdf}
                \caption{}
                \label{fig:lsndcontour}
            \end{subfigure}
            \caption{(a) The beam excess observed at LSND with the cut $R_\gamma > 10$. The red and green histograms are the expected beam-on backgrounds, and the blue histogram is the expected event rate with the best fit 3+1 oscillation hypothesis. (b) The best fit contours at the 90\% confidence level. Figures from Ref.~\cite{LSND:2001aii}}.
        \end{figure}

    \item[KARMEN \cite{KARMEN:2002zcm}] \hfill \\
        The \textbf{Ka}rlsruhe \textbf{R}utherford \textbf{M}edium \textbf{E}nergy \textbf{N}eutrino (KARMEN) experiment was another accelerator beam experiment similar to LSND, located at the spallation neutrino source ISIS at the Rutherford Laboratory in the UK, running 1997--2001. KARMEN searched for $\numubar \to \nuebar$ oscillations using a \SI{800}{\MeV} proton beam to produce a DAR neutrino beam like LSND. The detector was placed \SI{17.7}{\meter} away from the target and, unlike LSND, at an angle \ang{100} off the proton beam, reducing beam backgrounds. 

        With a background prediction of $15.8 \pm 0.5$ events, KARMEN observed 15 events, well within expectations and finding no evidence for oscillations.
        \Cref{fig:KARMEN} plots KARMEN's 90\% confidence level exclusion, compared with other experiments at the time. 
        The 3+1 results of our implementation of KARMEN is shown in \Cref{fig:KARMENfit}.

        \begin{figure}
            \centering
            \includegraphics[width=0.5\textwidth]{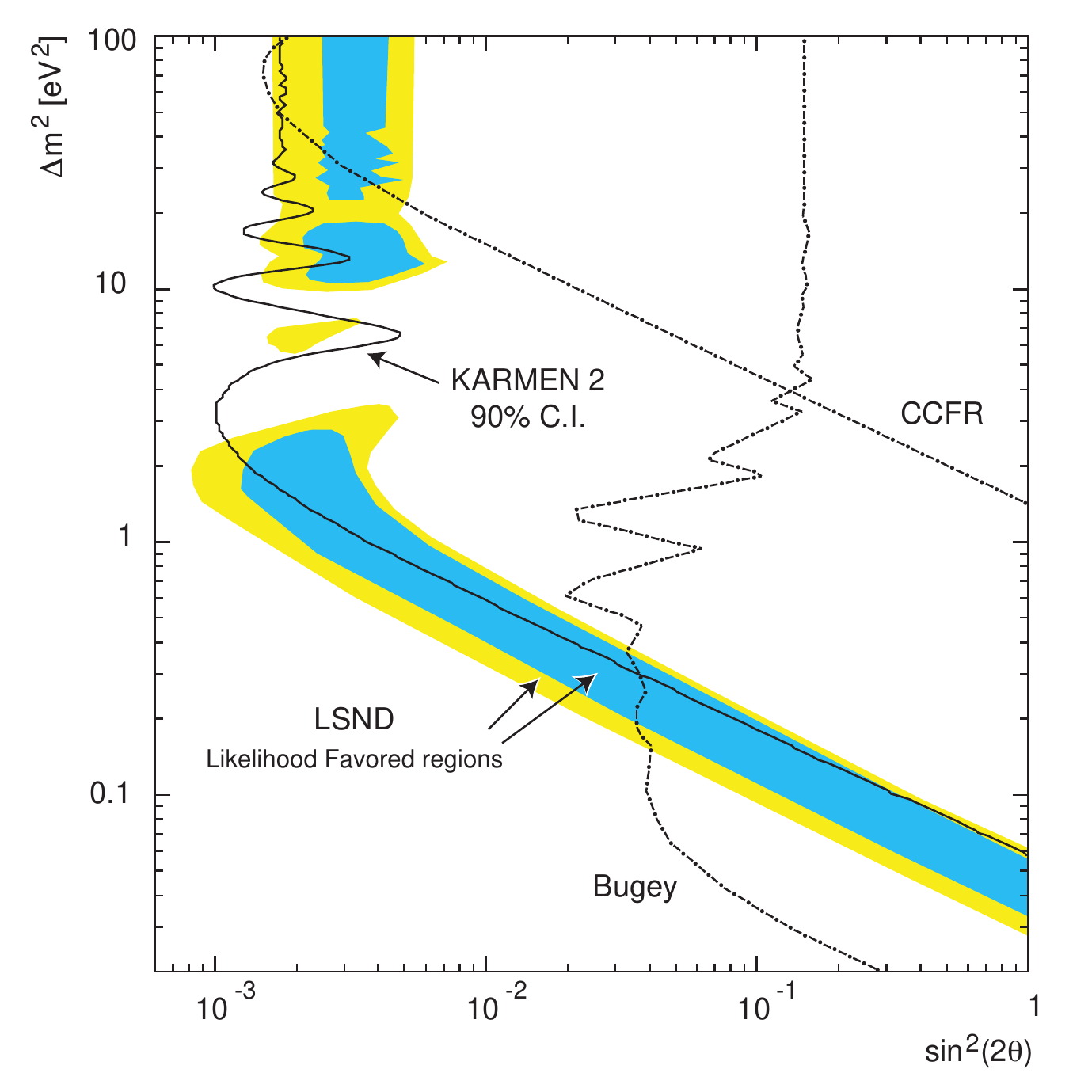}
            \caption{Comparison of KARMEN's confidence region at the 90\% confidence level with LSND's. Included in this plot are exclusions from two other experiments, CCFR and Bugey. We discuss these experiments below.
            Figure from Ref.~\cite{KARMEN:2002zcm}.}
            \label{fig:KARMEN}
        \end{figure}

    \item[MiniBooNE (BNB) \cite{MiniBooNE:2013uba,MiniBooNE:2020pnu}] \hfill \\
        The MiniBooNE experiment has already been described in detail in \Cref{chapter:miniboone}, and we simply refer to that chapter. The ``BNB'' in the experiment title refers to the Booster Neutrino Beam, which is the primary source of MiniBooNE's neutrino flux.

        The 3+1 results of our neutrino and antineutrino combined MiniBooNE fit is shown in \Cref{fig:MBfit}.

    \item[MiniBooNE (NuMI) \cite{MiniBooNE:2008hnl}] \hfill \\
        In addition to the BNB beam line, the MiniBooNE detector could also observe neutrinos from the NuMI beam line. 
        The NuMI beam produces neutrinos for the MINOS detectors by accelerating \SI{120}{\GeV} protons into a carbon target. 
        The MiniBooNE detector is located \SI{745}{\meter} from the NuMI production target, and at an angle 6.3\textdegree\ off the NuMI beam axis.
        Using data collected in 2005--2007, MiniBooNE searched for possible $\numu \to \nue$ oscillations from the NuMI target.
        The data was selected to be in the energy range $0.2 < E_\nu < 3\ \GeV$, and is shown in \Cref{fig:MBNuMI}.
        The observed data falls within the expectation, but with large systematic uncertainties in the expectation. 
        It's noted that in the low energy region $E_\nu < 0.9\ \GeV$ the data are systematically high at the $1.2\sigma$ level.

        The 3+1 results of our implementation of MiniBooNE-NuMI is shown in \Cref{fig:MB-NMfit}.

        \begin{figure}
            \centering
            \includegraphics[width=0.5\textwidth]{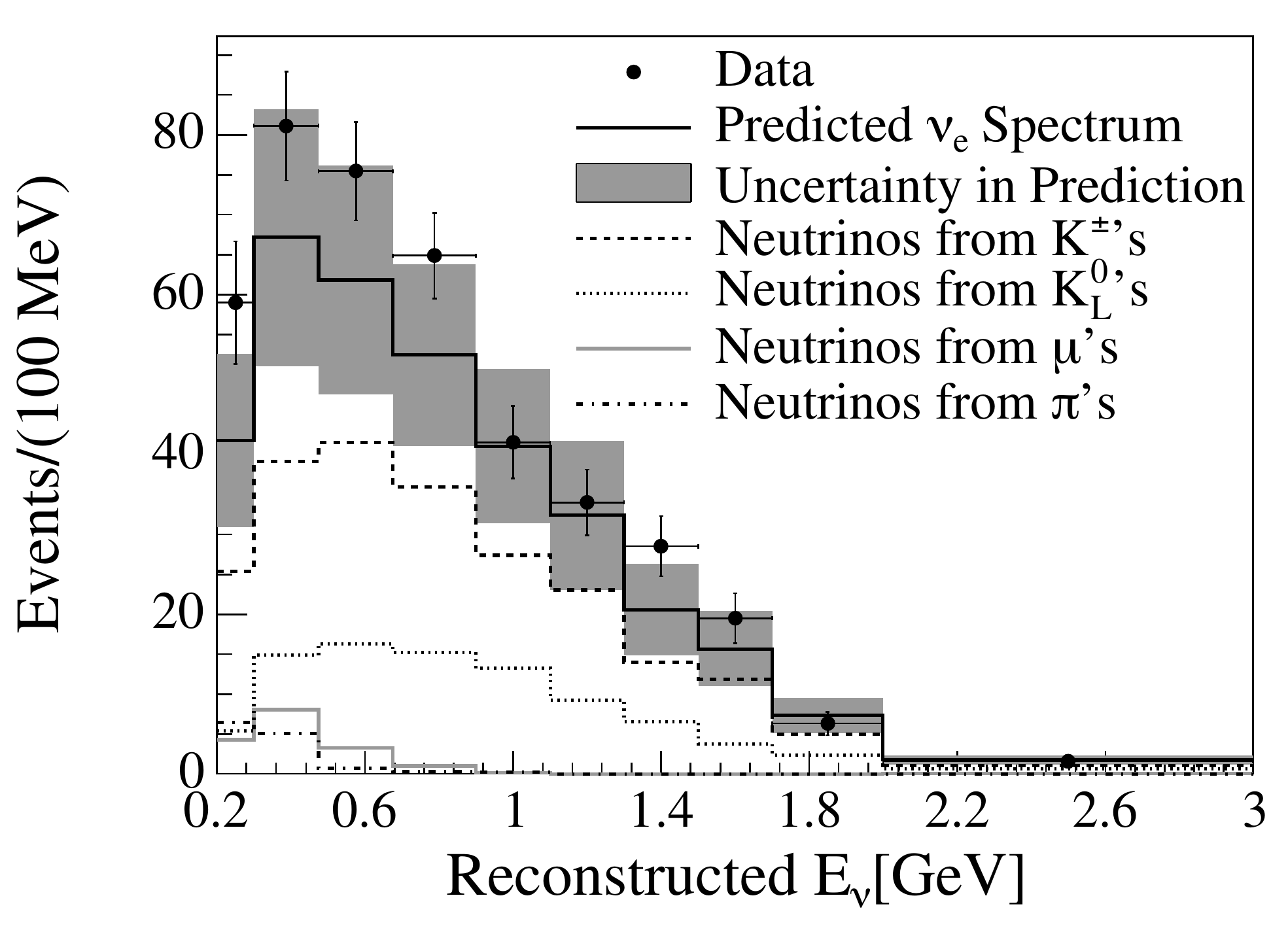}
            \caption{The observed $\nue$ event distribution in the MiniBooNE detector from the NuMI beam. While the observed data lies within the expectation, the expected distribution suffers from large systematic uncertainties. Figure from Ref.~\cite{MiniBooNE:2008hnl}.}
            \label{fig:MBNuMI}
        \end{figure}

    \item[NOMAD \cite{NOMAD:2003mqg}] \hfill \\
        The \textbf{N}eutrino \textbf{O}scillation \textbf{Ma}gnetic \textbf{D}etector (NOMAD) experiment was designed to search for $\numu \to \nutau$ oscillations using the neutrino beam produced by the \SI{450}{\GeV} proton synchrotron (SPS) at CERN. 
        The proton beam impinged a series of beryllium rods, producing secondary particles which were then focused by two magnetic lenses and led into a \SI{290}{\meter} decay tunnel. 
        The neutrinos, on average, traveled \SI{625}{\meter} before reaching the NOMAD detector.
        The detector was designed to identify electrons from $\tau^- \to e^- + \nuebar + \nutau$ decays.
        This allowed the detector to also search for $\numu \to \nue$ oscillations, motivated by the observations from LSND. 

        The NOMAD experiment conducted such a search using data collected 1995--1998. 
        In order to reduce systematic uncertainties, the experiment studied the ratio $R_{e\mu}$ of \nue to \numu CC interactions. 
        Additionally, the experiment took into account the energy and radial distribution of the neutrino beam. 
        The selected energy range extended up to \SI{300}{\GeV}, with a peak at \SI{\sim 40}{\GeV}. 
        This relatively large $\langle L \rangle/ \langle E \rangle \sim 0.02$ gave NOMAD sensitivity to a larger $\Dmq$ compared to LSND. 
        The data is binned as a function of visible energy, which is taken as an approximation of the neutrino energy.

        The observed ratios $R_{e\mu}$ are shown in \Cref{fig:NOMAD}.
        The observed data was consistent with the null hypothesis, and therefore excludes the LSND preferred parameter space at $\Dmq \gtrsim 10\ \eVq$.
        The exclusion is shown in \Cref{fig:NOMADexclusion}.

        The 3+1 results of our implementation of NOMAD is shown in \Cref{fig:NOMADfit}.

        \begin{figure}
            \begin{subfigure}{.59\textwidth}
                \centering
                \includegraphics[height=.4\textheight]{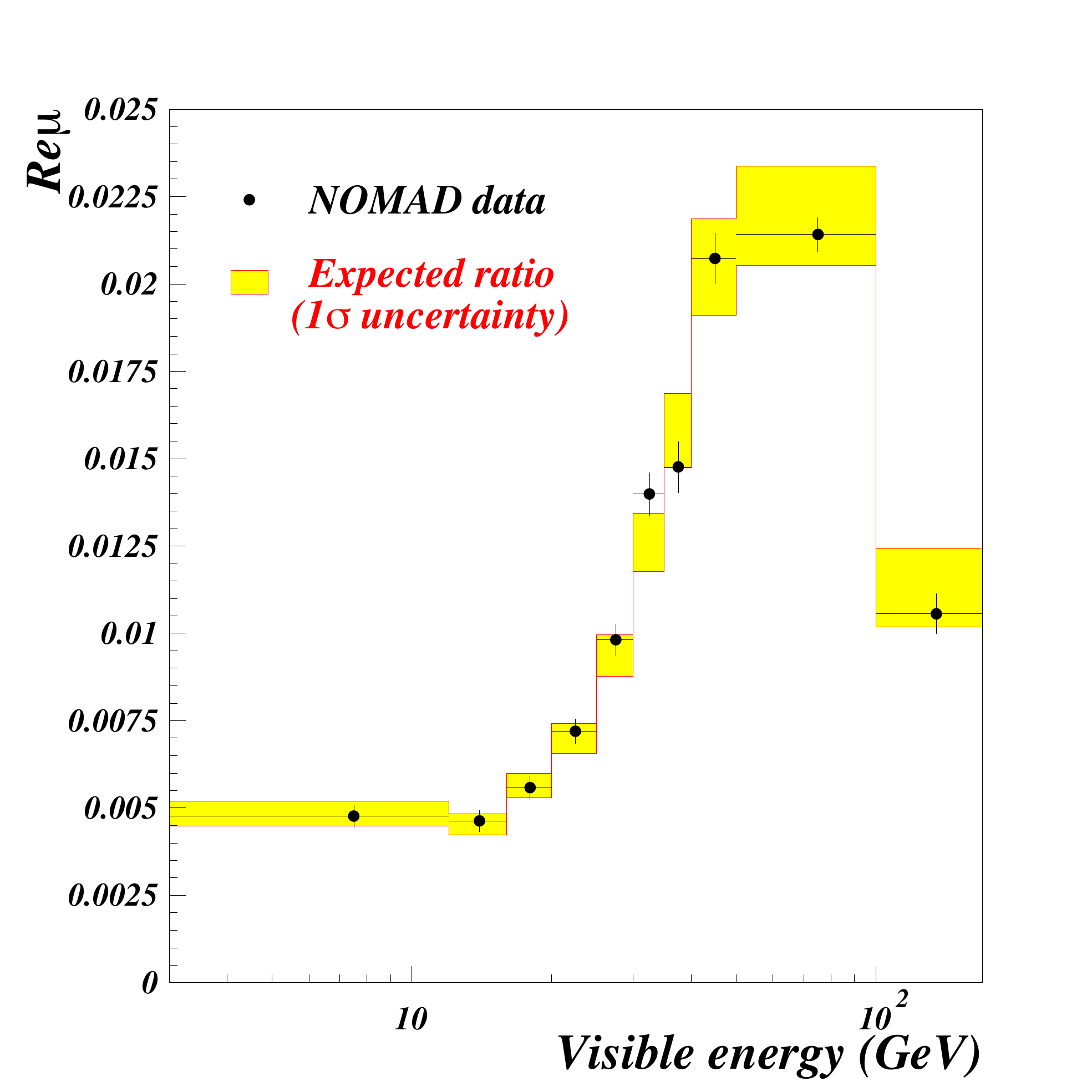}
                \caption{}
            \end{subfigure}
            \begin{subfigure}{.39\textwidth}
                \centering
                \includegraphics[height=.4\textheight]{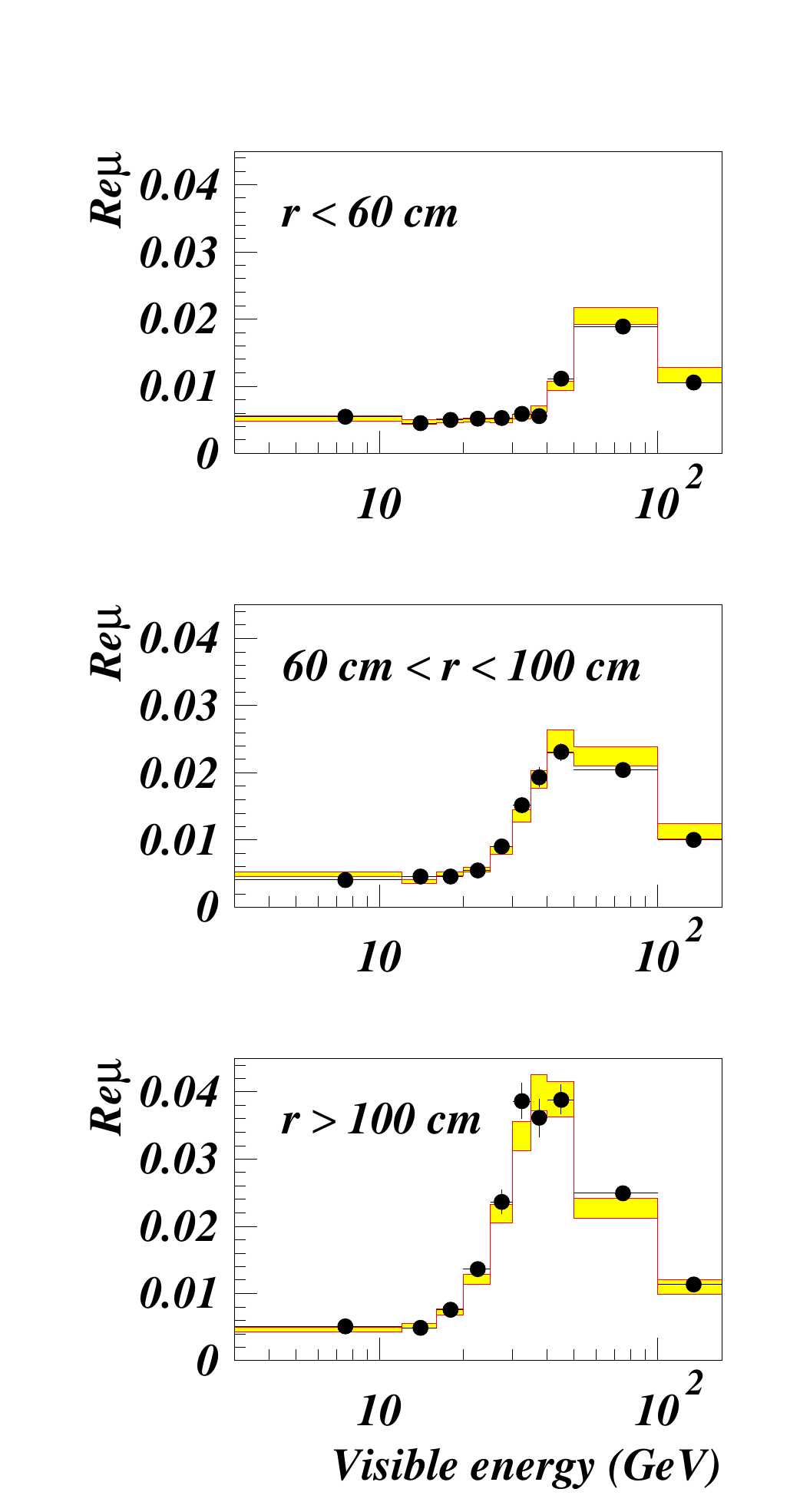}
                \caption{}
            \end{subfigure}
            \caption{(a) The observed ratios $R_{e\mu}$ versus expectation at the NOMAD detector, with $1\sigma$ bands in yellow. (b) The observed ratios $R_{e\mu}$ and expectation, separated by radial distribution. Figures from Ref.~\cite{NOMAD:2003mqg}.}
            \label{fig:NOMAD}
        \end{figure}

        \begin{figure}
            \centering
            \includegraphics[width=0.45\textwidth]{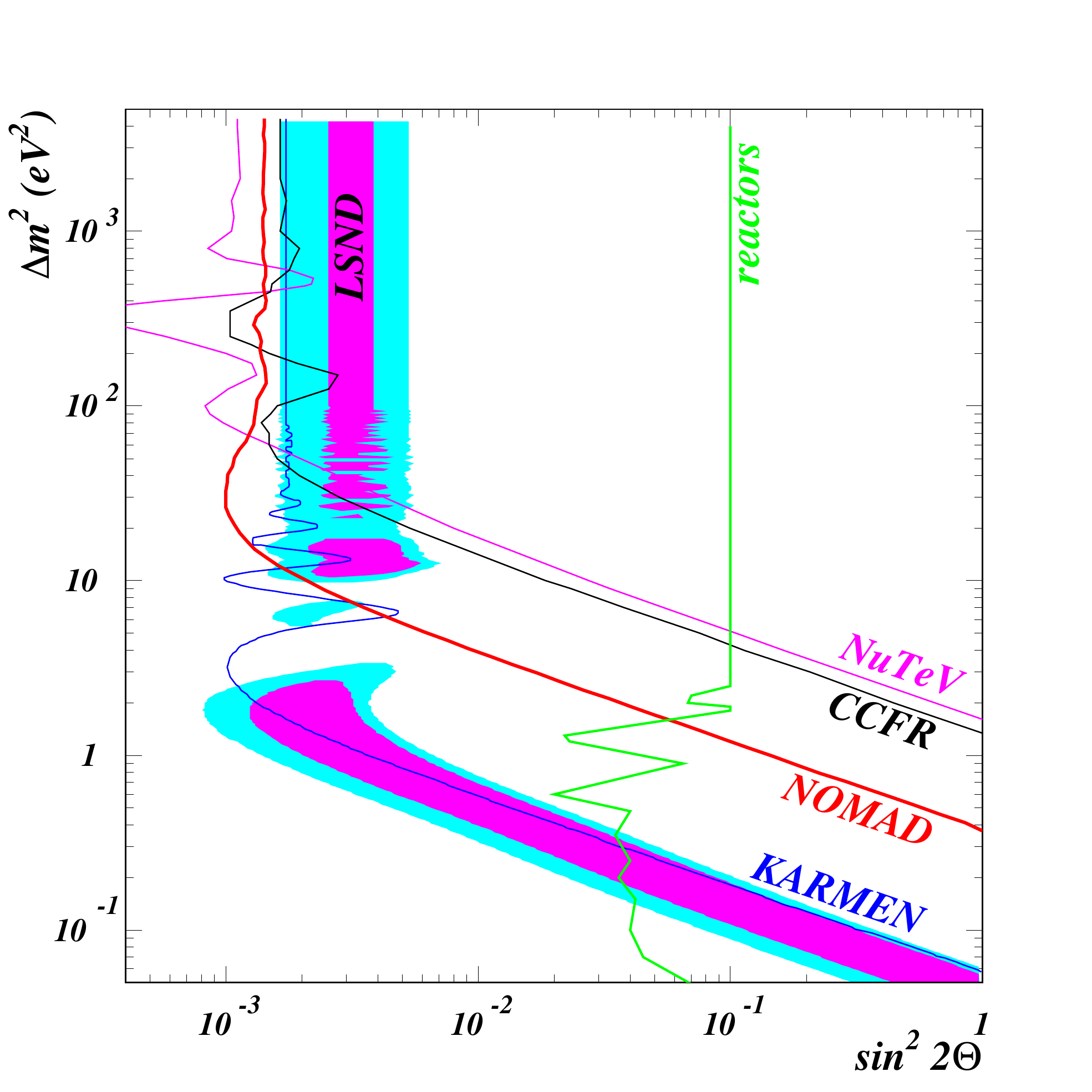}
            \caption{The 90\% NOMAD exclusion region, compared with other SBL experiments available at the time of the NOMAD analyis. We have discussed several of these experiments in this section. Figure from Ref.~\cite{NOMAD:2003mqg}.}
            \label{fig:NOMADexclusion}
        \end{figure}
\end{description}

\subsection{\texorpdfstring{$P(\nue \to \nue)$ \& $P(\nuebar \to \nuebar)$}{P(\nu e \rightarrow \nu e) \& P(\=\nu e \rightarrow \=\nu e)}}
\label{sec:nuedis}

\begin{description}

    \item[KARMEN/LSND (cross section) \cite{Conrad:2011ce}] \hfill \\
        In addition to the $\numubar \to \nuebar$ appearance analysis described above, both LSND and KARMEN conducted a measurement of the $\nue$ CC interaction cross section on \isotope[12]{C} \cite{KARMEN:1994xse,Armbruster:1998uk,LSND:2001fbw}. 
        Like \nuebar interactions, \nue interactions can be tagged by a coincident signal. 
        First, the incoming \nue undergoes the IBD interaction $\nue + \isotope[12]{C} \to \isotope[12]{N_\textrm{gs}} + e^-$. 
        Then, the $\isotope[12]{N}$ ground state decays like $\isotope[12]{N_\textrm{gs}} \to \isotope[12]{C} + e^+ + \nue$ with a Q-value of 16.3 MeV and a lifetime of 15.9 ms.  
        The observed $e^-$, followed by a $e^+$, allows the tagging of \nue events, and a cross section measurement can be made. 

        In practice, the measured cross section will be \textit{flux-averaged}, so that the measured quantity will depend on the flux knowledge. 
        If the $\nue$ flux is low, then the measured cross section will be low compared to theoretical predictions. 
        Therefore, the measured cross section can be used to place limits on the disappearance of the \nue flux by comparing the measured cross section versus expectation. 

        \begin{figure}
            \centering
            \begin{subfigure}{0.5\textwidth}
                \includegraphics[width=\textwidth]{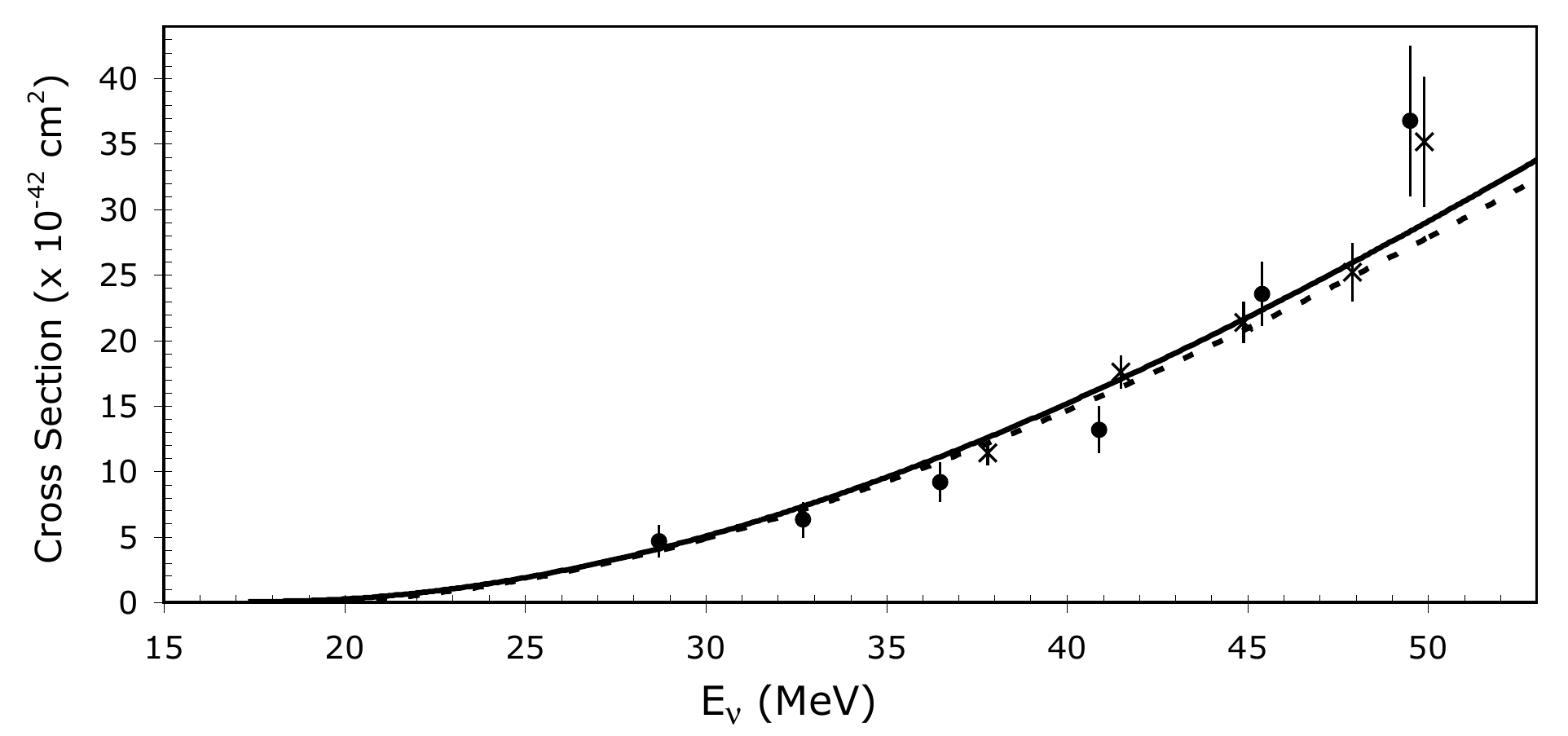}
                \caption{}
                \label{fig:KARMENLSNDmeasured}
            \end{subfigure}
            \begin{subfigure}{0.34\textwidth}
                \includegraphics[width=\textwidth]{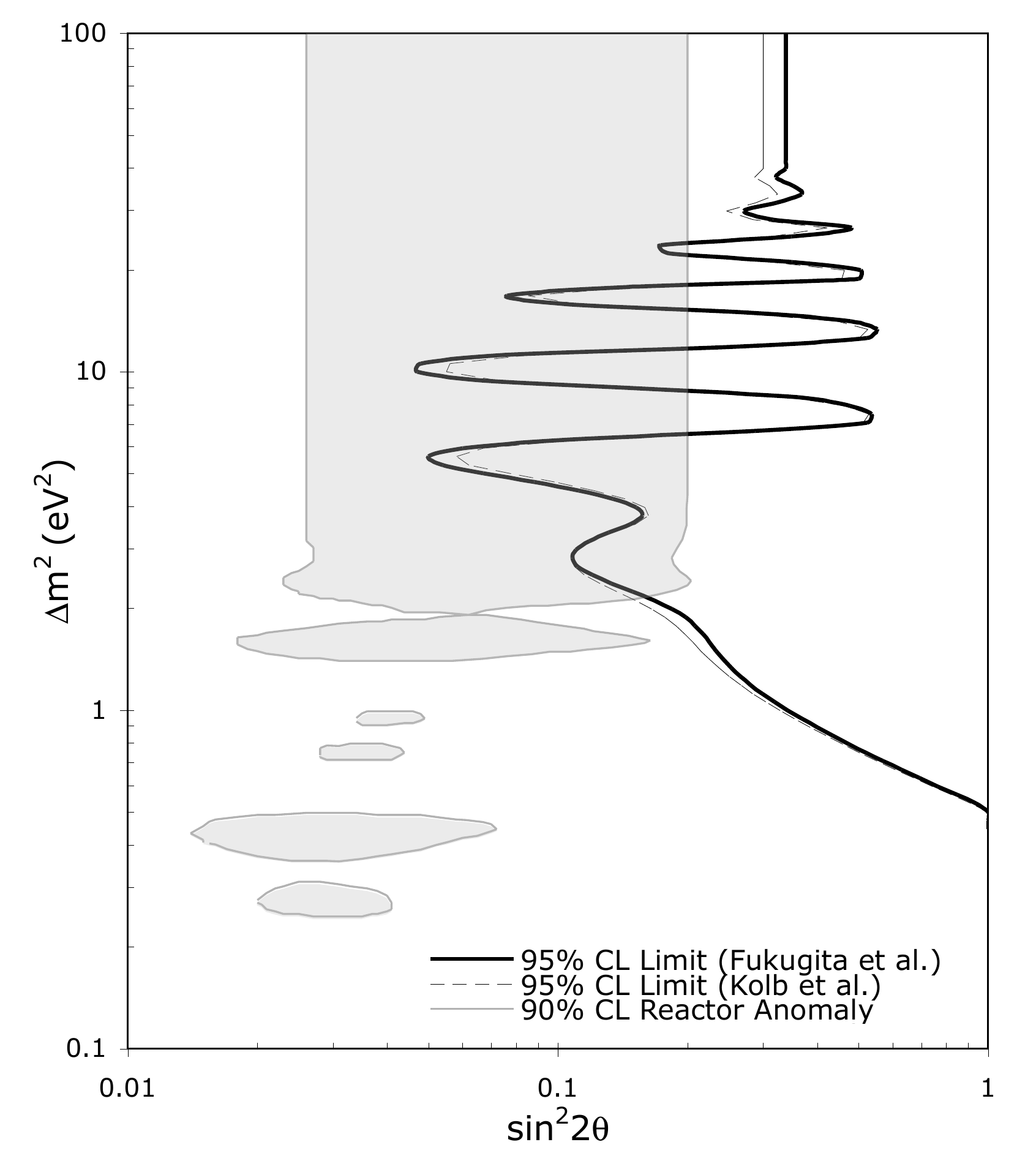}
                \caption{}
                \label{fig:KARMENTLSNDfits}
            \end{subfigure}
            \caption{(a) The measured $\nue + \isotope[12]{C} \to \isotope[12]{N_\textrm{gs}} + e^-$ cross section for LSND (crosses) and KARMEN (points). The multiple lines correspond to different cross section predictions. (b) The 95\% confidence level $\nue \to \nue$ disappearance limits. Each limit assumes a different interaction cross section model. The filled in grey contour corresponds to the RAA, which was discussed in \Cref{sec:RAA}. Figures from Ref.~\cite{Conrad:2011ce}}.
        \end{figure}

        The measured cross sections, compared to theoretical predictions, are shown in \Cref{fig:KARMENLSNDmeasured}. 
        No indication for oscillations is seen, and a limit is place on $\nue \to \nue$ disappearance. 
        \Cref{fig:KARMENTLSNDfits} shows the extracted limits. 

        The result of our 3+1 KARMEN/LSND cross section fit is shown in \Cref{fig:KARMENTLSNDfits}.
\end{description}

\begin{figure}
    \centering
    \begin{subfigure}{0.49\linewidth}
        \centering
        \includegraphics[width=\linewidth]{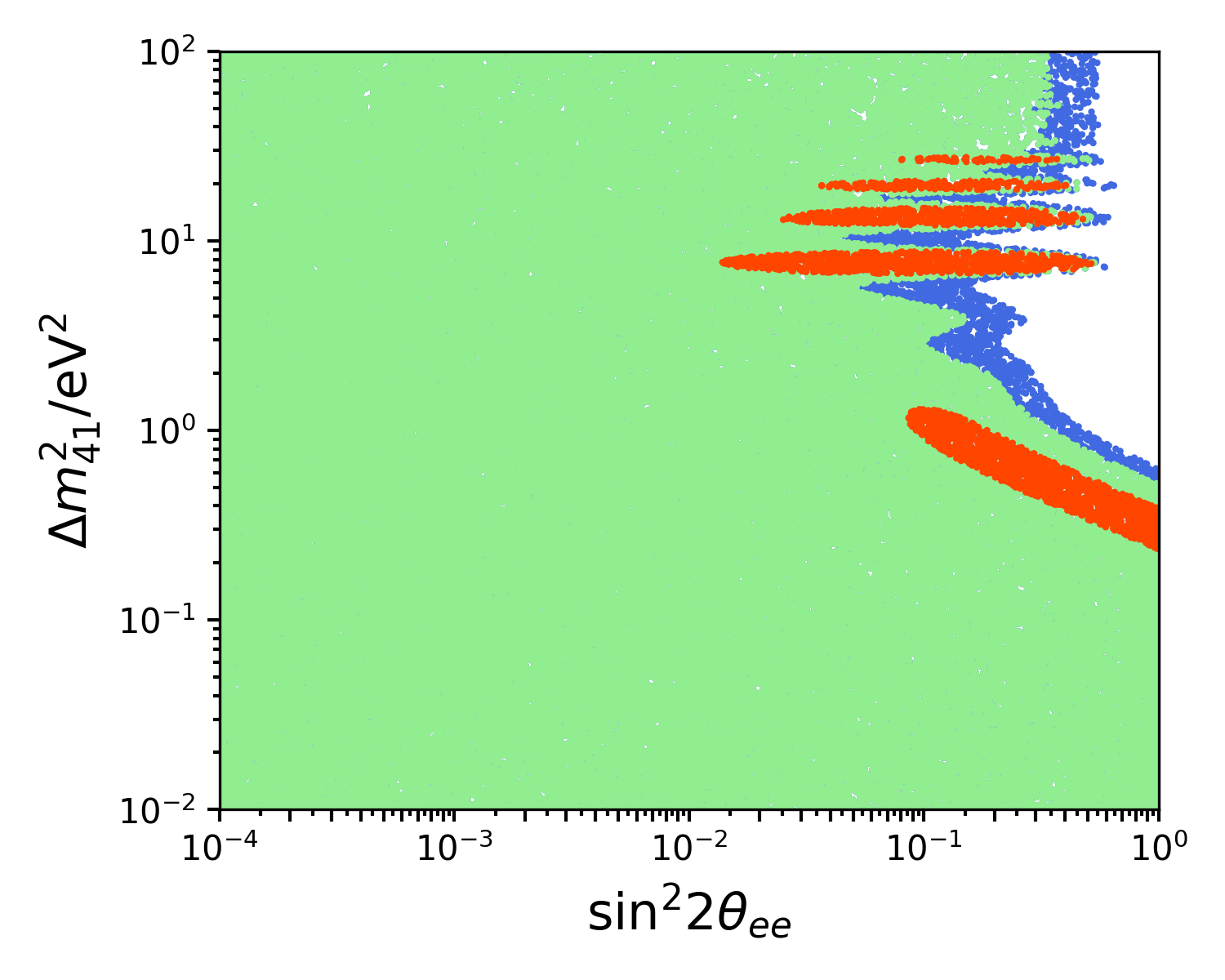}
        \caption{KARMEN/LSND (cross section)}
        \label{fig:KARMENLSNDxsecfit}
    \end{subfigure}
    \hfill
    \begin{subfigure}{0.49\linewidth}
        \centering
        \includegraphics[width=\linewidth]{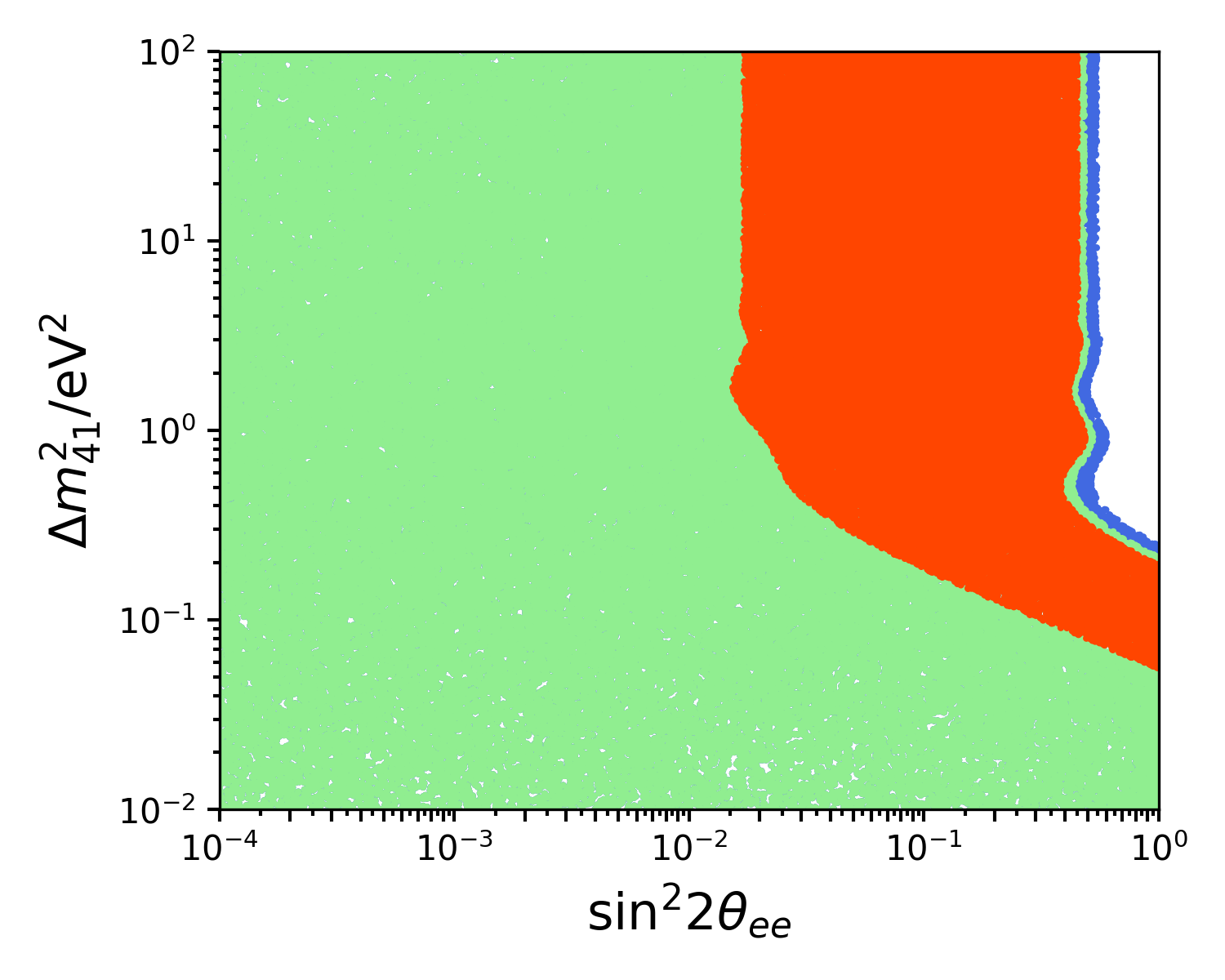}
        \caption{SAGE \& GALLEX}
        \label{fig:Galliumfit}
    \end{subfigure}

    \bigskip
    \begin{subfigure}{0.49\linewidth}
        \centering
        \includegraphics[width=\linewidth]{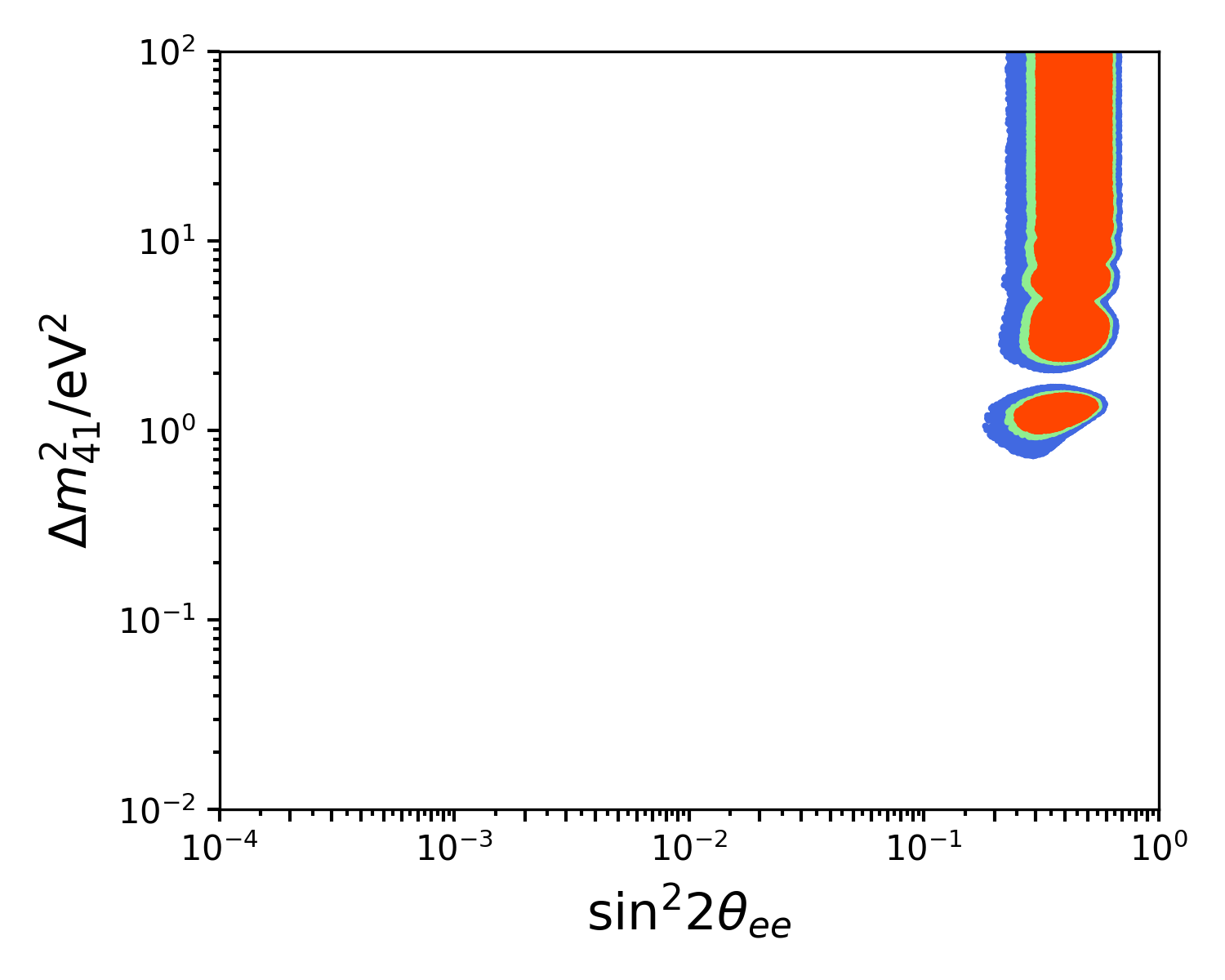}
        \caption{BEST}
        \label{fig:BESTfit}
    \end{subfigure}
    \caption{The 3+1 fits to the $\nue \to \nue$ disappearance oscillation data sets used in our global fits.}
    \label{fig:3+1galliumdisappearance}
\end{figure}

\begin{description}

    \item[SAGE \cite{SAGE:2009eeu} \& GALLEX \cite{Kaether:2010ag}] \hfill \\
        While we discussed the Gallium anomalies in \Cref{sec:gallium}, we will review the results again here.

        The \textbf{S}oviet-\textbf{A}merican \textbf{G}allium \textbf{E}xperiment (SAGE) and \textbf{Gall}ium \textbf{E}xperiment (GALLEX) were two \isotope[71]{Ga}-based detector experiments that measured solar neutrinos through the process $\nue + \isotope[71]{Ga} \to \isotope[71]{Ge} + e^-$.
        The produced \isotope[71]{Ge} would later be collected and counted. 
        Both experiments ran calibration tests by placing radioactive neutrino sources within the detector. 

        GALLEX conducted two calibration runs with \isotope[51]{Cr} sources. One run was in 1994, and the other 1995--1996. Through electron capture the source would emit four mono-energetic lines of \nue's with differing rates: 747 \keV (81.63\%), 427 \keV (8.95\%), 752 \keV (8.49\%), and 432 \keV (0.93\%) \cite{Barinov:2021asz}. 
        SAGE also conducted two callibration runs, first with \isotope[51]{Cr} (1994--1995) and then with \isotope[37]{Ar}. \isotope[37]{Ar} decays with two mono-energetic neutrino lines, one at \SI{811}{\keV} (90.2\%) and another at \SI{813}{\keV} (9.8\%) \cite{Abdurashitov:2005tb}. 

        Combined, the observed \isotope[71]{Ge} production was a factor of $R=0.87\pm0.05$ lower than expected. Collectively, these experiments are referred to as the ``Gallium'' experiments, and the anomalous data as the ``Gallium anomalies.'' The observed data are shown again in \Cref{fig:Galliumresults2} along with the results from BEST, which we discuss next. 

        \begin{figure}
        \centering
        \begin{subfigure}{0.45\textwidth}
            \includegraphics[width=\textwidth]{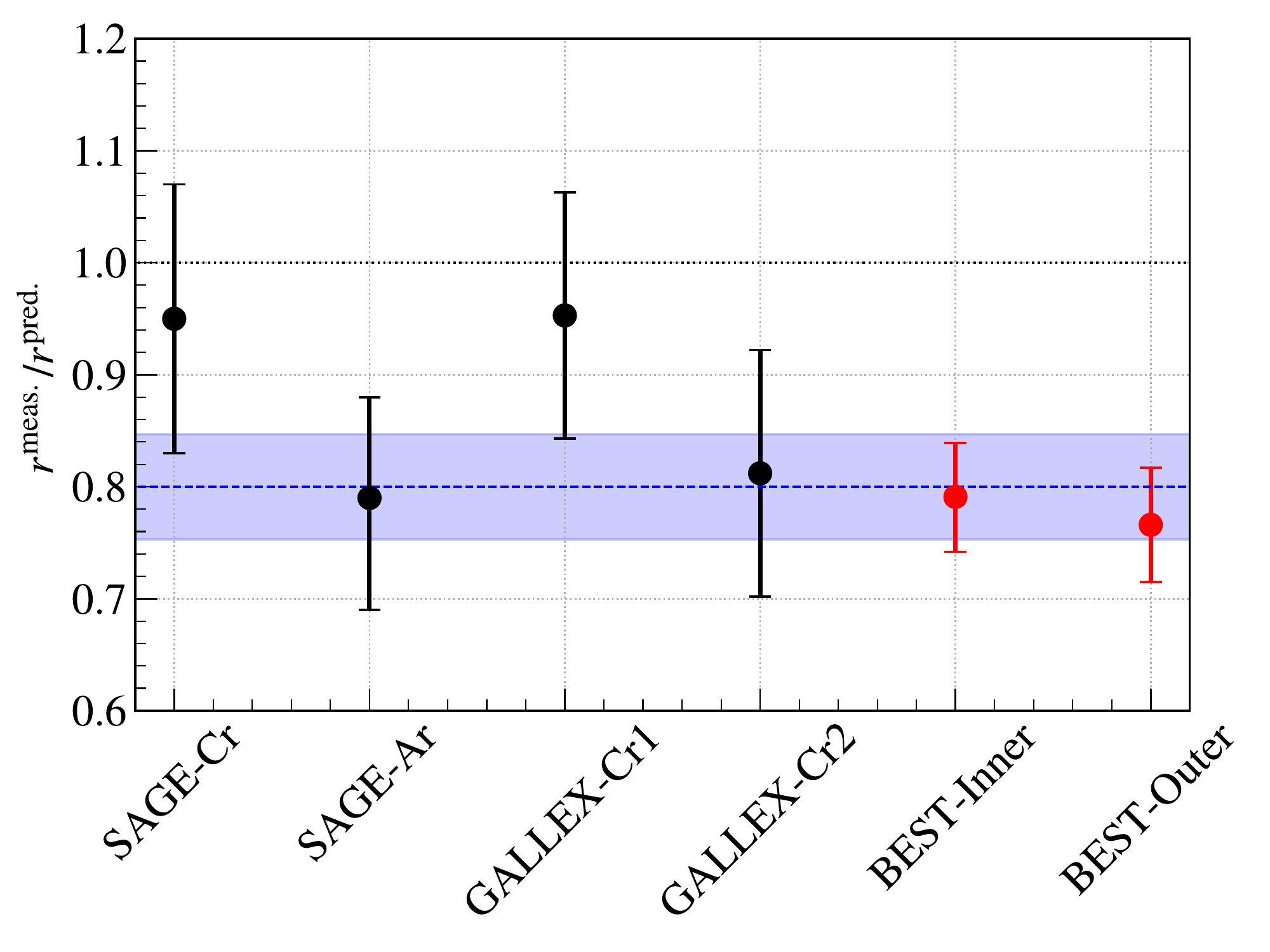}
            \caption{}
        \end{subfigure}
        \begin{subfigure}{0.45\textwidth}
            \includegraphics[width=\textwidth]{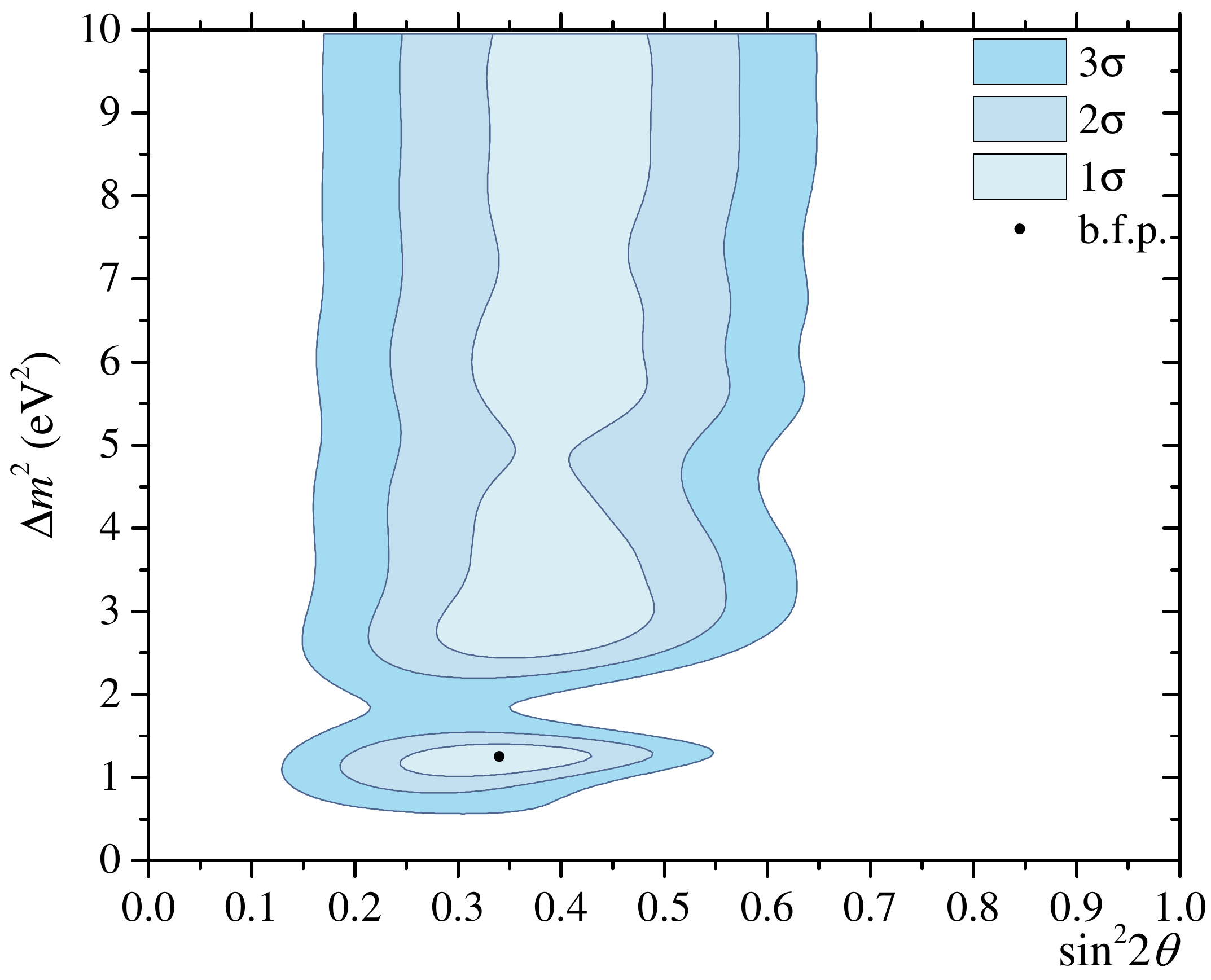}
            \caption{}
            \label{fig:GalliumBestexclusion}
        \end{subfigure}
        \caption{(a) The observed \isotope[71]{Ge} production rate over expectation for the various runs for SAGE, GALLEX, and BEST. Figure taken from Ref.~\cite{Barinov:2022wfh}. 
        (b) Allowed parameter regions of the combined SAGE, GALLEX, and BEST data for the 3+1 model. Figure taken from Ref.~\cite{Barinov:2021asz}.}
        \label{fig:Galliumresults2}
    \end{figure}

        The result of our SAGE and GALLEX 3+1 fit is shown in \Cref{fig:Galliumfit}.

    \item[BEST \cite{Barinov:2021asz,Barinov:2022wfh}] \hfill \\
        More recently, the \textbf{B}aksan \textbf{E}xperiment on \textbf{S}terile \textbf{T}ransitions (BEST) experiment ran to follow-up on the Gallium anomalies. In 2019, a $(3.414\pm0.008)$ MCi \isotope[51]{Cr} source was placed in the center of a dual volume gallium detector. The inner spherical voume of diameter 133.5 cm held 7.5 t of Ga, while the outer cylindrical volume with dimension $(h,\rho)=(234.5, 109) $ cm held 40.0 t. 

        Like the previous Gallium anomalies, BEST observed a deficit of \isotope[71]{Ge} production rates in both volumes, with ratios of $R_\textrm{in}=0.791\pm0.05$ for the inner volume and $R_\textrm{out}=0.766\pm0.05$ for the outer volume.
        The rate ratio between the two volumes is $0.97\pm0.07$, within unity. 
        Therefore, an overall deficit is observed, but not an oscillation between volumes. 

        Combining these results with the previous Gallium anomalies give the 3+1 fit results shown in \Cref{fig:GalliumBestexclusion}. In the oscillation hypothesis, a large mixing angle of $\sin^2 2\theta = 0.34$ is recovered for $\Dmq \gtrsim 1\ \eVq$. 

        The result of our 3+1 fit for BEST is shown in \Cref{fig:BESTfit}.
\end{description}

Before moving on to the reactor experiments, let's discuss how the approach of these experiments have changed since the author first began with their thesis work.

As discussed in \Cref{sec:RAA}, the Reactor Antineutrino Anomaly (RAA) has motivated the search for sterile neutrinos.
But the RAA refers to a deficit compared to models, and reactor models are known to be both difficult to derive and incorrect.
To avoid this limitation, modern reactor experiments try to measure oscillations over multiple baselines and compare the $\nuebar$ spectral shape as a function of distance; this eliminates the need for prior flux knowledge.
With a peak $\nuebar$ observed energy of \SI{\sim 5}{\MeV}, a multi-baseline detector would have to be placed \SI{\sim 5}{\m} from the reactor core.
This presents unique challenges, and we discuss these experiments below.

\begin{figure}
    \centering
    \begin{subfigure}{0.47\linewidth}
        \centering
        \includegraphics[width=\linewidth]{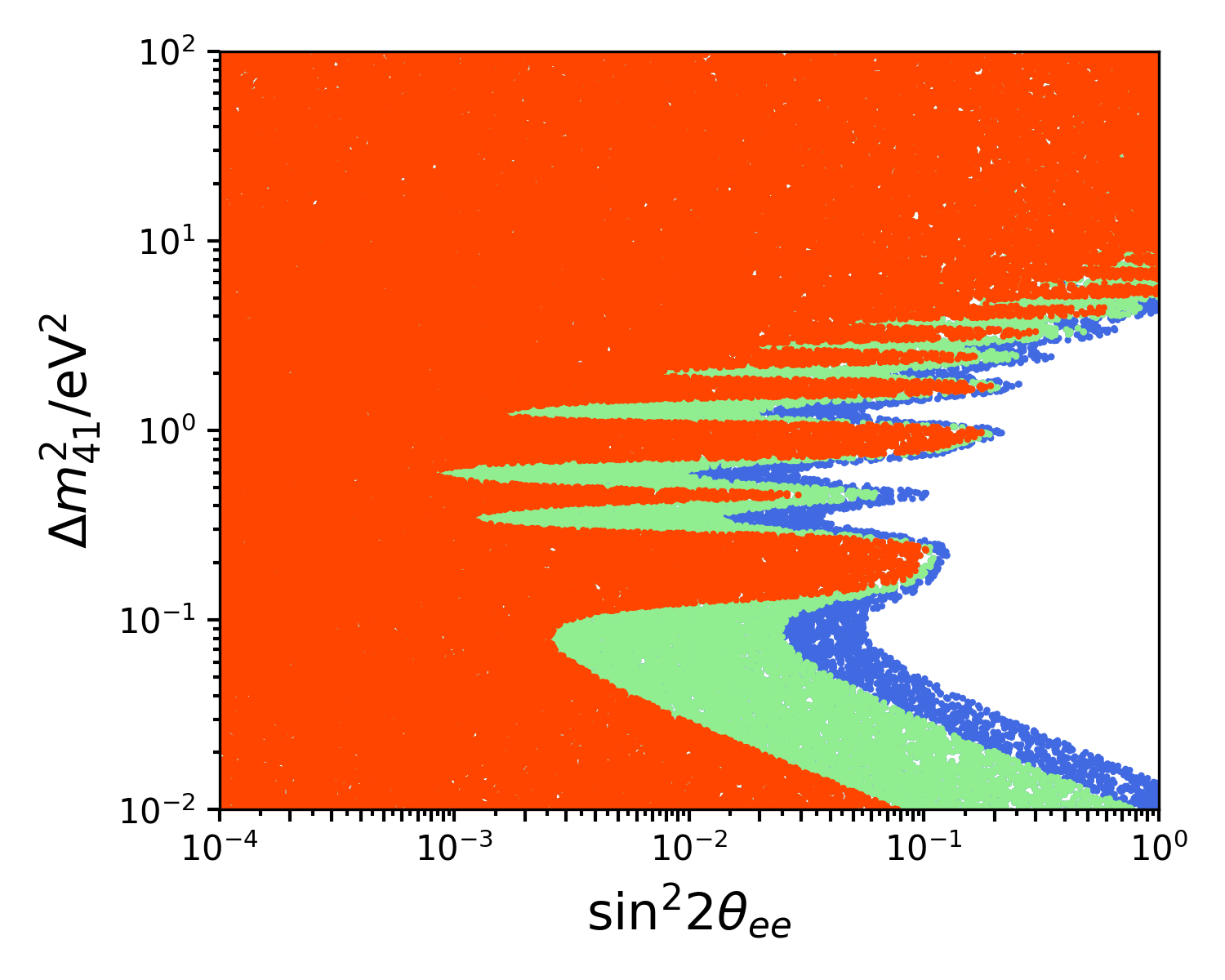}
        \caption{Bugey}
        \label{fig:Bugeyfit}
    \end{subfigure}
    \hfill
    \begin{subfigure}{0.47\linewidth}
        \centering
        \includegraphics[width=\linewidth]{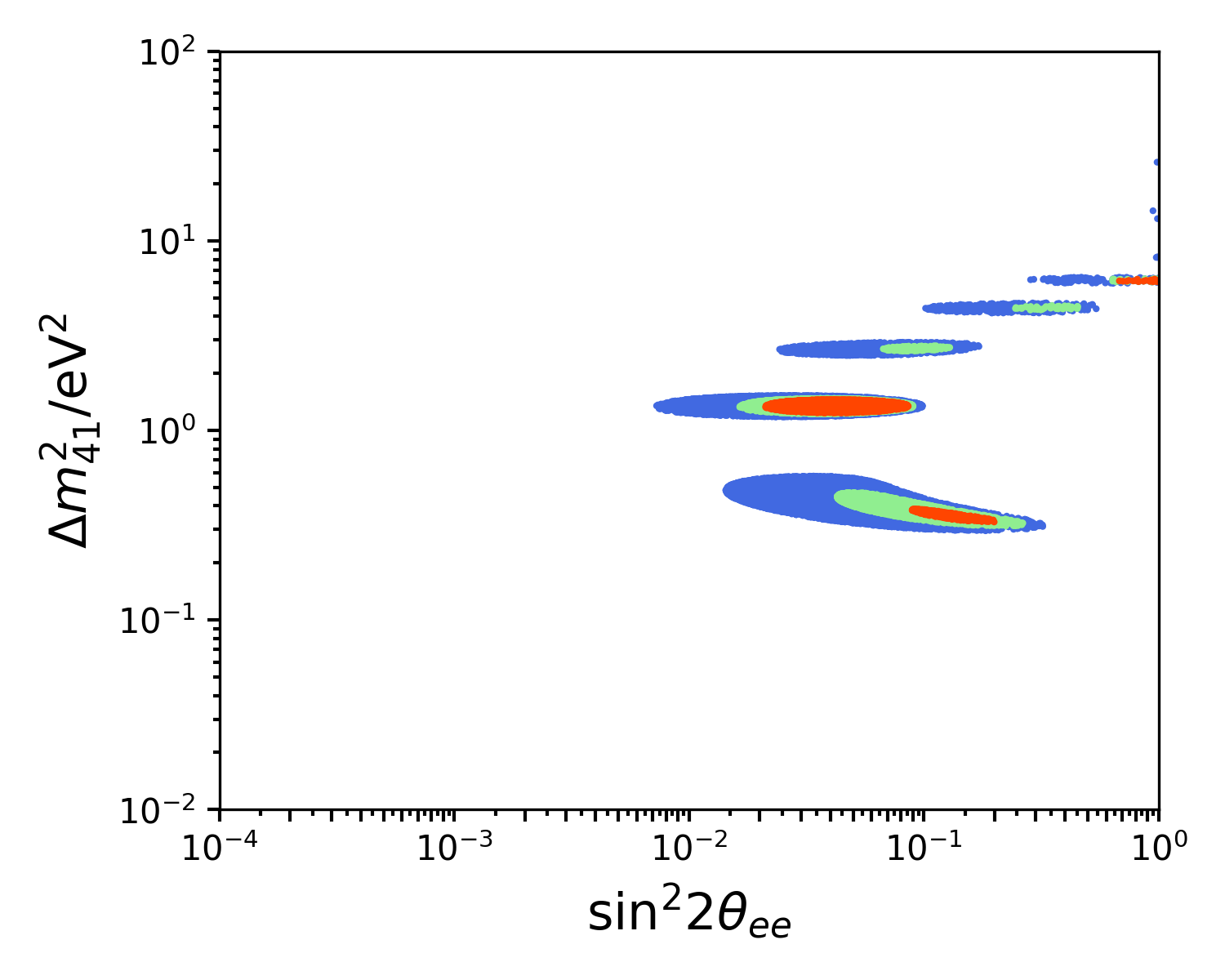}
        \caption{DANSS}
        \label{fig:DANSSfit}
    \end{subfigure}

    \begin{subfigure}{0.47\linewidth}
        \centering
        \includegraphics[width=\linewidth]{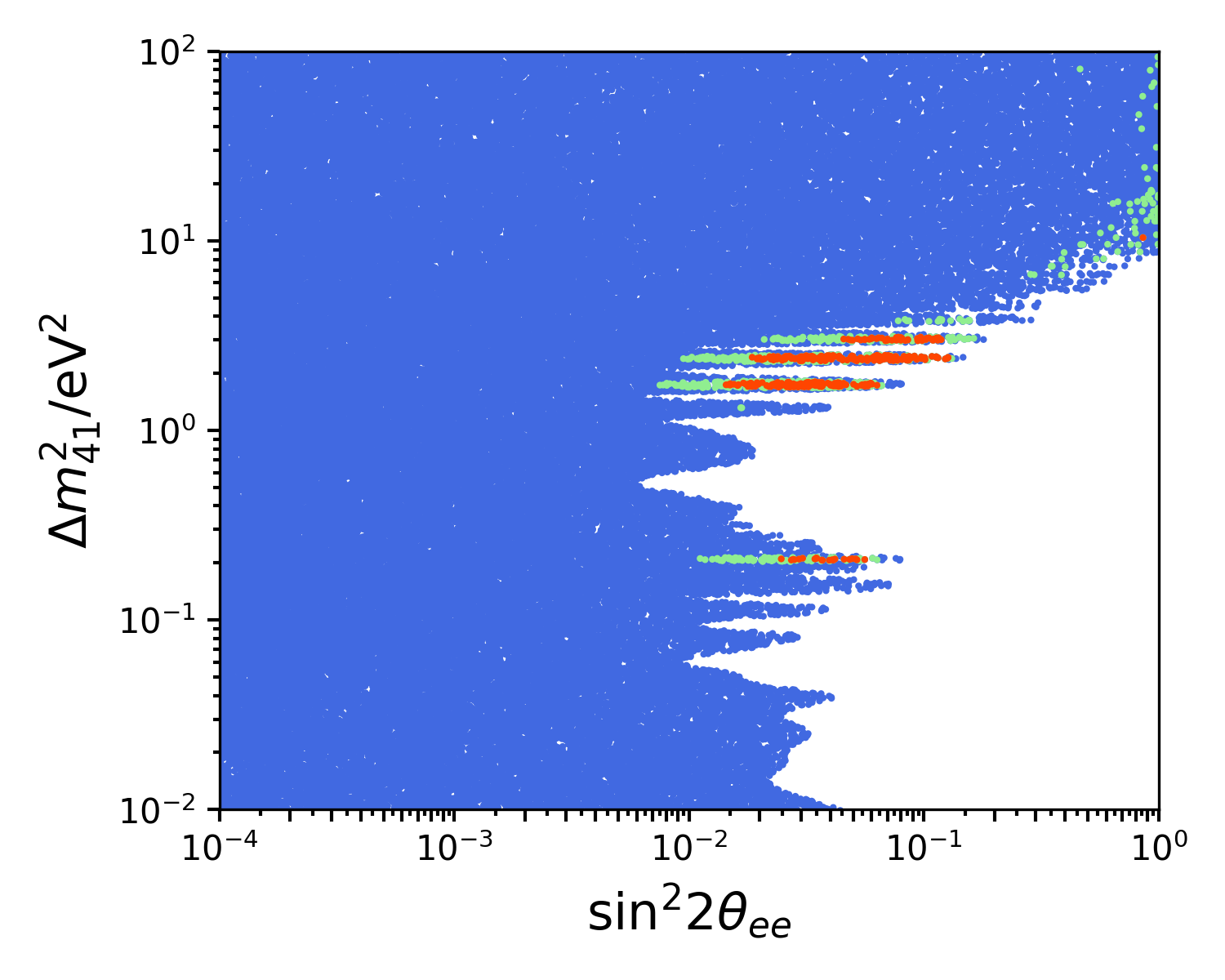}
        \caption{NEOS/RENO}
        \label{fig:NEOSRENOfit}
    \end{subfigure}
    \hfill
    \begin{subfigure}{0.47\linewidth}
        \centering
        \includegraphics[width=\linewidth]{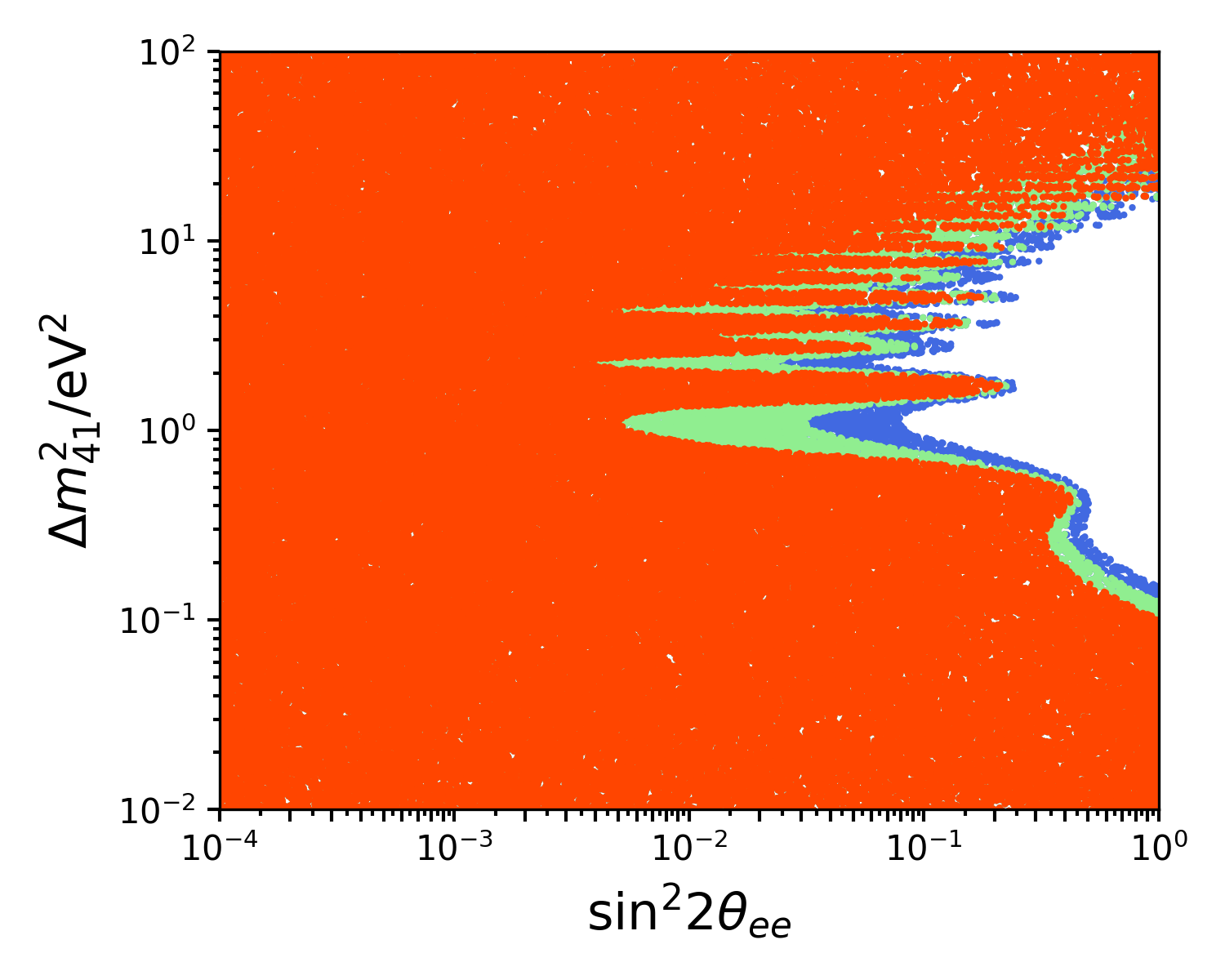}
        \caption{PROSPECT}
        \label{fig:PROSPECTfit}
    \end{subfigure}

    \begin{subfigure}{0.47\linewidth}
        \centering
        \includegraphics[width=\linewidth]{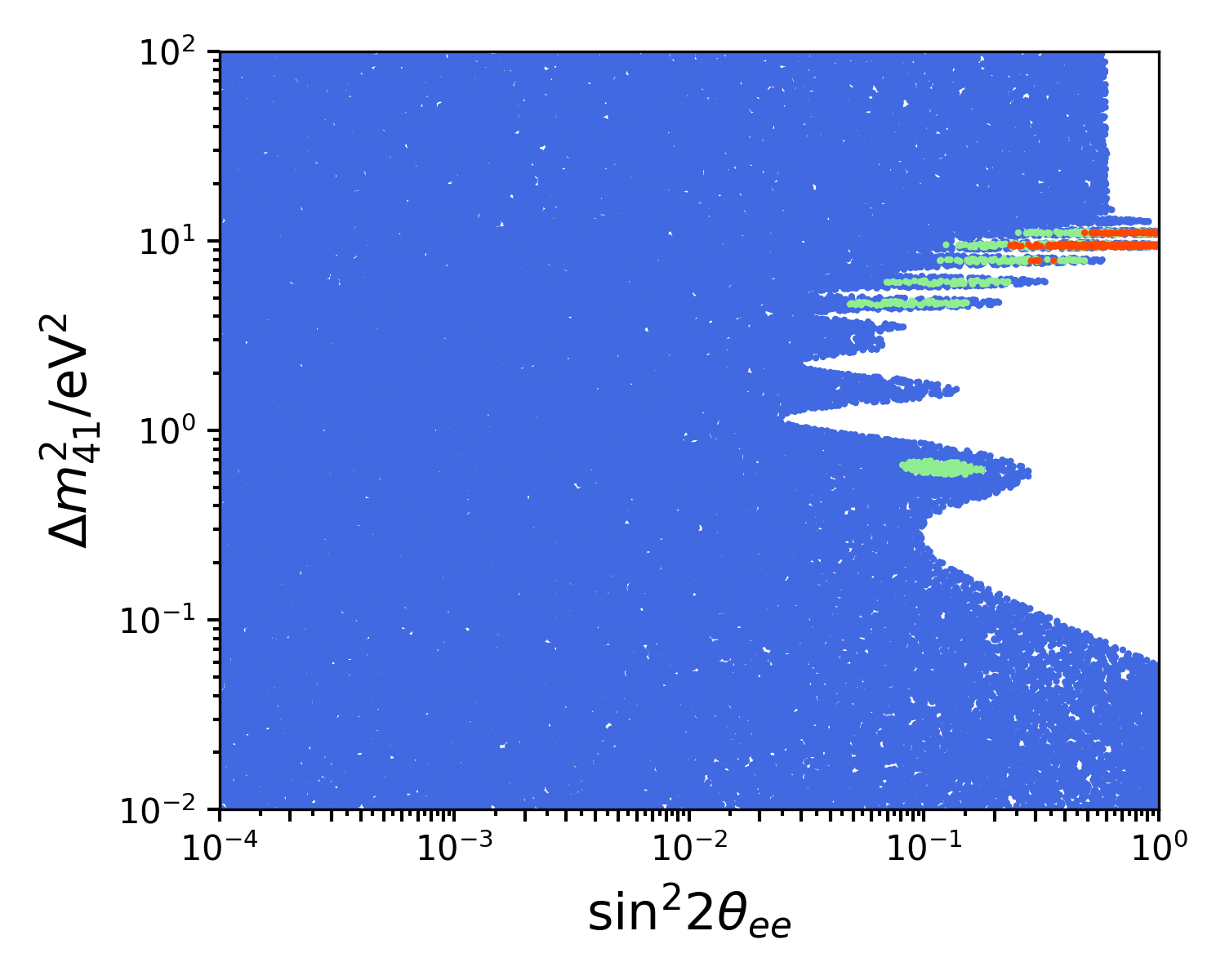}
        \caption{STEREO}
        \label{fig:STEREOfit}
    \end{subfigure}
    \hfill
    \begin{subfigure}{0.47\linewidth}
        \centering
        \includegraphics[width=\linewidth]{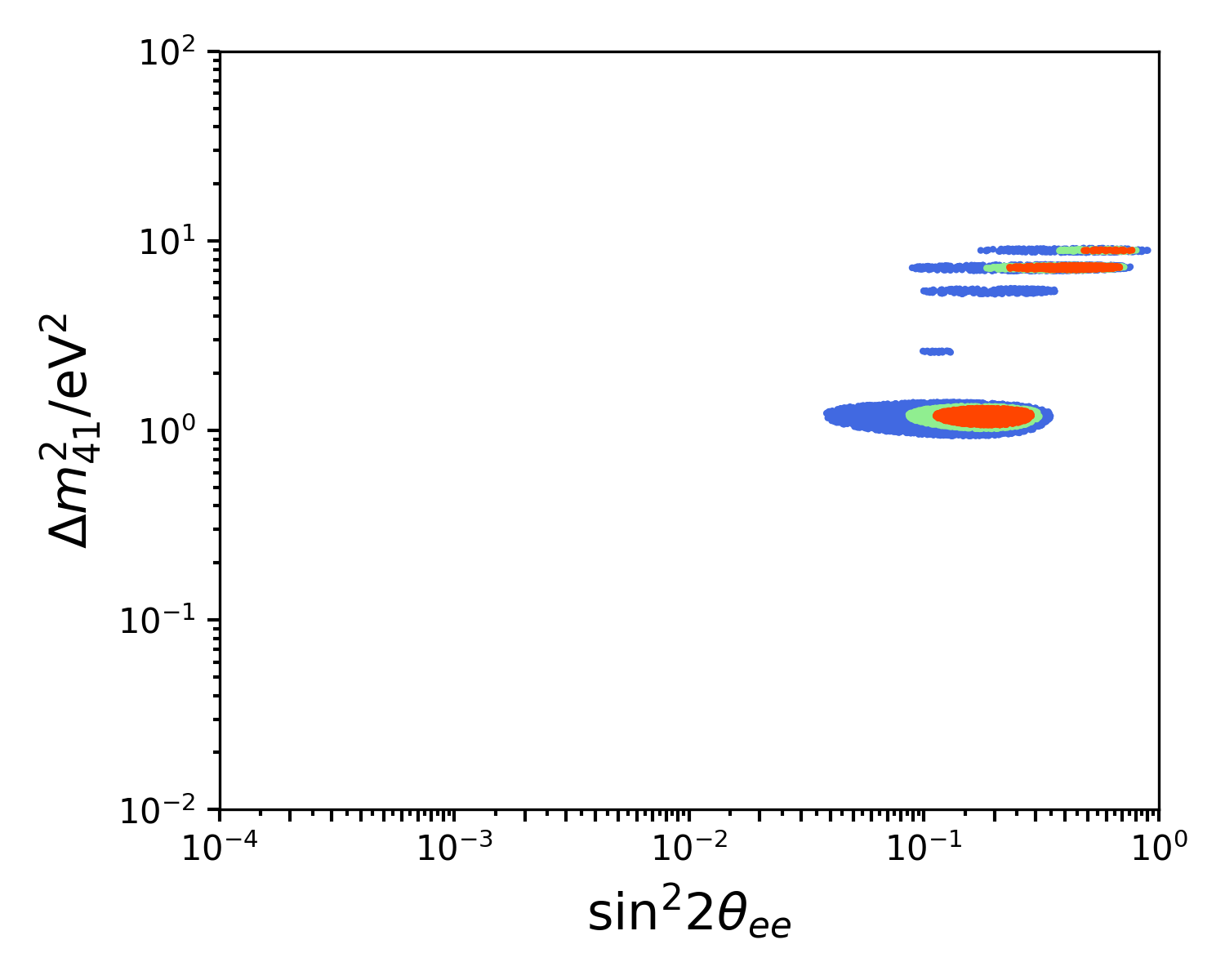}
        \caption{Neutrino-4}
        \label{fig:Neutrino-4fit}
    \end{subfigure}
    \caption{The 3+1 fits to the $\nuebar \to \nuebar$ disappearance oscillation data used in our global fits.}
    \label{fig:3+1reactor}
\end{figure}

The 3+1 fits for the reactor experiments, as we have implemented them, are shown in \Cref{fig:3+1reactor}.

\begin{description}

    \item[Bugey \cite{Declais:1994su}] \hfill \\
        A neutrino oscillation search was conducted at the Bugey reactor complex in France using three \isotope[6]{Li}-loaded liquid scintillator detectors at distances of 15, 40, and 95 m from the reactor core. 
        The collaboration did two analyses, one where the observed spectra was compared to nuclear models, and another where the spectra between baselines were compared.
        Previous publications \cite{Collin:2016rao,Collin:2016aqd} from our group used the first analysis, but we have changed to using the latter in recent publications \cite{Diaz:2019fwt,Vergani:2021tgc}.

        The ratios of the observed data between the various baselines are shown in \Cref{fig:Bugeydata}. The analysis finds no normalization difference between detectors nor spectral differences. The exclusions are shown in \Cref{fig:Bugeyexclusion}. Because the nearest detector was at \SI{15}{\meter}, Bugey was primarily sensitive to lower $\Dmq$.

        \begin{figure}
            \centering
            \begin{subfigure}{0.45\textwidth}
                \includegraphics[width=\textwidth]{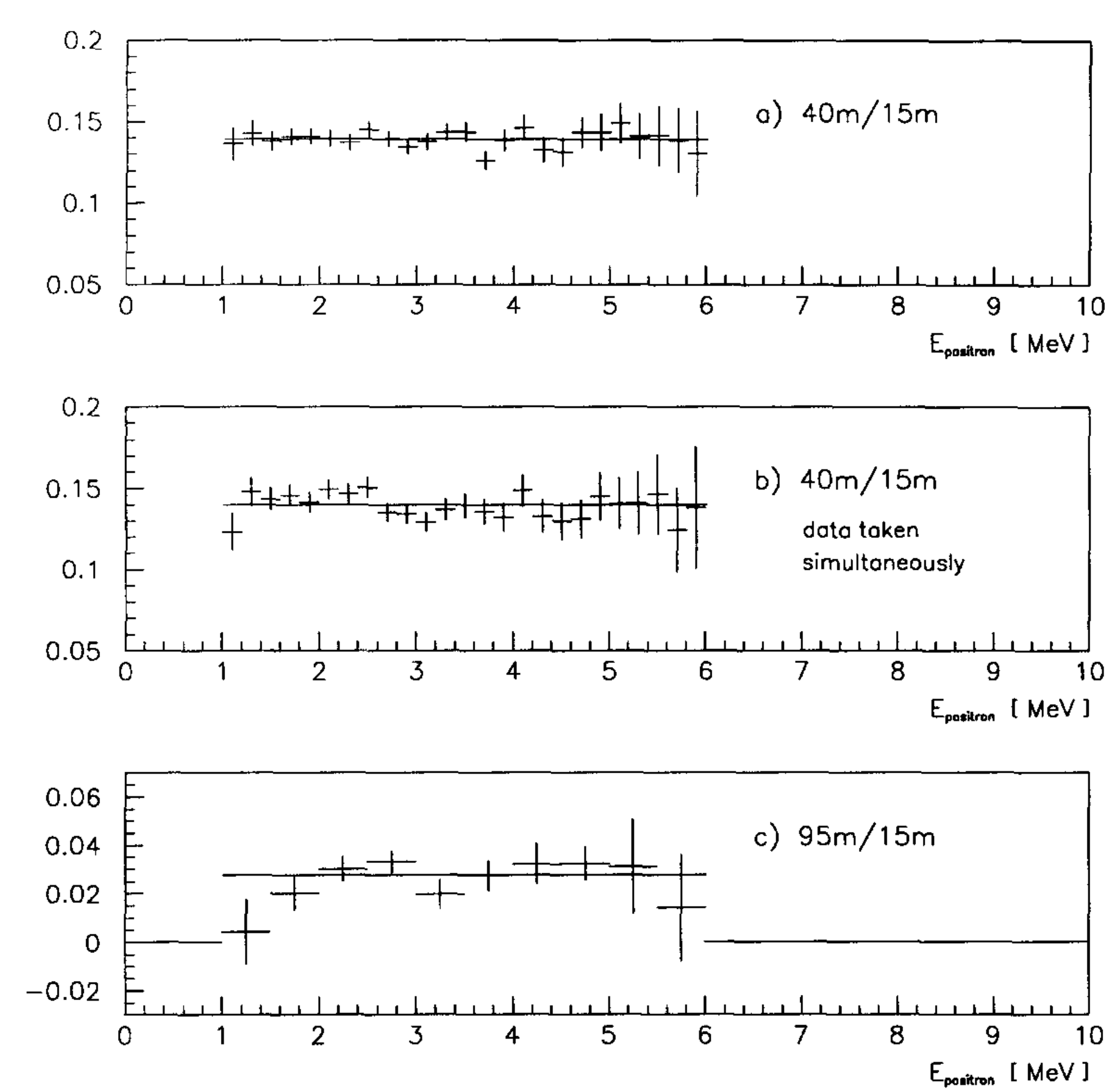}
                \caption{}
                \label{fig:Bugeydata}
            \end{subfigure}
            \begin{subfigure}{0.45\textwidth}
                \includegraphics[width=\textwidth]{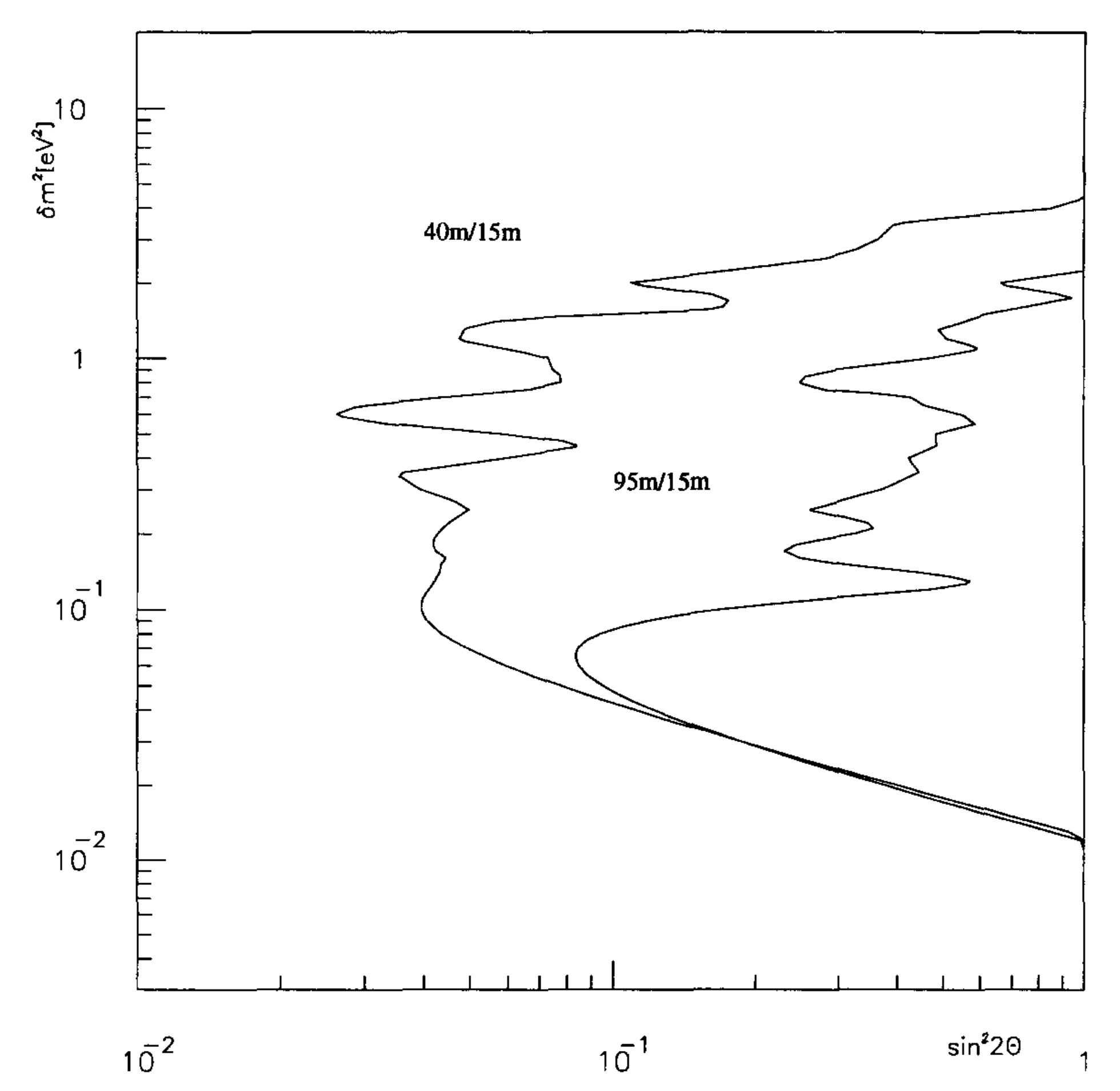}
                \caption{}
                \label{fig:Bugeyexclusion}
            \end{subfigure}
            \caption{(a) The ratios of the observed Bugey data between the various detector baselines. The expected ratios, in the absence of oscillations, would be approximately $(15/40)^{2} = 0.14$ for the 40m/15m comparison and $(15/95)^{2} = 0.025$ for the 95m/15m comparison. The black lines are a fit to a constant line. (b) The 90\% confidence level exclusion contours for two different detector comparisons. Figures from Ref.~\cite{Declais:1994su}.}
        \end{figure}

        Our 3+1 fit to Bugey is shown in \Cref{fig:Bugeyfit}.
        
    \item[DANSS \cite{DANSS:2018fnn}] \hfill \\
        The \textbf{D}etector of \textbf{A}nti\textbf{n}eutrino Based on \textbf{S}olid \textbf{S}cintillator (DANSS) experiment is an ongoing reactor neutrino experiment located at the Kalinin Nuclear Power Plant in Russia. 
        The detector is a highly segmented scintillator detector with a volume of \SI{\sim 1}{\m\cubed}, placed on a movable platform so that the \nuebar spectra is measured at three distances, \SI{10.7}{\m}, \SI{11.7}{\m}, and \SI{12.7}{\m} from the reactor core center. 
        The platform is placed under the reactor and moves vertically, so that the ``top'' position is nearer to the reactor, and the ``bottom'' position is further. 
        The reactor core, in turn, is quite large: a cylindrical shape with dimensions $(h, \rho) = (3.7, 1.6)$ m.

        DANSS conducts a shape-only analysis, where the spectra between the different positions are normalized before the ratios are taken. Therefore, DANSS searches for a spectral distortion from oscillations, without relying on reactor models. 

        For DANSS data taken 2016--2018, the results are shown in \Cref{fig:DANSSexclusion}. 
        A best fit oscillation point is found at $(\Dmq, \sin^2 2\theta) = (1.4\ \eVq, 0.05)$  with a $\Delta \chi^2 = 13.1$. 
        The collaboration has yet to publish an analysis with a complete uncertainty treatment, so the significance of the measurement is still being studied. 
        In \Cref{fig:DANSSexclusion}, an exclusion curve is published.

        \begin{figure}
            \centering
            \begin{subfigure}{0.45\textwidth}
                \includegraphics[width=\textwidth]{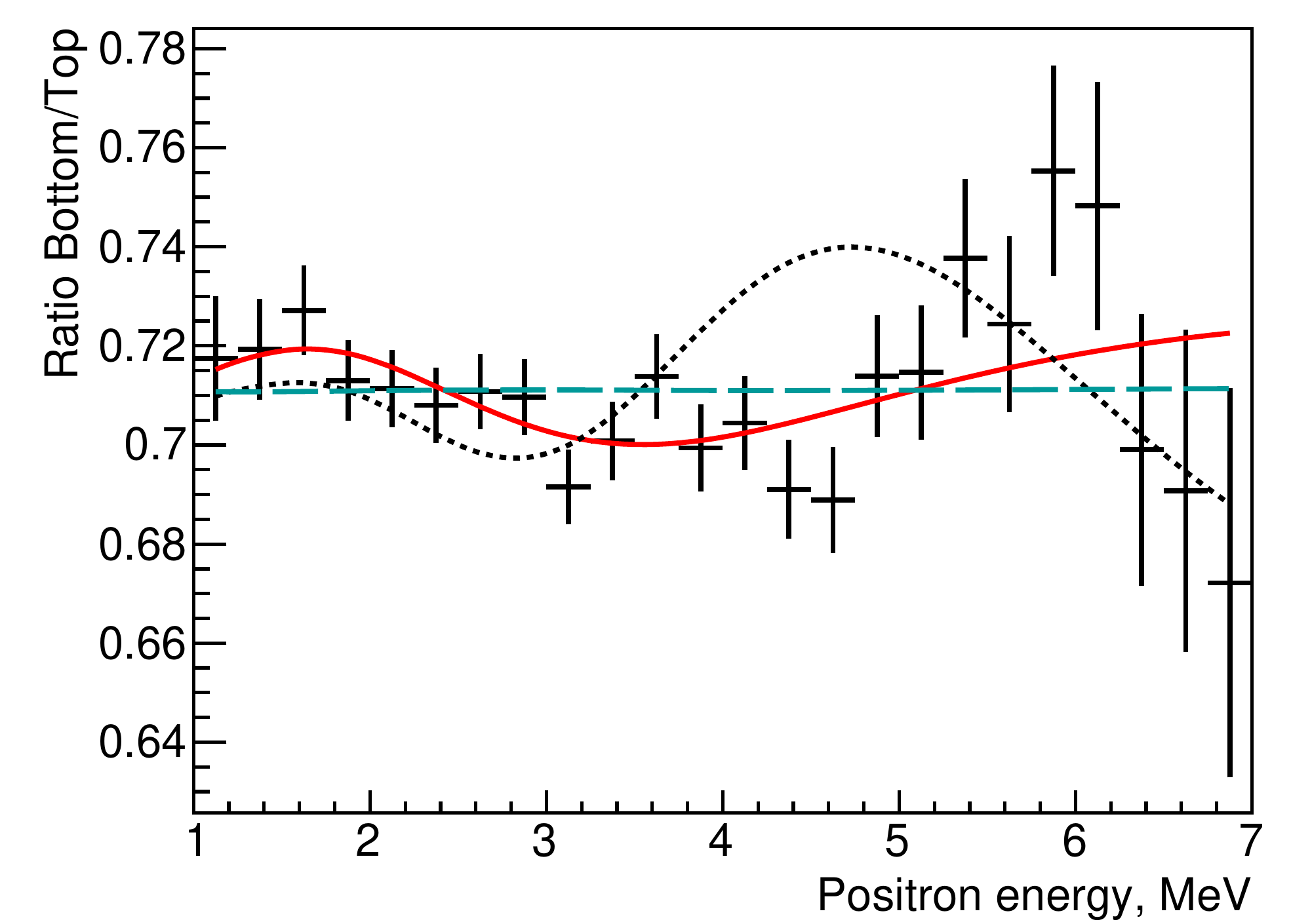}
                \caption{}
                \label{fig:DANSSdata}
            \end{subfigure}
            \begin{subfigure}{0.45\textwidth}
                \includegraphics[width=\textwidth]{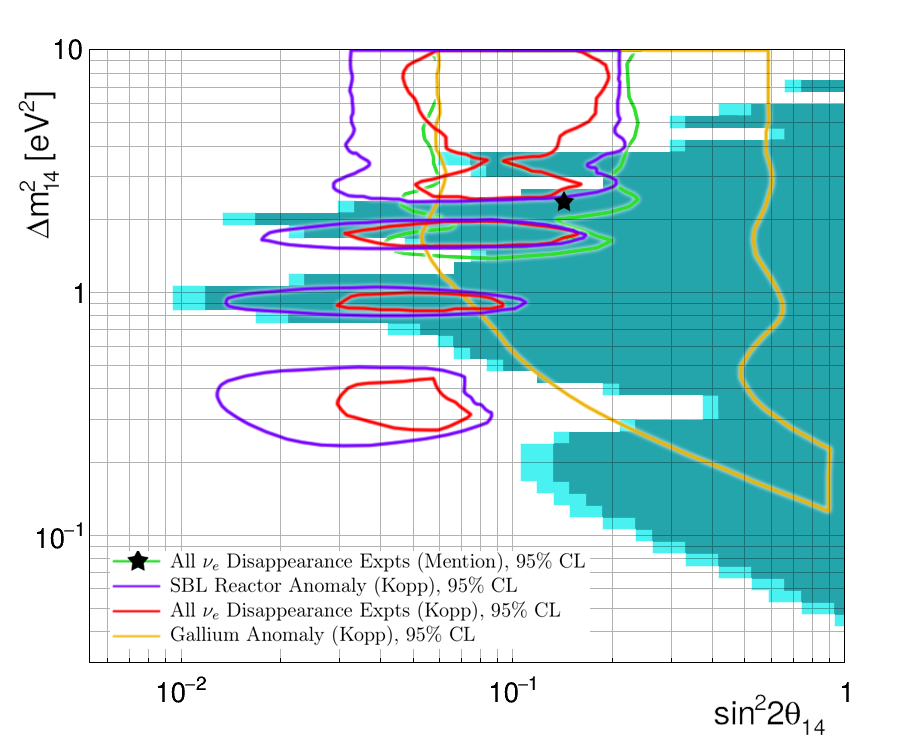}
                \caption{}
                \label{fig:DANSSexclusion}
            \end{subfigure}
            \caption{(a) The ratio of the observed positron energy spectra between the bottom (further) and top (nearer) detector positions. The dashed curve is the no-oscillation hypothesis, which is taken to be the ratio of the observed total rates between the bottom and top positions. The solid curve is the expectation from DANSS's best fit point for oscillation: $\sin^2 2\theta = 0.05$ and $\Delta m^2 = 1.4\ \eVq$. The dotted curve is the expectation at DANSS from a fit to the RAA and Gallium (SAGE \& GALLEX only) anomaly. (b) The 90\% (cyan) and 95\% (dark cyan) confidence level exclusion region for DANSS.
            Figures from Ref.~\cite{DANSS:2018fnn}.}
        \end{figure}

        We show our 3+1 fit to DANSS in \Cref{fig:DANSSfit}.
        
    \item[NEOS/RENO \cite{RENO:2020hva}] \hfill \\
        The \textbf{N}eutrino \textbf{E}xperiment for \textbf{O}scillation at \textbf{S}hort Baseline (NEOS) experiment is an ongoing experiment at the Hanbit Nuclear Power Complex in Korea.
        The cylindrical liquid scintillator detector has dimensions $(h, \rho) = (1.21, 0.515)$ m and sits $23.7\pm0.3$ m from the reactor core. The core, also cylindrical, has dimensions $(h, \rho) = (3.8, 1.55)$ m. 

        Unlike DANSS, the NEOS detector is a single volume and at a static position. To avoid systematic uncertainties from nuclear models, the analysis compares its data with a ``reference flux'' from a different experiment. 
        This reference flux would ideally not contain spectral features from a $\Dmq \sim \SI{1}{\eV\squared}$ e.g. if the flux was measured at a distance far beyond the oscillation length.
        Initially, NEOS used Daya Bay's unfolded \nuebar flux measurement \cite{NEOS:2016wee} as their reference flux, but has moved to a joint analysis with the \textbf{R}eactor \textbf{E}xperiment for \textbf{N}eutrino \textbf{O}scillation (RENO) collaboration. 
        This is an improvement as the RENO detector lies in the same reactor complex as the NEOS detector, reducing systematic uncertainties relating to reactor complexes and reactor cores.
        The RENO detector is far enough from the reactor core (294 m) so that no shape information from a $\Dmq \sim 1\ \eVq$ mass splitting would be discernible.

        Current results from NEOS use 180 days of reactor-on and 45 days of reactor-off data collected in 2015--2016. 
        The data from the joint analysis with RENO is shown in \Cref{fig:NEOSdata}, and confidence regions in \Cref{fig:NEOSexclusion}. 

        The best fit point is found at $(\Dmq, \sin^2 2\theta) = (2.41\ \eVq, 0.08)$, with a p-value of 8.2\%. Therefore, NEOS does not see a significance signature for oscillations, but does have an allowed region at the $1\sigma$ level. 

        \begin{figure}
            \centering
            \begin{subfigure}{0.45\textwidth}
                \includegraphics[width=\textwidth]{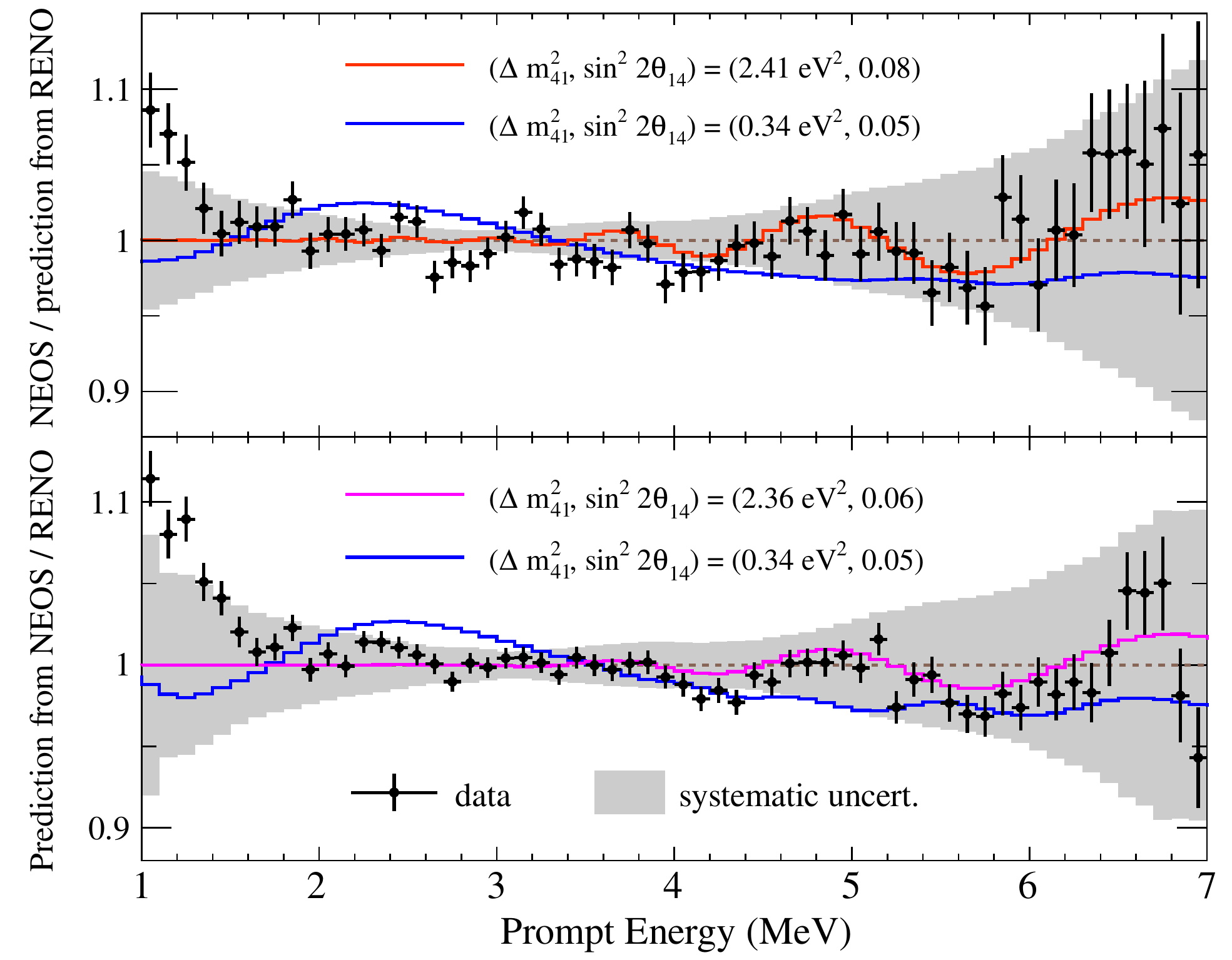}
                \caption{}
                \label{fig:NEOSdata}
            \end{subfigure}
            \begin{subfigure}{0.45\textwidth}
                \includegraphics[width=\textwidth]{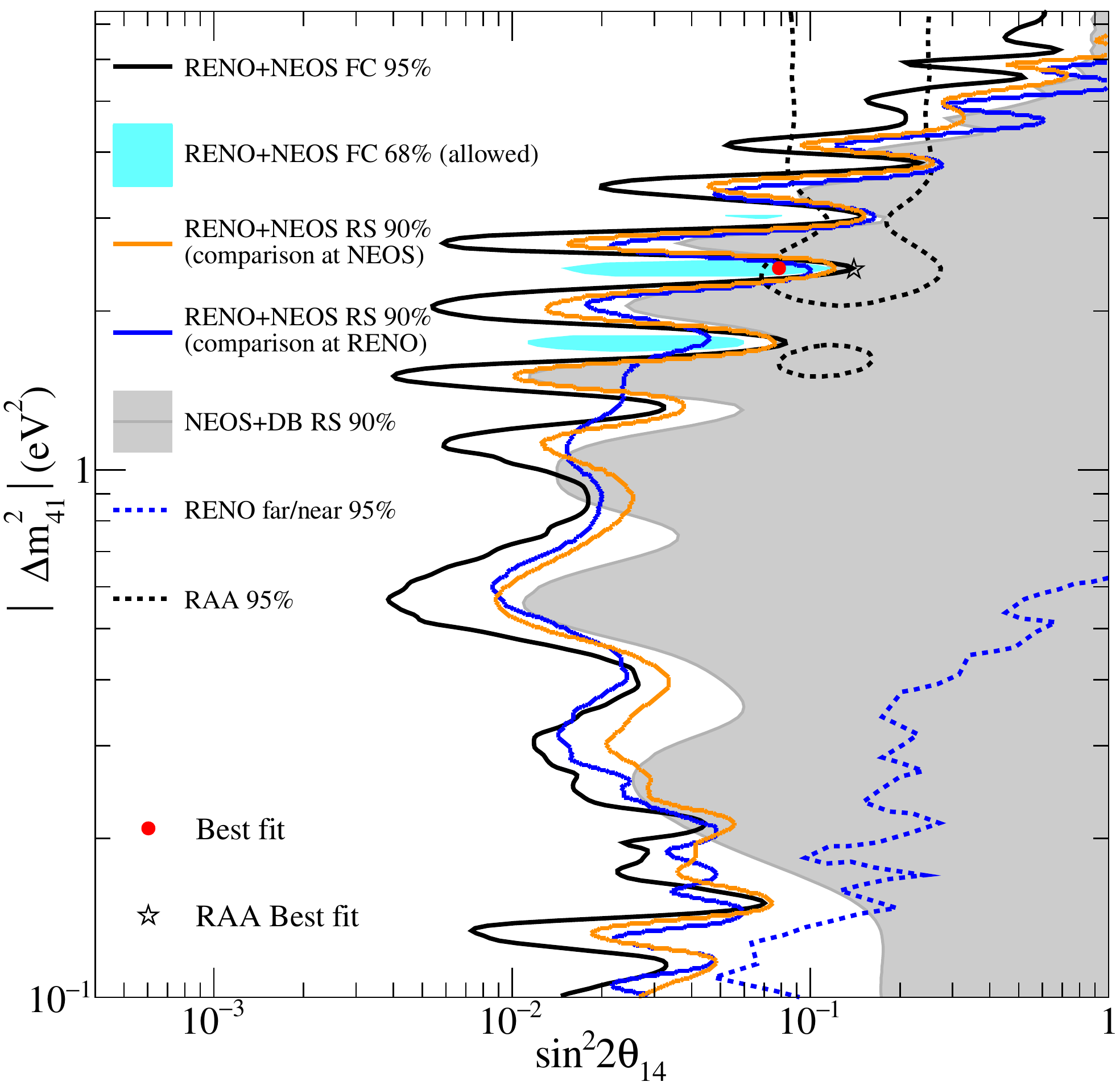}
                \caption{}
                \label{fig:NEOSexclusion}
            \end{subfigure}
            \caption{(a) Comparisons of the observed prompt energy spectra versus the expectation. In the upper plot, the NEOS data are compared to an expectation from the unfolded \nuebar spectra from RENO measurements. The lower plot shows the reverse: the RENO prompt spectra compared to an expectation from the unfolded \nuebar spectra from NEOS measurements. (b) Various exclusion limits by the NEOS collaboration. We only note the 95\% (black line) and 68\% (cyan filled) confidence levels of the NEOS/RENO joint analysis. The ``NEOS+DB'' contour is the exclusion obtained when NEOS used the unfolded Daya Bay \nuebar flux as their reference flux \cite{NEOS:2016wee}. 
            Figures from Ref.~\cite{RENO:2020hva}.}
        \end{figure}

        We show our 3+1 fit to NEOS/RENO in \Cref{fig:NEOSRENOfit}.
        
    \item[PROSPECT \cite{PROSPECT:2020sxr}] \hfill \\
        The \textbf{P}recision \textbf{R}eactor \textbf{O}scillation and \textbf{Spect}rum Experiment (PROSPECT) is an ongoing reactor neutrino experiment located near the High Flux Isotope Reactor (HFIR) at Oak Ridge National Laboratory. 
        Unlike the previous reactor experiments described, HFIR is a highly enriched uranium research reactor. 
        This offers two benefits.
        First, the reactor core is compact, with dimensions $(h, \rho) = (0.508, 0.2175)$ m. This reduces the uncertainties in \nuebar propagation distances, and allows the detector to be placed closer to the reactor core. 
        Second,  the fission fraction of HFIR is always kept above 99\% \isotope[235]{U}. This simplifies the modeling of the reactor core, and reduces uncertainties that would arise from the reactor core's composition changing with time. 

        The PROSPECT detector is a rectangular volume with dimensions $2.0 \times 1.6 \times 1.2$ \si{\m\cubed}, subdivided into 154 optically isolated rectangular segments. 
        The detector sits very close to the reactor core, with a center-to-center distance of $7.9 \pm 0.1$ m. 
        This gives the detector a large baseline range compared to the center-to-center distance, ranging 6.7--9.2 m.

        In 2018, PROSPECT collected 96 days of reactor-on data, and 73 days of reactor-off data. 
        The results are shown in \Cref{fig:PROSPECTdata}.
        In the analysis, the data are divided into 10 baseline bins and 16 prompt energy bins. For a given \textit{energy} bin, the predicted \nuebar spectra is normalized (across baseline bins) to the total observed rate in that energy bin.
        Therefore, the analysis does not depend on the spectral shape of the true \nuebar reactor flux.

        A best fit is found at $(\Dmq, \sinsqtth) = (1.78\ \eVq, 0.11)$ with a $\Delta \chi^2 = 4$.
        MC simulations showed that this value of $\Delta \chi^2$ has a p-value of $0.57$ with respect to the null hypothesis. 
        Therefore, no significant evidence for oscillation is observed. 
        \Cref{fig:PROSPECTexclusion} plots the 95\% confidence level of the data. 

        \begin{figure}
            \centering
            \begin{subfigure}{0.45\textwidth}
                \includegraphics[width=\textwidth]{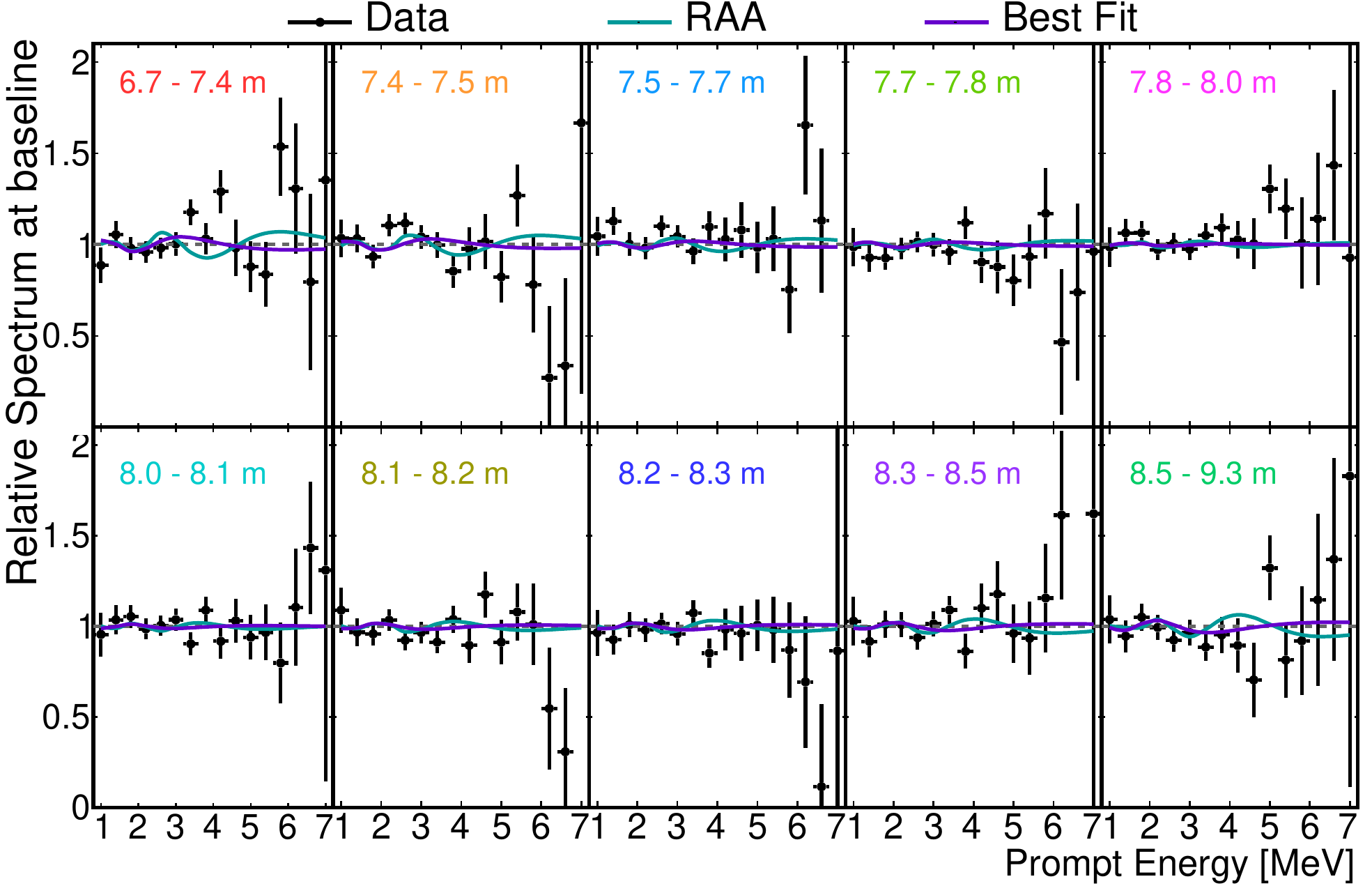}
                \caption{}
                \label{fig:PROSPECTdata}
            \end{subfigure}
            \begin{subfigure}{0.45\textwidth}
                \includegraphics[width=\textwidth]{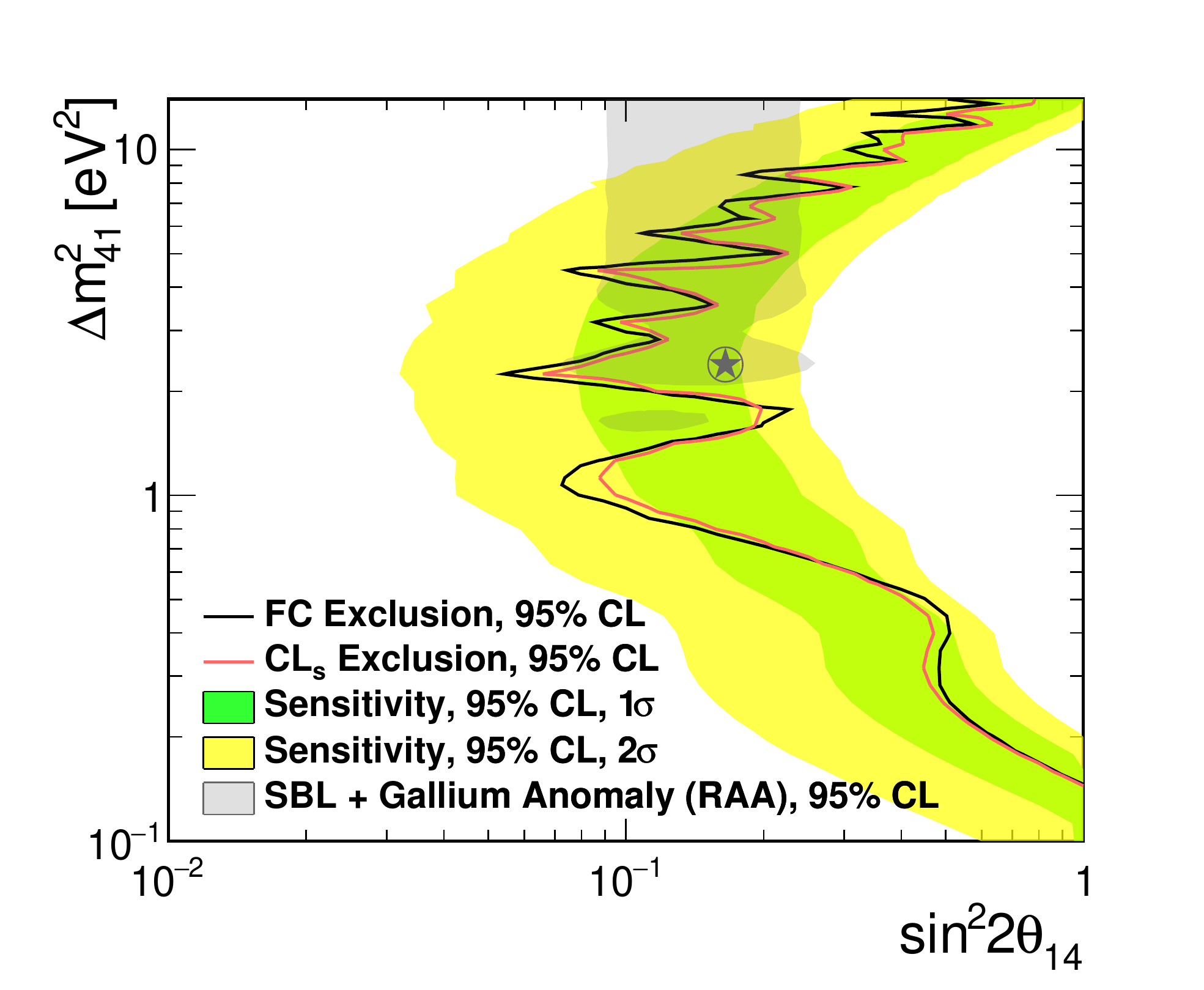}
                \caption{}
                \label{fig:PROSPECTexclusion}
            \end{subfigure}
            \caption{(a) The observed prompt energy spectra compared to expectation at PROSPECT. 
            The prediction is normalized such that the spectra for a given energy bin is normalized (across baseline bins) to the data. 
            (b) The 95\% exclusion and sensitivities derived from PROSPECT. 
            Two methods of calculating the exclusions and sensitivities are displayed.
            Figures from Ref.~\cite{PROSPECT:2020sxr}.}
        \end{figure}

         We show our 3+1 fit to PROSPECT in \Cref{fig:PROSPECTfit}.
         
    \item[STEREO \cite{STEREO:2019ztb}] \hfill \\
        The STEREO experiment is an ongoing reactor neutrino experiment at the Institut Laue-Langevin (ILL) research center in Grenoble, France.
        Like other research reactors, the STEREO's reactor is compact and composed of highly enriched \isotope[235]{U} (93\%).
        The STEREO detector is composed of six opitcally separated cells filled with liquid scintillator. 
        The distinct cells allow the measurement of the \nuebar spectrum over baselines 9.4-11.2 m from the reactor core. 

        The STEREO experiment ran in two phases, with 179 days of total reactor-on time. Phases I and II were treated as two independent experiments, with the results of phase II shown in \Cref{fig:STEREOdata}. 
        Compared to pseudoexperiments, the observed p-value is 9\%.
        Therefore, the no-sterile hypothesis cannot be rejected. 
        The observed exclusion is shown in \Cref{fig:STEREOexclusion}.

        \begin{figure}
            \centering
            \begin{subfigure}{0.45\textwidth}
                \includegraphics[width=\textwidth]{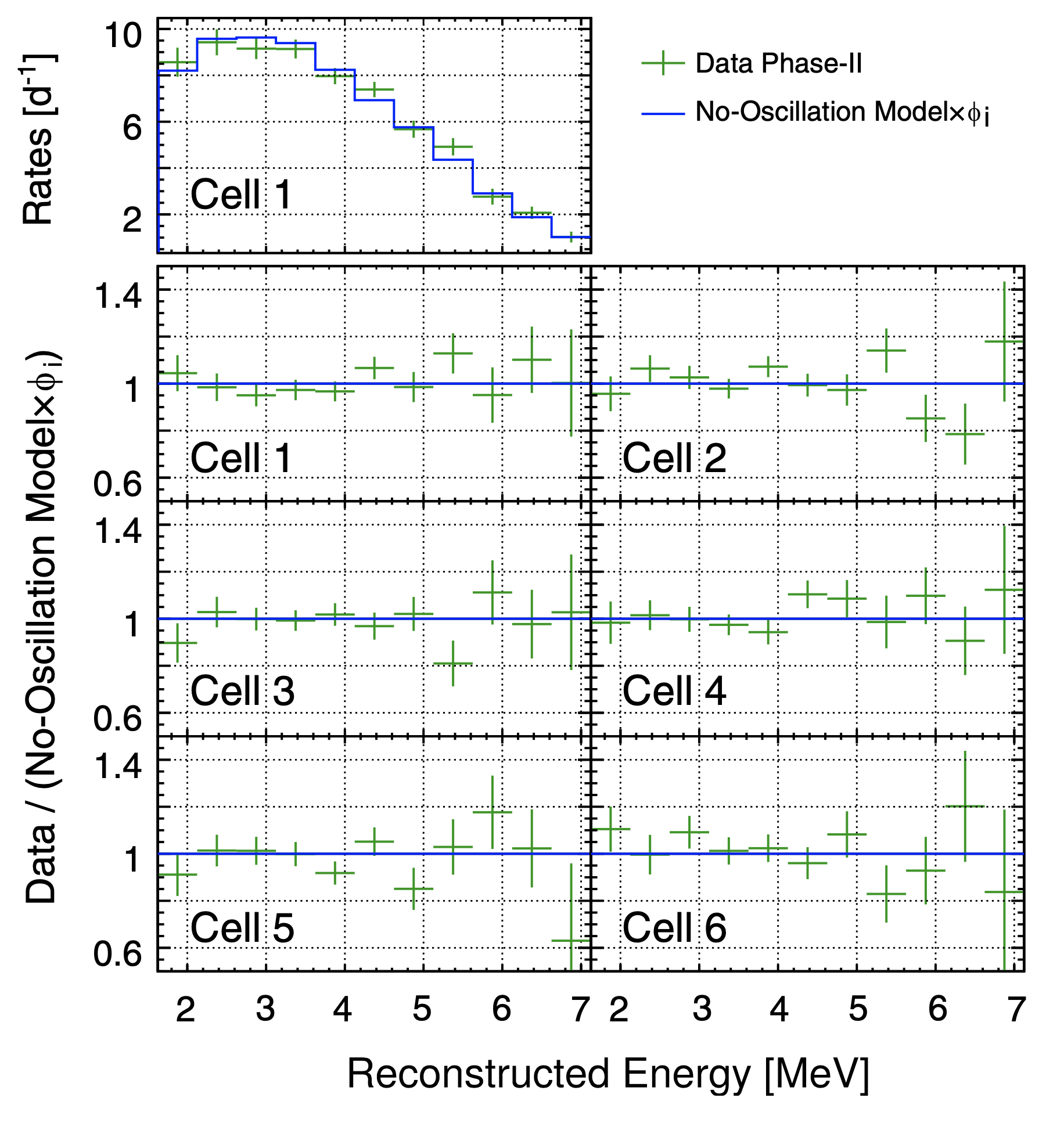}
                \caption{}
                \label{fig:STEREOdata}
            \end{subfigure}
            \begin{subfigure}{0.45\textwidth}
                \includegraphics[width=\textwidth]{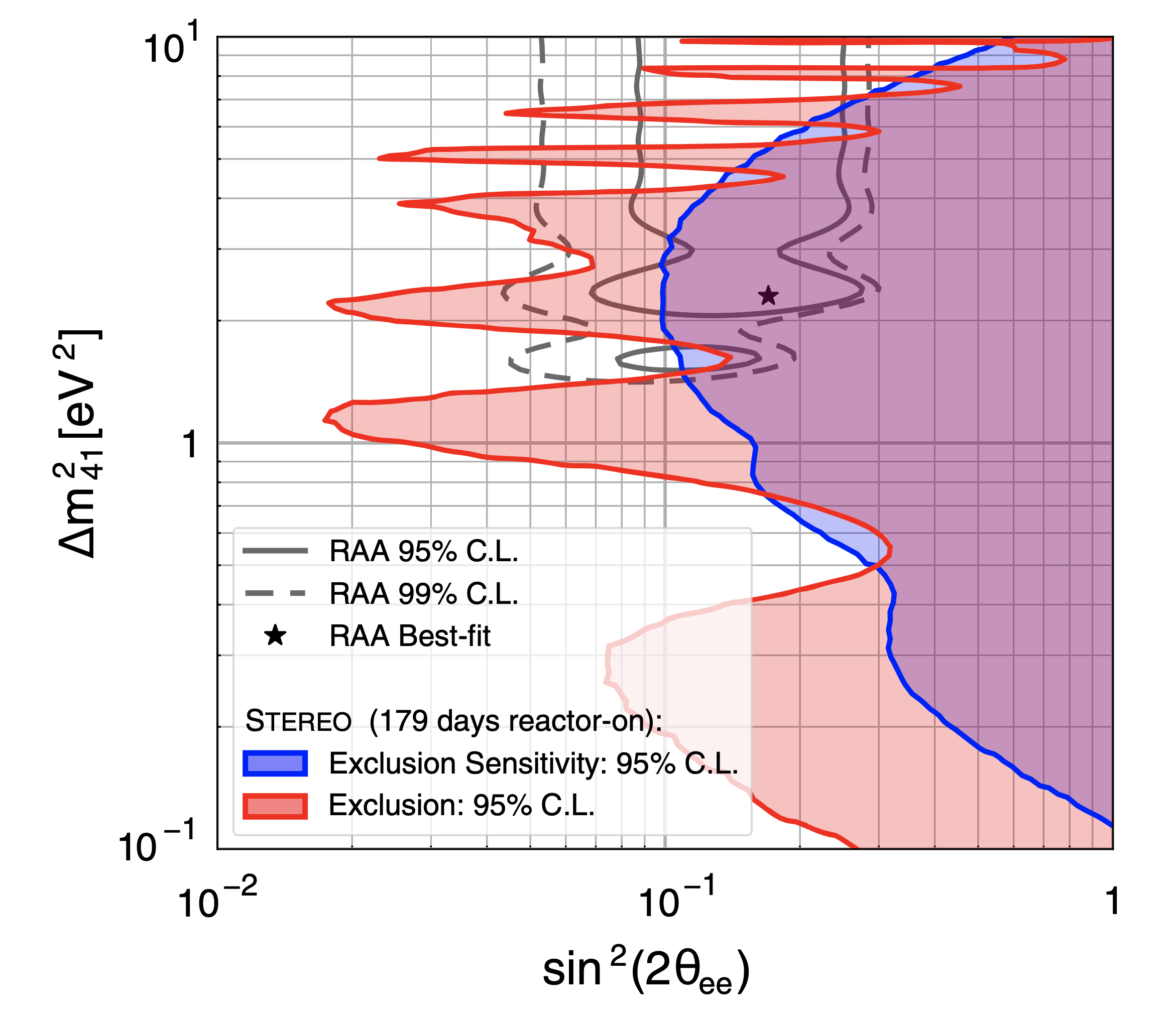}
                \caption{}
                \label{fig:STEREOexclusion}
            \end{subfigure}
            \caption{(a) The top-most plot shows the absolute comparison between the observed data in Phase II and the null hypothsis in the first cell. The remaining six plots show the relative comparison of the measured rates versus expectation for each cell in the detector. The normalization for each energy bin common across all cells is allowed to float. (b) The exclusion sensitivity and observed exclusion at the 95\% confidence level is shown. Figures are taken from Ref.~\cite{STEREO:2019ztb}.}
        \end{figure}

         We show our 3+1 fit to STEREO in \Cref{fig:STEREOfit}.
         
    \item[Neutrino-4 \cite{Serebrov:2020kmd}] \hfill \\
        Neutrino-4 is an ongoing reactor neutrino experiment located near the SM-3 research nuclear reactor in Russia. 
        Being a research reactor, the core is primarily \isotope[235]{U} and compact, with dimensions $0.25 \times 0.42 \times 0.42\ \textrm{m}^3$.

        The detector, a 1.8 m$^3$ volume of liquid scintillator, is divided into $5 \times 10$ segments of dimensions $0.225 \times 0.225 \times 0.85\ \textrm{m}^3$ each.  
        Further, the detector as a whole is placed on rails so that total baseline range sampled is 6--12 m from the reactor core. 
        This also allows multiple subsegments of the detector to be placed at the same distance from the core, reducing detector calibration systematics.

        In 2016--2020, Neutrino-4 recorded data for 720 days of reactor-on and 860 days of reactor-off. 
        The results are shown in \Cref{fig:Neutrino4}. 
        Neutrino-4 claims a significant signal for oscillations, with a best fit at $(\Delta m^2, \sinsqtth) = (7.3\ \eVq, 0.36)$ at a significance of $2.9\sigma$.
        \begin{figure}
            \centering
            \includegraphics[width=0.95\textwidth]{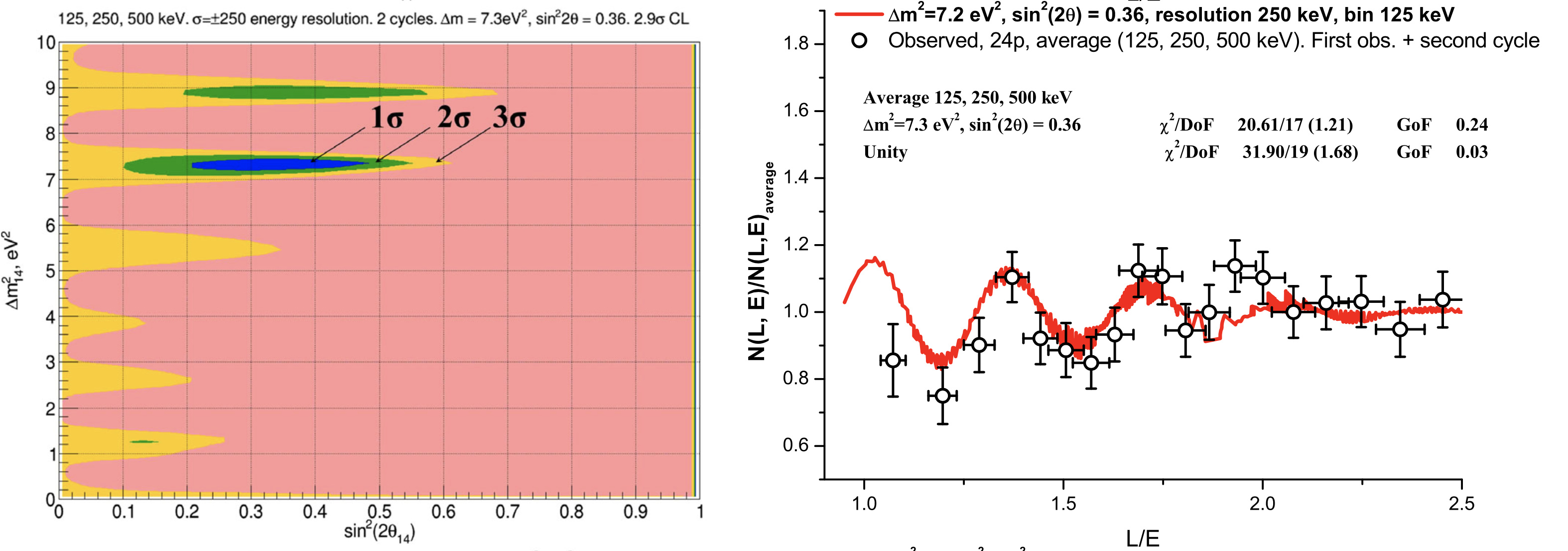}
            \caption{\textbf{Left}: The best fit contours reported by Neutrino-4. The best fit is found at $(\Delta m^2, \sinsqtth) = (7.3\ \eVq, 0.36)$ with a $2.9\sigma$ significance. \textbf{Right}: The ratio of data versus expectation observed by Neutrino-4. The red line gives the expected signal at the best fit point. 
            Figures from Ref.~\cite{Serebrov:2020kmd}.
            }
            \label{fig:Neutrino4}
        \end{figure}    

         We show our 3+1 fit to Neutrino-4 in \Cref{fig:Neutrino-4fit}.
\end{description}

\subsection{\texorpdfstring{$P(\numu \to \numu)$ \& $P(\numubar \to \numubar)$}{P(\nu \mu \rightarrow \nu \mu) \& P(\=\nu \mu \rightarrow \=\nu \mu)}}

\begin{description}

    \item[CDHS \cite{DYDAK1984281}] \hfill \\ The CDHS collaboration conducted a $\numu \to \numu$ disappearance search using the CERN Super Proton Synchrotron (SPS) neutrino beam.
        The SPS impinged a 19.2 GeV proton beam onto a beryllium target, producing neutrinos with a peak flux at 1 GeV.
        The $\numu$'s would then be observed by two detectors, placed at 130 m and 885 m from the target. 
        The detectors were composed of alternating planes of iron plates and scintillators.

        Unlike the other experiments listed in this section, CDHS did not bin their events by energy.
        Instead, CDHS sorted their events by the length traveled by the observed muons, acting as a proxy for $\numu$ energy. 

        The observed ratios between the two detectors are shown in \Cref{fig:CDHSdata}. 
        No evidence for $\numu$ disappearance was found between the two detectors. 
        The extracted exclusion can be seen in \Cref{fig:CDHSexclusion}. 

        \begin{figure}
            \centering
            \begin{subfigure}{0.55\textwidth}
                \includegraphics[height=.3\textheight]{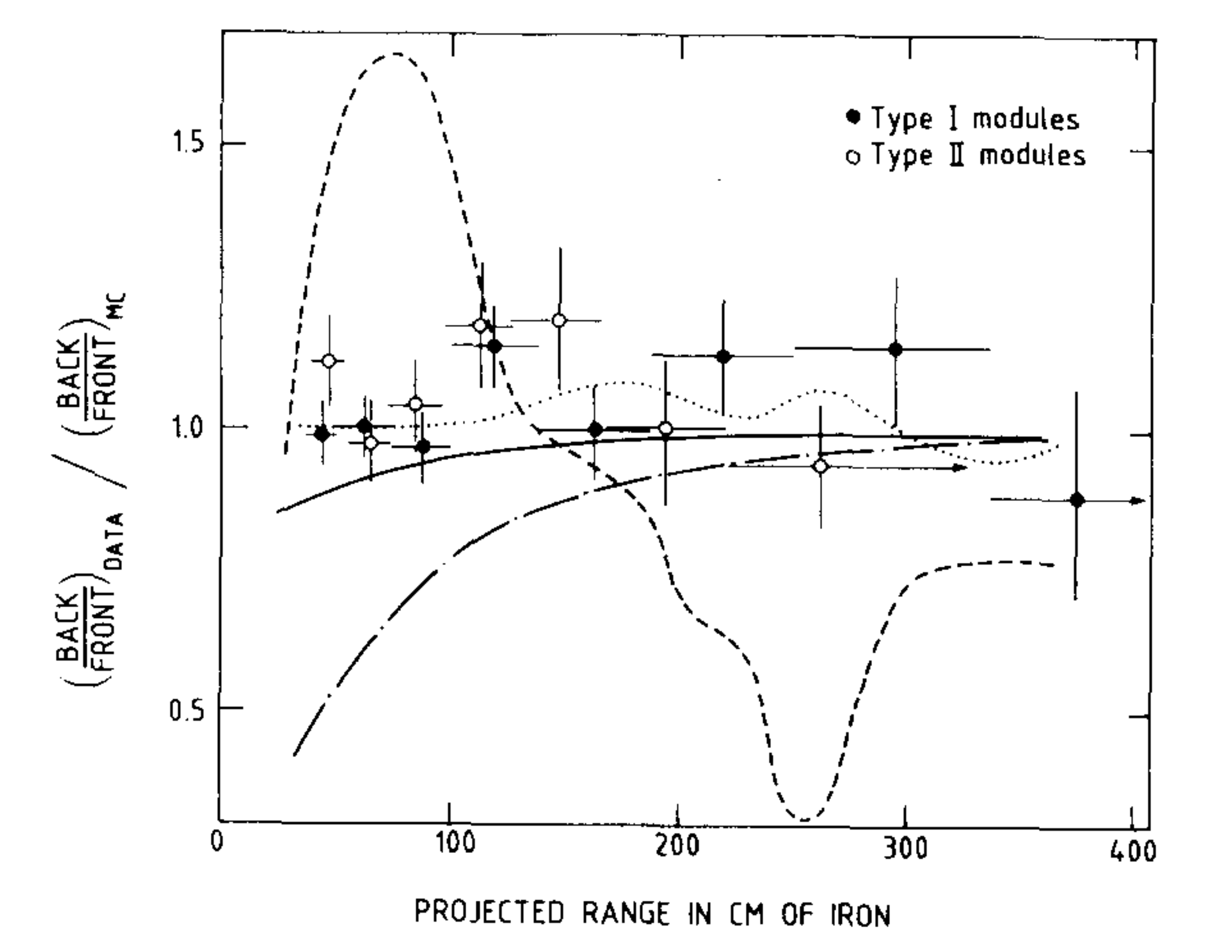}
                \caption{}
                \label{fig:CDHSdata}
            \end{subfigure}
            \begin{subfigure}{0.35\textwidth}
                \includegraphics[height=.3\textheight]{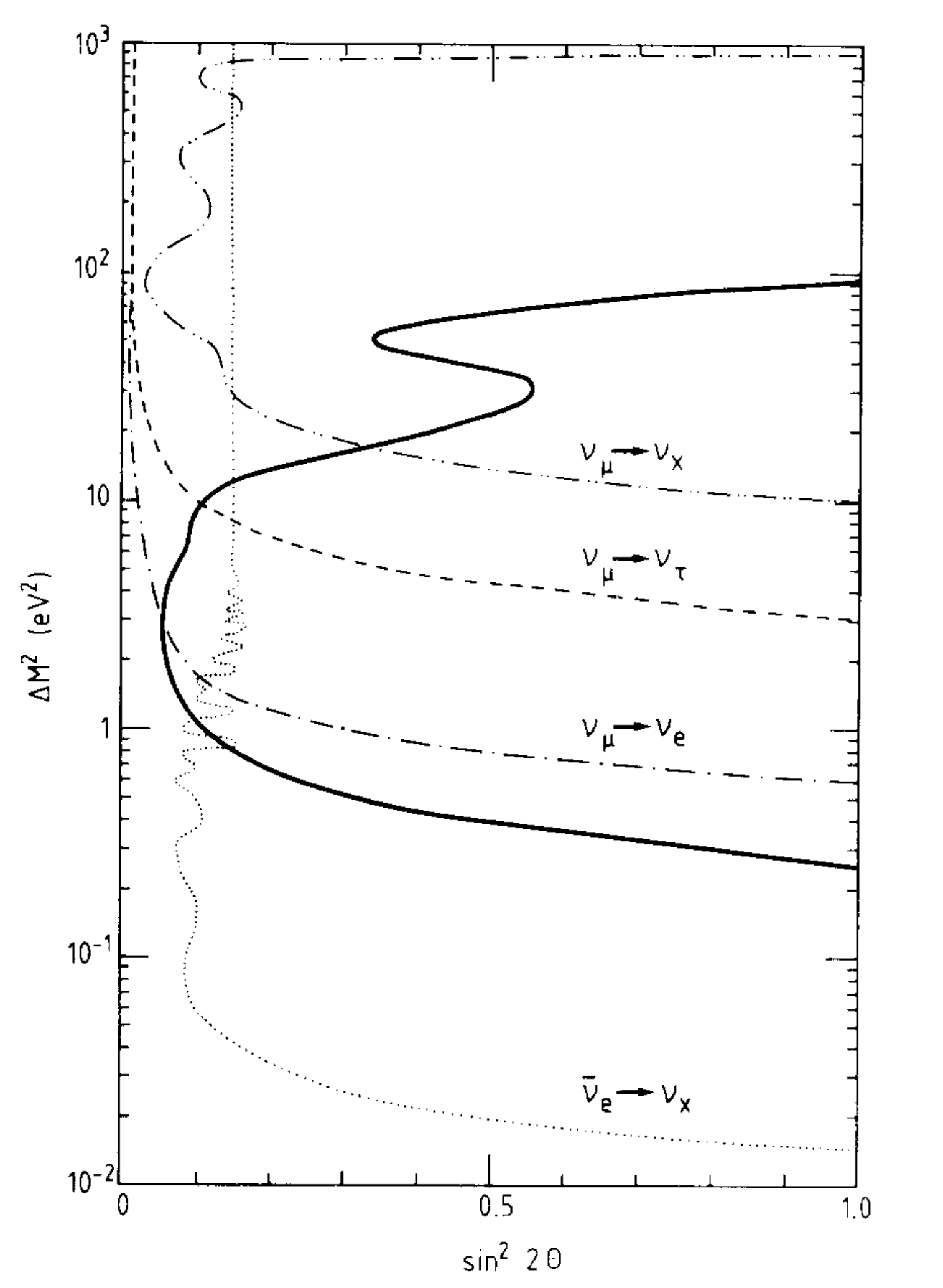}
                \caption{}
                \label{fig:CDHSexclusion}
            \end{subfigure}
            \caption{(a) The observed ratios of $\numu$ events between the two CDHS detectors, as a function of muon track length. The different colored dots correspond to different subdetector types in the CDHS detectors. The different lines correspond to expectations for different sterile neutrino hypotheses. (b) The 90\% confidence level from the CDHS observations, shown in the solid line. The remaining lines were the best limits, at the time, for other oscillation channels.
            Figures from Ref.~\cite{DYDAK1984281}.}
            \label{fig:CDHS}
        \end{figure}
        
        We show the results of our CDHS 3+1 fit in \Cref{fig:CDHSfit}.
        
    \item[CCFR84 \cite{PhysRevLett.52.1384}] \hfill \\ 
        The CCFR collaboration conducted a $\numu \to \numu$ and $\numubar \to \numubar$ disappearance search using two detectors in the Fermilab narrow band neutrino beam. 
        The beam was tuned to provide data at five meson momentum settings (100, 140, 165, 200, and 250 GeV) for $\pi^+$s and $K^+$s, providing neutrinos between 40 and 230 GeV. 
        The two detectors were placed \SI{715}{\m} and \SI{1116}{\m} from the midpoint of the decay pipe. 

        The observed ratios between the two detectors, for $\numu$ and $\numubar$ data, are shown in \Cref{fig:CCFRdata}. 
        No evidence for oscillation was observed, and the 90\% exclusion curves are shown in \Cref{fig:CCFRexclusion}.

        \begin{figure}
            \centering
            \begin{subfigure}{0.50\textwidth}
                \includegraphics[width=\textwidth]{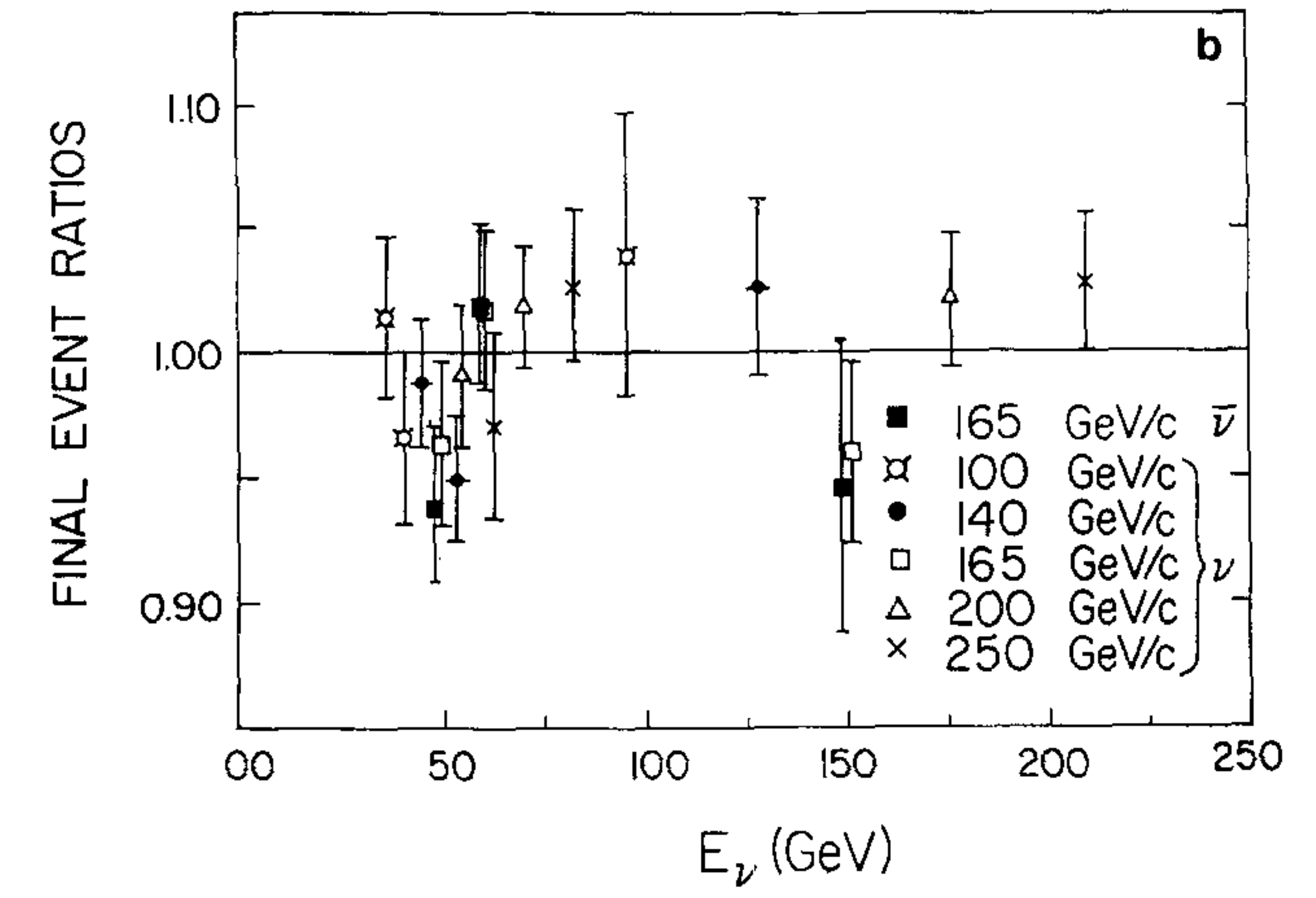}
                \caption{}
                \label{fig:CCFRdata}
            \end{subfigure}
            \begin{subfigure}{0.30\textwidth}
                \includegraphics[width=\textwidth]{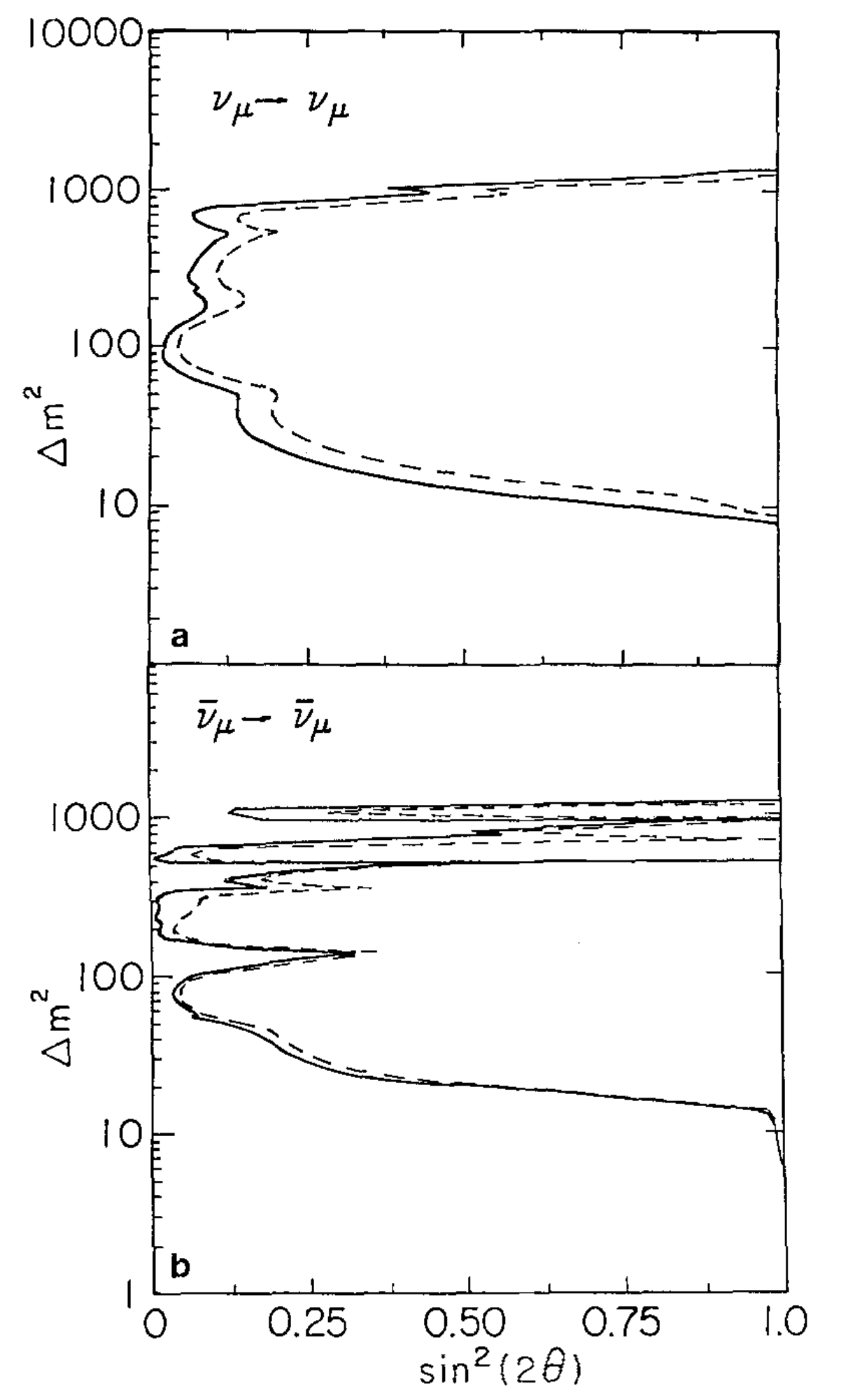}
                \caption{}
                \label{fig:CCFRexclusion}
            \end{subfigure}
            \caption{(a) The observed ratios of events seen between the far and near CCFR detectors. The plot shows both $\numu$ and $\numubar$ data. (b) The 90\% confidence level limits shown for $\numu$ (top) and $\numubar$ (bottom) disappearance. The two lines in each plot corresponds to two different methods to draw the exclusions.
            Figures from Ref.~\cite{PhysRevLett.52.1384}.}
            \label{fig:CCFR}
        \end{figure}

        We show the results of our CCFR 3+1 fit in \Cref{fig:CCFRfit}.
        
    \item[MiniBooNE/SciBooNE \cite{SciBooNE:2011qyf,MiniBooNE:2012meu}] \hfill \\ 
        In addition to the MiniBooNE $\numu \to \nue$ and $\numubar \to \nuebar$ appearance analyses described earlier, MiniBooNE conducted  $\numu \to \numu$ and $\numubar \to \numubar$ disappearance analyses jointly with the SciBooNE detector. 
        The SciBooNE detector was located \SI{100}{\meter} from the BNB neutrino production target, sharing the same neutrino flux as MiniBooNE. 
        The SciBooNE detector was composed of three sub-detectors: a highly segmented scintillator tracker (SciBar), an electromagnetic calorimeter, and a muon range detector (MRD). 

        In the $\numu \to \numu$ analysis, events were collected in three $\numu$ samples: SciBar-stopped events, MRD-stopped events, and MiniBooNE events. 
        These three samples were fit simultaneously to an oscillation model. The data verses expectation can be seen in \Cref{fig:SBMBnumu}.
        The data gave a p-value over 50\%, showing no evidence for oscillations. 
        The exclusions are shown in \Cref{fig:SBMBnumuexclusion}. 

        \begin{figure}
            \centering
            \includegraphics[width=0.95\textwidth]{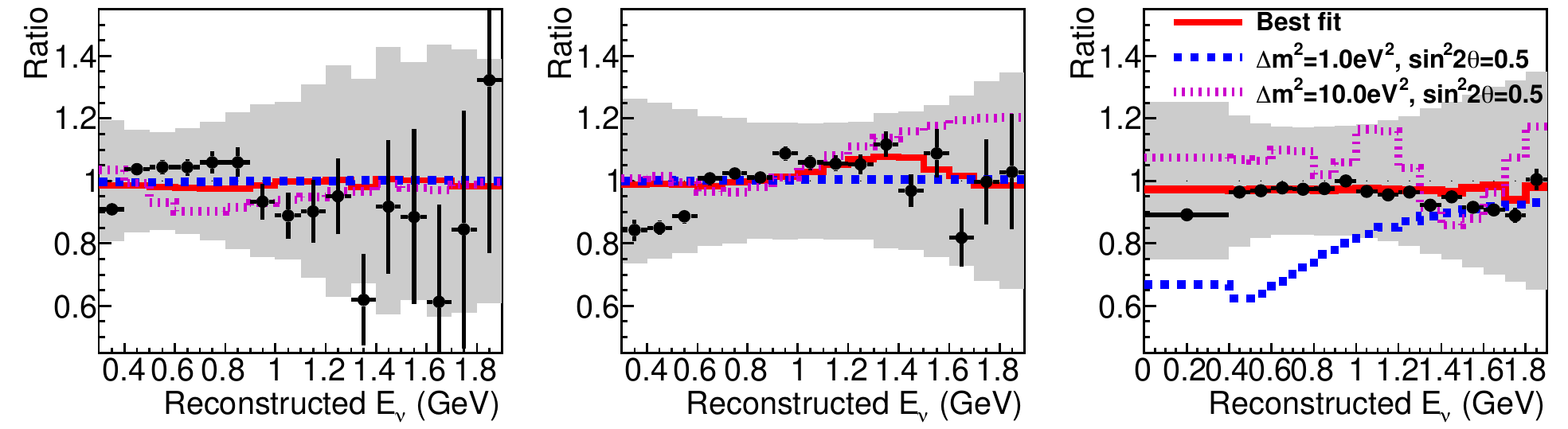}
            \caption{The ratio of observed $\numu$ rates over expectation for, left to right, SciBar-stopped, MRD-stopped, and MinibooNE samples. Figures from Ref.~\cite{SciBooNE:2011qyf}.}
            \label{fig:SBMBnumu}
        \end{figure}    

        \begin{figure}
            \centering
            \includegraphics[width=0.45\textwidth]{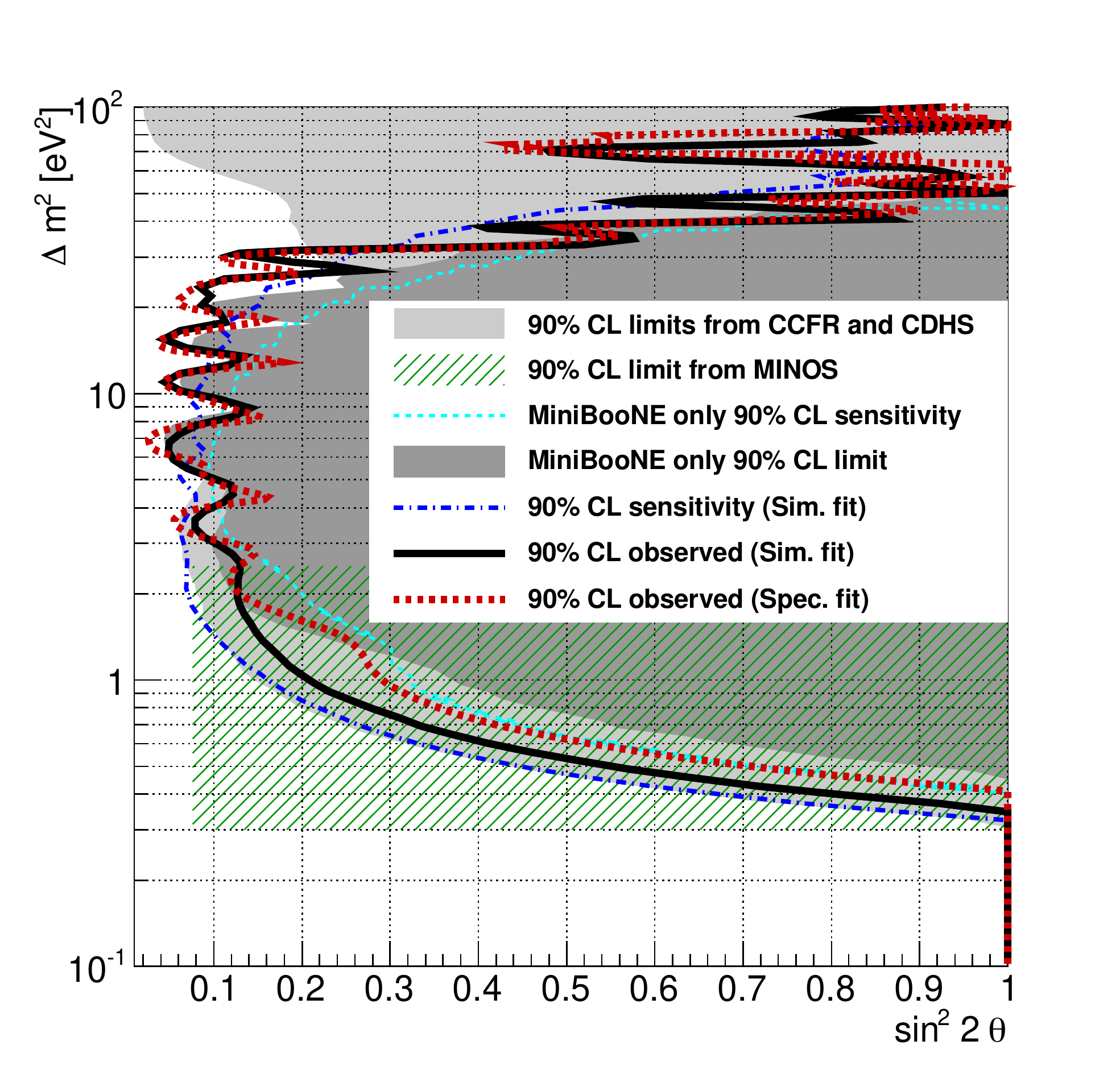}
            \caption{The 90\% confidence level limit for the MiniBooNE/SciBooNE $\numu$ disappearance joint fit. Figure from Ref.~\cite{SciBooNE:2011qyf}.}
            \label{fig:SBMBnumuexclusion}
        \end{figure}    

        A similar analysis was conducted for $\numubar$ disappearance, using MiniBooNE data taken 2006--2012 and SciBooNE data taken 2007--2008. 
        Like the $\numu$ analysis, the $\numubar$ analysis used SciBar-stopped and MRD-stopped events. 
        The data can be seen in \Cref{fig:MBSBnumubardata}. 
        Both detectors observed an excess compared to expectation, so no evidence for oscillations between the detectors was observed.
        The 90\% exclusion is shown in \Cref{fig:SBMBnumubarexclusion}.

        \begin{figure}
            \centering
            \begin{subfigure}{0.45\textwidth}
                \includegraphics[width=\textwidth]{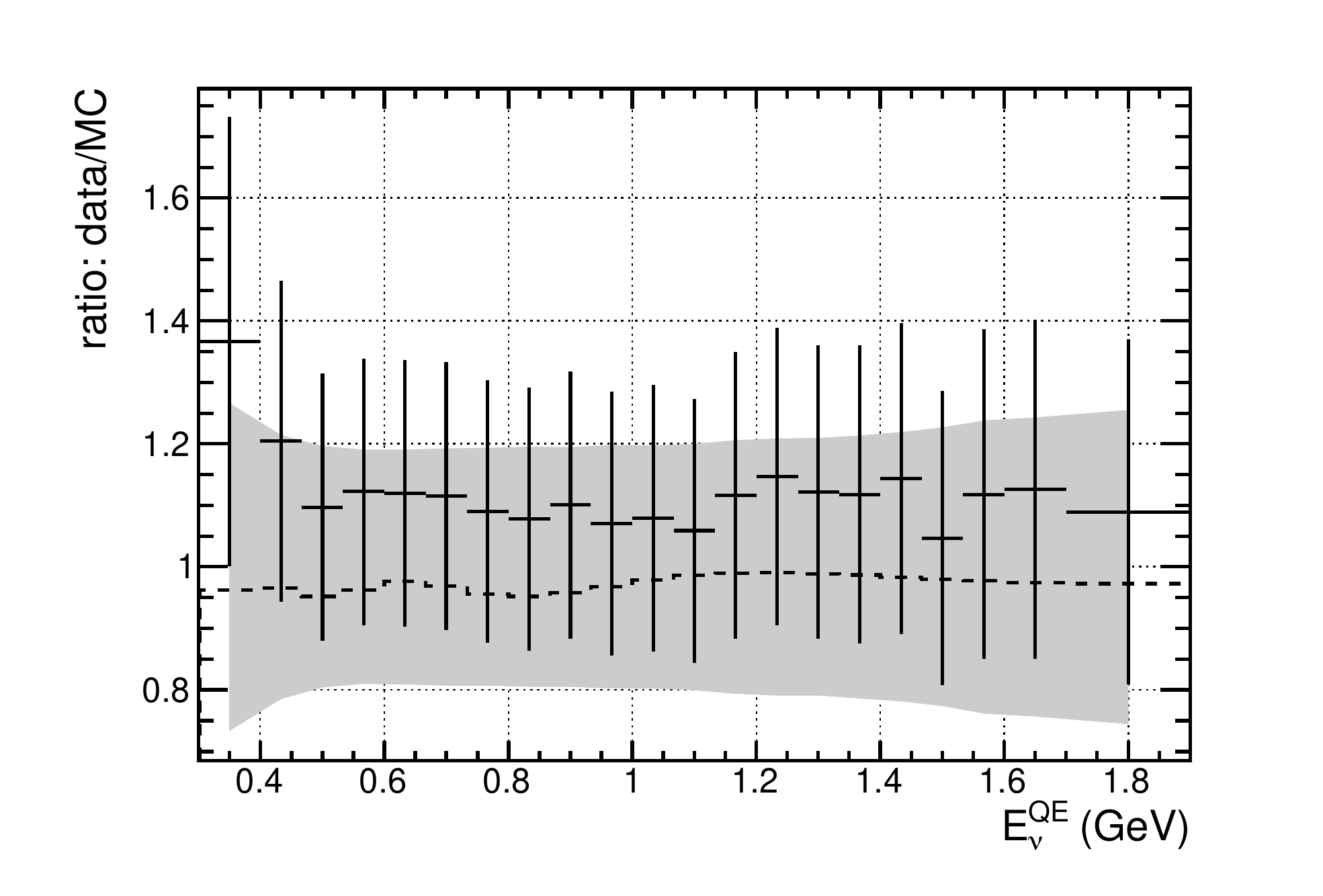}
                \caption{MiniBooNE}
                \label{fig:MBnumubardata}
            \end{subfigure}
            \begin{subfigure}{0.45\textwidth}
                \includegraphics[width=\textwidth]{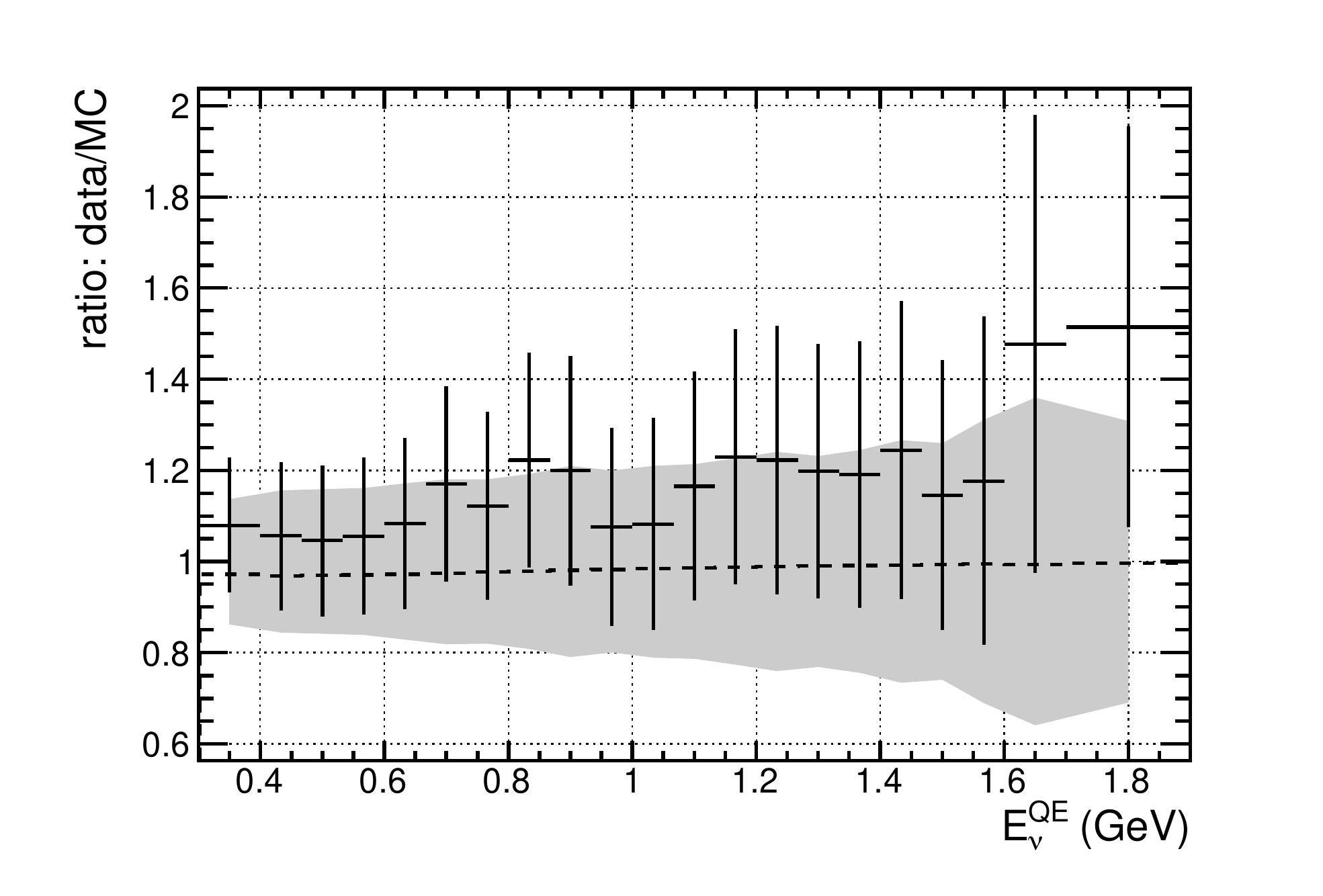}
                \caption{SciBooNE}
                \label{fig:SBnumubardata}
            \end{subfigure}
            \caption{The ratios of observed over expected $\numubar$ events in MiniBooNE (left) and SciBooNE (right). No oscillation deficit was observed. Note that the y-axis does not start at 0.
            Figures from Ref.~\cite{MiniBooNE:2012meu}.}
            \label{fig:MBSBnumubardata}
        \end{figure}

        \begin{figure}
            \centering
            \includegraphics[width=0.45\textwidth]{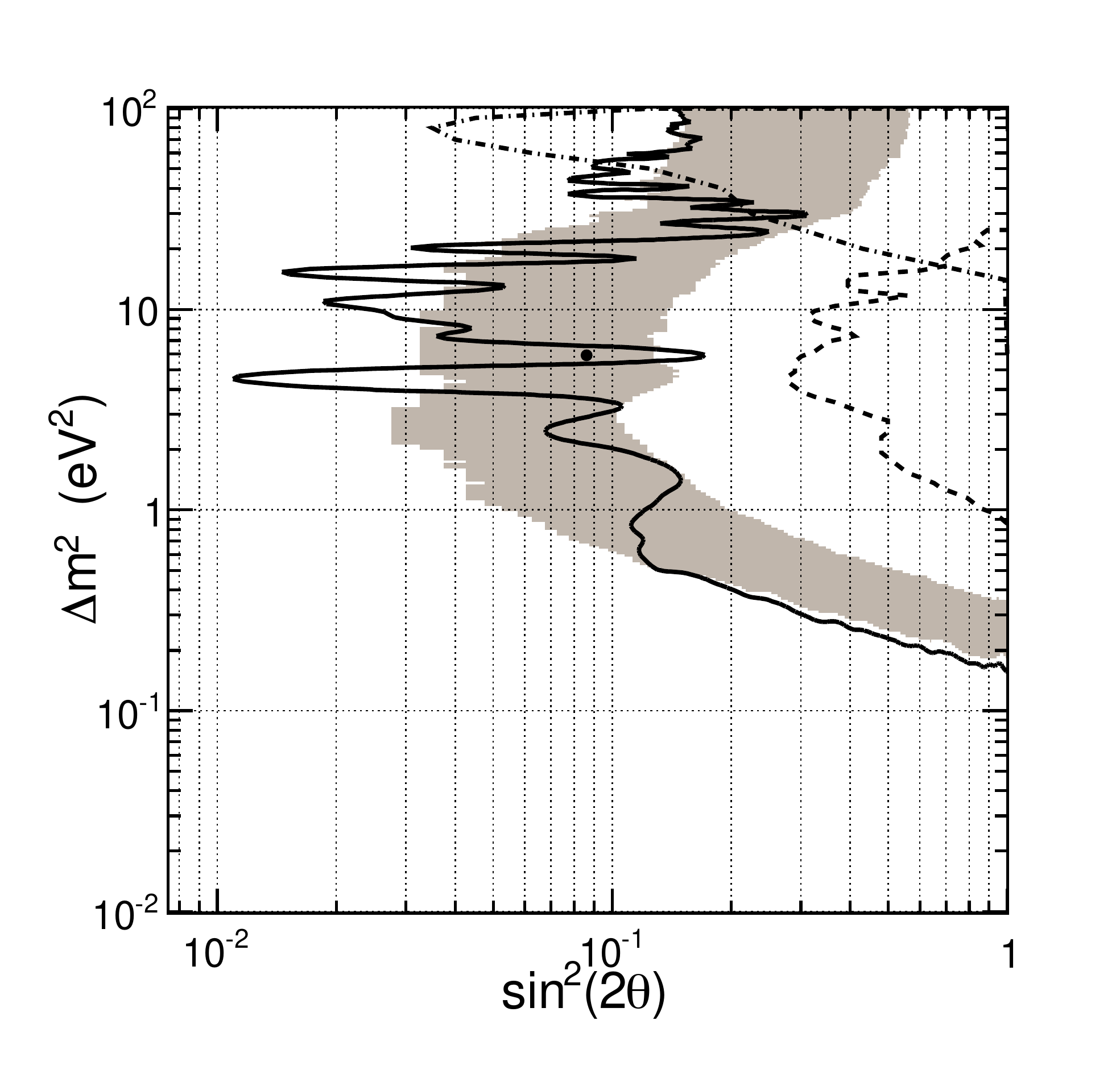}
            \caption{The 90\% confidence level for the MiniBooNE/SciBooNE joint $\numubar$ disappearance analysis, shown in the solid line. 
            The dashed line is the 90\% confidence level from the 2009 MiniBooNE disappearance analysis \cite{MiniBooNE:2009ozf} and the dot-dashed line is the 90\% for CCFR.  
            Figure from Ref.~\cite{MiniBooNE:2012meu}.}
            \label{fig:SBMBnumubarexclusion}
        \end{figure}    

        We show the results of our MiniBooNE/SciBooNE joint analysis 3+1 fit in \Cref{fig:MB-SBfit}.
        
    \item[MINOS-CC \cite{MINOS:2011xqg,MINOS:2012dbe,MINOS:2016viw}] \hfill \\ 
        The \textbf{M}ain \textbf{I}njector \textbf{N}eutrino \textbf{O}scillation \textbf{S}earch (MINOS) experiment was built to measure the standard model neutrino oscillation parameters using \numu and \numubar oscillations.
        MINOS detected the neutrino flux from the NuMI beamline, with a near detector 1.04 km from the beam target and a far detector at 734 km. 
        The detectors were magnetized, such that it could differentiate $\mu^{-}$'s from $\mu^{+}$'s.
        The beam peaked at \SI{3}{\GeV}. 

        In our fits, we use both $\numubar \to \numubar$ and $\numu \to \numu$ oscillation channels  from three different MINOS data sets. 
        The first is the $\numubar \to \numubar$ oscillation data set, collected in two phases in 2009--2011 using the $\numubar$-enhanced beam. The second data set is a $\numubar \to \numubar$ oscillation analysis using the 7\% wrong-signed neutrinos in the $\numu$ configuration. 
        The last data set is the $\numu \to \numu$ oscillation analysis from 2016. 
        In all three data sets, the observations in the near detector is used to predict the flux at the far detector given some model. 
        
        The three data sets are shown in \Cref{fig:MINOSdata}.
        The results of our MINOS fits are shown in \Cref{fig:MINOSfit}.

        \begin{figure}
            \centering
            \begin{subfigure}{0.3\textwidth}
                \includegraphics[width=\textwidth]{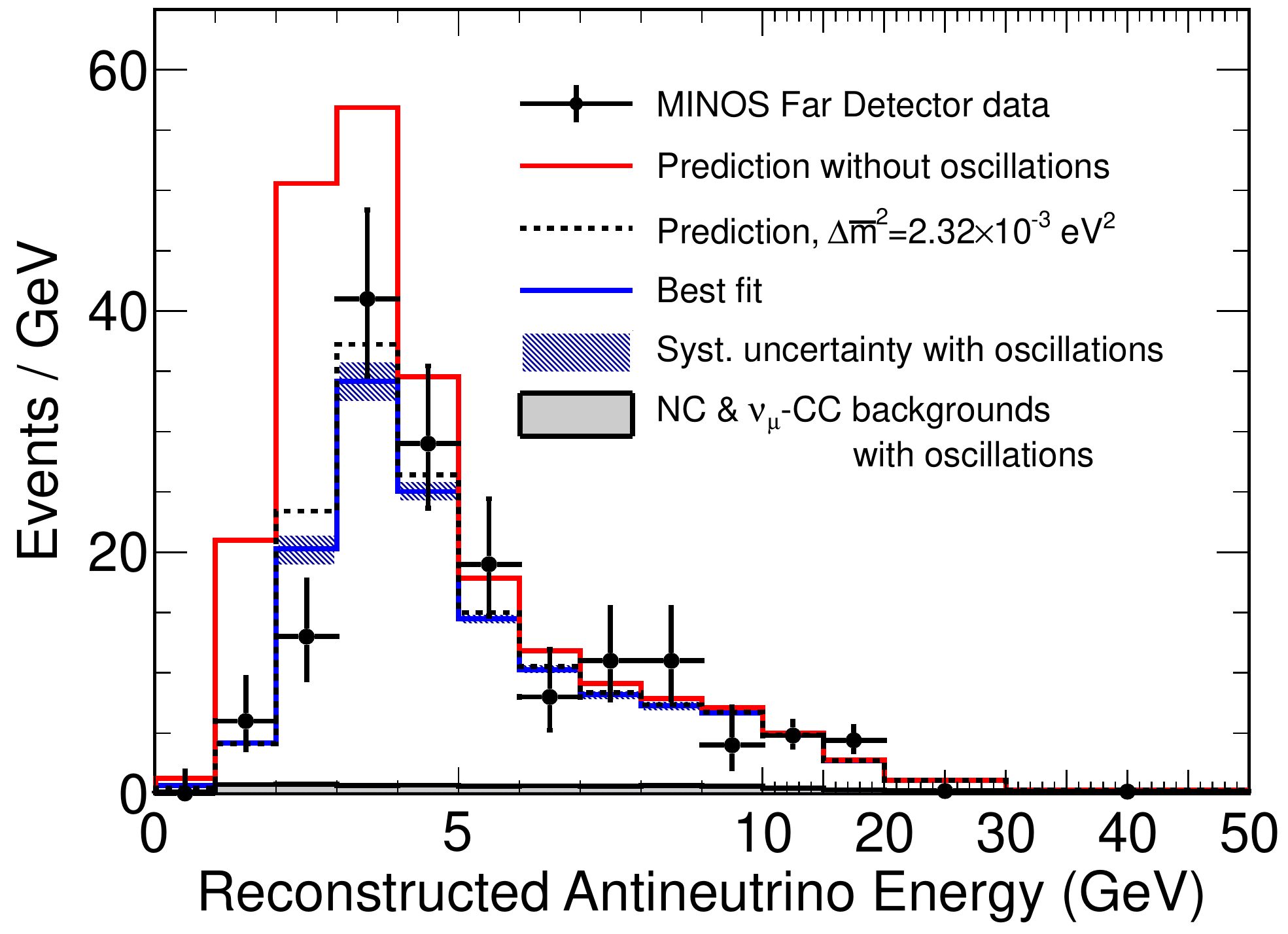}
                \caption{}
            \end{subfigure}
            \begin{subfigure}{0.3\textwidth}
                \includegraphics[width=\textwidth]{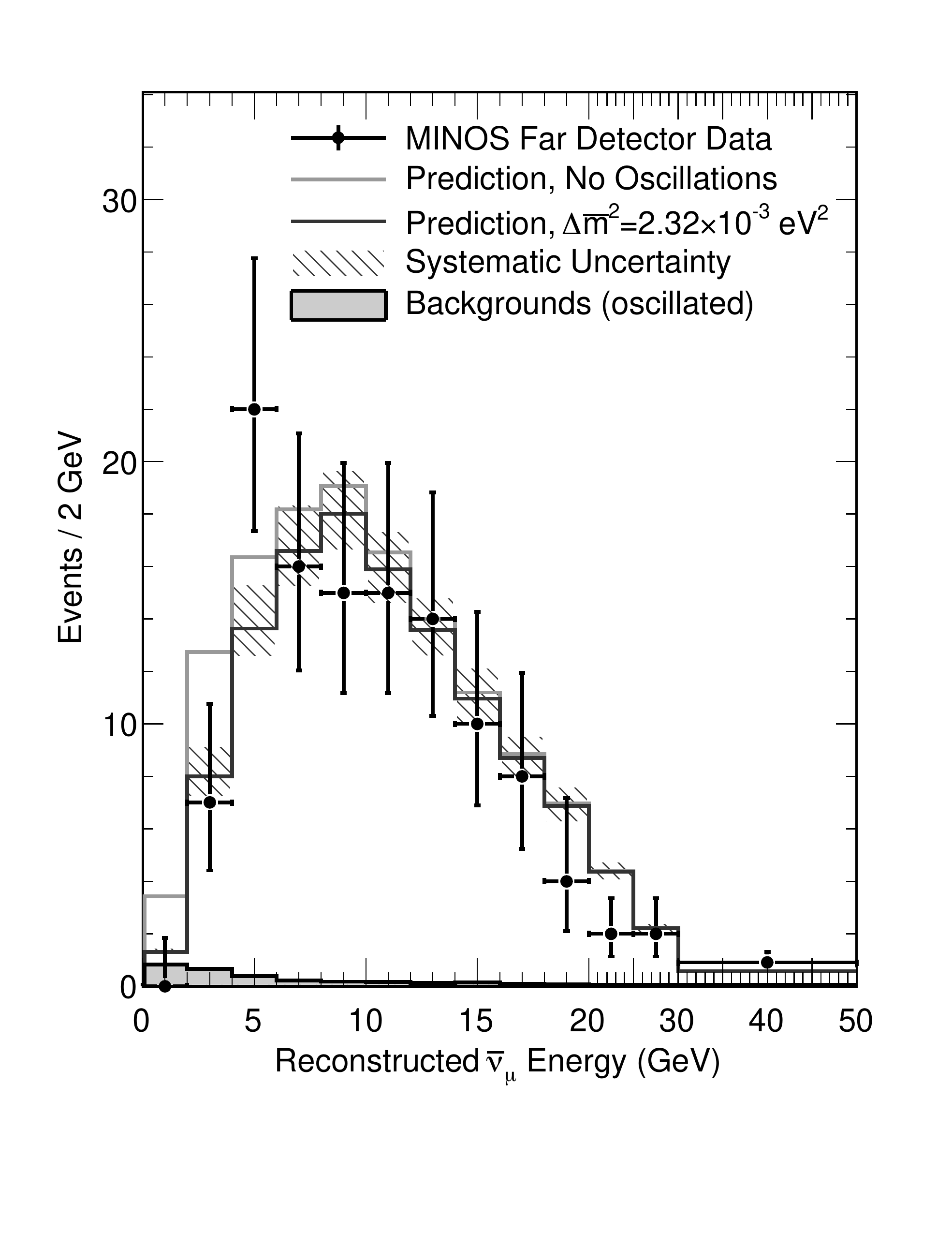}
                \caption{}
            \end{subfigure}
            \begin{subfigure}{0.3\textwidth}
                \includegraphics[width=\textwidth]{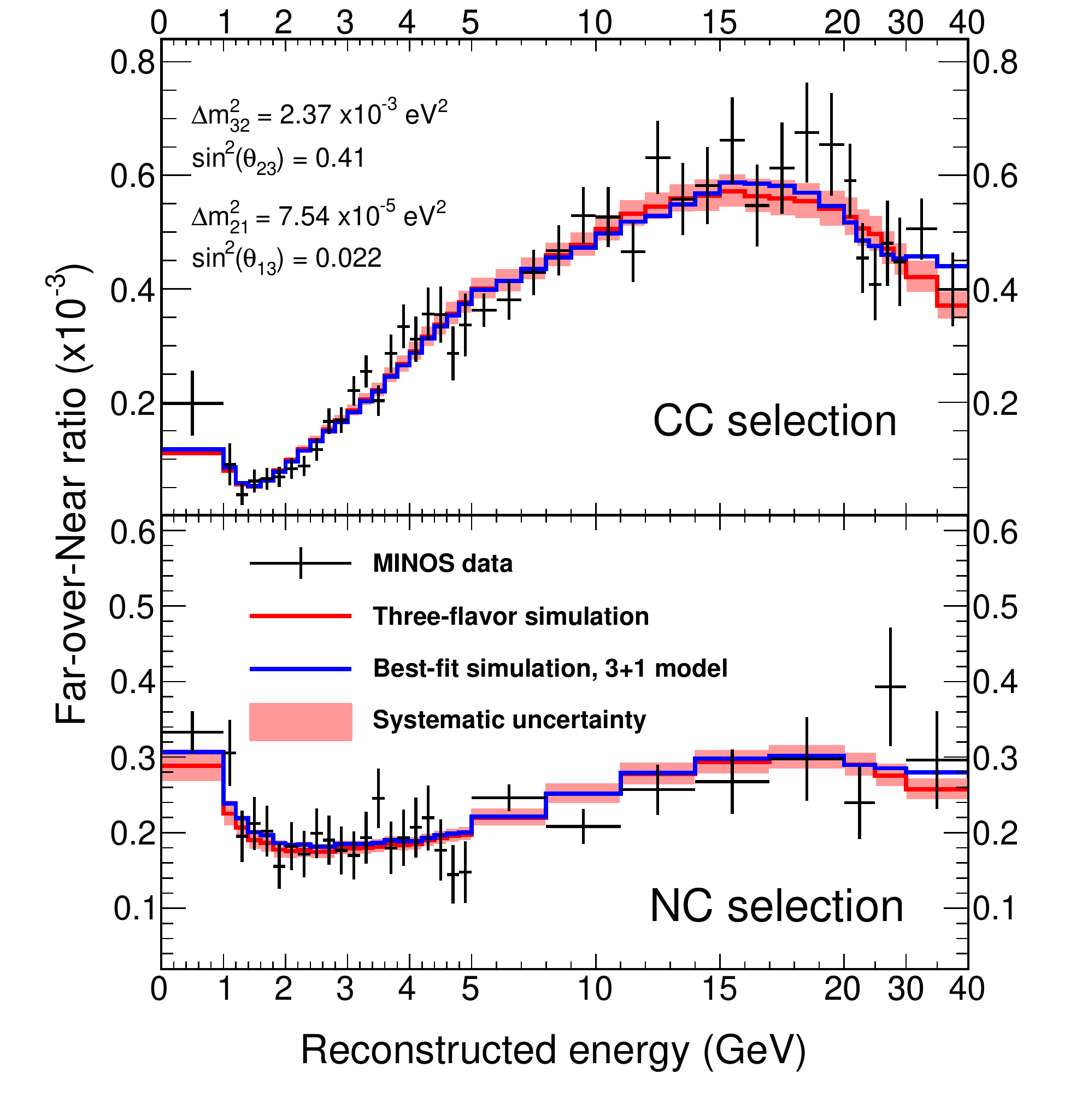}
                \caption{}
            \end{subfigure}
            \caption{(a) The observed $\numubar$ spectra at the far detector in the $\numubar$-enhanced beam configuration. In MINOS's analysis, the standard model parameters are fitted, not the sterile parameters. Figure from Ref.~\cite{MINOS:2012dbe}. (b) The observed wrong-signed $\numubar$ spectra at the far detector in the $\numu$-enhanced beam configuration. Figure from Ref.~\cite{MINOS:2011xqg}. (c) The observed $\numu$ spectra ratio between the far and near detector. We only consider the CC sample in our fits. Figure from Ref.~\cite{MINOS:2016viw}.}
            \label{fig:MINOSdata}
        \end{figure}

        \begin{figure}
            \centering
            \begin{subfigure}{0.49\linewidth}
                \centering
                \includegraphics[width=\linewidth]{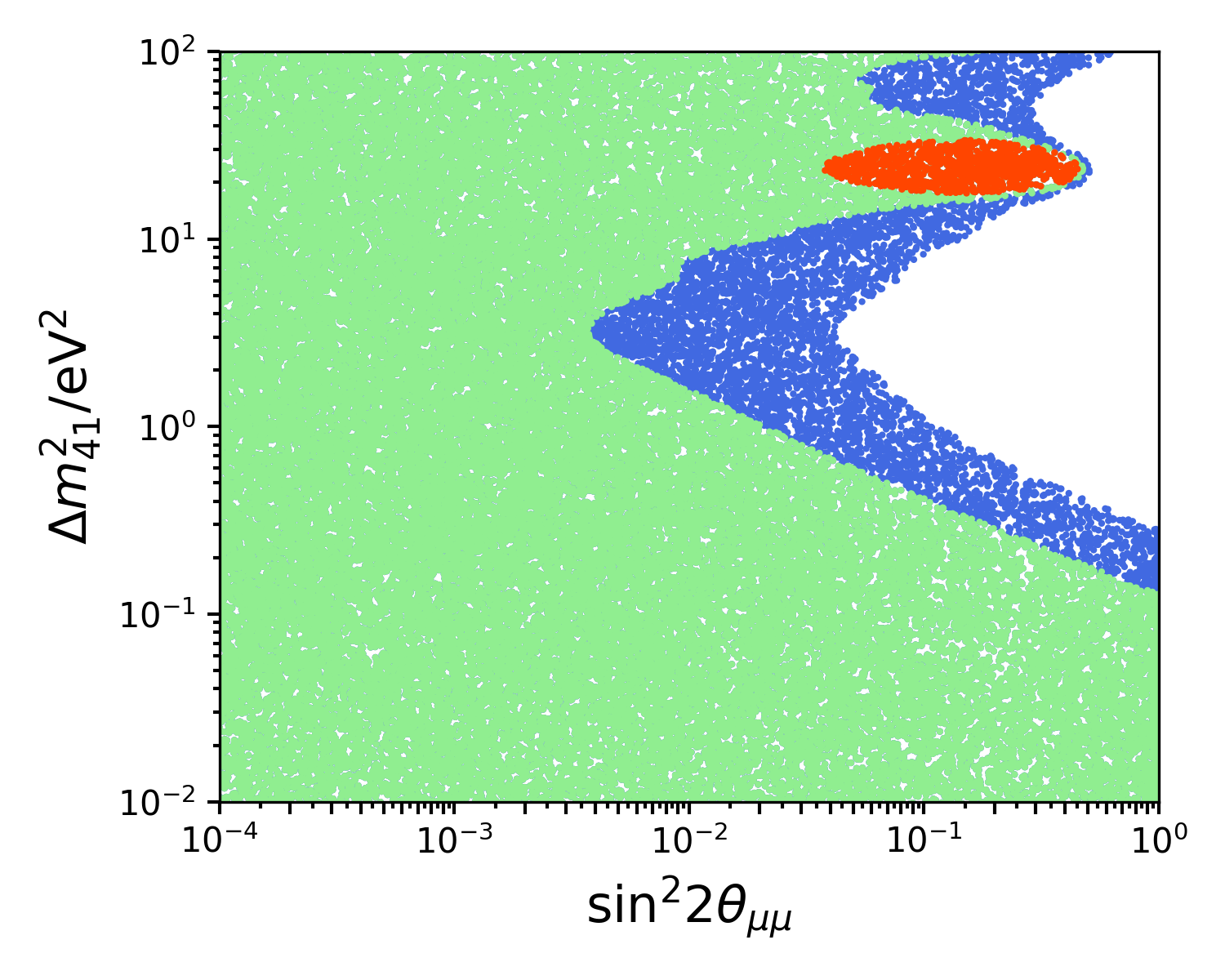}
                \caption{CDHS}
                \label{fig:CDHSfit}
            \end{subfigure}
            \hfill
            \begin{subfigure}{0.49\linewidth}
                \centering
                \includegraphics[width=\linewidth]{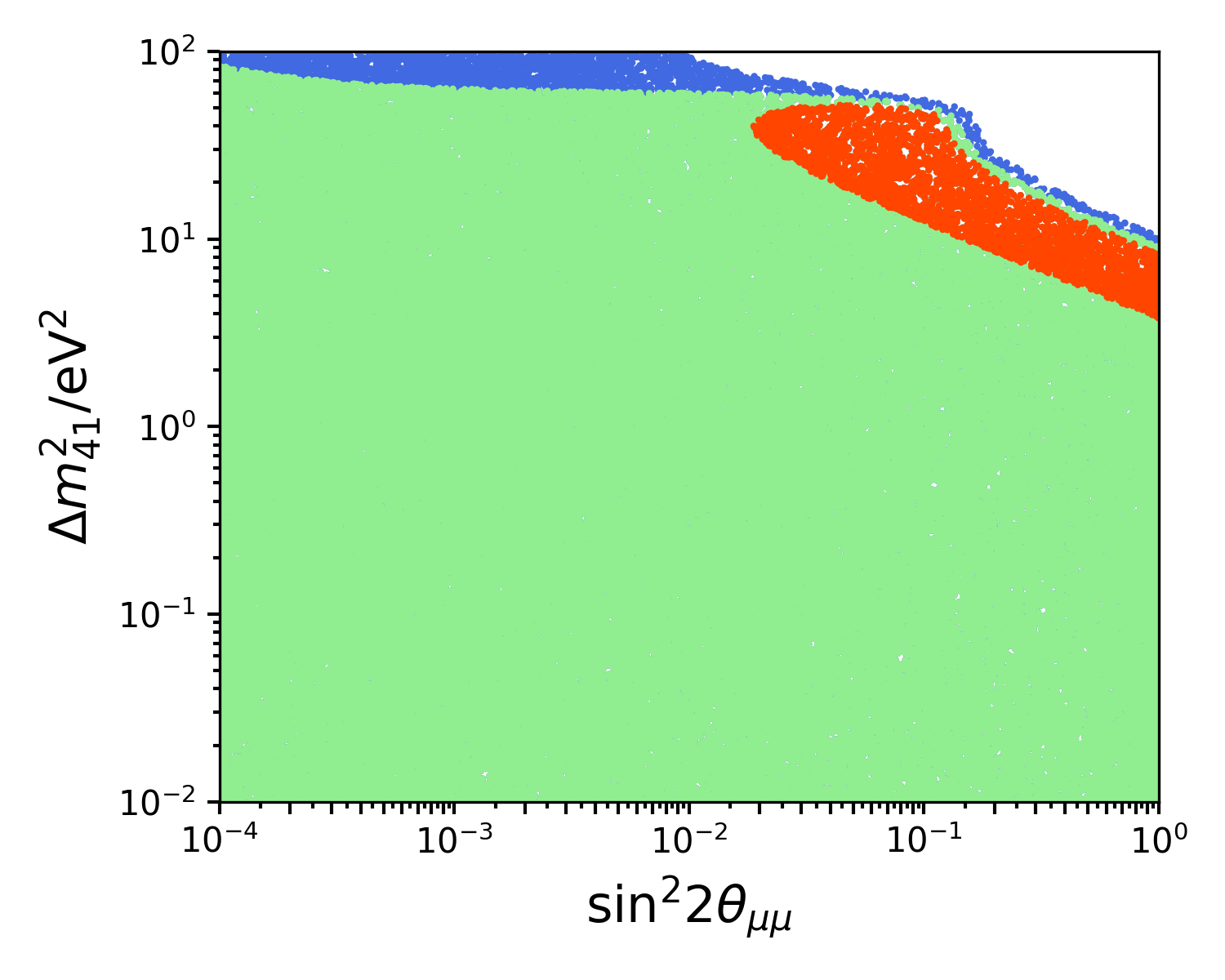}
                \caption{CCFR}
                \label{fig:CCFRfit}
            \end{subfigure}

            \bigskip
            \begin{subfigure}{0.49\linewidth}
                \centering
                \includegraphics[width=\linewidth]{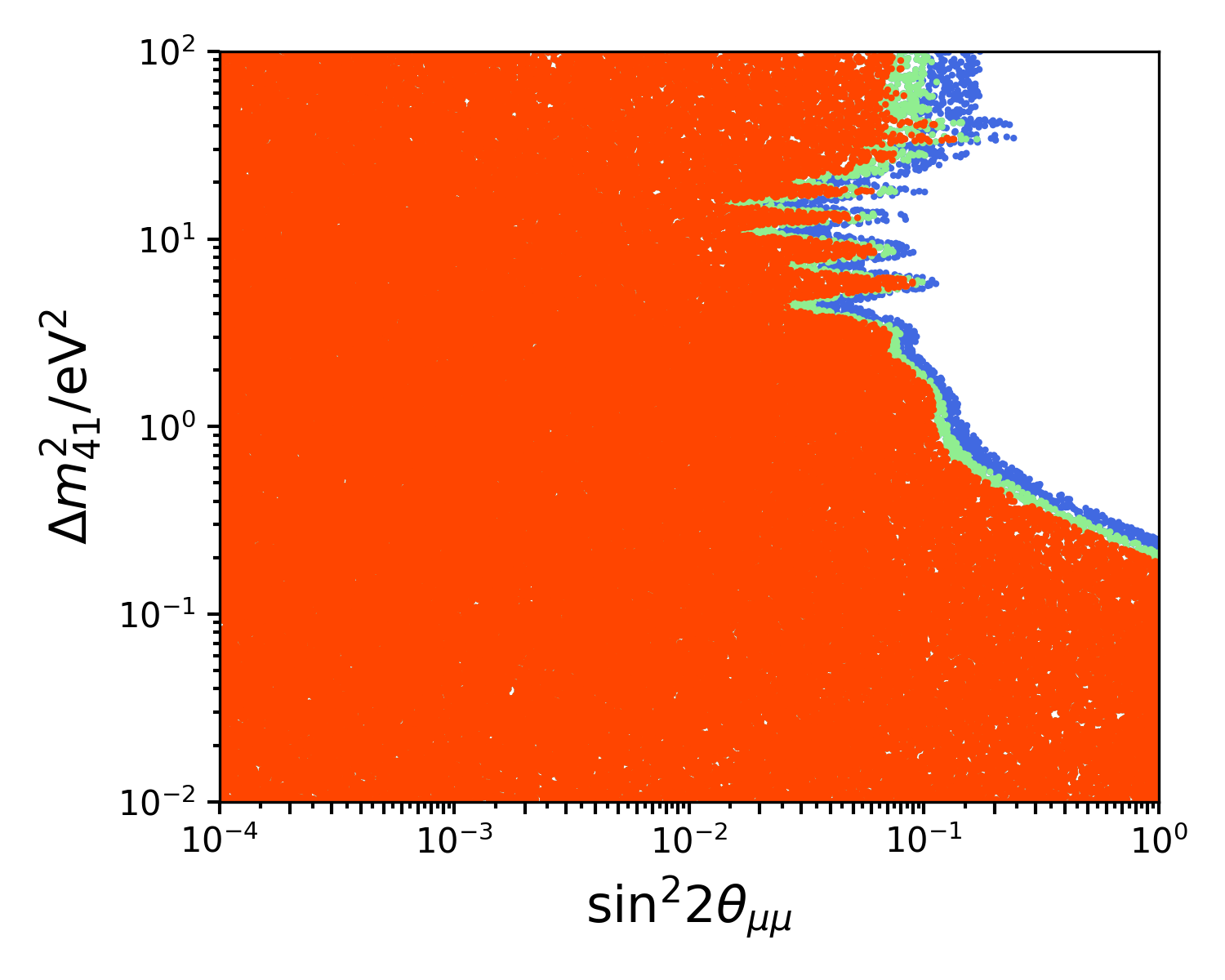}
                \caption{MiniBooNE/SciBooNE}
                \label{fig:MB-SBfit}
            \end{subfigure}
            \hfill
            \begin{subfigure}{0.49\linewidth}
                \centering
                \includegraphics[width=\linewidth]{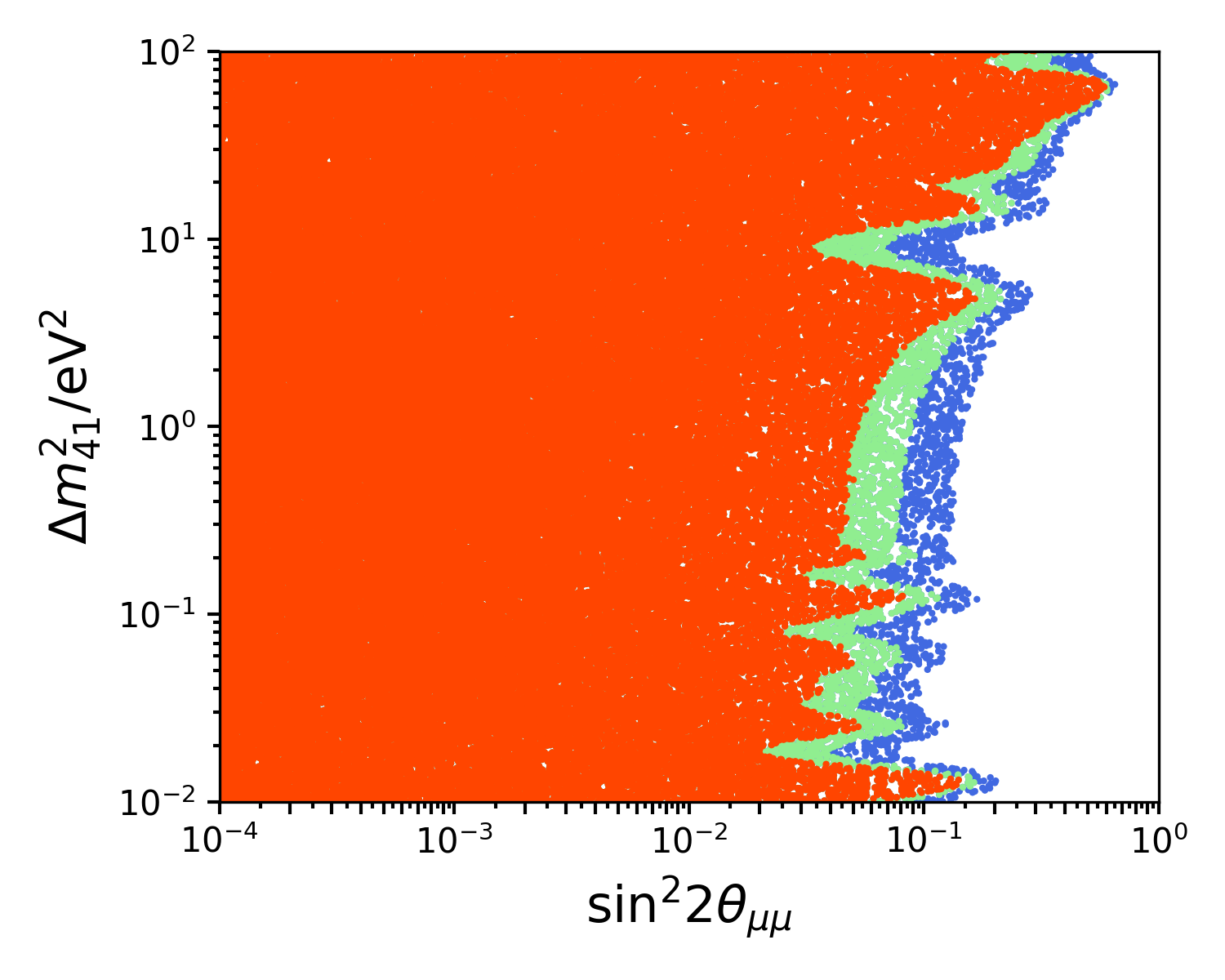}
                \caption{MINOS}
                \label{fig:MINOSfit}
            \end{subfigure}
            \caption{The 3+1 fits to the $\numu\to\numu$ and $\numubar\to\numubar$ disappearance experiments used in our global fits.}
            \label{fig:numufits}
        \end{figure}
\end{description}

\section{Methodology}
\label{sec:methodology}

A thorough description of the methodology of our fits can be found in Ref.~\cite{Diaz:2019fwt}, but we summarize the significant points here. 

For a particular model, a Markov Chain Monte Carlo (MCMC) is used to explore the parameter space. 
The algorithm follows that used by the \texttt{emcee} Python package \cite{Foreman-Mackey:2012any}, but with our own implementation in C++.
The sampled parameters, for the models described in \Cref{sec:sterilemodels}, are: $\Dmqfo$, $\Uef$, and $\Umuf$ for the 3+1 model; $\Dmqfo$, $\Dmqfiveo$, $\Uef$, $\Uefive$, $\Umuf$, $\Umufive$, and $\phi_{\mu e}$ for the 3+2 model; and $\Dmqfo$, $\Uef$, $\Umuf$, and $\Gamma$ for the 3+1+Decay model.
For each model, we enforce the unitarity conditions $\sum_{\alpha} |U_{\alpha i}|^2 < 1$ for each mass index $i$, and $\sum_{i} |U_{\alpha i}|^2 < 1$ for each flavor index $\alpha$.
For each model, $\alpha \in \{e, \mu\}$; and  $i \in \{4\}$ for both the 3+1 and 3+1+Decay models, while $i \in \{4,5\}$ for the 3+2 model.
We also impose the constraint $\Dmqfo < \Dmqfiveo$ for the 3+2 model.

At each sampled point in the parameter space, two values are recorded: a $\chi^2$ value and a log-likelihood. 
For the frequentist fits, the test statistic $\Delta \chi^2 = \chi^2(\vec{\theta}) - \chi^2_{\mathrm{min}}$ is used, where $\chi^2(\vec{\theta})$ is the $\chi^2$ at some parameter set $\vec{\theta}$ and $\chi^2_{\mathrm{min}}$ is the minimum $\chi^2$ found in the parameter space.
$\Delta \chi^2$ is assumed to follow a $\chi^2$ distribution with degrees of freedom equal to the difference in degrees of freedom between the null model and the sterile model under consideration. 
When drawing two dimensional confidence regions, as have been shown in this chapter, the $\Delta \chi^2$ is profiled over the remaining dimensions, and the contours are drawn assuming two degrees of freedom unless otherwise stated.

The log-likelihoods serve a dual purpose.
First, the MCMC explores the parameters space guided by the log-likelihood.
This allows a more efficient exploration of the parameter space, which would otherwise be computationally prohibitive if we were to scan in a grid over multiple dimensions (e.g. 7 dimensions for the 3+2 model).
Second, the MCMC naturally samples the posterior, which allows the drawing of Bayesian credible regions. 
In our analysis, we use the python package \texttt{corner.py} \cite{corner} to draw these regions. 

In addition to searching for the sterile parameters that best fit the data, we would also like to test the internal consistency of such a model. 
For example, in the 3+1 model, the three oscillation channels studied here, $\numu \to \nue$, $\nue \to \nue$, $\numu \to \numu$ (and their antineutrino analog), probe three different oscillation equations that depend on the mixing parameters differently:

\begin{align}
    P(\numu \to \nue )
    & =  
    \sin^{2} 2\theta_{\mu e} \sin^{2}\left( 1.27 \Delta m_{41}^2  \frac{L }{E } \right)\\
    P(\nue \to \nue )
    & =  
    1 - \sin^{2} 2\theta_{ee} \sin^{2}\left( 1.27 \Delta m_{41}^2  \frac{L }{E } \right)\\
    P(\numu \to \numu )
    & =  
    1 - \sin^{2} 2\theta_{\mu \mu} \sin^{2}\left( 1.27 \Delta m_{41}^2  \frac{L }{E } \right),
\end{align}
where the effective mixing angles $\sin^{2} 2\theta_{\alpha \beta}$ are given by 
\begin{align}
    \sin^{2} 2\theta_{\mu e} &= 4 |U_{\mu 4}|^{2} |U_{e 4}|^{2}\\
    \sin^{2} 2\theta_{e e} &= 4 |U_{e 4}|^{2} (1- |U_{e 4}|^{2})\\
    \sin^{2} 2\theta_{\mu \mu } &= 4 |U_{\mu 4}|^{2} (1-|U_{\mu 4}|^{2}).
\end{align}
We can see that the three effective mixing angles depend on only two different mixing elements, $|U_{e 4}|^{2}$ and $|U_{\mu 4}|^{2}$.
Therefore, the mixing angles are not independent and we can test if the different data sets provide consistent values. 

We test this by splitting the data sets into two groups, an appearance and disappearance data set. 
The appearance data set would be sensitive to the product $|U_{\mu 4}|^{2} |U_{e 4}|^{2}$, while the disappearance data set would be composed of experiments that are sensitive to either $|U_{e 4}|^{2}$ or $|U_{\mu 4}|^{2}$. 
We then apply the Parameter Goodness of Fit (PG) test \cite{Maltoni:2003cu} on these data sets. 
We perform separate fits on the two data subsets, along with the fit to the global data set.
We use the three minimum $\chi^2$'s, $\chi^2_{\mathrm{glob}}$, $\chi^2_{\mathrm{app}}$, $\chi^2_{\mathrm{dis}}$, to construct an effective
\begin{equation}
    \chi^2_{\mathrm{PG}} = \chi^2_{\mathrm{glob}} - (\chi^2_{\mathrm{app}} + \chi^2_{\mathrm{app}})
\end{equation}
with an effective number of degrees of freedom
\begin{equation}
    N_{\mathrm{PG}} = (N_{\mathrm{app}} + N_{\mathrm{dis}}) - N_{\mathrm{glob}},
\end{equation}
where $N_{\textrm{x}}$ are the number of degrees of freedom for each subset.
This $\chi^2_{\mathrm{PG}}$ is assumed to follow a $\chi^2$ distribution with $N_{\mathrm{PG}}$ degrees of freedom, and the resulting p-value tells us the probability for the difference between the subsets to arise from chance if the underlying physics were consistent.

\section{Results}
\label{sec:globalfitresults}

\subsection{\texorpdfstring{3+1}{3+1} Model}

\begin{figure}
    \centering
    \begin{subfigure}{0.49\textwidth}
        \includegraphics[width=\textwidth]{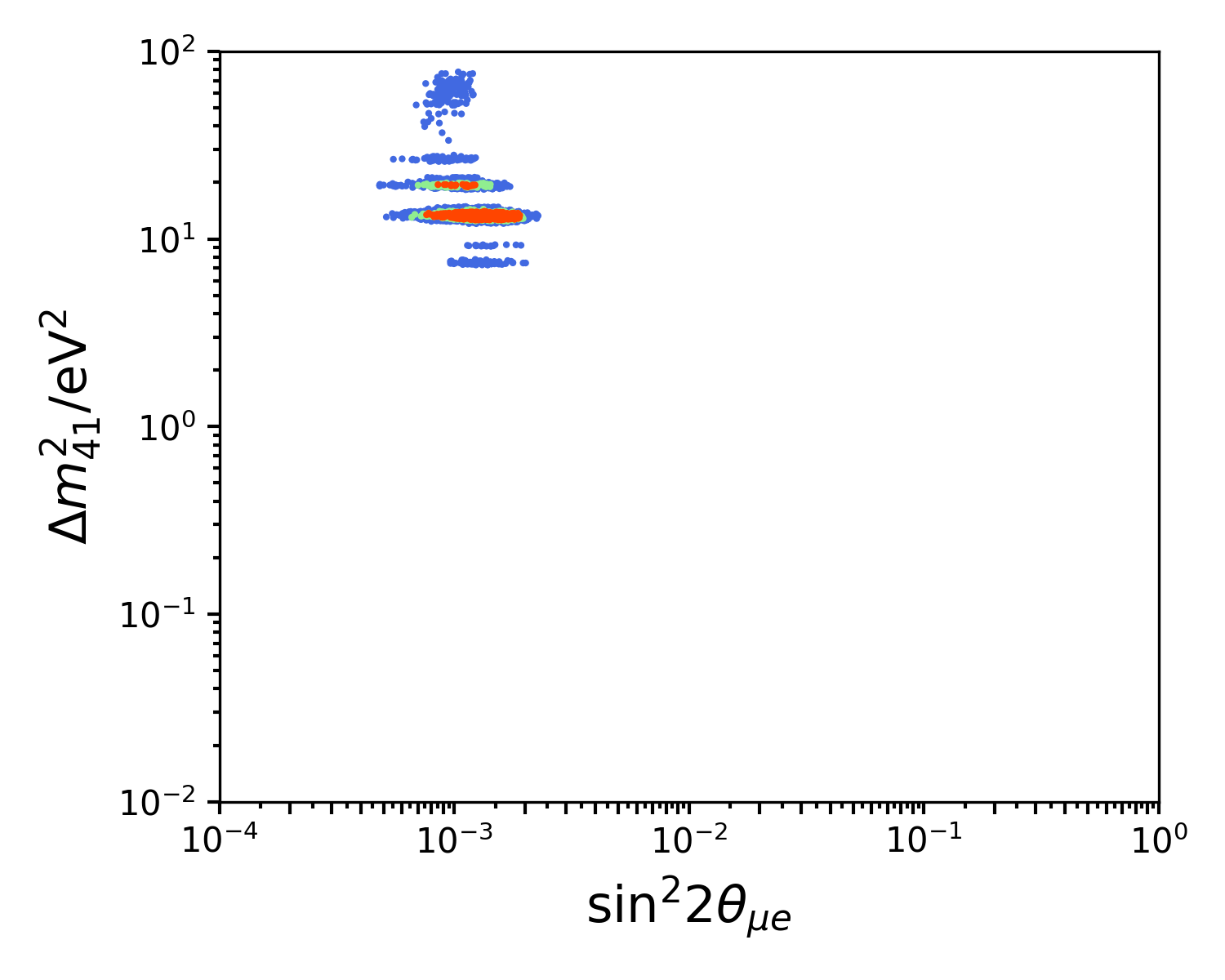}
        \caption{}
        \label{fig:sin22thmueglobalfreqresults}
    \end{subfigure}
    \begin{subfigure}{0.49\textwidth}
        \includegraphics[width=\textwidth]{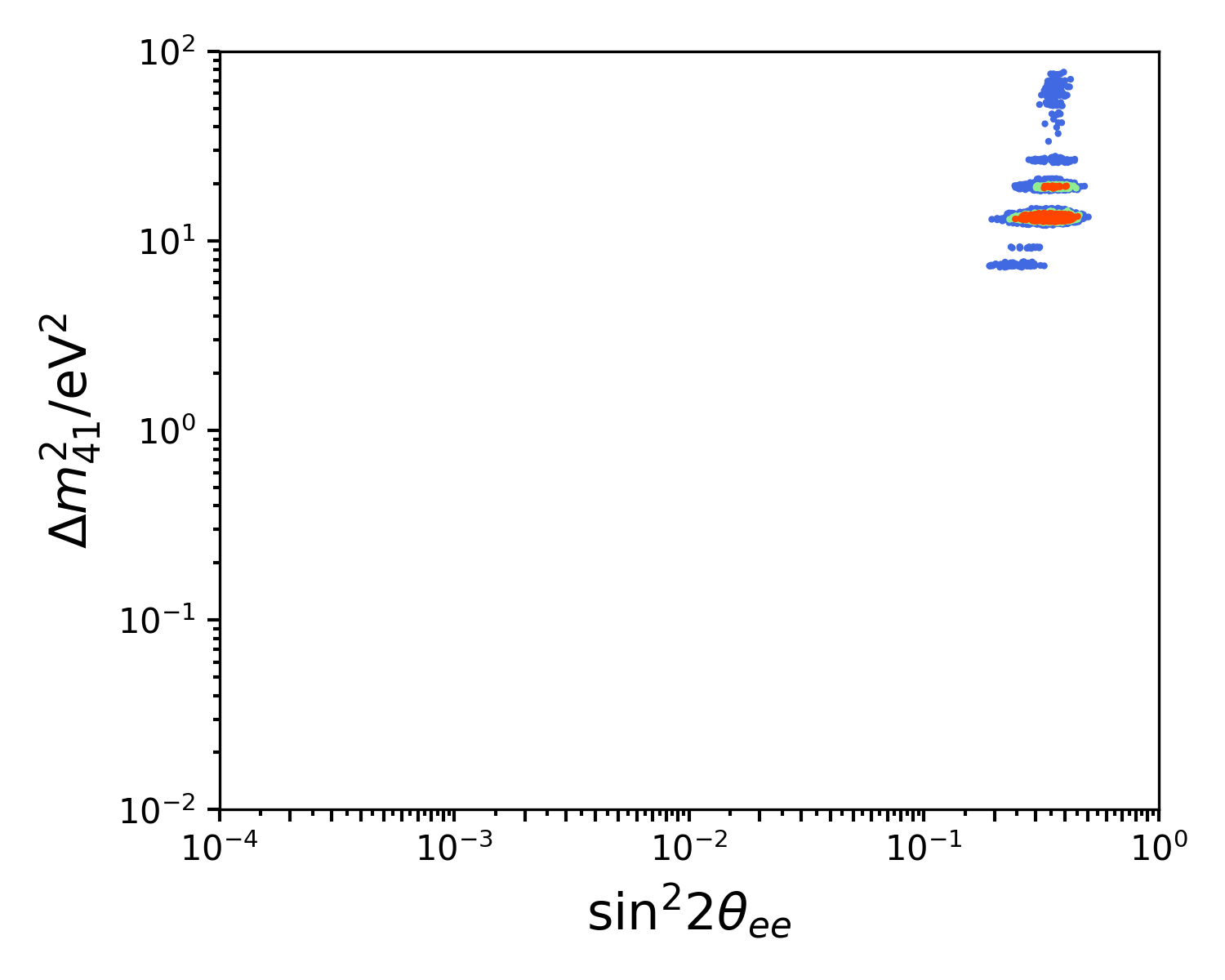}
        \caption{}
    \end{subfigure}

    \bigskip
    \begin{subfigure}{0.49\textwidth}
        \includegraphics[width=\textwidth]{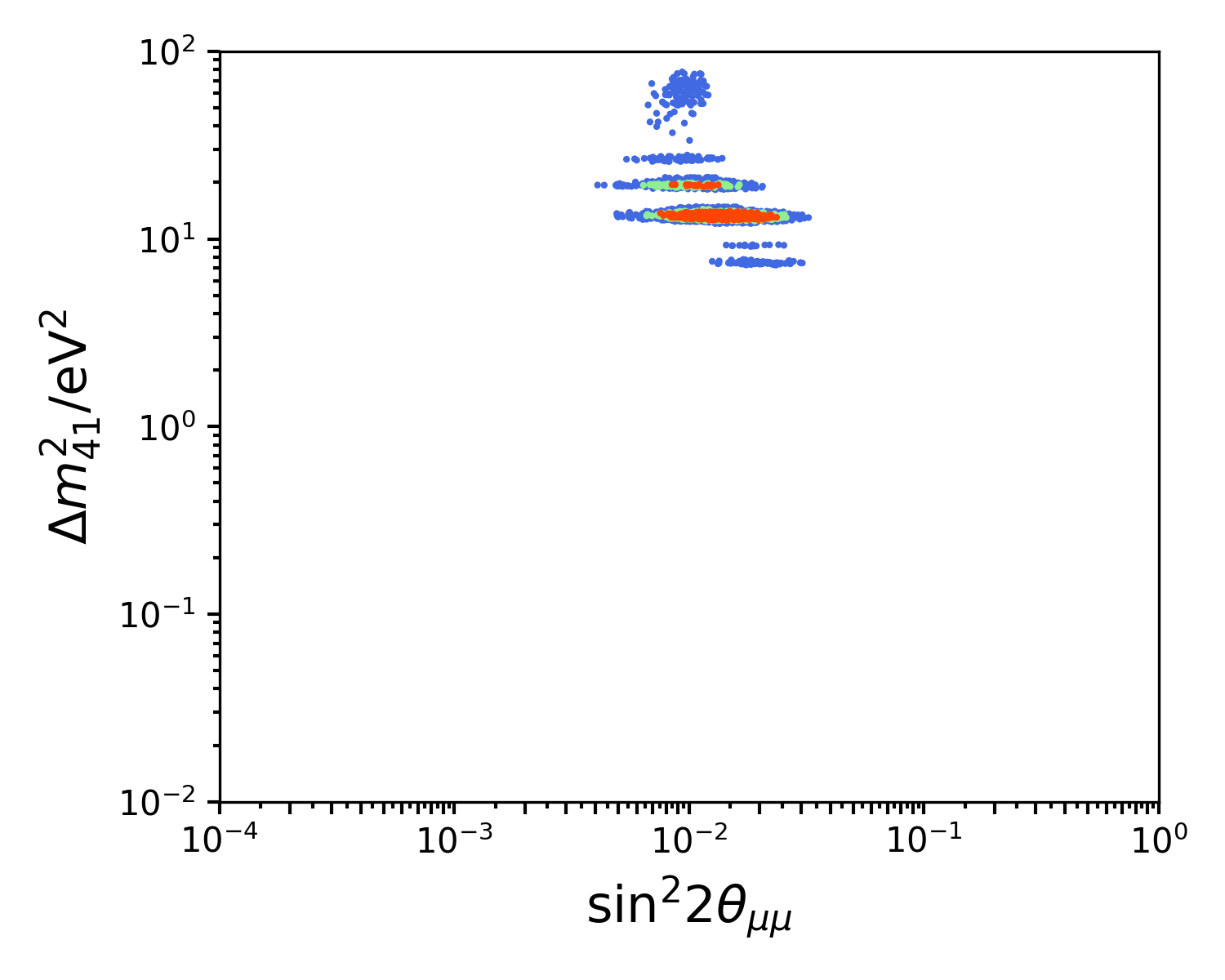}
        \caption{}
    \end{subfigure}
    \caption{Results of the 3+1 global fits. We plot the confidence regions in terms of three different, but not independent, mixing parameters: (a) $\sin^2 2\theta_{\mu e}$, (b) $\sin^2 2\theta_{e e}$, and (c) $\sin^2 2\theta_{\mu \mu}$. The confidence regions correspond to 90\%, 95\%, and 99\% in red, green, and blue, respectively.}
    \label{fig:globalfreqresults}
\end{figure}

We first fit the experiments listed in \Cref{sec:experiments} to the 3+1 sterile model which we reviewed in \Cref{sec:3plus1}. 
The results of this global fit are shown in \Cref{fig:globalfreqresults}. 
The best fit mass-squared splitting is found at $\Dmqfo = 13.1\ \eVq$, with mixing parameters $\Uef = 0.30$
and $\Umuf = 0.065$.
The best fit mixing parameters can also be written in terms of the effective mixing parameters $\sin^2 2\theta_{\mu e} = 0.0015$, $\sin^2 2\theta_{e e} = 0.32$, $\sin^2 2\theta_{\mu \mu} = 0.017$.

The improvement of the 3+1 model compared to the null is found to be $\Delta \chi^2 = 51$, with the addition of only 3 degrees of freedom. 
This substantial improvement has a p-value of $p=4.9 \times 10^{-11}\ (6.6 \sigma)$. 
Therefore, the experiments included in our fit strongly prefer a model like sterile neutrinos.

\begin{figure}
    \centering
    \begin{subfigure}{0.49\textwidth}
        \includegraphics[width=\textwidth]{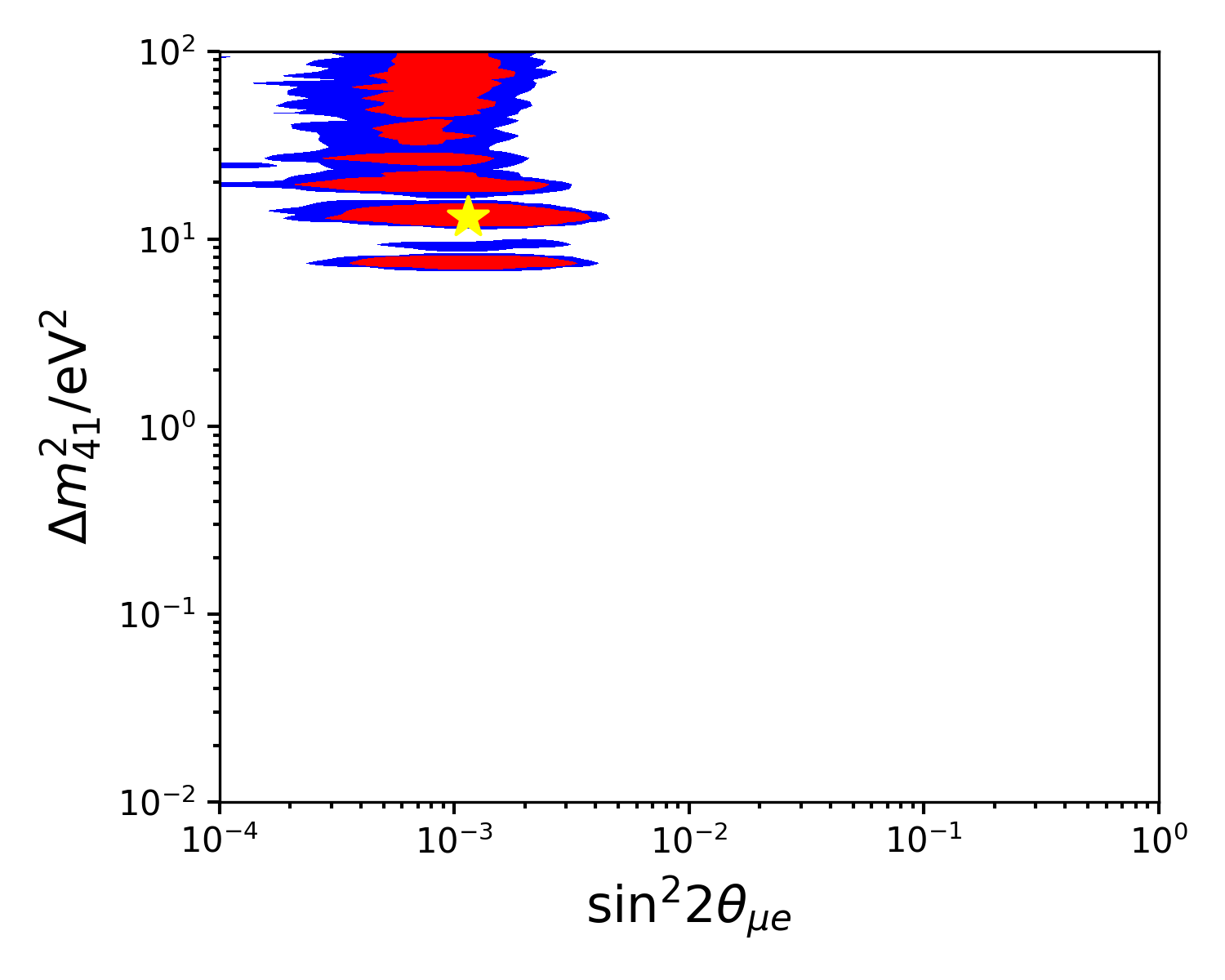}
        \caption{}
    \end{subfigure}
    \begin{subfigure}{0.49\textwidth}
        \includegraphics[width=\textwidth]{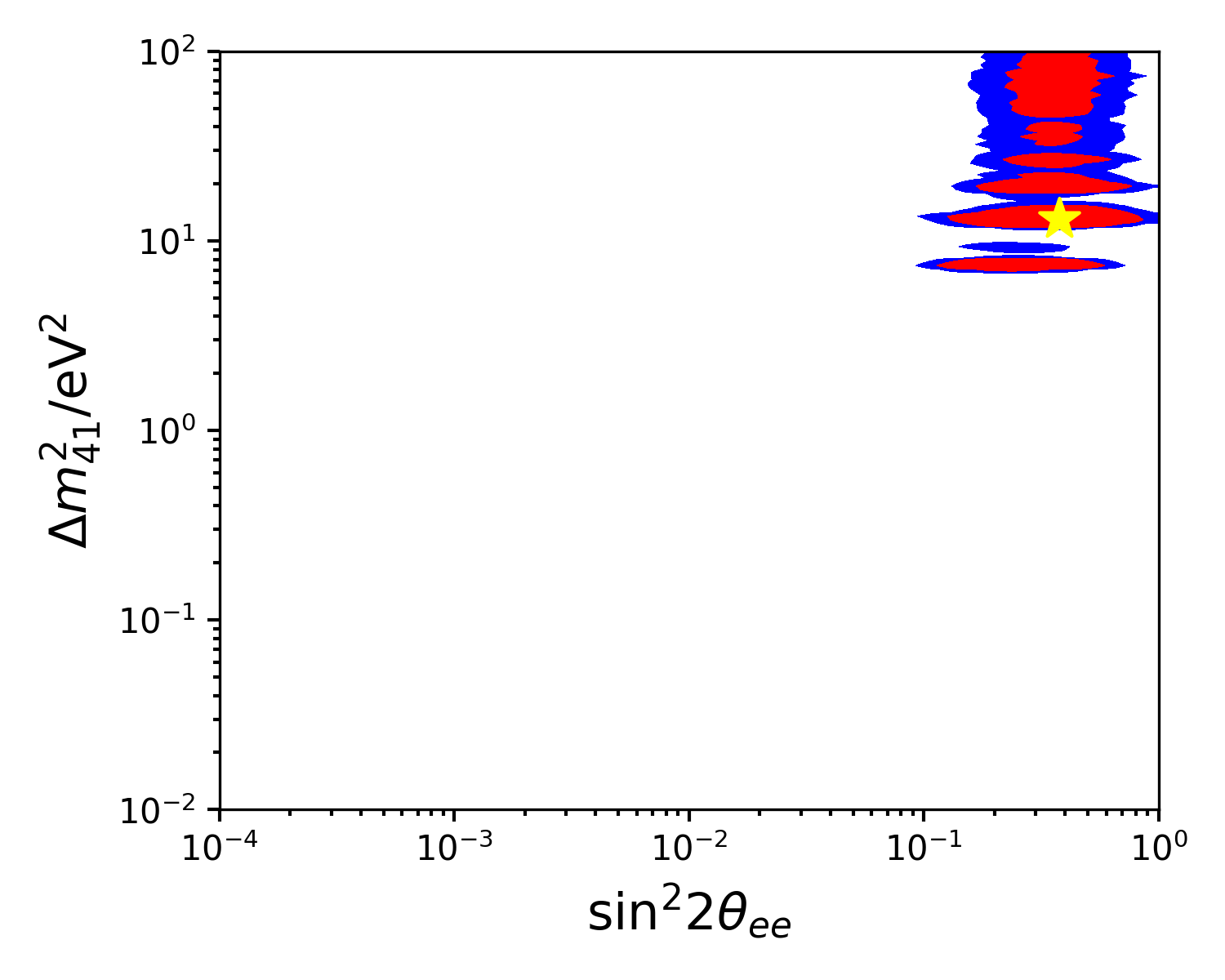}
        \caption{}
    \end{subfigure}

    \bigskip
    \begin{subfigure}{0.49\textwidth}
        \includegraphics[width=\textwidth]{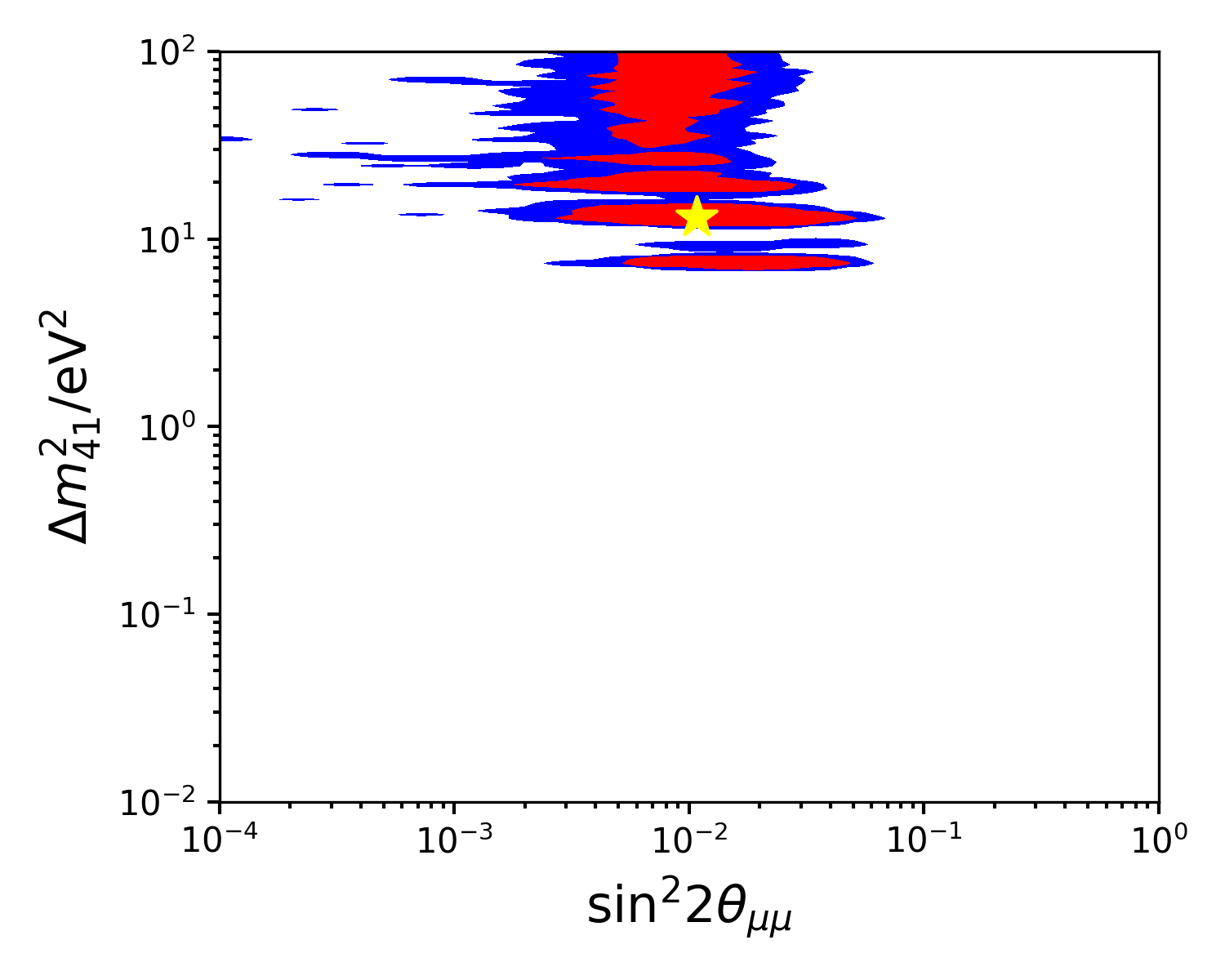}
        \caption{}
    \end{subfigure}
    \caption{Results of the Bayesian 3+1 global fits. We plot the confidence regions in terms of three different, but not independent, mixing parameters: (a) $\sin^2 2\theta_{\mu e}$, (b) $\sin^2 2\theta_{e e}$, and (c) $\sin^2 2\theta_{\mu \mu}$. The credible regions correspond to 90\% in red and 99\% in blue.}
    \label{fig:globalbayesresults}
\end{figure}

The results of the Bayesian fit can also be seen in \Cref{fig:globalbayesresults}. 
Compared to the frequentist fit in \Cref{fig:globalfreqresults}, we see good overlap between the regions, with the Bayesian contours being wider. 

\begin{figure}
    \centering
    \begin{subfigure}{0.49\textwidth}
        \includegraphics[width=\textwidth]{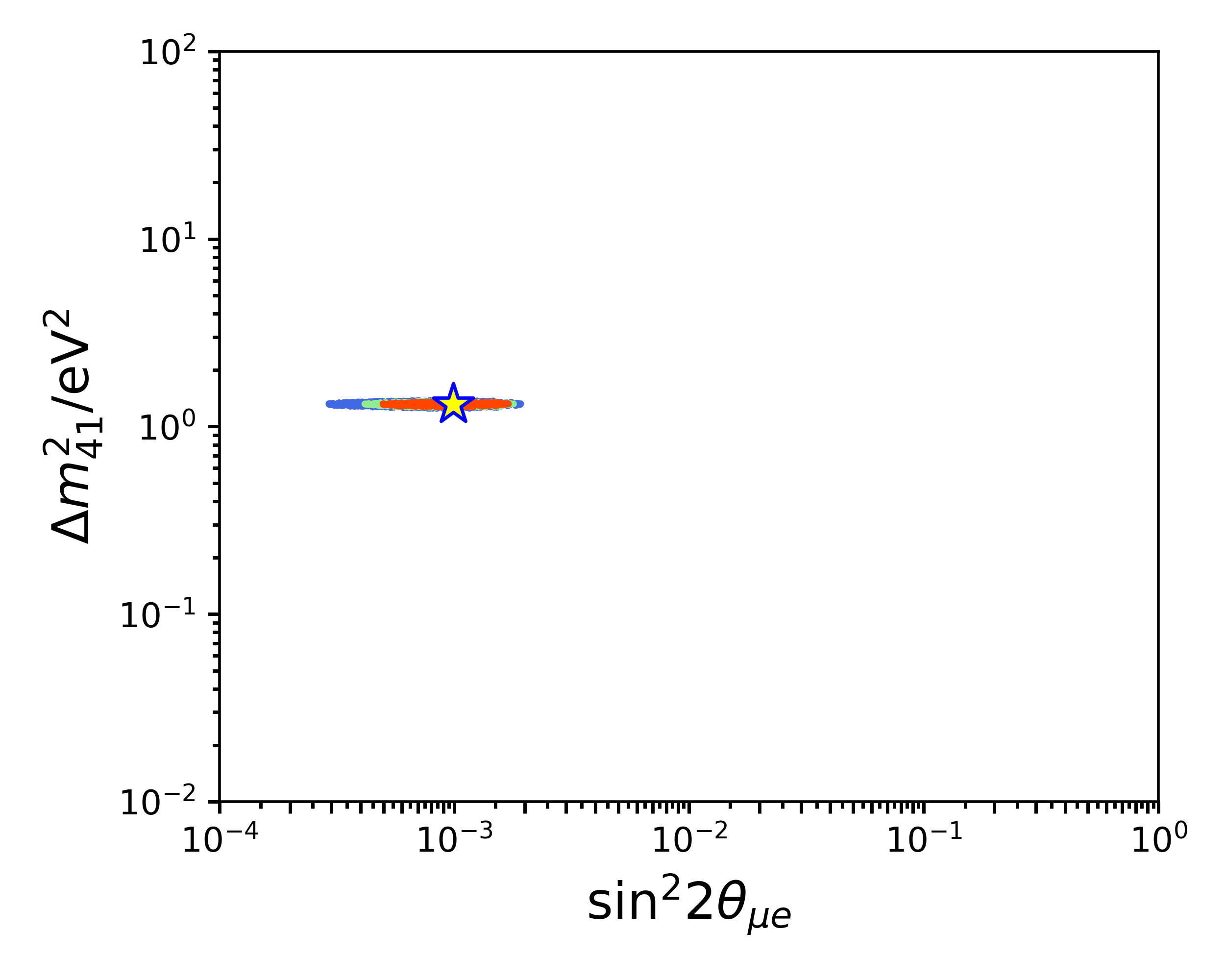}
        \caption{}
    \end{subfigure}
    \begin{subfigure}{0.49\textwidth}
        \includegraphics[width=\textwidth]{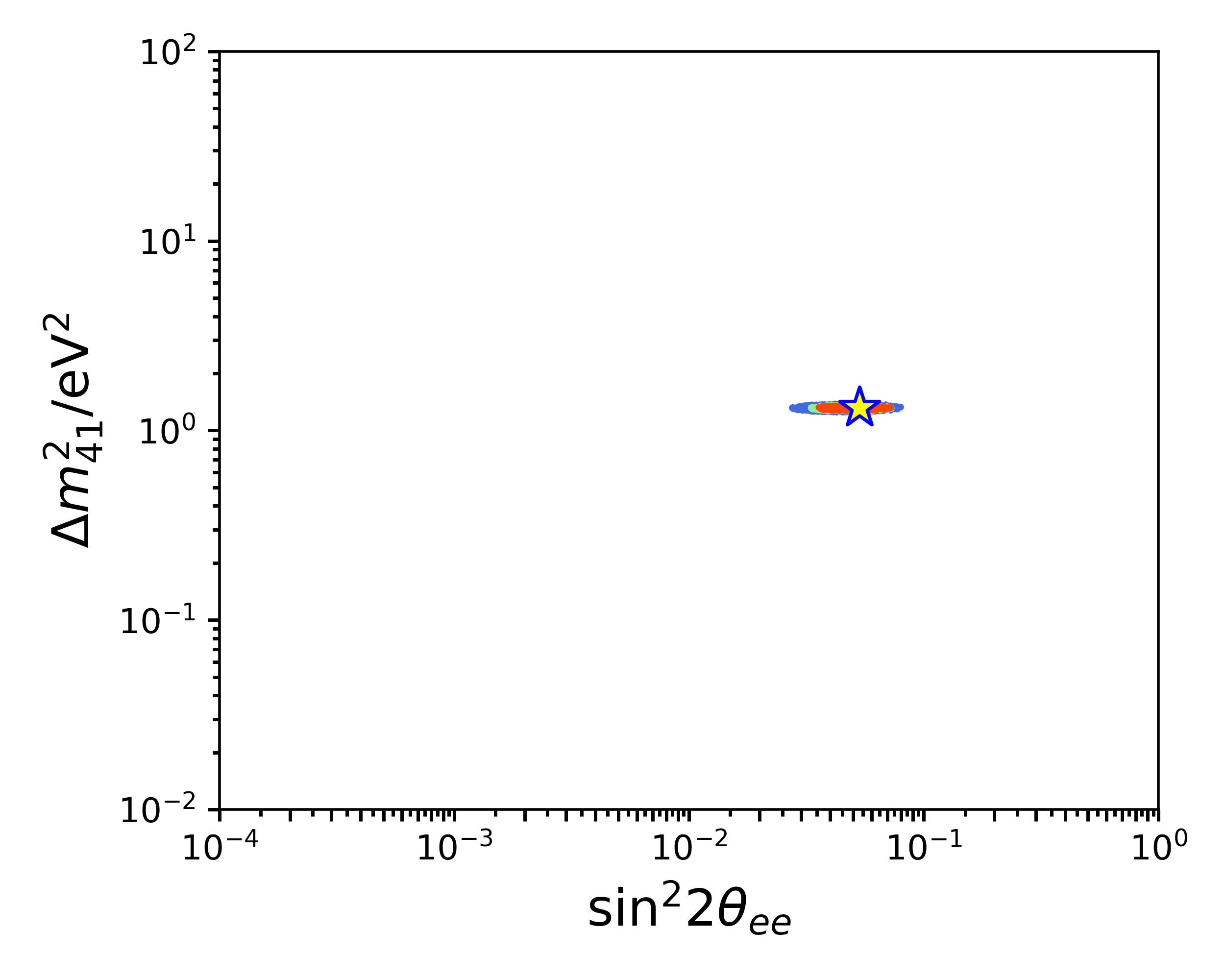}
        \caption{}
        \label{fig:physicsreportglobalfreqresultsee}
    \end{subfigure}

    \bigskip
    \begin{subfigure}{0.49\textwidth}
        \includegraphics[width=\textwidth]{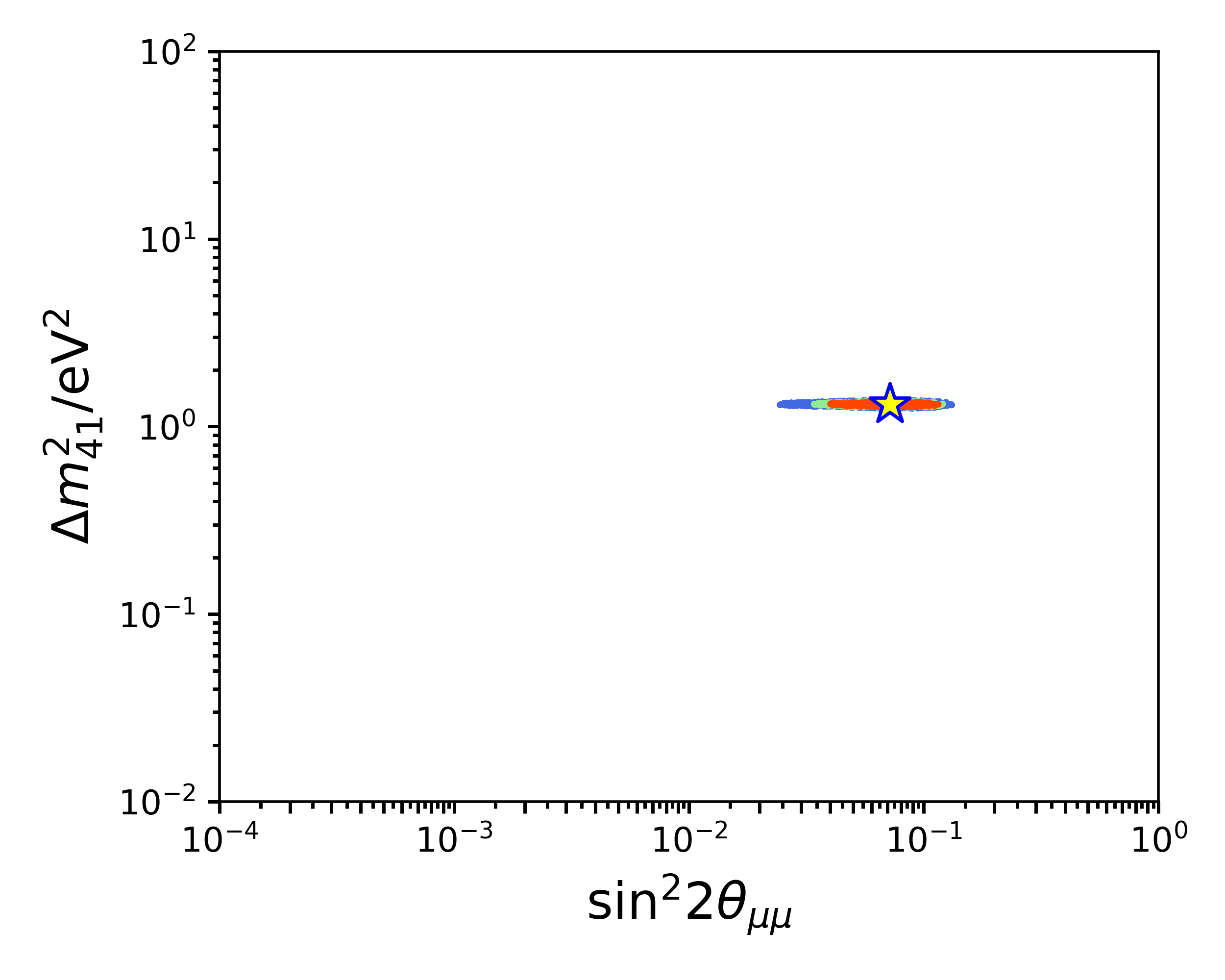}
        \caption{}
    \end{subfigure}
    \caption{The frequentist global fit results from our previous review in Ref.~\cite{Diaz:2019fwt}.}
    \label{fig:physicsreportglobalfreqresults}
\end{figure}

\begin{figure}
    \centering
    \includegraphics[width=0.5\textwidth]{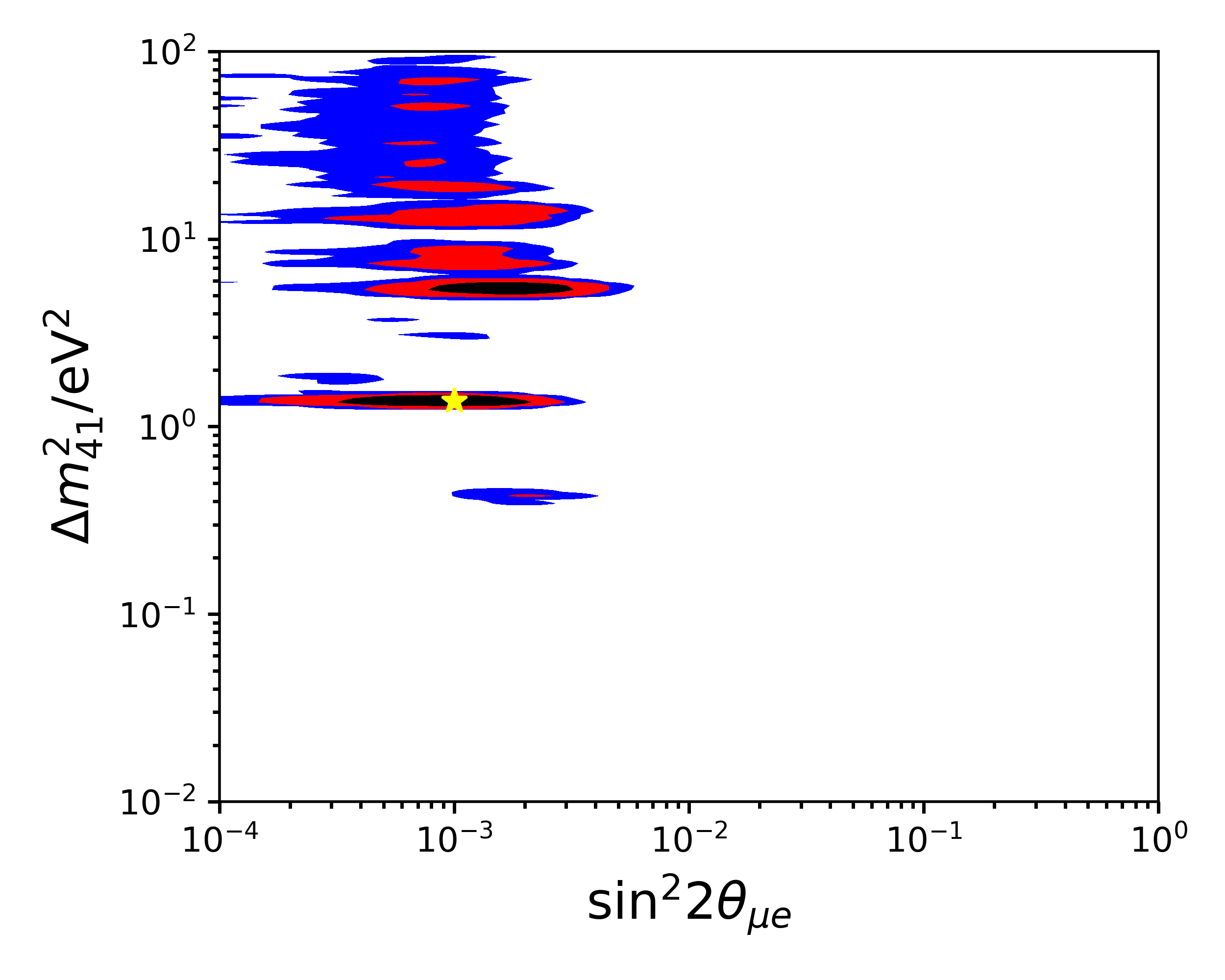}
    \caption{The Bayesian global fit results from our previous review in Ref.~\cite{Diaz:2019fwt}.}
    \label{fig:physicsreportglobalbayesresults}
\end{figure}

Before reporting on the tension, we would like to compare our current results with that from our previous review in Ref.~\cite{Diaz:2019fwt}.
We show in \Cref{fig:physicsreportglobalfreqresults} the frequentist 3+1 results from that analysis, and in \Cref{fig:physicsreportglobalbayesresults} the Bayesian results. 
Comparing our current frequentist results in \Cref{fig:globalfreqresults} and the previous results in \Cref{fig:physicsreportglobalfreqresults}, we find a substantial difference in the allowed $\Dmqfo$ values. 
We explain this change as being due to the addition of BEST, which had observed a $4\sigma$ deviation from the null model \cite{Barinov:2022wfh}. 
The previous best fit region is incompatible with the very strong signal observed by BEST, as can be seen by comparing \Cref{fig:physicsreportglobalfreqresultsee} and \Cref{fig:BESTfit}. 
Therefore, the $\Dmqfo \approx 1.32\ \eVq$ best fit island found in Ref.~\cite{Diaz:2019fwt} becomes disfavored. 
While no other islands were found in the previous frequentist fits, the Bayesian results from the previous fits, displayed in \Cref{fig:physicsreportglobalbayesresults}, revealed higher mass splittings which the current fits are compatible with. 

\begin{figure}
    \centering
    \begin{subfigure}{0.49\textwidth}
        \includegraphics[width=\textwidth]{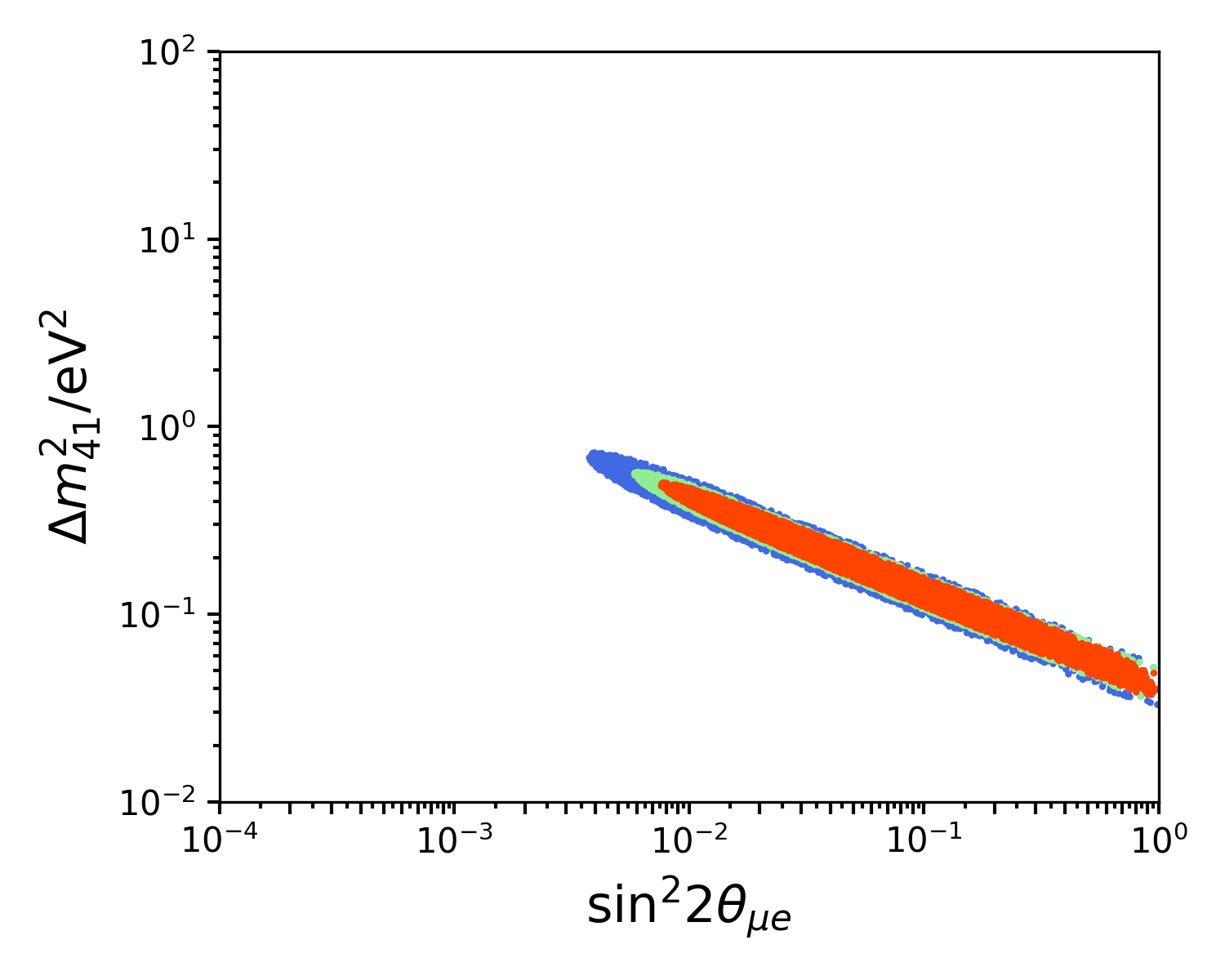}
        \caption{}
    \end{subfigure}
    \begin{subfigure}{0.49\textwidth}
        \includegraphics[width=\textwidth]{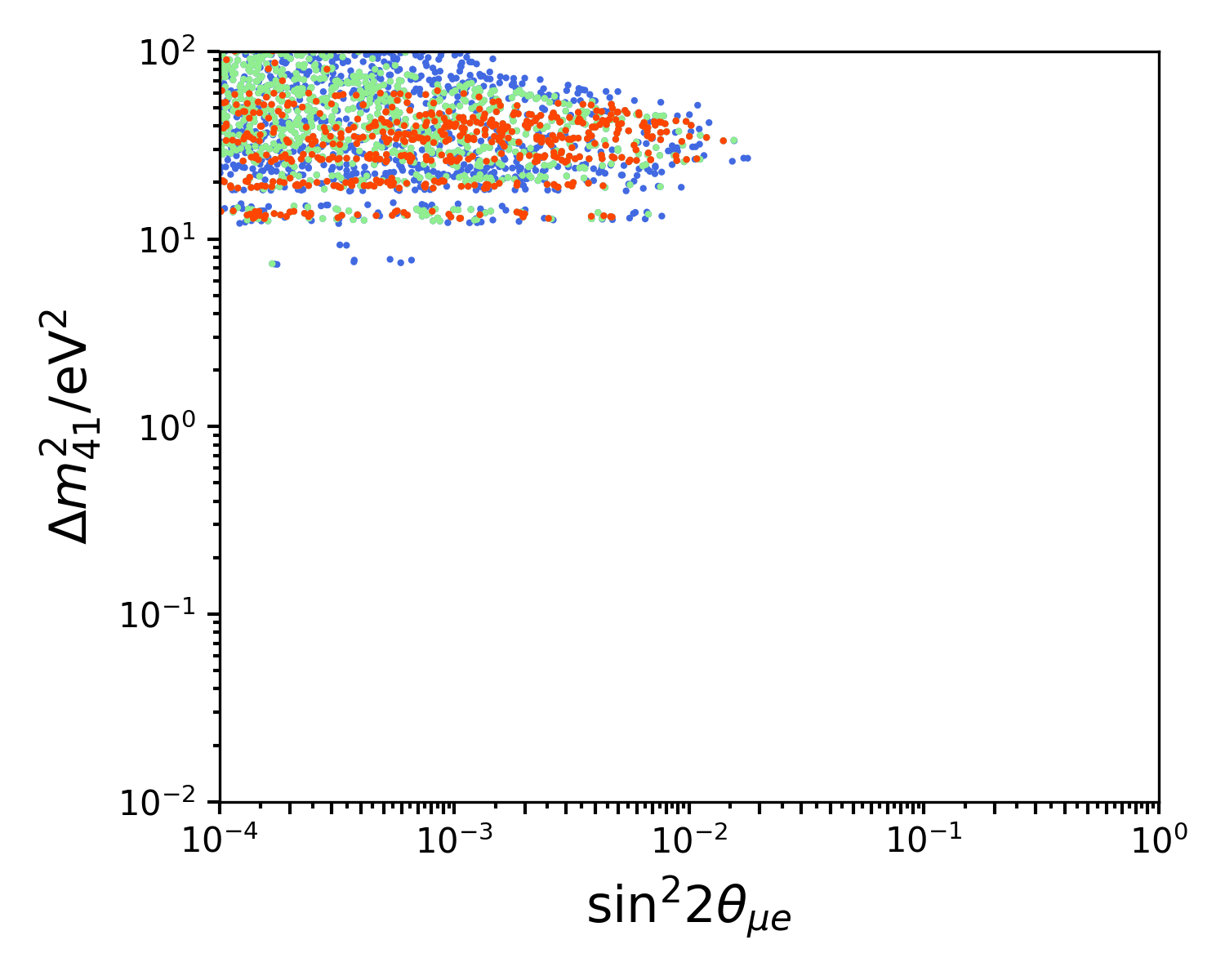}
        \caption{}
    \end{subfigure}

    \caption{3+1 fit results for (a) the appearance-only data sets and (b) the disappearance-only data sets.}
    \label{fig:3plus1appearanceanddisappearancefits}
\end{figure}

To test the internal consistency of this model, we calculate the tension by separating the data sets into two groups, as described above: the appearance and disappearance data sets. 
The appearance data are sensitive to the product $|U_{e4}||U_{\mu4}|$, while the disappearance data are individually sensitive to $|U_{e4}|^2$ or $|U_{\mu4}|^2$. 
The appearance-only and disappearance-only 3+1 fits are shown in \Cref{fig:3plus1appearanceanddisappearancefits}.
Visually, we can already see that these two subsets of the data do not agree in parameter space. 
To quantify this tension, we will use the PG test.
We find a test statistic value of $\chi^2_{\mathrm{PG}} = \chi^2_{\mathrm{glob}} - (\chi^2_{\mathrm{app}} + \chi^2_{\mathrm{dis}}) = 728 - (79+619) = 30$, with degrees of freedom $k=(2+3)-3=2$.
This gives a p-value of $p=3.1\times10^{-7} (5.1\sigma)$.
Clearly, there exists an internal inconsistency within the 3+1 model despite the overall preference that the data has for the 3+1 model over the null model.
This motivates the exploration of models more complex than the minimal 3+1 sterile neutrino model.

\subsection{\texorpdfstring{3+2}{3+2} Model}

We now consider the expanded 3+2 model, where we now have two sterile neutrino mass and weak states.
In addition to the three parameters introduced in the 3+1 model, (\Dmqfo, \Uef, \Umuf), the parameters \Dmqfiveo, \Uefive, \Umufive, and $\phi_{\mu e}$ are added in the 3+2 model, for a total of seven parameters.

\begin{figure}
    \centering
    \begin{subfigure}{0.49\textwidth}
        \includegraphics[width=\textwidth]{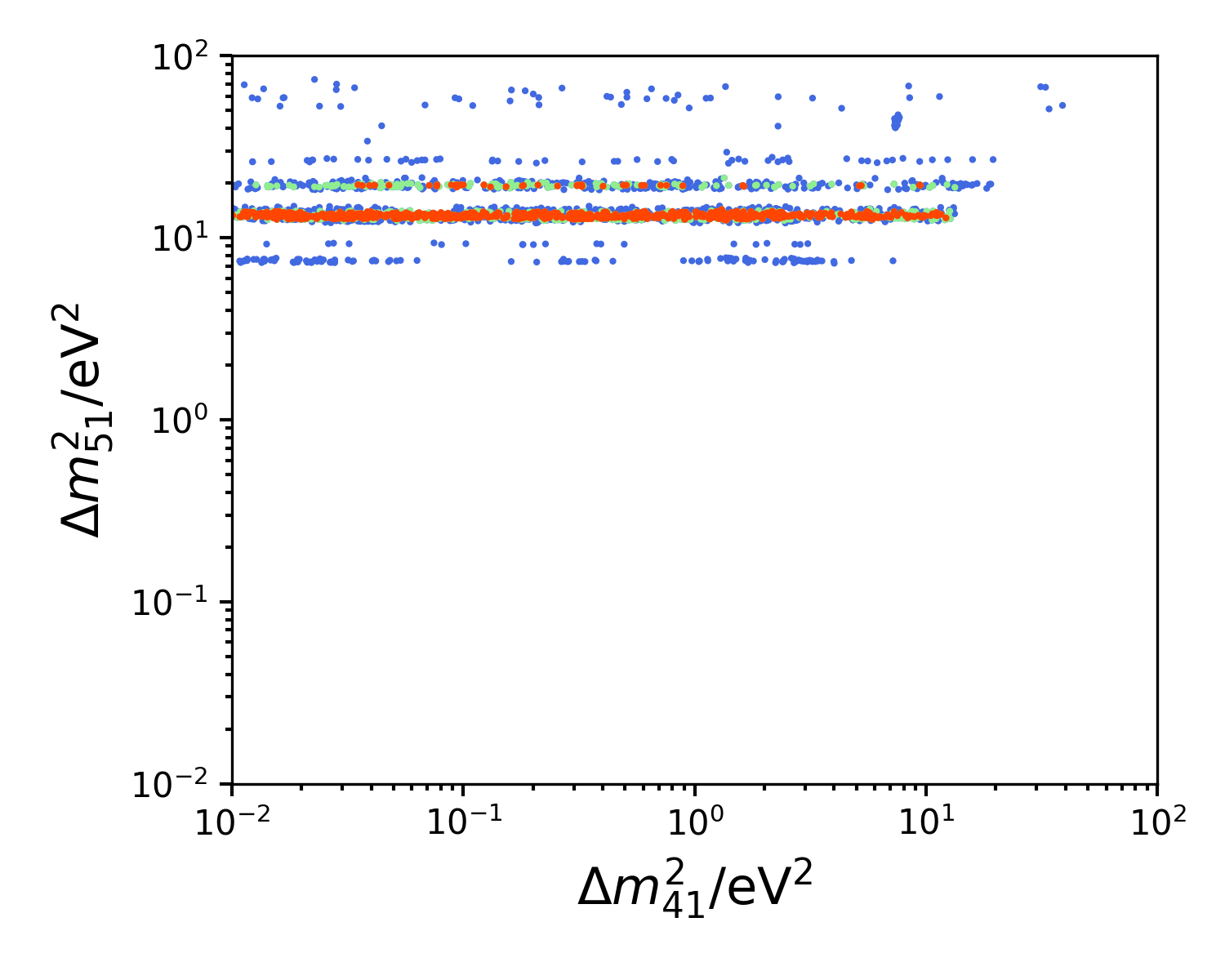}
        \caption{}
         \label{fig:3plus2dm241vsdm251}
    \end{subfigure}
    \begin{subfigure}{0.49\textwidth}
        \includegraphics[width=\textwidth]{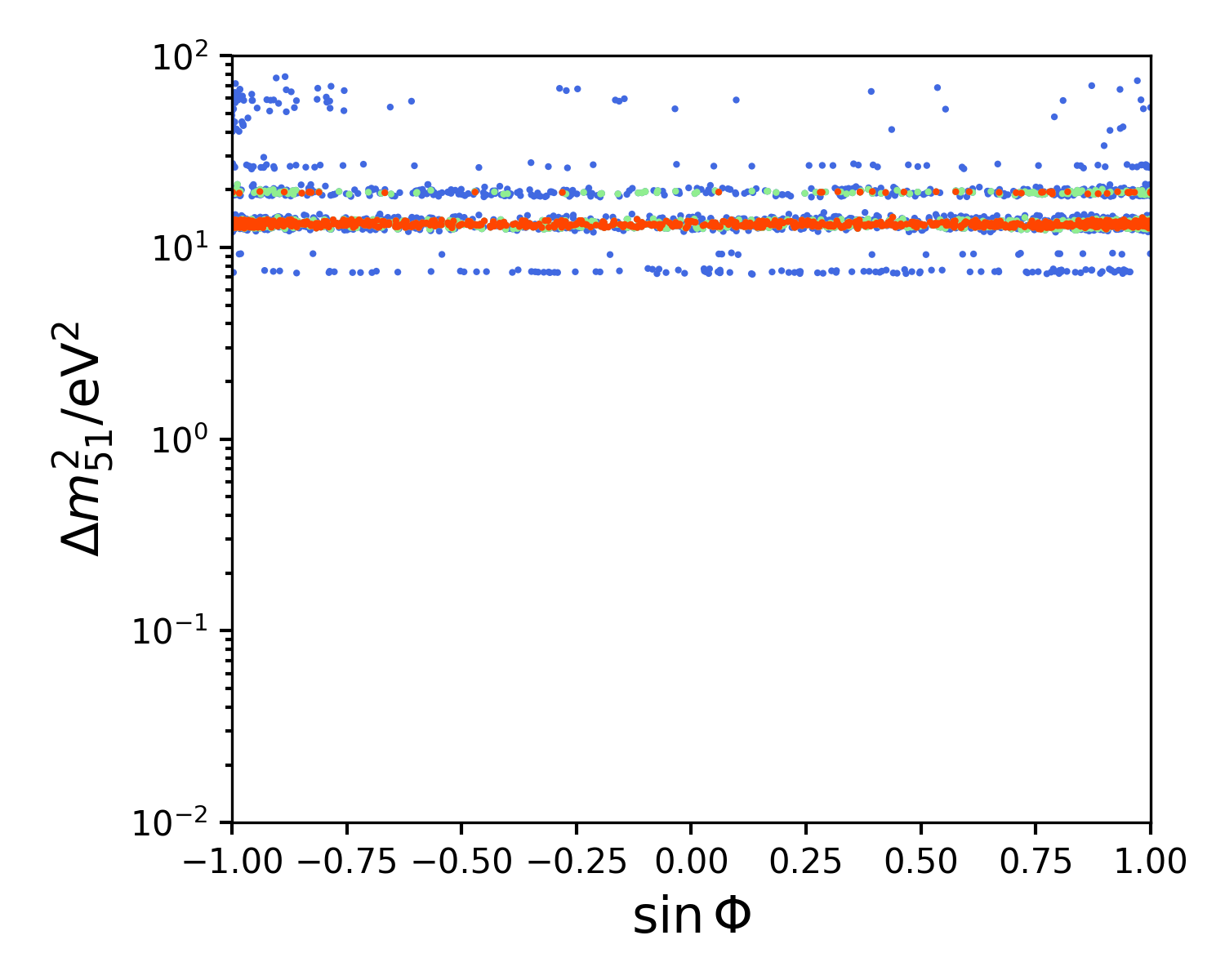}
        \caption{}
        \label{fig:3plus2dm2vssinphi}
    \end{subfigure}
    \caption{Results of the 3+2 global fits. (a) The best fit region in the \Dmqfo vs \Dmqfiveo plane. We can see that the fit finds preferred values of \Dmqfiveo, but leaves \Dmqfo unconstrained. The fitted values of \Dmqfiveo are consistent with the values of \Dmqfo found the in the 3+1 model. (b) The best fit region in the $\sin(\phi_{\mu e})$ vs \Dmqfiveo plane. We see no sensitivity to the additional CP-violating parameter $\phi_{\mu e}$.}
    \label{fig:3plus2fits}
\end{figure}

Our fit finds the following best fit parameters: 
$\Dmqfo = \SI{2.2e-3}{\eV\squared}, 
\Uef = 0.18, 
\Umuf = \num{3.2e-4}, 
\Dmqfiveo = \SI{13.1}{\eV\squared}, 
\Uefive = 0.30, 
\Umufive = 0.054$ and 
$\phi_{\mu e} = 0.78\pi$.
At these parameter points, we find the improvement of the model compared to the null to be at $\Delta \chi^2=51$.
We find, then, that the 3+2 model provides minimal improvement to the data, compared to the 3+1 model.
In \Cref{fig:3plus2dm241vsdm251}, we show the best fit regions in the \Dmqfo vs \Dmqfiveo plane.
We can see that the 3+2 model ends up fitting \Dmqfiveo to the \Dmqfo values found in the 3+1 fit in \Cref{fig:sin22thmueglobalfreqresults}, but leaves the other mass-squared splitting unconstrained. 
A similar feature is seen in \Cref{fig:3plus2dm2vssinphi}, where we plot the best fit region in the $\sin(\phi_{\mu e})$ vs \Dmqfiveo  plane.
Here, the data seems to be insensitive to the additional parameter $\phi_{\mu e}$.
We conclude, therefore, that the 3+2 model provides negligible improvement to the data compared to the 3+1 model.

We use the PG test to test the consistency of the 3+2 model, following the same procedure as for the 3+1 model.
We find a test statistic value of $\chi^2_{\mathrm{PG}} = \chi^2_{\mathrm{glob}} - (\chi^2_{\mathrm{app}} + \chi^2_{\mathrm{dis}}) = 728 - (75+613) = 40$, with degrees of freedom $k=(5+6)-7=4$.
This gives a p-value of $p=4.3\times10^{-8}\ (5.5\sigma)$.
Thus, we find that the 3+2 model actually worsens the tension compared to the 3+1 model. 

\subsection{3+1+Decay Model}

We now consider the 3+1+Decay model described in \Cref{sec:3plus1plusdecay}.
Here we have four dimensions to fit over. 
The first three are the same as the 3+1 case (\Dmqfo, \Uef, \Umuf), with the fourth being the decay width $\Gamma$ introduced in \Cref{sec:3plus1plusdecay}.

For this model, we show the results under two different conditions. 
In the first, we show the results with no bounds on $\Gamma$, to provide a fit that makes no model assumptions on the lifetime of $\nu_{4}$. 
In the second, we assume the decay width $\Gamma$ is given specifically by \Cref{eq:decayrate}, and apply the condition that $g^2 < 4\pi$. 
This is to ensure that, in that particular model of $\Gamma$,  we remain in the perturbative regime and that unitarity is preserved.
This leads to the restriction that $\gamma \leq m_{4}/4$, or $\tau \geq 4/m_{4}$.

\begin{figure}
	\centering
	\begin{subfigure}{0.32\linewidth}
		\centering
		\includegraphics[width=\linewidth]{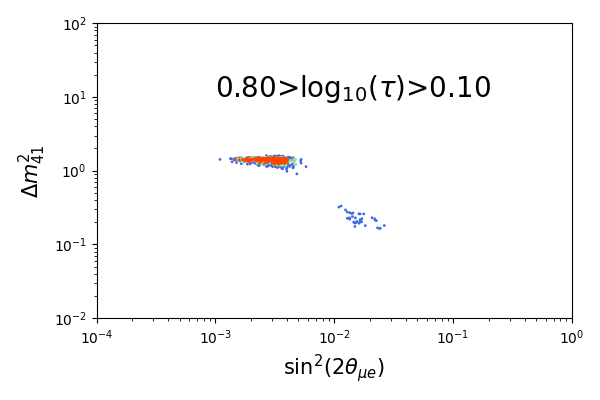}
	\end{subfigure}
	\hfill
	\begin{subfigure}{0.32\linewidth}
		\centering
		\includegraphics[width=\linewidth]{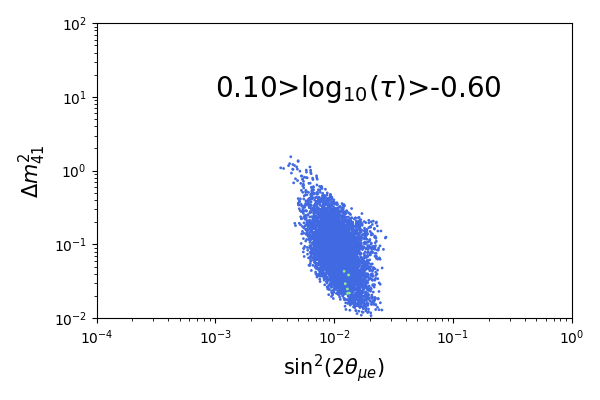}
	\end{subfigure}
	\hfill
	\begin{subfigure}{0.32\linewidth}
		\centering
		\includegraphics[width=\linewidth]{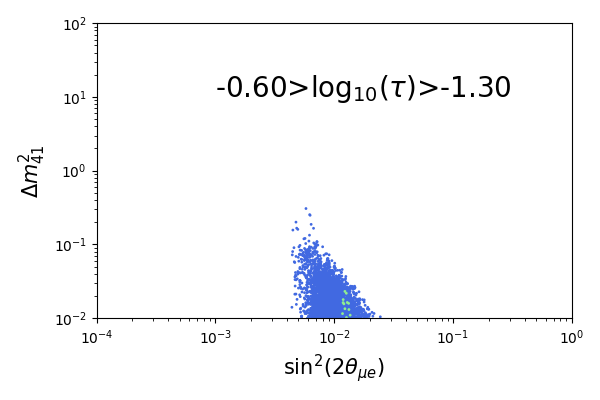}
	\end{subfigure}
	
	\bigskip
	\begin{subfigure}{0.32\linewidth}
	\centering
	\includegraphics[width=\linewidth]{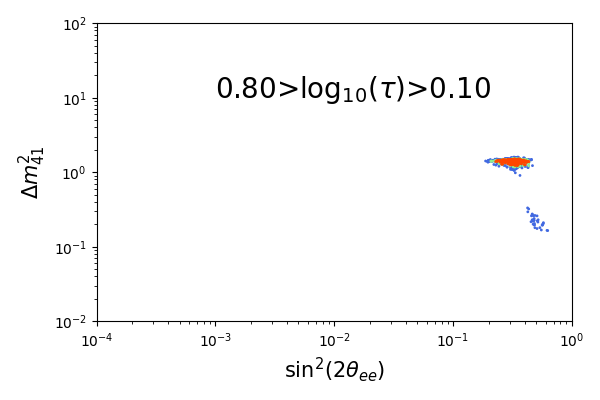}
	\end{subfigure}
	\hfill
	\begin{subfigure}{0.32\linewidth}
		\centering
		\includegraphics[width=\linewidth]{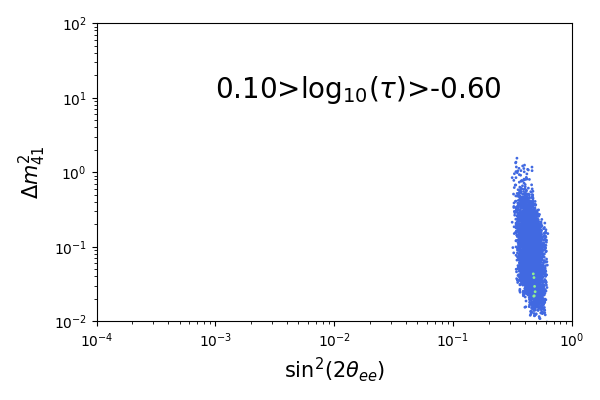}
	\end{subfigure}
	\hfill
	\begin{subfigure}{0.32\linewidth}
		\centering
		\includegraphics[width=\linewidth]{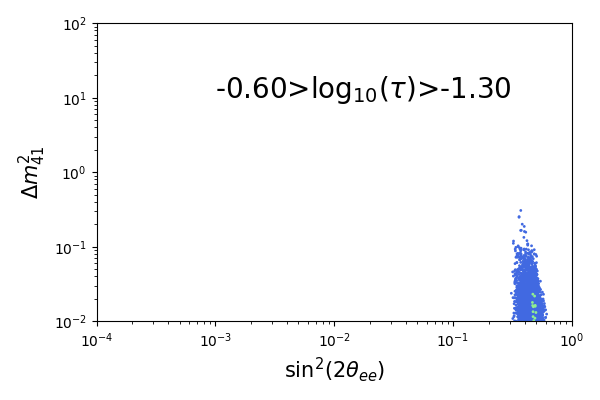}
	\end{subfigure}
	
	\bigskip
	\begin{subfigure}{0.32\linewidth}
		\centering
		\includegraphics[width=\linewidth]{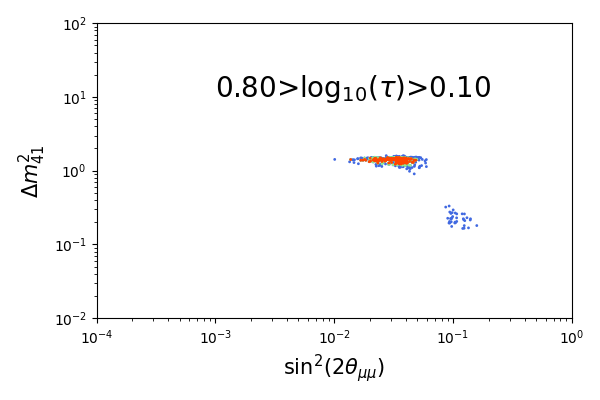}
	\end{subfigure}
	\hfill
	\begin{subfigure}{0.32\linewidth}
		\centering
		\includegraphics[width=\linewidth]{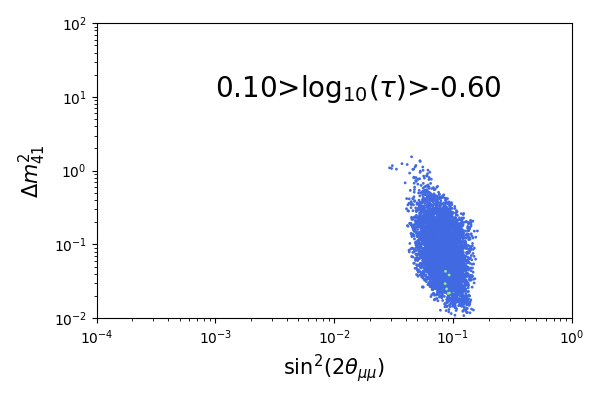}
	\end{subfigure}
	\hfill
	\begin{subfigure}{0.32\linewidth}
		\centering
		\includegraphics[width=\linewidth]{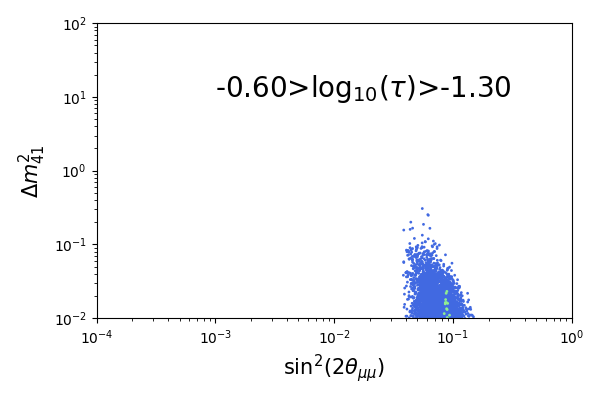}
	\end{subfigure}
	\caption{The results of the 3+1+Decay fits with no constraints on $g^2$. The first row shows the changing confidence regions for $\sin^2 2\theta_{\mu e}$, the second for $\sin^2 2\theta_{e e}$, and the third for $\sin^2 2\theta_{\mu \mu}$. The columns are sliced into different intervals of lifetimes $\tau$.}
	\label{fig:globalfitsdecay}
\end{figure}

For the first case, we find a best fit at $\Dmqfo = \SI{1.4}{\eV\squared}$, $\Uef = 0.3$, $\Umuf=0.09$, and $\tau = \SI{2.7}{\per\eV}$. 
Written in terms of mixing angles, the best fit is found at $\sin^2 2\theta_{\mu e} = 0.0027$, $\sin^2 2\theta_{e e} = 0.34$, $\sin^2 2\theta_{\mu \mu} = 0.030$.
The best fit confidence regions are shown in \Cref{fig:globalfitsdecay}, sliced in different intervals of $\tau$.
The contours are drawn assuming Wilks' theorem with three degrees of freedom. 
The first feature to notice is how different the preferred parameter space looks like when compared to the 3+1 case in \Cref{fig:globalfreqresults}. 
In particular, the mass splitting drops down nearly an order of magnitude. 
Interestingly, this brings the $\Dmqfo$ to a value near that which was found in the previous 3+1 fit shown in \Cref{fig:physicsreportglobalfreqresults}, but shifted to larger mixing angles. 
Another interesting feature is that the contour does not extend beyond $\tau > \SI{0.8}{\per\eV}$; therefore, there is a preference for a decaying sterile neutrino model versus a non-decaying sterile neutrino model.

\begin{figure}
	\centering
	\begin{subfigure}{0.49\linewidth}
		\centering
		\includegraphics[width=\linewidth]{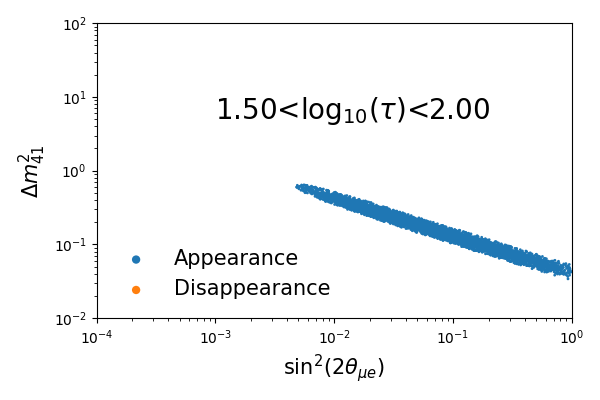}
	\end{subfigure}
	\hfill
	\begin{subfigure}{0.49\linewidth}
		\centering
		\includegraphics[width=\linewidth]{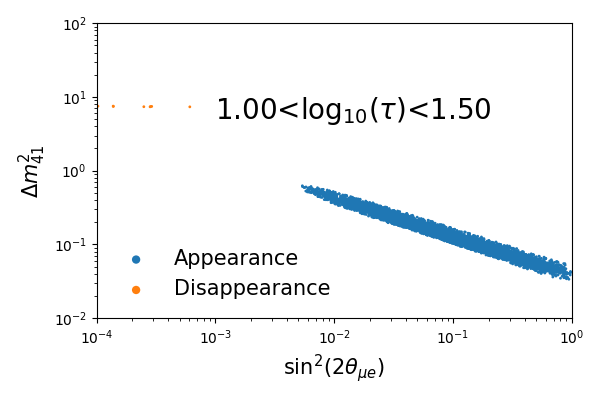}
	\end{subfigure}

	\bigskip
	\begin{subfigure}{0.49\linewidth}
		\centering
		\includegraphics[width=\linewidth]{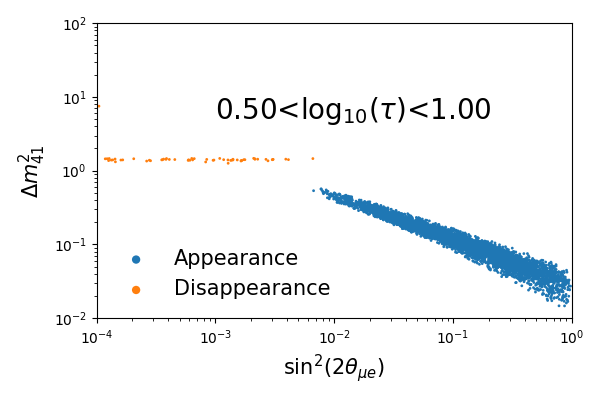}
	\end{subfigure}
	\hfill
	\begin{subfigure}{0.49\linewidth}
		\centering
		\includegraphics[width=\linewidth]{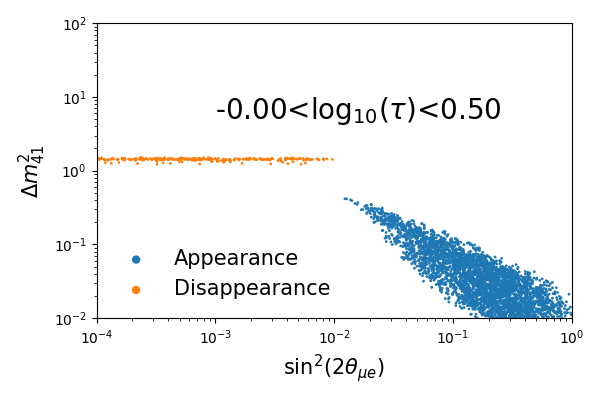}
	\end{subfigure}

	\bigskip
	\begin{subfigure}{0.49\linewidth}
		\centering
		\includegraphics[width=\linewidth]{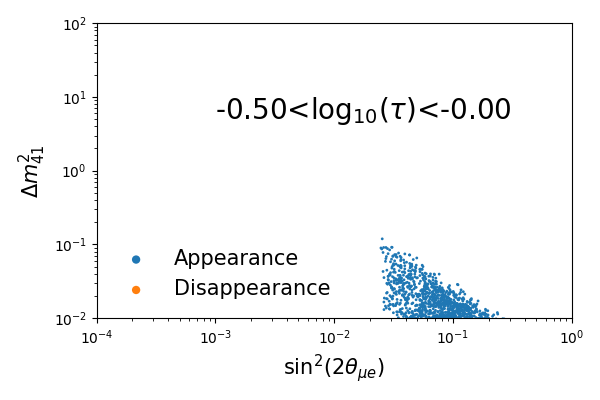}
	\end{subfigure}
	\caption{The confidence regions for the appearance and disapearance fits with the 3+1+Decay model at the 95\% confidence level. Here, we place no restrictions on $\Gamma$.}
	\label{fig:decaytension}
\end{figure}

 To test the tension in this model, we once again utilize the PG test by separating the experiments into an appearance data set and a disappearance data set. 
 We find a test statistic value of $\chi^2_{\mathrm{PG}} = \chi^2_{\mathrm{glob}} - (\chi^2_{\mathrm{app}} + \chi^2_{\mathrm{dis}}) = 710 - (79+611) = 19$, with degrees of freedom $k=(3+4)-4=3$.
 Compared to the 3+1 model, the tension is reduced from a $\chi^2_{\mathrm{PG}}$ of 30 to 19 with the 3+1+Decay model. 
 This reduced tension corresponds to a p-value of $p=2.7\times10^{-4}\ (3.6\sigma)$. 
 While this is a substantial improvement compared to the tension for the 3+1 model, this tension is nonetheless troublesome. 
 The confidence regions for the appearance and disappearance fits are shown in \Cref{fig:decaytension} for the 95\% confidence level.
 
\begin{figure}
        \centering
        \begin{subfigure}{0.49\textwidth}
                \includegraphics[width=\textwidth]{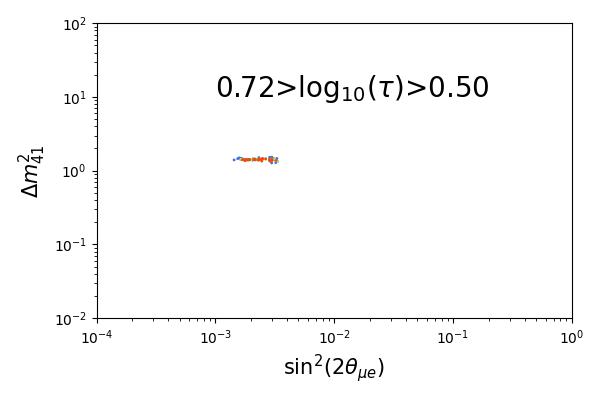}
        \end{subfigure}
        \begin{subfigure}{0.49\textwidth}
                \includegraphics[width=\textwidth]{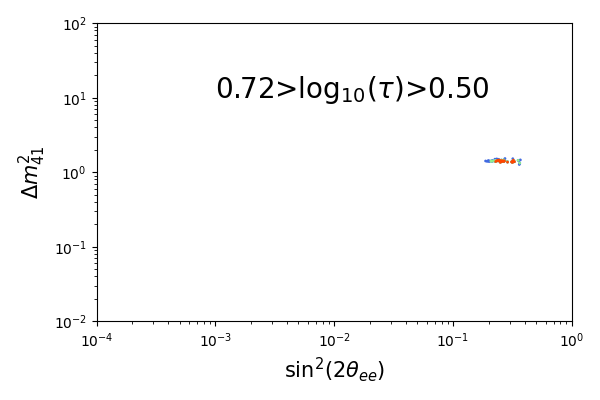}
        \end{subfigure}

        \bigskip
        \begin{subfigure}{0.49\textwidth}
                \includegraphics[width=\textwidth]{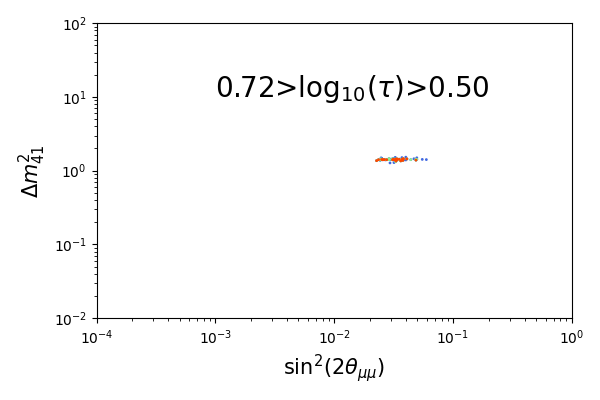}
        \end{subfigure}
        \caption{The best fit contours for the 3+1+Decay model when the constraint $g^2<4\pi$ is applied to the decay width $\Gamma$ given in \Cref{eq:decayrate}.}
        \label{fig:globalfitsdecayg2lessthan4pi}
\end{figure}

 For the case that we assume the specific decay width $\Gamma$ as given in \Cref{eq:decayrate} and restrict $g^2 < 4 \pi$, we obtain the best fit point  $\Dmqfo = \SI{1.35}{\eV\squared}$, $\Uef = 0.3$, $\Umuf=0.09$, and $\tau = \SI{3.5}{\per\eV}$. 
Written in terms of mixing angles, the best fit is found at $\sin^2 2\theta_{\mu e} = 0.0029$, $\sin^2 2\theta_{e e} = 0.32$, $\sin^2 2\theta_{\mu \mu} = 0.033$.
We show in \Cref{fig:globalfitsdecayg2lessthan4pi} the best fit contours of this model with the coupling constant constraint.
This time, only a single island exists, again with a preference for a finite lifetime.

\begin{figure}
        \centering
        \begin{subfigure}{0.49\textwidth}
                \includegraphics[width=\textwidth]{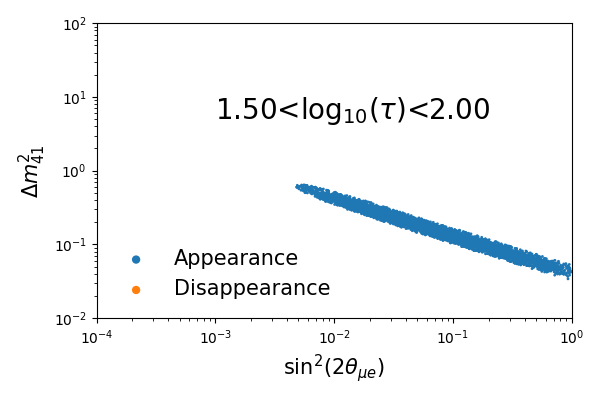}
        \end{subfigure}
        \begin{subfigure}{0.49\textwidth}
                \includegraphics[width=\textwidth]{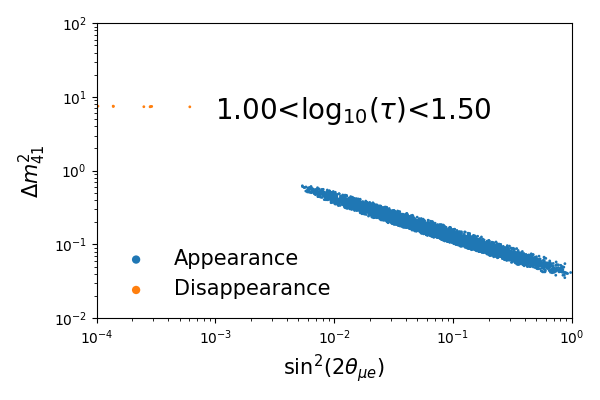}
        \end{subfigure}

        \bigskip
        \begin{subfigure}{0.49\textwidth}
                \includegraphics[width=\textwidth]{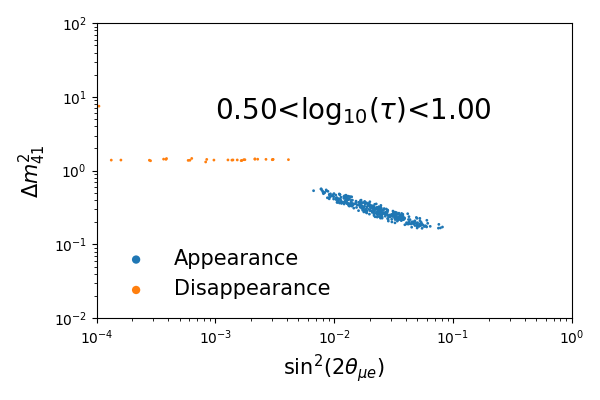}
        \end{subfigure}
        \caption{The confidence regions for the appearance and disappearance fits with the 3+1+Decay model at the 95\% confidence level. The restriction $g^{2} < 4\pi$ is placed here.}
        \label{fig:globalfitsdecayg2lessthan4pitension}
\end{figure}

We find a test statistic value of $\chi^2_{\mathrm{PG}} = \chi^2_{\mathrm{glob}} - (\chi^2_{\mathrm{app}} + \chi^2_{\mathrm{dis}}) = 711 - (79+612) = 19$, with degrees of freedom $k=(3+4)-4=3$. This gives the same p-value as the case with the unrestricted $\Gamma$.
So while the preferred parameter space is significantly restricted when we place the bound $g^2<4\pi$, the best fit point remains similar and the relief in tension is the same.
A comparison of the appearance and disappearance fits can be seen in \Cref{fig:globalfitsdecayg2lessthan4pitension} for the 95\% confidence level.

\section{Discussion}

The results of our global fits above give us a very confusing picture of the sterile neutrino model.
We find that the data observed strongly prefer a minimal sterile neutrino mode, the  3+1 model, versus the SM picture; but irreconcilable tension exists within that model.
Adding a second sterile state to the model, the 3+2 model, provides negligible improvement to the fit and worsens the tension.
Expanding the picture into a more exotic model, the 3+1+Decay, provides some relief to the tension, but not enough to give us ease.

While simple sterile neutrino models are not able to give us a consistent picture, the various phenomena that can be explained by sterile neutrinos continues to encourage the development of novel models and new experimental techniques.
In particular, we notice that while there exists experiments that observe something like $\numu\to\nue$ and $\nue\to\nue$ oscillations, there still has yet to be an experiment that observes $\numu\to\numu$ oscillations.
Further, all the experiments listed above conduct measurements with vacuum oscillations.
To continue exploring the sterile neutrino hypothesis, it would be interesting to search in unique ways.
In the remaining chapters, we present an expansion of a sterile neutrino analysis that performs its search at a substantially higher energy than previous sterile neutrino searches and utilizing non-vacuum phenomena.

    \chapter{Summary of the Previous Sterile Neutrino Search in IceCube}
\label{ch:IceCube}

\section{IceCube in a Nutshell}

The IceCube Neutrino Observatory is a gigaton-scale neutrino detector embedded within the antarctic ice at 1450--2450 m below the surface \cite{IceCube:2016zyt}.
The flagship purpose of IceCube is to search for point-sources of neutrinos outside of our solar system.
For this thesis, though, we will restrict the discussion to the detector itself and the sterile neutrino analysis conducted with IceCube.

\begin{figure}
    \centering
    \includegraphics[height=.35\textheight]{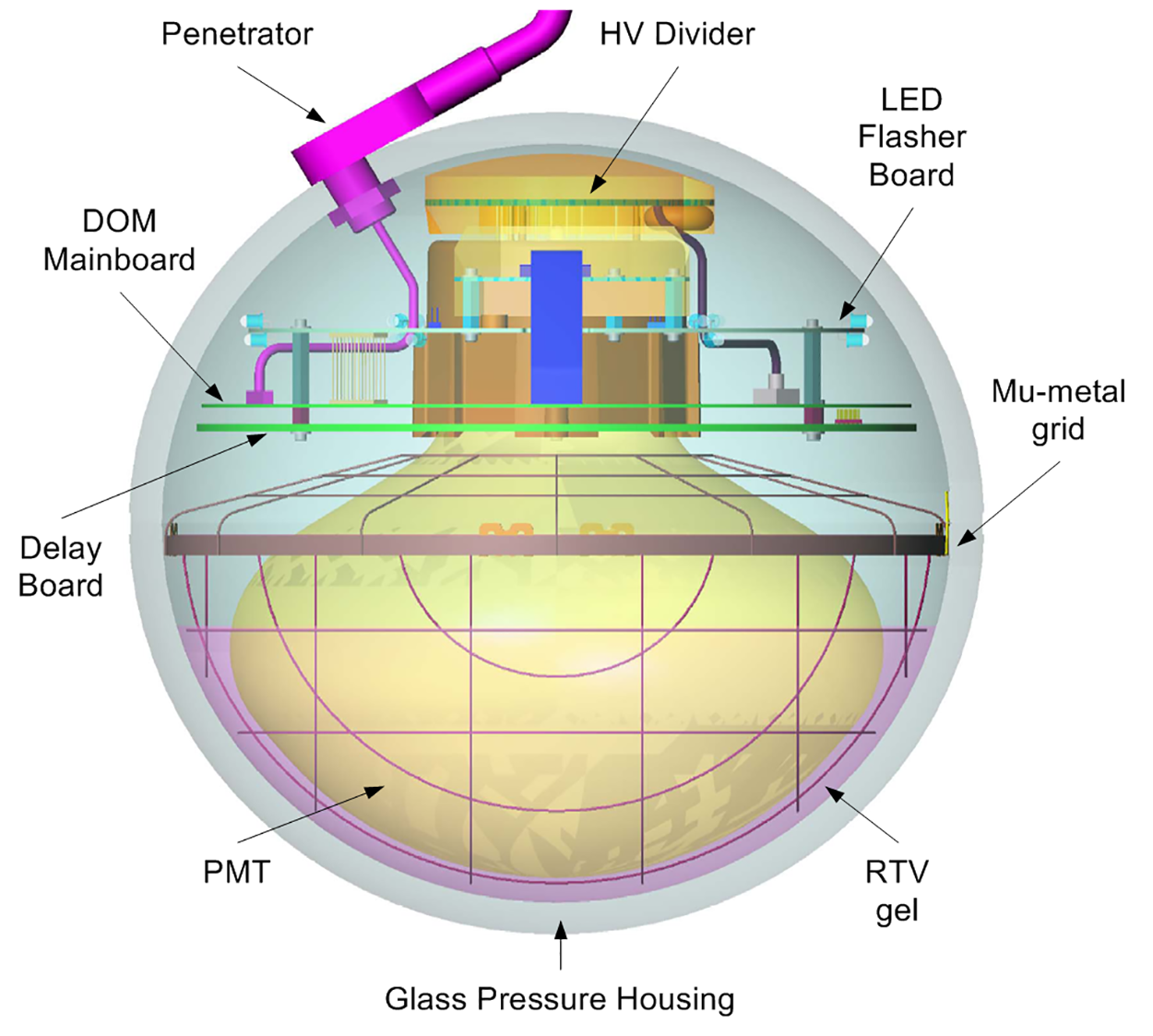}
    \caption{A schematic of one of the 5160 DOMs in IceCube. Note that the PMT points downwards.}
    \label{fig:DOM}
\end{figure}

\begin{figure}
    \centering
    \includegraphics[height=0.49\textheight]{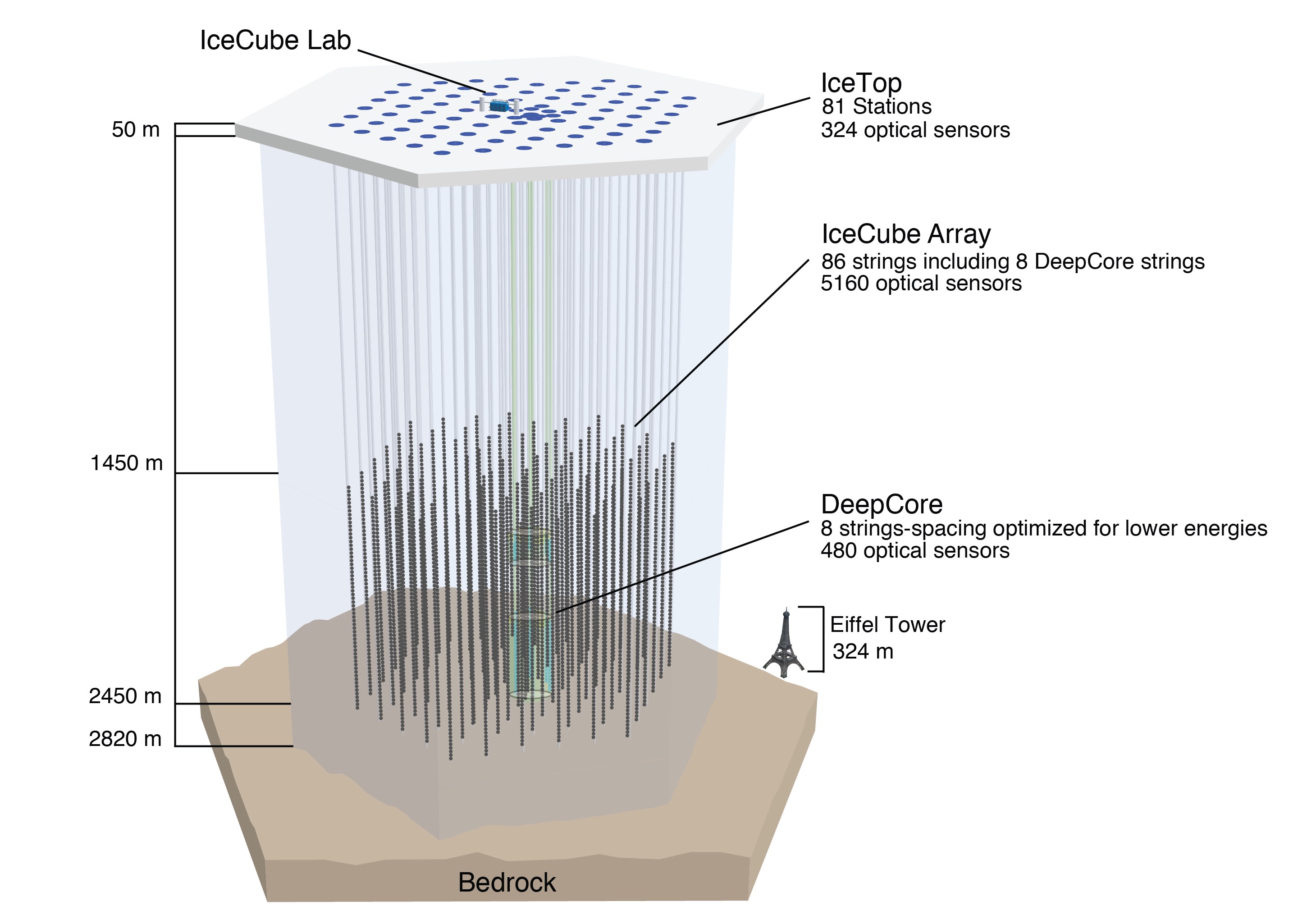}
    \caption{A diagram of the IceCube detector showing the distribution of the 86 strings and 5160 DOMs.}
    \label{fig:ICdetector}
\end{figure}

The detector is composed of 5160 digital optical modules (DOMs), which are the detector units embedded within the ice.
Each DOM contains a photomultiplier tube (PMT) which points downwards, as well as a signal digitizer board.
A schematic is shown in \Cref{fig:DOM}.
These DOMs are placed on 86 vertical strings, with 60 DOMs on each. 
The primary array of strings (78 strings) are arranged in an approximately triangular grid with \SI{125}{\meter} horizontal spacing, and a vertical spacing of \SI{17}{\meter} between DOMs. 
A subset of DOMs (8 strings), called DeepCore, are placed closer together, with an average inter-string spacing of \SI{72}{\meter} and vertical DOM separation between 7 and \SI{10}{\meter}. 
The dimensions of the detector were optimized to search for high-energy low-flux astrophysical neutrinos.
A diagram of the detector is shown in \Cref{fig:ICdetector}.

\section{Sterile-Induced Neutrino Oscillation in Matter}
\label{sec:sterileneutrinooscillationinmatter}

In addition to astrophysical neutrinos, IceCube also detects neutrinos that are produced in the Earth's atmosphere and later interact near the detector.
As will be discussed in the next section, these atmospheric neutrinos are used to conduct a sterile neutrino search.
This search utilizes the fact that these high-energy atmospheric neutrinos can travel through the Earth's matter before reaching the detector, and that the presence of a sterile neutrino can modify matter-propagating neutrino oscillations beyond the modification expected from the SM (as in \Cref{sec:matteroscillation}).

In this section, we discuss how these matter oscillations can be modified by the existence of a sterile neutrino.
For this discussion, we again assume that the neutrinos are traveling through a medium of constant density.

Like in \Cref{sec:matteroscillation}, we start with the effective Hamiltonian in the flavor basis, 
\begin{equation}
    \mathcal{H}_{F}=\frac{1}{2E}(U \mathbb{M}^{2} U^{\dagger} + \mathbb{A}).
\end{equation}
Here, 
\begin{equation}
    \mathbb{M}^{2} =
    \begin{pmatrix}
        0 & 0 & 0 & 0\\
        0 & \Delta m_{21}^{2} & 0 & 0 \\
        0 & 0 & \Delta m_{31}^{2} & 0 \\
        0 & 0 & 0 & \Delta m_{41}^{2}
    \end{pmatrix}, 
    \quad
    \mathbb{A} =
    \begin{pmatrix}
        A_{\textrm{CC}} + A_{\textrm{NC}} & 0 & 0 & 0\\
        0 & A_{\textrm{NC}} & 0 & 0\\
        0 & 0 & A_{\textrm{NC}} & 0\\ 
        0 & 0 & 0 & 0
    \end{pmatrix},
\end{equation}
where 
\begin{equation}
    A_\textrm{CC} \equiv 2 E V_\textrm{CC} = 2 \sqrt{2} E G_{F} N_e,
    \quad
    A_\textrm{NC} \equiv 2 E V_\textrm{NC} = -\sqrt{2} E G_{F} N_n,
\end{equation}
and $N_n$ is the neutron density.
In $\mathbb{A}$, we kept the NC terms. Note that the sterile component has neither CC nor NC terms. 

In a two neutrino model, where we are considering only $\numu-\nu_s$ oscillations, we can simplify to 
\begin{equation}
    \mathbb{M}^{2} =
    \begin{pmatrix}
        0 & 0 \\
        0 & \Dmq \\
    \end{pmatrix}, 
    \quad
    \mathbb{A} =
    \begin{pmatrix}
        A_{\textrm{NC}} & 0 \\
        0 & 0
    \end{pmatrix}.
\end{equation}
We can see that our Hamiltonian ends up looking nearly identical to that derived in \Cref{sec:matteroscillation}, so that the derived oscillation parameters can be obtained by making the replacement $A_{\textrm{CC}}\to A_{\textrm{NC}}$, or $N_e\to - N_n/2$, in \Crefrange{eq:229}{eq:234}.

In this sterile-enhanced matter oscillation scenario, the resonant energy $E_{\nu}^{\textrm{R}}$ would be found at 
\begin{equation}
    E_{\nu}^{\textrm{R}}= 
    -\frac{ \Delta m^{2} \cos 2 \theta} {\sqrt{2} G_\textrm{F} N_n }.
\end{equation}
If we assume that $\Dmq>0$ and $\theta<\pi/4$, then we get a negative value for $E_{\nu}^{\textrm{R}}$. What this means is that the muon neutrino resonance can only be observed for antineutrinos, and not neutrinos. 

In reality, a full oscillation treatment for four neutrinos propagating through varying density is required.
For the work here and the remaining chapters, the neutrino propagation through the Earth is numerically calculated using the open-source neutrino oscillation calculator \texttt{nuSQuIDS} \cite{Arguelles:2021twb}, which we describe more of later. 

\section{8-year Sterile Neutrino Search}
\label{sec:MEOWS}

As a result of the effect of matter oscillations discussed in \Cref{sec:sterileneutrinooscillationinmatter}, a sterile neutrino search can be conducted with IceCube that would not be possible with vacuum oscillations. 
Such an analysis has already been started and published \cite{IceCube:2020phf,IceCube:2020tka}, which we will summarize in this section. 
In \Cref{ch:MEOWSplusth34,ch:ICresults}, we will discuss the continuation of this work and the final work of this thesis. 

As discussed in the previous section, the existence of a sterile neutrino affects the oscillation of the active neutrinos as they propagate through matter. 
As a reminder: regardless of the vacuum values of the mixing angles and mass-squared splittings, there exists a resonant energy for a given matter density that would result in maximal mixing for either neutrinos or antineutrinos. 
Refs.~\cite{IceCube:2020phf,IceCube:2020tka} exploits this at IceCube, using the atmospheric muon antineutrinos produced around the Earth and which propagate through the Earth's matter towards IceCube. 
That search is called Matter Enhanced Oscillations With Steriles (MEOWS).
While ``MEOWS'' is not an official name, we will refer to the analysis as such in this thesis.

\subsection{\texorpdfstring{\numu}{Muon Neutrino} Flux}

\begin{figure}
\centering
    \includegraphics[width=0.57\textwidth]{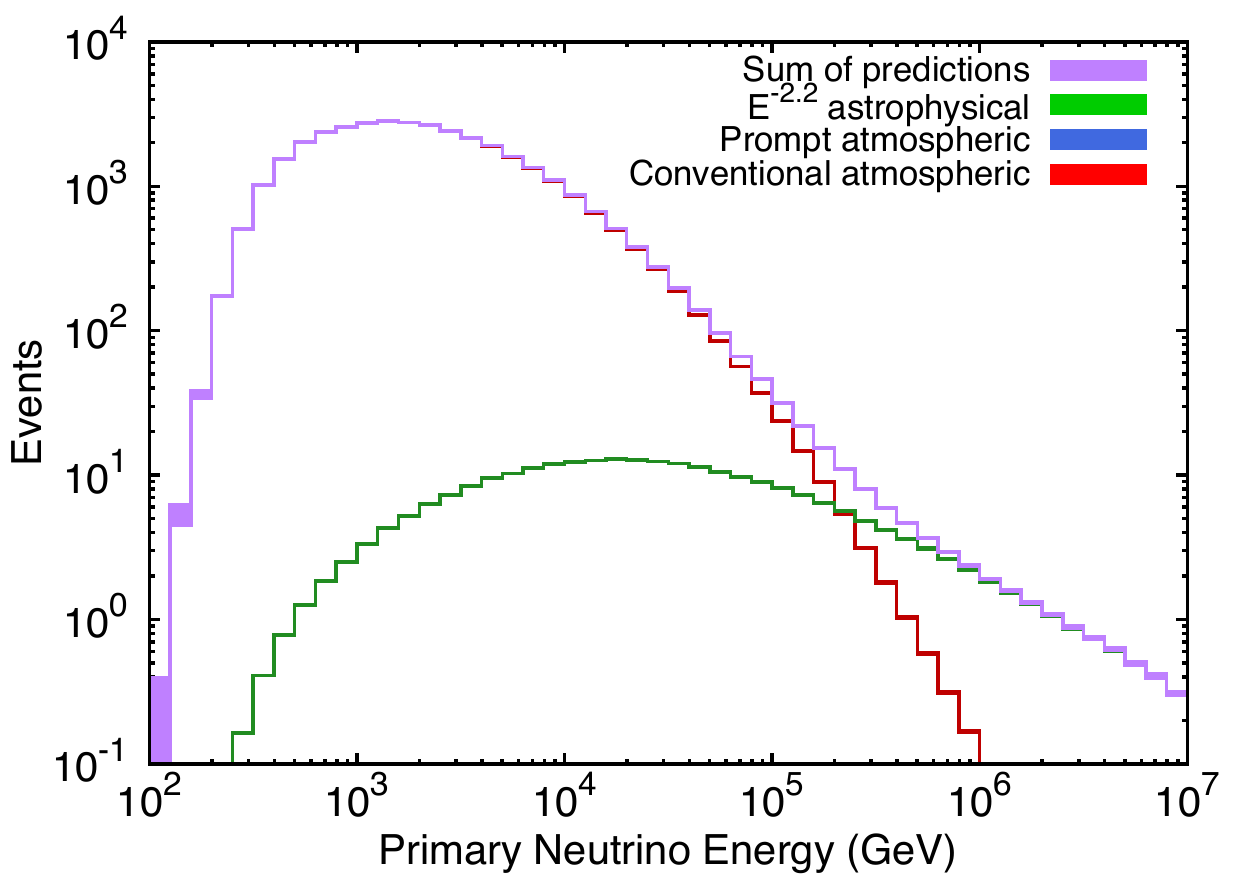}
    \caption{The best fit flux templates in the northern sky astrophysics \numu search in Ref.~\cite{IceCube:2015qii}. The atmospheric component is the sum of the conventional and prompt components, which we define later in \Cref{sec:atmoshpericneutrinos}. The prompt component fitted to zero. Supplemental figure from Ref.~\cite{IceCube:2015qii}.}
    \label{fig:eventrate}
\end{figure}

\Cref{fig:eventrate} shows the best fit template event rates for a northern sky astrophysical muon neutrino search conducted at IceCube \cite{IceCube:2015qii}. 
The atmospheric muon neutrino event distributions (in red), are seen to peak at $\SI{\sim 1}{\TeV}$.

\begin{figure}
    \centering
    \includegraphics[width=.9\textwidth]{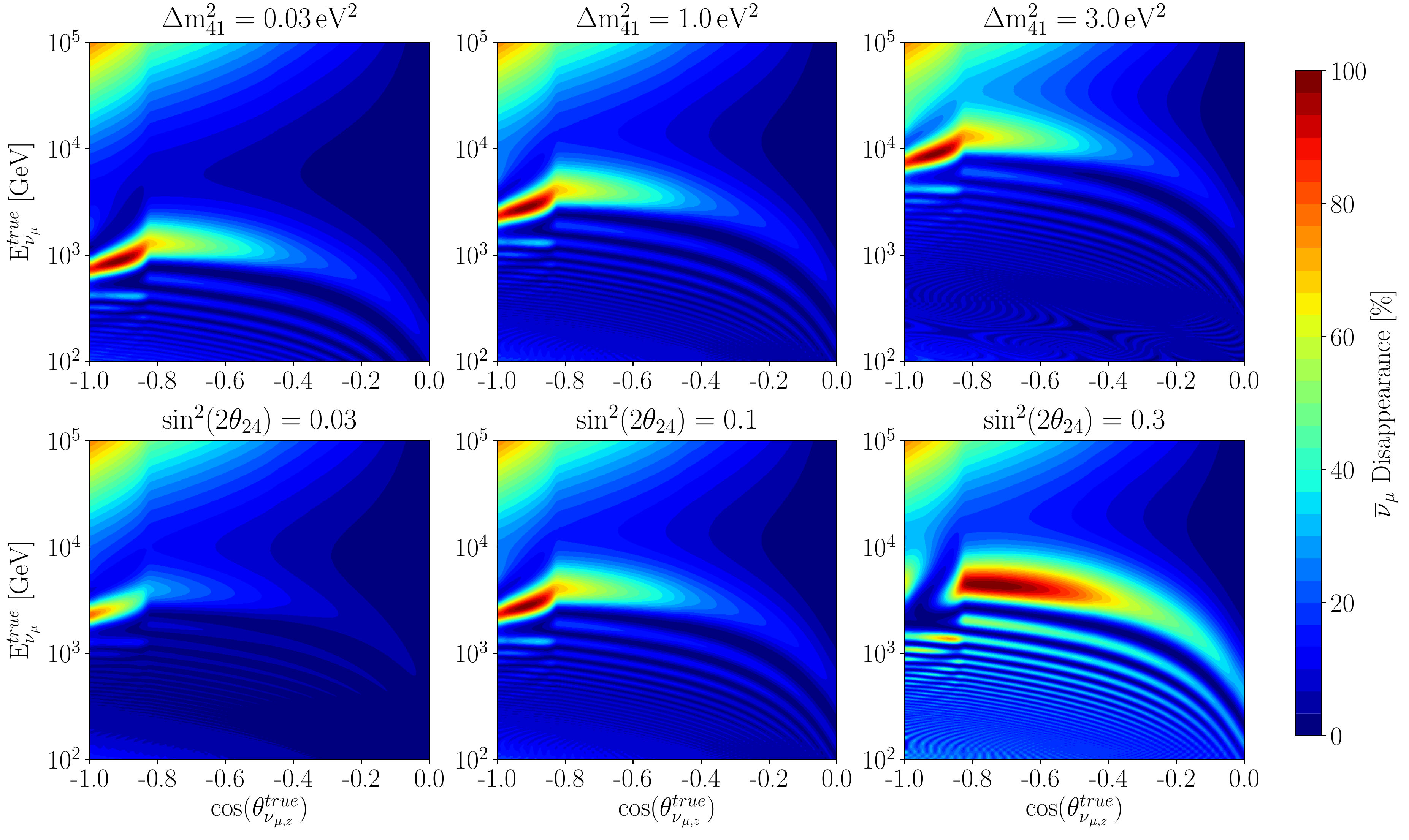}
    \caption{The expected disappearance of atmospheric \numubar at the IceCube detector for various sterile neutrino parameters \Dmqfo and \sinsqtthtf. The top row keeps the mixing angle $\sinsqtthtf = 0.1$ fixed with increasing \Dmqfo from left to right. The bottom row keeps $\Dmqfo= 1\ \eVq$ fixed with increasing \sinsqtthtf from left to right. Figure taken from Ref.~\cite{IceCube:2020tka}.}
    \label{fig:MEOWSoscillograms}
\end{figure}

In \Cref{fig:MEOWSoscillograms}, we show a series of oscillograms.
Each plot shows the disappearance probability of atmospheric $\numubar$ after traversing through `the Earth's matter, for some sterile parameters \Dmqfo and \sinsqtthtf.
The x-axis gives the direction from which the neutrino is coming: $\cos \theta = -1$ refers to neutrinos that come from directly below the detector and traverse the Earth's core, while $\cos \theta = 0$ refer to neutrinos from the horizon. 
For each plot, the disappearance in the upper left region is due to SM neutrino interactions at high energies with the Earth's matter. 
The large disappearance seen in the $10^3-10^4\ \GeV$ range is the resonant disappearance that is the result of the existence of a sterile neutrino state. 
This is the target signal for the sterile neutrino analysis in IceCube. 
As we can see, for typical sterile parameters considered, the resonant disappearance occurs at an energy near the peak muon neutrino flux seen in \Cref{fig:eventrate}.

\subsection{\texorpdfstring{\numu}{Muon Neutrino} Interactions Near the Detector}
\label{sec:numuinteractions}

\begin{figure}
    \centering
    \includegraphics[width=.50\textwidth]{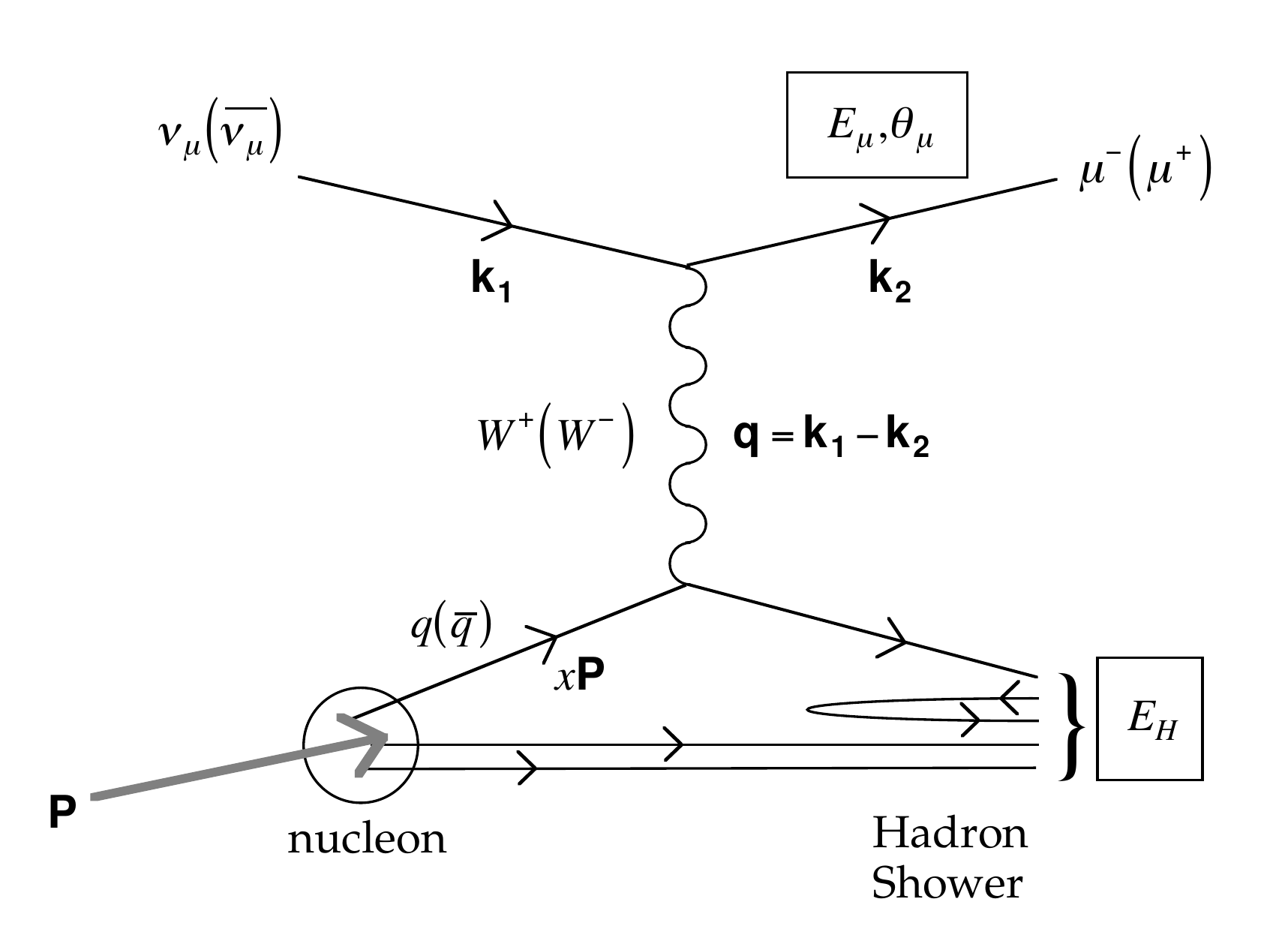}
    \caption{A deep inelastic scattering event of a \numu/\numubar interacting with the quarks of a nucleon. The final products are a $\mu^{-}/\mu^{+}$ and a hadronic shower. Figure taken from Ref.~\cite{Conrad:1997ne}.}
    \label{fig:DIS}
\end{figure}

When a \numu or \numubar is near the detector, it can interact with the ice or bedrock. 
At the energies of interest, neutrinos undergo Deep Inelastic Scattering (DIS) \cite{Formaggio:2012cpf}.  
Here, the neutrinos are of high enough energy that they can resolve the quarks individually in the nucleon. 
A tree-level diagram of a DIS scattering is shown in \Cref{fig:DIS}. 
We refer to Ref.~\cite{Formaggio:2012cpf} for details on neutrino scattering. 

\begin{figure}
    \centering
    \includegraphics[height=.85\textheight]{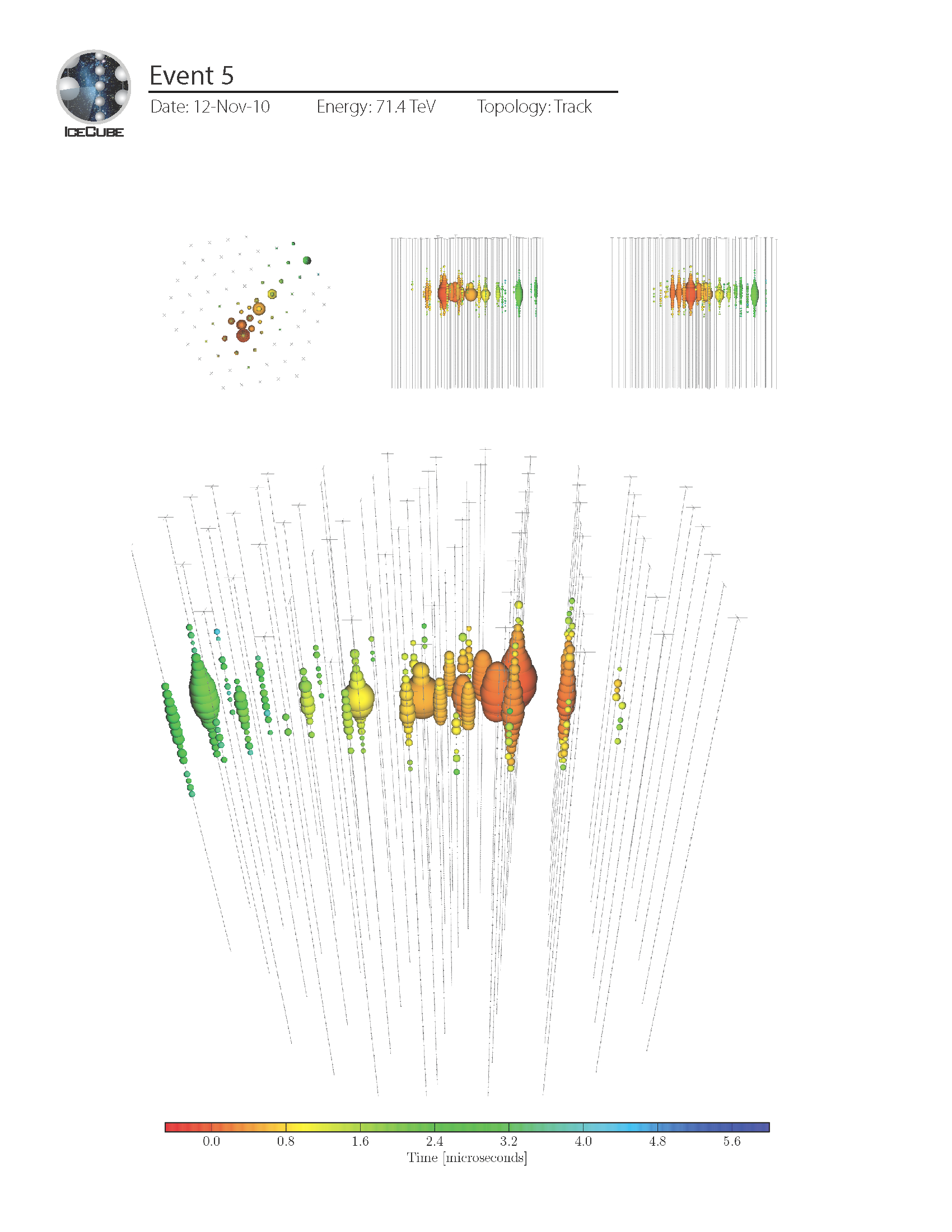}
    \caption{A \numu event observed in IceCube. Each colored sphere is a DOM that was hit. The size of the sphere corresponds to the energy deposited, and the color corresponds to the time when the DOM was hit. The event started near the center of the detector (red spheres), and moved left until exiting the detector. }
    \label{fig:eventdisplay}
\end{figure}

With muons travelling at $\SI{\sim 1}{\TeV}$ through matter, they undergo various kinds of interactions.
With the muon critical energy in ice being at \SI{1.03}{\TeV}, our muons of interest go through both ionization and radiative interactions.
Examples of the latter are: bremsstrahlung, $e^{+}e^{-}$ pair production, and photonuclear interactions \cite{Groom:2001kq}.
The muon energy loss can be written as 
\begin{equation}
    \langle -dE/dx \rangle = a(E) + b(E) E,
\end{equation}
where $a(E)$ is the ionization energy loss and $b(E)$ is the sum of the pair production, bremsstahlung, and photonuclear contributions. 
An approximation of the average propagation distance can be calculated by integrating
\begin{equation}
    R(E) = \int dE' [a(E') + b(E')E']^{-1}.
\end{equation}
This approximated range is called the ``continuous-slowing-down-approximation'' (CSDA) range.
At higher energies, fluctuations in energy losses makes the CSDA range of limited use, but it can still provide an order of magnitude approximation for the distance travelled by a charged particle in a medium. 
For a $\mu$ of energy \SI{1}{\TeV}, the CSDA returns an average range of \SI{2.4}{\km}.
Up at \SI{10}{\TeV}, the average range becomes \SI{7.8}{\km}.
Therefore, a muon produced near IceCube will traverse a very long distance, over a kilometer long and frequently longer than the length of the detector. 
An IceCube event display is shown in \Cref{fig:eventdisplay}.
There, a \numu undergoes a CC interaction near the center of the detector and the outgoing $\mu$ travels hundreds of meters to the left before exiting the detector. 

In IceCube, a $\mu^+$ track is indistinguishable to a $\mu^-$ track.
Therefore, we cannot distinguish between a \numu CC event and a \numubar CC event.
In the context of observed events, we will thus use ``\numu'' to refer both \numu and \numubar events. 

\subsection{Results}

We summarize here the results of the previous sterile neutrino analysis in IceCube \cite{IceCube:2020phf,IceCube:2020tka}.
Over a live-time of 7.634 years, \num{305735} up-going \numu events were observed.
These events were binned in terms of reconstructed energy and direction, and the sterile parameters \Dmqfo and \sinsqtthtf were fitted.
Two analyses were performed, a frequentist and a Bayesian analysis. 
The details of the frequentist analysis can be found in \Cref{sec:freqanalysis} in the context of the updated analysis, and the details for the Bayesian analysis can be found in \Cref{sec:bayesanalysis}.

\begin{figure}
    \centering
    \includegraphics[width=0.54\textwidth]{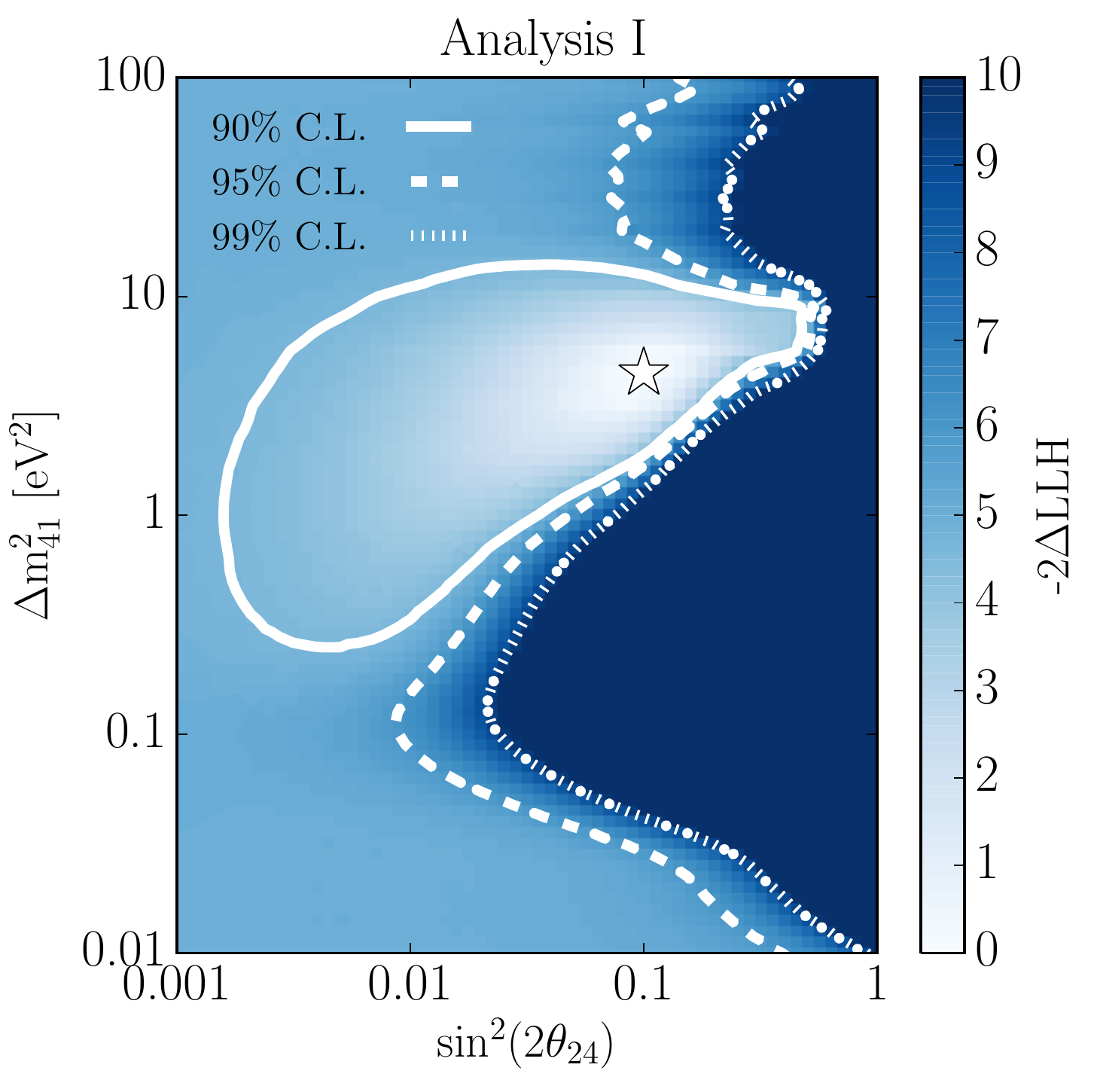}
    \caption{The best fit contour for the previous sterile neutrino analysis at IceCube. The best fit point was found at $\Dmqfo=4.5\ \eVq$ and $\sinsqtthtf = 0.10$, with a p-value of 8\%. Figure taken from Ref.~\cite{IceCube:2020tka}.}
    \label{fig:MEOWSresult}
\end{figure}

The result of the frequentist analysis is shown in \Cref{fig:MEOWSresult}.
The best fit sterile parameters were found at $\Dmqfo=4.5\ \eVq$ and $\sinsqtthtf = 0.10$, with a p-value of 8\%. 
Therefore, no significant preference for sterile neutrinos was found.

\begin{figure}
    \centering
    \includegraphics[width=0.54\textwidth]{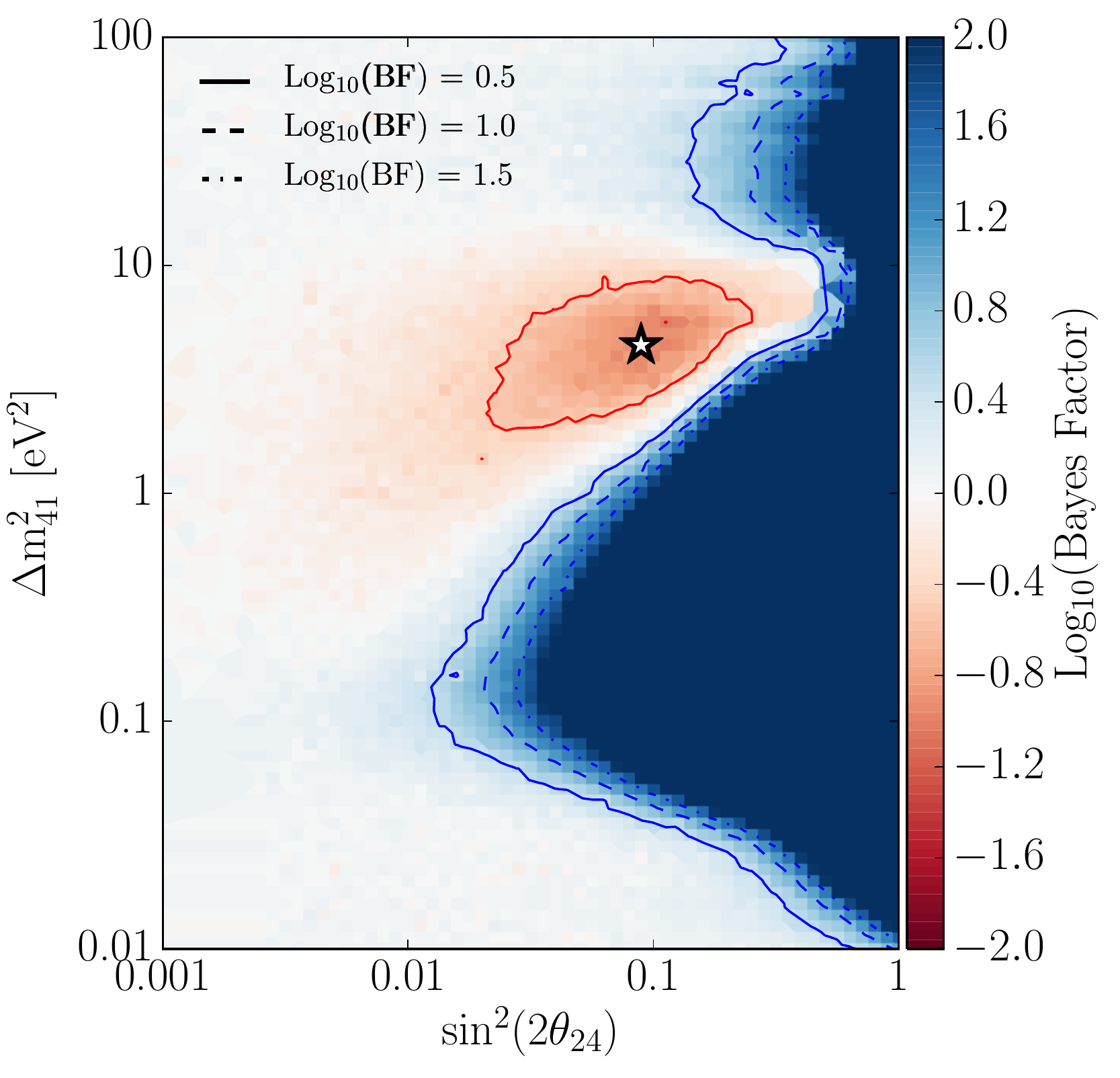}
    \caption{The result of the Bayesian fit in the previous sterile neutrino analysis at IceCube. At each point, the value of the Bayes factor relative to the null hypothesis is plotted. The point with the largest evidence was found at $\Dmqfo=4.5\ \eVq$ and $\sinsqtthtf = 0.10$, with a Bayes factor of $\log_{10}K=-1.03$ relative to the no-sterile hypothesis. Figure taken from Ref.~\cite{IceCube:2020phf}.}
    \label{fig:MEOWSbayesresult}
\end{figure}

The result of the Bayesian analysis is shown in \Cref{fig:MEOWSbayesresult}.
The analysis was conducted by calculating the Bayes factor, as described in \Cref{sec:bayesanalysis}.
At the best fit point of $\Dmqfo=4.5\ \eVq$ and $\sinsqtthtf = 0.10$, a Bayes factor of $\log_{10}K=-1.03$ relative to null was found.
According to Jeffreys' scale, this constitutes a ``strong'' preference for the sterile model compared to the no-sterile model.

    \chapter{MEOWS+\texorpdfstring{$\theta_{34}$}{\texttheta34}}
\label{ch:MEOWSplusth34}

The MEOWS analysis is unique amongst the sterile neutrino analyses in that it studies matter induced oscillations, as opposed to vacuum oscillations that the experiments listed in \Cref{sec:experiments} explored.
This provides an independent manner to search for sterile neutrinos, possibly shedding light on the difficulties with the sterile models discussed in \Cref{sec:globalfitresults}. 

As we noted in \Cref{ch:globalfits}, a $\numubar\to\numubar$ vacuum disappearance experiment would only be sensitive to \Umufsq and \Dmqfo for a 3+1 model.
In matter oscillations, on the other hand, $\numubar\to\numubar$ depends on each of the new sterile neutrino parameters: \Uefsq, \Umufsq, \Utaufsq, \Dmqfo, $\delta_{14}$, and $\delta_{24}$.
The mixing matrix elements can also be written in terms of the mixing angles $\theta_{14}, \theta_{24}, \theta_{34}$.
The relationship is given by 
\begin{align}
    \Uefsq &= \sin^2 \theta_{14} \label{eq:usandthetas1} \\
    \Umufsq &= \sin^2 \theta_{24} \cos^2 \theta_{14} \\ 
    \Utaufsq &= \sin^2 \theta_{34} \cos^2 \theta_{24} \cos^2 \theta_{14}.
    \label{eq:usandthetas3}
\end{align}

In the MEOWS analysis described in \Cref{sec:MEOWS}, a few simplifications were made. 
First, $\theta_{14}$ has a negligible effect on the atmospheric \numubar disappearance through the Earth \cite{Razzaque:2011ab,Esmaili:2012nz}, so $\theta_{14}$ was set to $0$ (equivalently, $\Uefsq = 0$), which made $\delta_{14}$ negligible as well. 
This leaves $\theta_{24}, \theta_{34}, \Dmqfo$, and $\delta_{24}$ as the relevant sterile parameters. 
Finally, a choice was made to set $\theta_{34} = 0$ (equivalently, $\Utaufsq = 0$), which also made $\delta_{24}$ negligible.
This was done for computational reasons, with the justification that $\theta_{34} = 0$ is a conservative estimate. 

The goal of our present work is to expand the MEOWS analysis to study the effects of a non-zero $\theta_{34}$ using the collected MEOWS neutrino sample. 
In fact, few experiments have placed limits on $\theta_{34}$, and these existing limits remain weak.
For example, the MINOS/MINOS+ experiments place a limit, at a fixed 
$\Dmqfo = 0.5\ \eVq$, of only
$\sin^2 \theta_{34} < 0.49\ (\theta_{34} < \ang{44})$
at the 95\% CL \cite{MINOS:2017cae}. 

\section{Oscillograms}

\begin{figure}
    \centering
    \begin{subfigure}{0.49\linewidth}
        \centering
        \includegraphics[width=\linewidth]{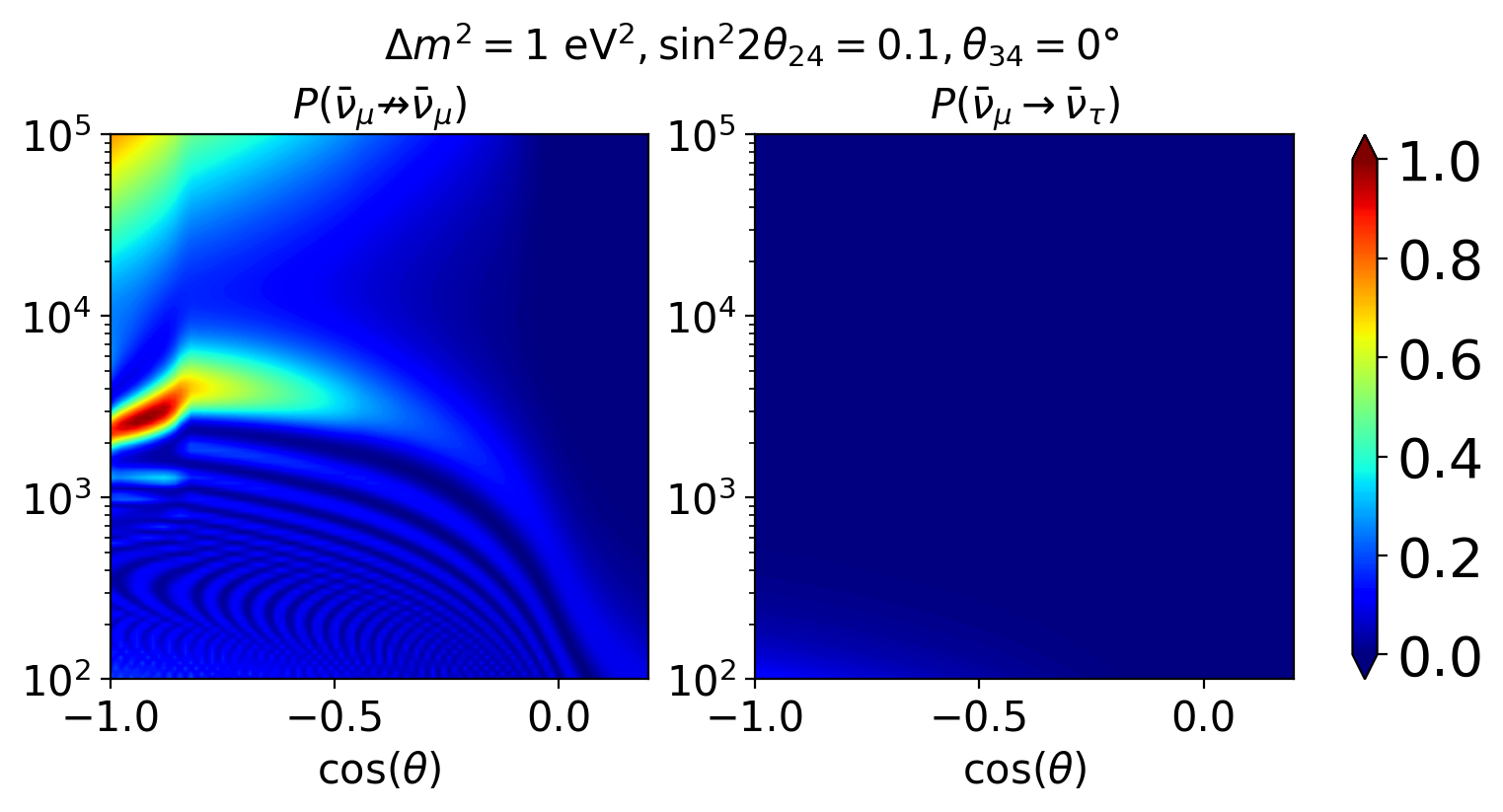}
    \end{subfigure}
    \hfill
    \begin{subfigure}{0.49\linewidth}
        \centering
        \includegraphics[width=\linewidth]{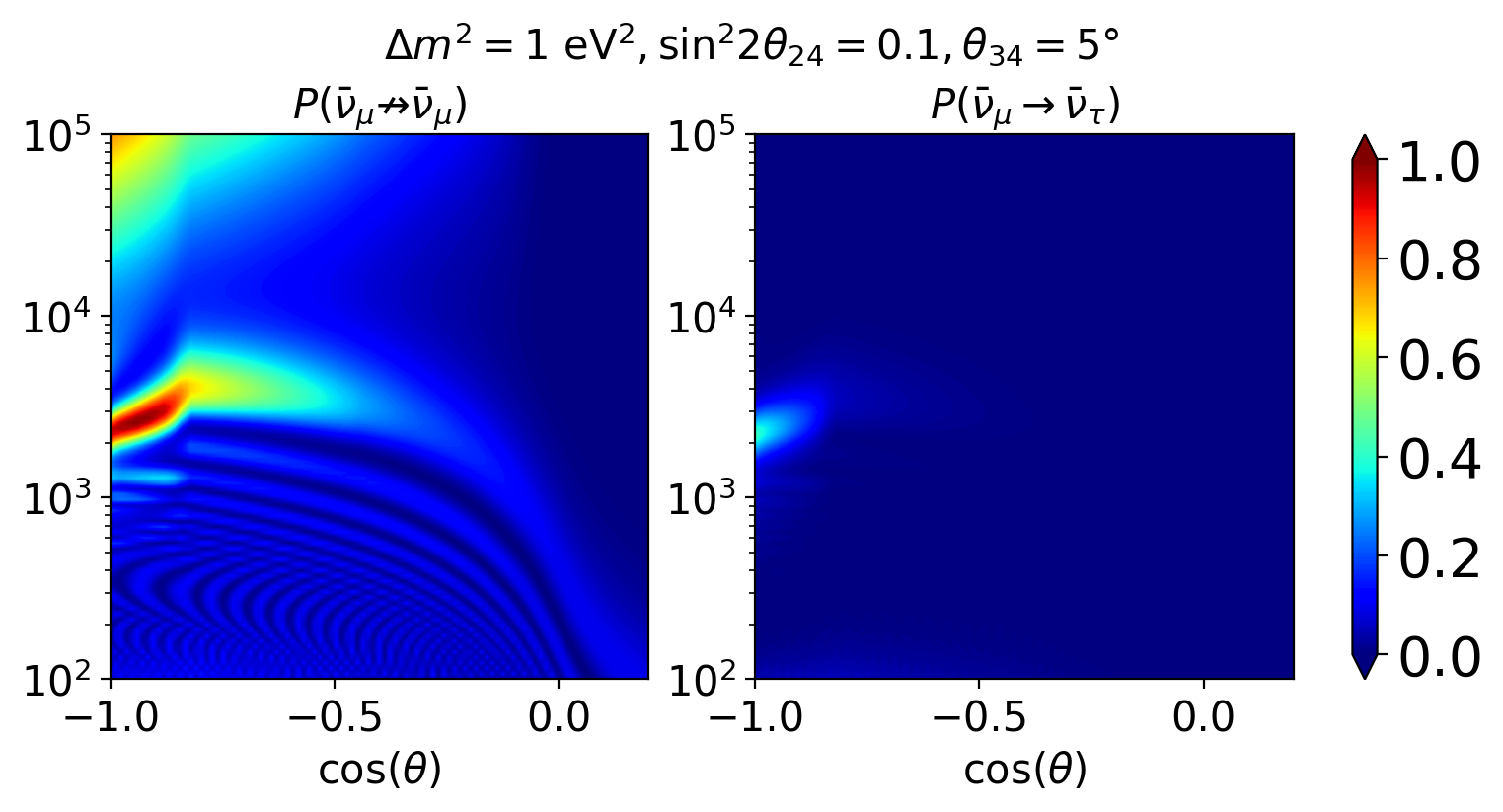}
    \end{subfigure}

    \begin{subfigure}{0.49\linewidth}
        \centering
        \includegraphics[width=\linewidth]{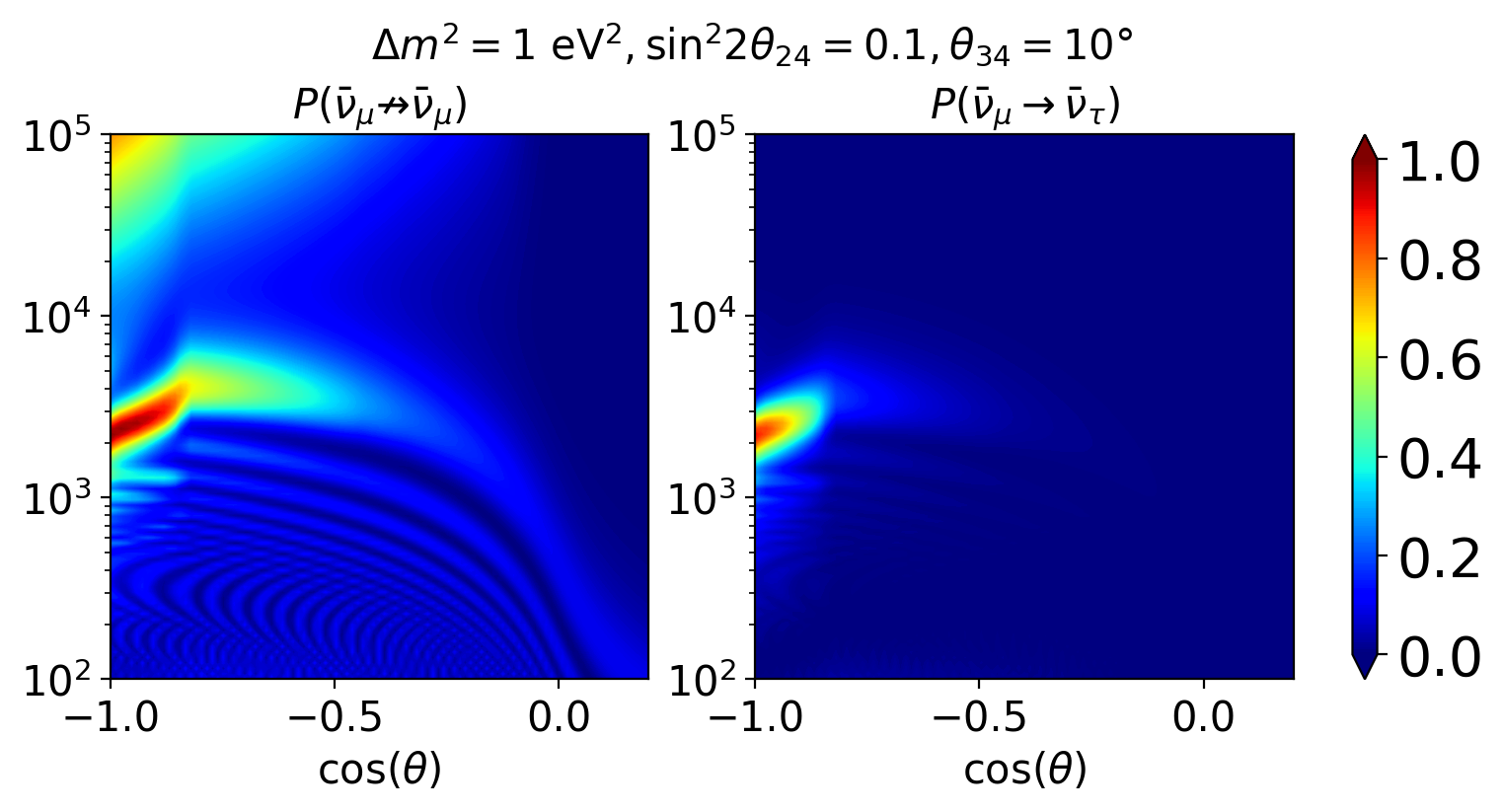}
    \end{subfigure}
    \hfill
    \begin{subfigure}{0.49\linewidth}
        \centering
        \includegraphics[width=\linewidth]{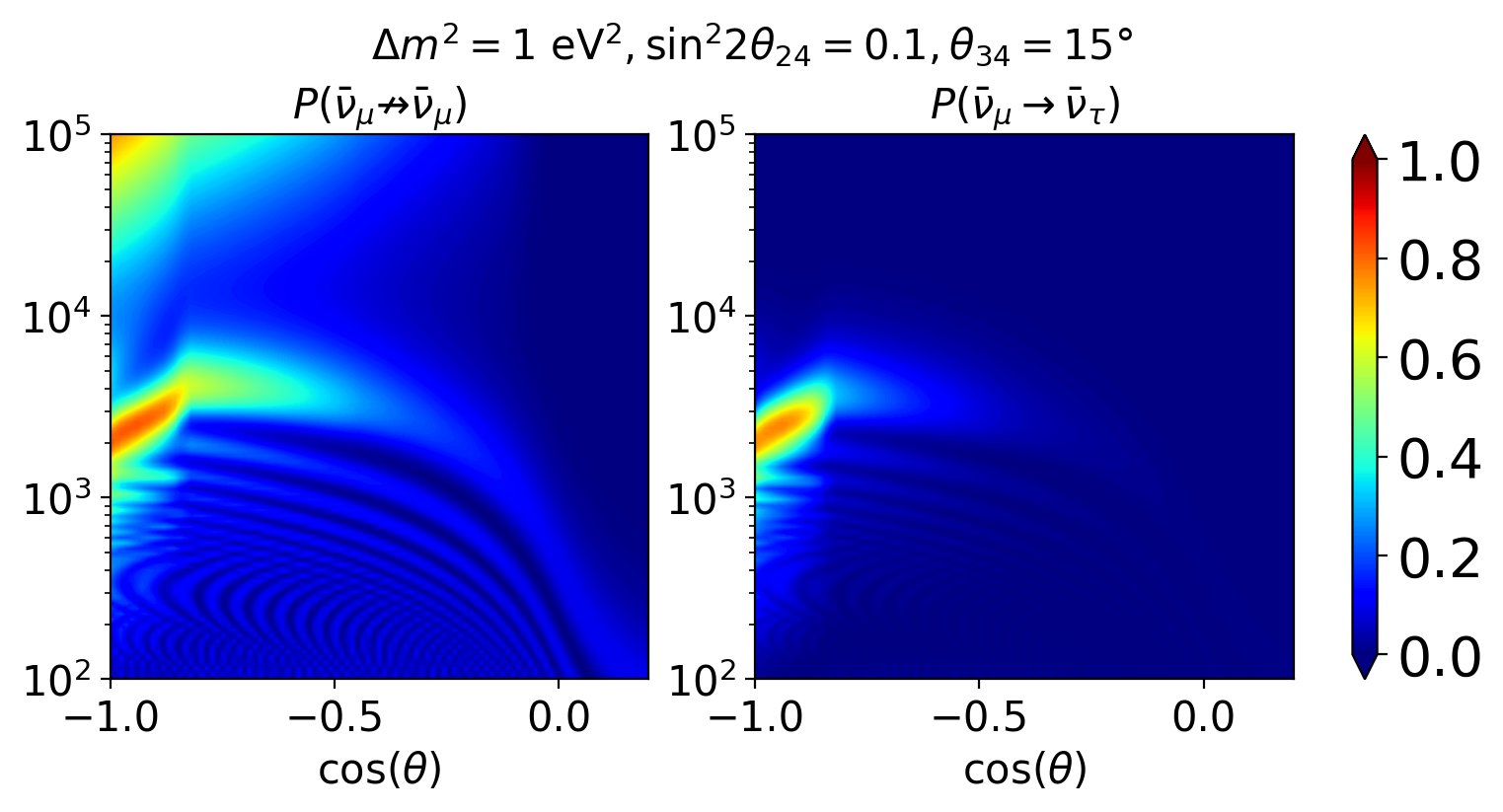}
    \end{subfigure}

    \begin{subfigure}{0.49\linewidth}
        \centering
        \includegraphics[width=\linewidth]{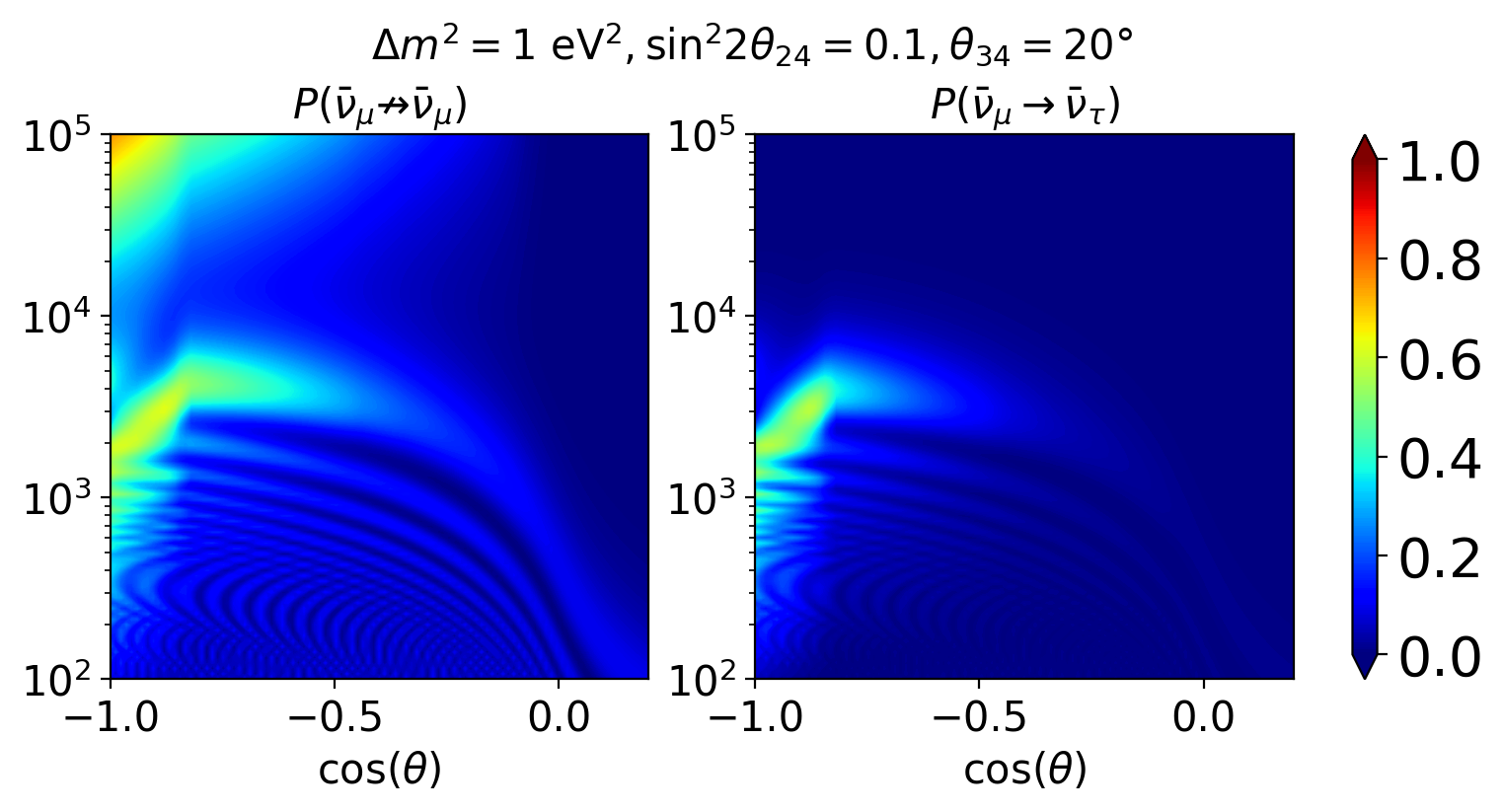}
    \end{subfigure}
    \hfill
    \begin{subfigure}{0.49\linewidth}
        \centering
        \includegraphics[width=\linewidth]{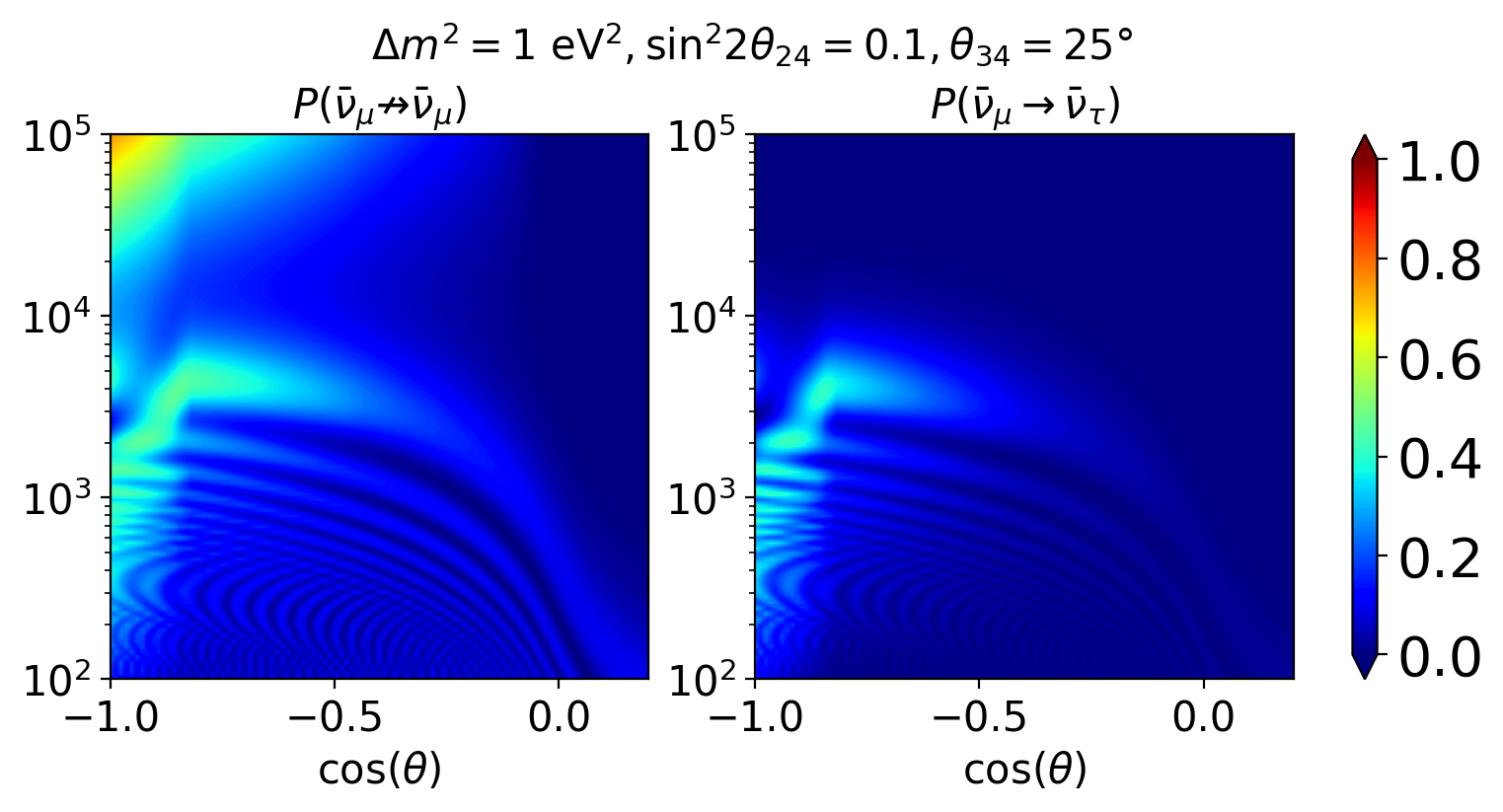}
    \end{subfigure}
    \caption{A series of oscillograms, where we show the expected $\numubar \to \numubar$ disappearance and $\numubar \to \nutaubar$ appearance for different values of $\theta_{34}$.
    In each plot, we keep $\Dmqfo = \SI{1}{\square\eV}$ and $\sinsqtthtf=0.1$ constant.}
    \label{fig:MEOWSth34oscillograms}
\end{figure}

Before discussing details of the analysis, we will see how $\theta_{34}$ affects \numubar oscillations through the Earth. 
\Cref{fig:MEOWSth34oscillograms} shows a series of oscillograms with increasing $\theta_{34}$. 
$\Dmqfo$ and $\sinsqtthtf$ are left constant at some representative value.
For each pair of oscillograms, the left plot shows the \numubar disappearance.
We can see how the resonance substantially changes with increasing $\theta_{34}$, broadening and smearing into lower energies.
Therefore, while $\theta_{34} = 0$ is the conservative selection, a proper fit to a 3+1 sterile model in IceCube has to take into account $\theta_{34}$ if it wants to accurately measure the hypothetical parameters.

Alongside each \numubar disappearance plot, we also show the $\numubar \to \nutaubar$ appearance probability.
With $\theta_{34}=0$, \nutaubar appearance is negligible.
But with increasing $\theta_{34}$, \nutaubar appearance becomes significant.
This appearance must be taken into account when doing our analysis.

\section{Neutrino Sources}

\subsection{Atmospheric Neutrinos}
\label{sec:atmoshpericneutrinos}

\begin{figure}
    \centering
    \includegraphics[height=.4\textheight]{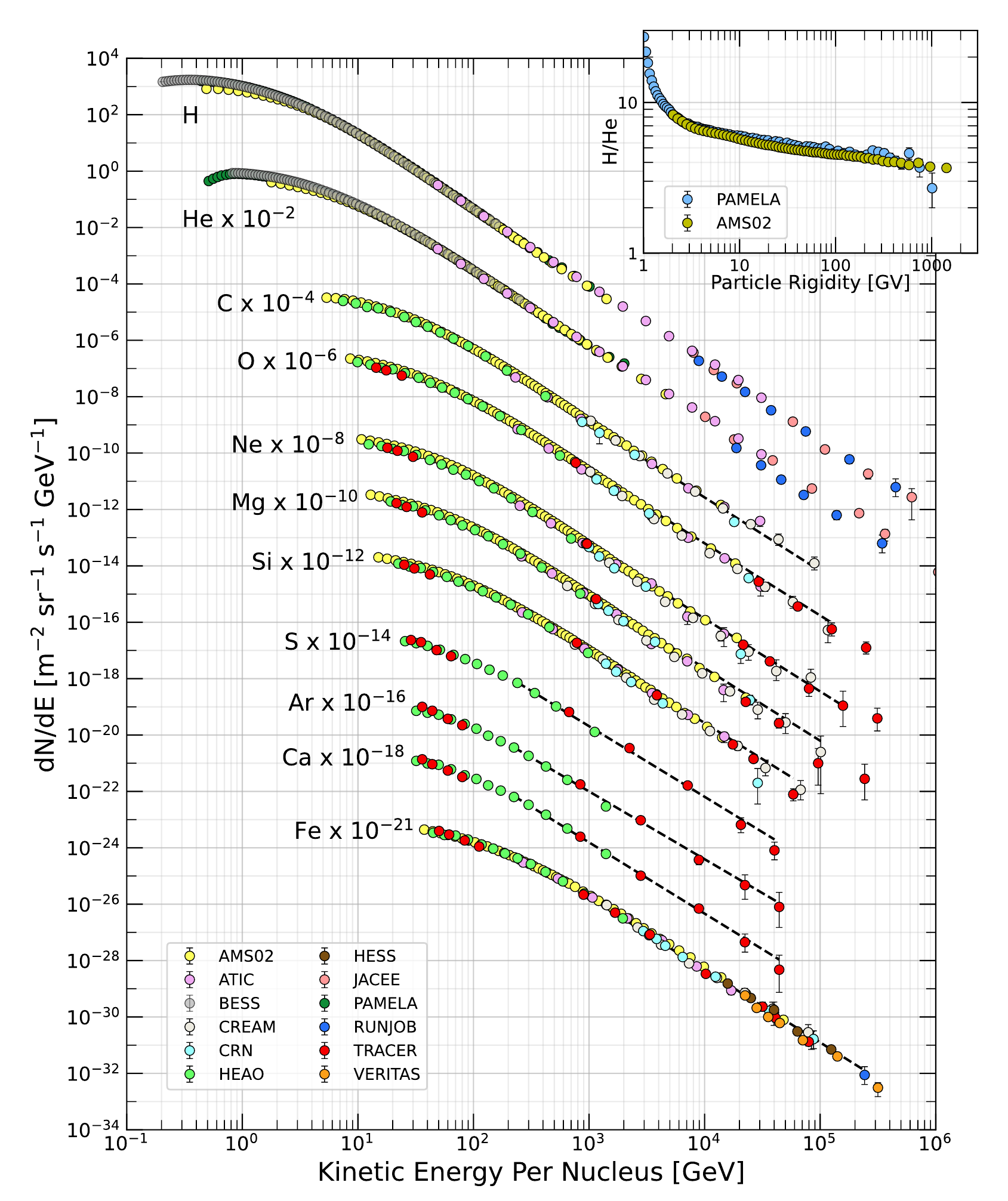}
    \caption{Fluxes of cosmic ray nuclei as a function of kinetic energy per nucleus. Figure taken from Ref.~\cite{Workman:2022ynf}.}
    \label{fig:cosmicrayflux}
\end{figure}

The neutrinos that the MEOWS analysis primarily detects are atmospheric muon neutrinos.
These neutrinos are produced by the interactions of cosmic rays with the Earth's atmosphere.
Cosmic rays are primarily composed of hydrogen and helium nuclei, with a smaller proportion of heavier elements.
\Cref{fig:cosmicrayflux} shows the flux of various cosmic ray nuclei experimentally measured.

The spectra of these cosmic ray approximately follow a steeply falling power-law spectrum.
Within our energy range of interest (several \GeV up to 100 \TeV), the intensity of primary nucleons is approximately given by 
\begin{equation}
    I_{N}(E) \approx 1.8 \times 10^{4} (E/1\ \GeV)^{-\gamma}\ \si{\nucleon\per\square\meter\per\second\per\steradian\per\GeV},
\end{equation}
where $E$ is the energy per nucleon and $\gamma \approx 2.7$.
These cosmic rays then collide with the Earth's atmosphere, producing mesons that later decay into neutrinos (similar to how accelerating a proton beam into a target produces neutrinos at LSND and MiniBooNE).
The flux evolution of cosmic rays and their daughter particles through the atmosphere are described by coupled cascade equations.
Of interest for our analysis are the neutrino fluxes and the muon fluxes, the former of which is our source and the latter of which is a background.

The atmospheric neutrino flux is divided into two components a ``conventional'' component and a ``prompt'' component.
The conventional component originates from the decay of pions, kaons, and muons.
The prompt component, on the other hand, originates from the decay of higher-mass charmed mesons.
These charmed mesons decay quickly, before they have a chance to lose substantial energy interacting with the atmosphere.
Therefore, the prompt neutrino component is in a higher energy range than the conventional component.

For the MEOWS analysis, the conventional flux is derived using the \texttt{Matrix Cascade Equation (MCEq)} package \cite{Fedynitch:2015zma}. 
As inputs, \texttt{MCEq} takes in an initial cosmic ray model, a hadronic interaction model, and an atmospheric density profile.
For the cosmic ray model, the Hillas-Gaisser 2012 H3a model \cite{Gaisser:2011klf} is used. 
As the cosmic rays collide with the atmosphere, the evolution of the secondary particles is guided by a hadronic interaction model.
We use the SYBLL 2.3c model \cite{Riehn:2017mfm}.
Finally, the evolution also depends on the atmospheric density profile of the Earth.
Data from the Atmospheric Infrared Sounder (AIRS) on NASA's Aqua satellite is used \cite{AIRS}.

The prompt component is taken to be the one calculated in Ref.~\cite{Bhattacharya:2015jpa}. 

\subsection{Astrophysical Neutrinos}

In addition to atmospheric neutrinos, astrophysical neutrinos are taken into account.
While their origin is unknown, IceCube has established the existence of a diffuse flux of astrophysical neutrinos.
For this analysis, the astrophysical neutrino flux is taken to be isotropic with a falling power-law spectrum
\begin{equation}
    \frac{dN_\nu}{dE} = \Phi_{\textrm{astro}} \times \left( \frac{E_\nu}{\SI{100}{\TeV}} \right)^{-\gamma_{\textrm{astro}}},
    \label{eq:astroflux}
\end{equation}
with a nominal normalization $\Phi_{\textrm{astro}} = \SI{0.787e-18}{\per\GeV\per\steradian\per\second\per\square\cm}$ and a nominal spectral index of $\gamma_{\textrm{astro}} = 2.5$.

\section{Neutrino Propagation Through the Earth}

After calculating an initial neutrino flux at the Earth's surface, we have to propagate this neutrino flux through the Earth. 
As mentioned earlier, this calculation is very difficult to do analytically taking into account the varying density profile of the Earth and the multiple neutrino flavors. 
We therefore use the \texttt{nuSQuIDS} package to numerically propagate the neutrinos through the Earth, given some sterile neutrino hypothesis.
In addition to the difficult calculation of matter oscillations, \texttt{nuSQuIDS} also takes into account non-coherent interactions. 
These include flux attenuation from neutrinos interacting in the Earth; neutrino energy losses due to neutral current interactions; \nue and \numu production from $\tau$ decays; and neutrino production from $W^{-}$ decays in Glashow resonances. 

\texttt{nuSQuIDS} also takes in as input the density profile of the Earth.
Here, we use the Preliminary Reference Earth Model (PREM model) \cite{Dziewonski:1981xy}, which assumes a spherically symmetric Earth with density varying as a function of radius.

After propagating the neutrinos through the Earth with some sterile neutrino model, \texttt{nuSQuIDS} provides the final neutrino flux at the IceCube detector.

\section{Neutrino Interactions Near the Detector}

When a neutrino interacts in or near the detector, its interaction products are visible to the IceCube detector.

For a \numu CC interaction, the interaction products are an outgoing $\mu$ and a hadronic shower starting at the interaction point.
As described in \Cref{sec:numuinteractions}, a $\mu$ can travel long distances in the ice.
If the muon traverses through the detector, it leaves behind a ``track'' of hit DOMs, as seen in \Cref{fig:eventdisplay}.
For \numu CC events that occur outside the detector, the hadronic shower will not be visible and the detector would only be able to see the $\mu$ if it traverses through the detector. 

\begin{figure}
    \centering
    \includegraphics[height=.90\textheight]{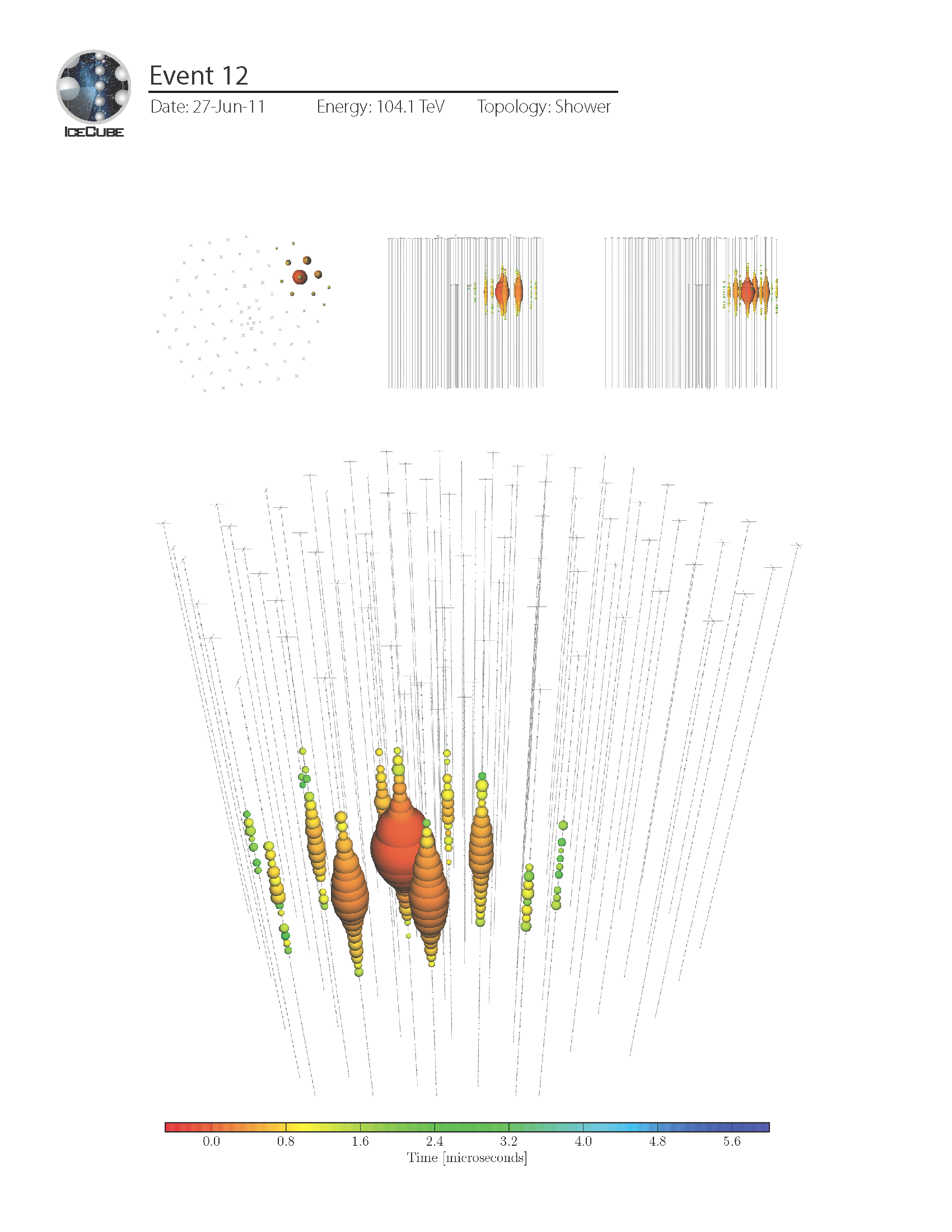}
    \caption{An example of a cascade event in IceCube. Unlike a track event, cascade events deposit their energy in an approximately spherical volume centered at the interaction point.}
    \label{fig:eventdisplaycascade}
\end{figure}

For a \nue CC interaction, the products are an outgoing $e$ and a hadronic shower.
The outgoing $e$ will produce an electromagnetic shower which, like a hadronic shower, will remain close to the interaction point.
This interaction will produce a roughly spherically symmetric distribution of hit DOMs, which is referred to as a ``cascade.''
An example of a cascade is shown in \Cref{fig:eventdisplaycascade}.

For a NC interaction of any neutrino flavor $\nu_\alpha$, the interaction products are the $\nu_\alpha$ and a hadronic shower.
The outgoing $\nu_\alpha$ is invisible to the detector, so the only signature is a cascade from the hadronic shower. 

In summary: as seen by the detector, $\numu$ CC events produce ``tracks'' from the $\mu$ traversing the detector, while \nue CC and all-flavor NC events produce ``cascades.''

The signature of a \nutau CC interaction will depend on how the outgoing $\tau$ decays.
Of interest for us is when the $\tau$ decays leptonically by $\tau \to \mu + ...$, as this has the same event signature as a $\numu$ event.
This decay has a branching ratio of 18\%.
As seen in \Cref{fig:MEOWSth34oscillograms}, a non-zero $\theta_{34}$ can lead to significant $\numu\to\nutau$ appearance, so these events have to be taken into account as well. 

Simulated neutrino interaction points are chosen using the program \texttt{LeptonInjector} \cite{IceCube:2020tcq}. The injected energy follows some nominal flux.
A related program, \texttt{LeptonWeighter} \cite{IceCube:2020tcq}, allows the reweighting of events to some arbitrary flux after the simulation.
This lets us use a single simulation set, which can be then reweighted for any neutrino flux at IceCube, i.e. any sterile neutrino hypothesis.

When a neutrino interaction point is selection, the secondary products are propagated through the rock and ice using the \texttt{PROPOSAL} software \cite{Koehne:2013gpa}. 
The deposited energy is accounted for by \texttt{PROPOSAL}, and the resulting photons are then propagated using the IceCube software \texttt{CLSim}.
The photons are propagated until they are either absorbed or collected by a DOM. 

For the \numu simulations, we use the same simulation set as the previous MEOWS analysis, corresponding to about 500 years of live time. 

For the work in this thesis, we also generated a new set of \nutau simulations. 
For this, we modified \texttt{PROPOSAL} to properly take into account the $\tau$ polarization when decaying into a $\mu$. 
\texttt{PROPOSAL} typically propagates and decays particles assuming that they are unpolarized. 
$\tau^{-}$'s ($\tau^{+}$'s), though, are very short-lived, and the weak interaction guarantees that they are produced in a left-(right-)handed chirality state. 
Therefore, they will decay polarized. 
The rest frame cross section for the $\mu^{\pm}$ energy and direction from a $\tau^{\pm}$ decay is \cite{Workman:2022ynf,Lipari:1993hd}
\begin{equation}
    \frac{d^{2}\Gamma}{dx d\cos\vartheta}
\propto
[3-2x\pm |\mathbf{P}_{\tau}| \cos \vartheta (2x -1)]x^{2},
\label{eq:taudecay}
\end{equation}
where $x \equiv 2 E_{\mu}/m_{\tau}$, $\vartheta$ is the angle between the muon momentum and the $\tau$ spin, and $|\mathbf{P}_{\tau}|$ is the degree of $\tau$ polarization. 
The standard \texttt{PROPOSAL} code neglects the $\cos \vartheta$ dependent term, but we introduce it for our analysis. 
Looking at \Cref{eq:taudecay}, we can see that for left-handed (LH) $\tau^{-}$ and right-handed (RH) $\tau^{-}$ the muon is emitted preferentially in the direction of motion.
Boosting to the lab frame, the $\mu^{\pm}$ energy distribution from a $\tau^{\pm}$ decay is given by \cite{Lipari:1993hd}
\begin{equation}
    F_{\tau^{\pm}\to\mu^{\pm}}(x) =   
    \left( \frac{5}{3} - 3 x^{2} + \frac{4}{3} x^{3} \right)
    \mp
    \mathbf{P}_{\tau} \left( \frac{1}{3} - 3 x^{2} + \frac{8}{3} x^{3} \right).
\end{equation}

\begin{figure}
    \centering
    \includegraphics[width=.55\textwidth]{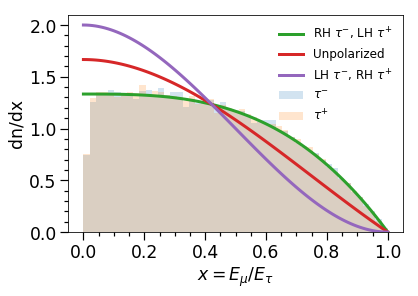}
    \caption{The expected ratio $x$ of the energies between a $\mu$ and a $\tau$, when the $\tau$ decays leptonically as $\tau \to \mu + ...$. 
    \texttt{PROPOSAL} typically assumes that all particles are unpolarized, so $x$ would follow the red line.
    We added a correction to \texttt{PROPOSAL} to take into account that a $\nutau$ CC event would create polarized $\tau$'s, and these $\tau$'s would in turn decay polarized.}
    \label{fig:taudecay}
\end{figure}

The $x$ distribution in the lab frame is shown in \Cref{fig:taudecay}.
The green line corresponds to $\mu$'s from completely polarized LH $\tau^{-}$ and RH $\tau^{+}$ decays, and the purple line corresponds to $\mu$'s from RH $\tau^{-}$ and LH $\tau^{+}$ decays.
The red line is the $x$ distribution if we assumed that the $\tau$'s were unpolarized when decaying. 
The histogram shows the results of our simulation after producing the muons as in \Cref{eq:taudecay} and boosting to the lab frame.
Without our correction, the simulation muon energy distribution would incorrectly follow the red line.

\section{Muon Reconstruction}

The observable of interest are muons traversing the detector, which produce tracks.
The MEOWS analysis selected for tracks whether they begin inside or outside the detector.
Tracks that originate outside of the detector are referred to as ``through-going'' tracks, while tracks that begin inside the detector are called ``starting'' tracks.
While through-going muons provide worse energy reconstruction since the hadronic shower and initial track occur outside of the detector, they greatly increase the available statistics.
Further contributing to the poor energy resolution is that these muons are likely to escape the detector, so the total energy of the muon is rarely contained entirely within the detector.

\begin{figure}
    \centering
    \includegraphics[width=0.65\textwidth]{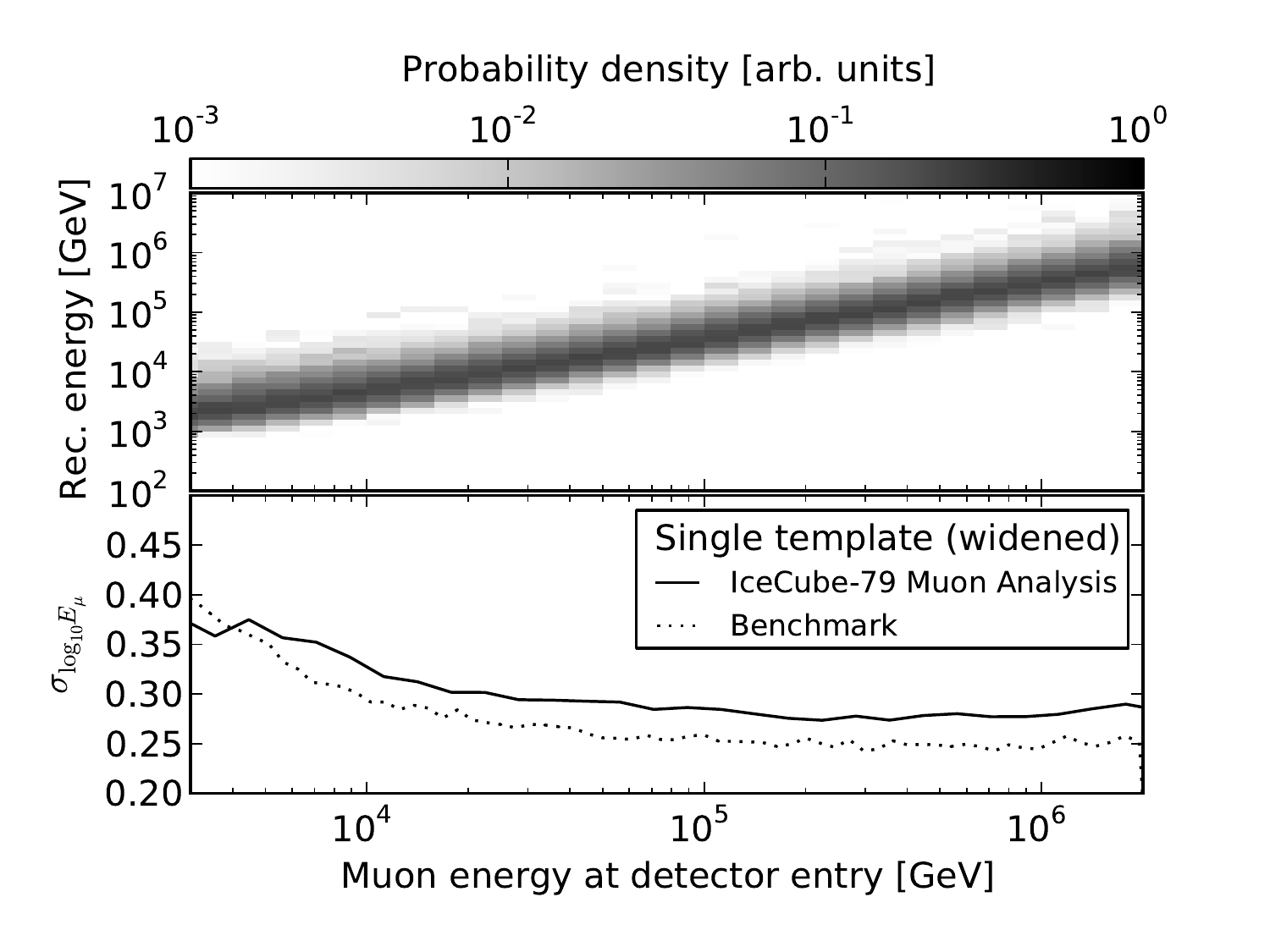}
    \caption{Energy reconstruction of muon tracks, as a function of the true muon energy when it entered the detector. Figure taken from Ref.~\cite{IceCube:2013dkx}.}
    \label{fig:muexenergyresolution}
\end{figure}

\begin{figure}
    \centering
    \begin{subfigure}{0.49\textwidth}
        \includegraphics[width=\textwidth]{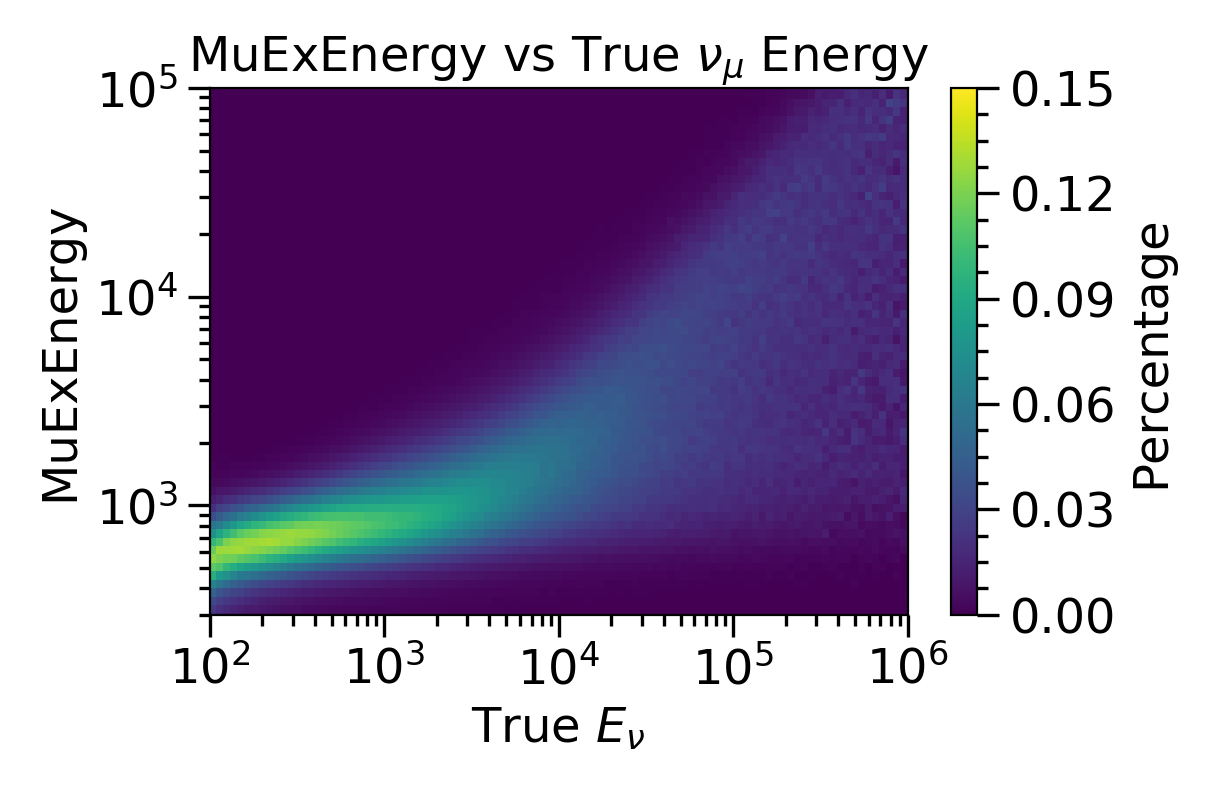}
        \caption{}
    \end{subfigure}
    \hfill
    \begin{subfigure}{0.49\textwidth}
        \includegraphics[width=\textwidth]{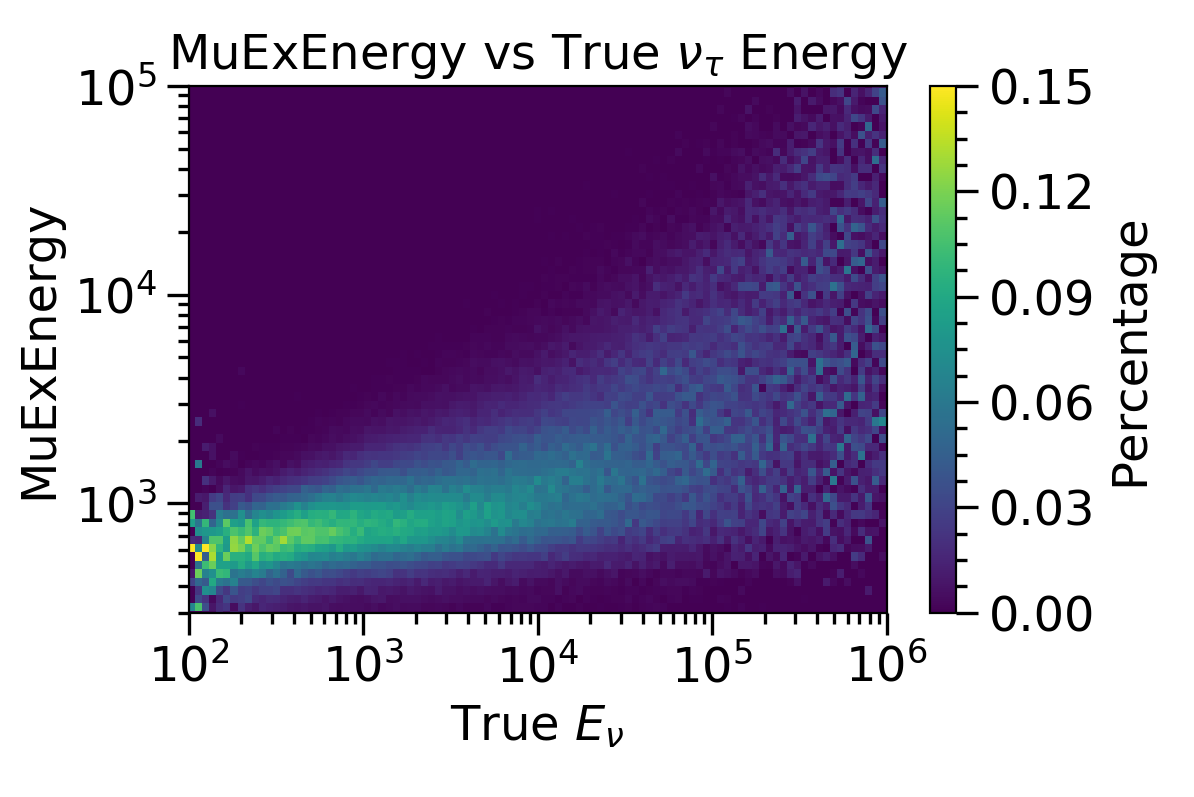}
        \caption{}
    \end{subfigure}

    \begin{subfigure}{0.49\textwidth}
        \includegraphics[width=\textwidth]{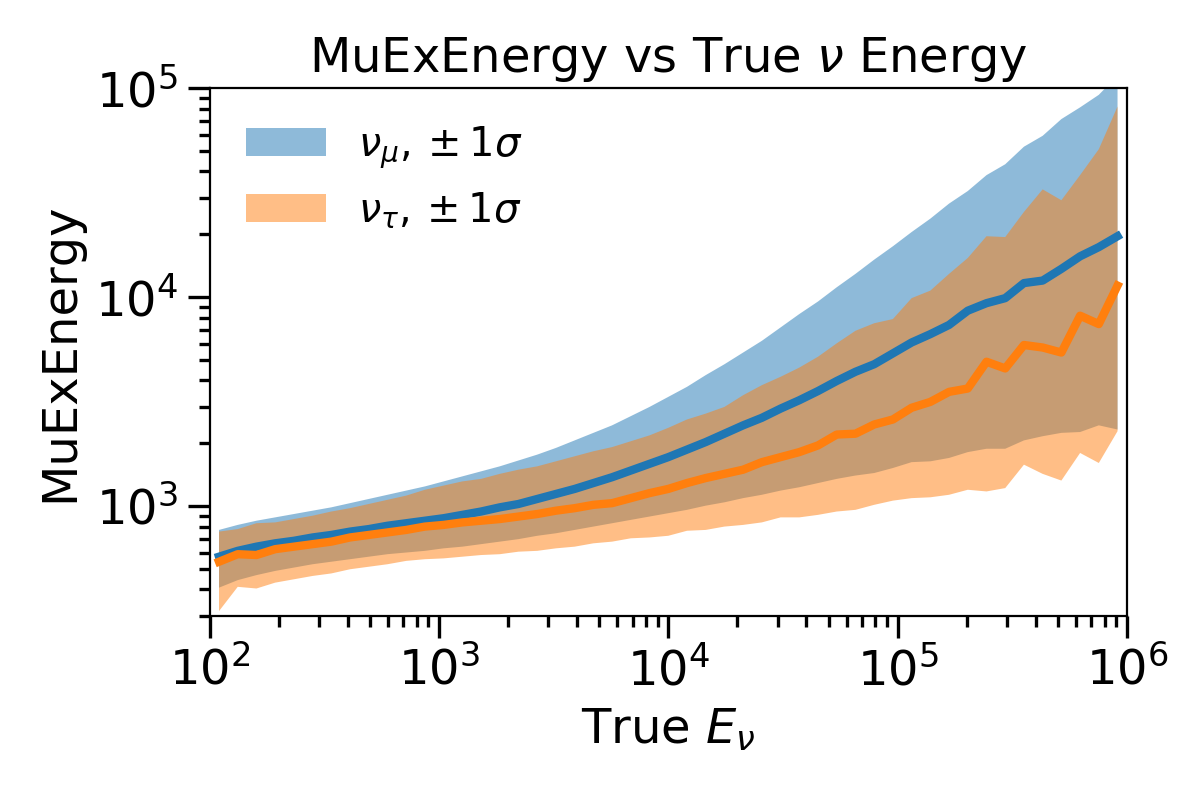}
        \caption{}
    \end{subfigure}
    \caption{(a) The energy reconstruction of muons from $\numu$ CC events as a function of true \numu energy. (b) The energy reconstruction of muons from \nutau CC events followed by $\tau \to \mu ...$ decays, as a function of true \nutau energy. (c) A comparison of the muon energy reconstruction between \numu CC and \nutau CC events. All plots use events that pass the MEOWS event selection.}
    \label{fig:energyreconstructions}
\end{figure}

The energy reconstruction algorithm used is an internal IceCube algorithm called \texttt{MuEX}, used in an astrophysical neutrino search analysis \cite{IceCube:2015qii} and described in more detail in Refs.~\cite{IceCube:2013dkx,Weaver:2015bja}.
\Cref{fig:muexenergyresolution} shows the reconstructed muon energy distribution as a function of the true muon energy when it enters the detector; the energy resolution is quite poor.
Relative to the true neutrino energy on interaction, the energy resolution is even worse.
We show in \Cref{fig:energyreconstructions} the reconstruction muon energy distribution compared to true $\numu$ energy in the MEOWS simulation sample, and the reconstructed muon energy distribution compared to true $\nutau$ energy from our simulation.

The muon direction, on the other hand, is reconstructed well.
Due to muons traversing such a large distance through IceCube, the large lever arm allows good reconstruction of the direction.
This can be seen in \Cref{fig:eventdisplay}.
Using the internal \texttt{MPEfit} algorithm, the direction is reliably reconstructed to better than $\ang{1}$.

\section{Event Selection}

In this analysis we use the same event selection as the MEOWS analysis.
The event selection was designed to obtain a very pure sample of $\numu$ CC events.
The IceCube detector, while being \SI{1.5}{\km} under ice, still triggers 3000 times per second \cite{IceCube:2015wro} due to penetrating atmospheric muons.
These atmospheric muons constitute the largest backgrounds in the analysis.
A detailed description of the cuts used are summarized in Ref.~\cite{IceCube:2020tka}, but we will summarize here the more significant cuts. 

At IceCube, the majority of the neutrino flux is composed of muon neutrinos.
As discussed, \numu CC events produce a track in the detector, a very distinct signal compared to the cascades produced by \nue and NC events. 
Therefore, the first step is IceCube's muon filter, which removes non-track events. 
\nutau CC events where the $\tau$ decays into a muon produces a near identical signal as a \numu CC event, so those will not be cut out. 

Track events will then have their energies and direction fitted.
A cut is applied so that the reconstructed energies fall within 500--9976 \GeV.
Outside of this range, the event count falls quickly, and the less understood astrophysical flux begins to dominate at higher energies.

At the horizon ($\cos \theta = 0$), IceCube already has \SI{157}{\km} of water-equivalent shielding.
While muons are highly penetrating, they are not so penetrating that they could traverse thousands of kilometers through the Earth.
Therefore, any track reconstructed as coming from below the horizon ($\cos \theta < 0$) will very likely come from a \numu event instead of an atmospheric muon, and a cut is applied as such.
If an above-horizon ($\cos \theta > 0$) atmospheric muon were to be misreconstructed as $\cos \theta < 0$, its fit would be poor and therefore still removable.

\begin{figure}
    \begin{subfigure}{.49\textwidth}
    \includegraphics[width=\textwidth]{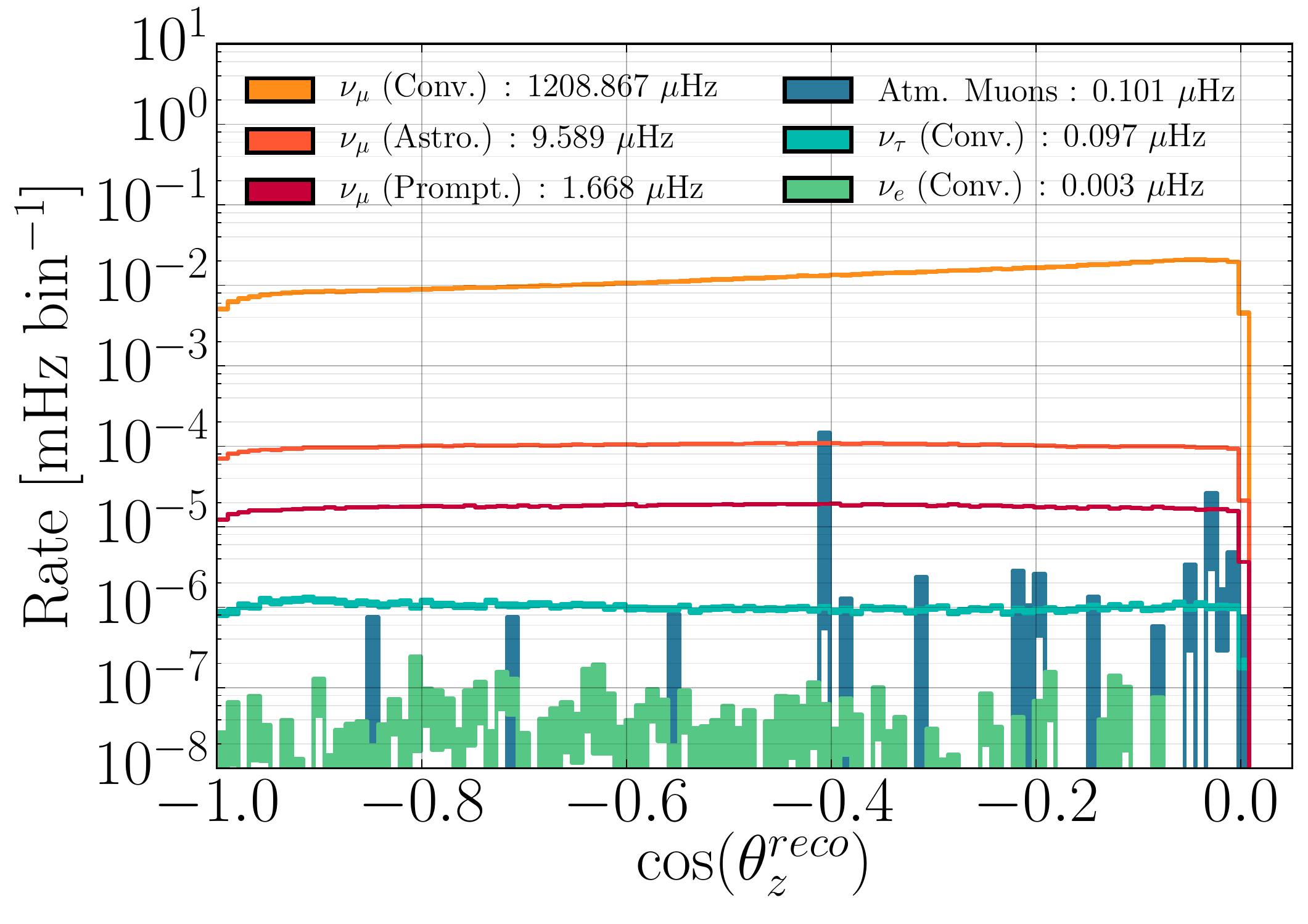}
    \caption{}
    \end{subfigure}
    \hfill
    \begin{subfigure}{.49\textwidth}
    \includegraphics[width=\textwidth]{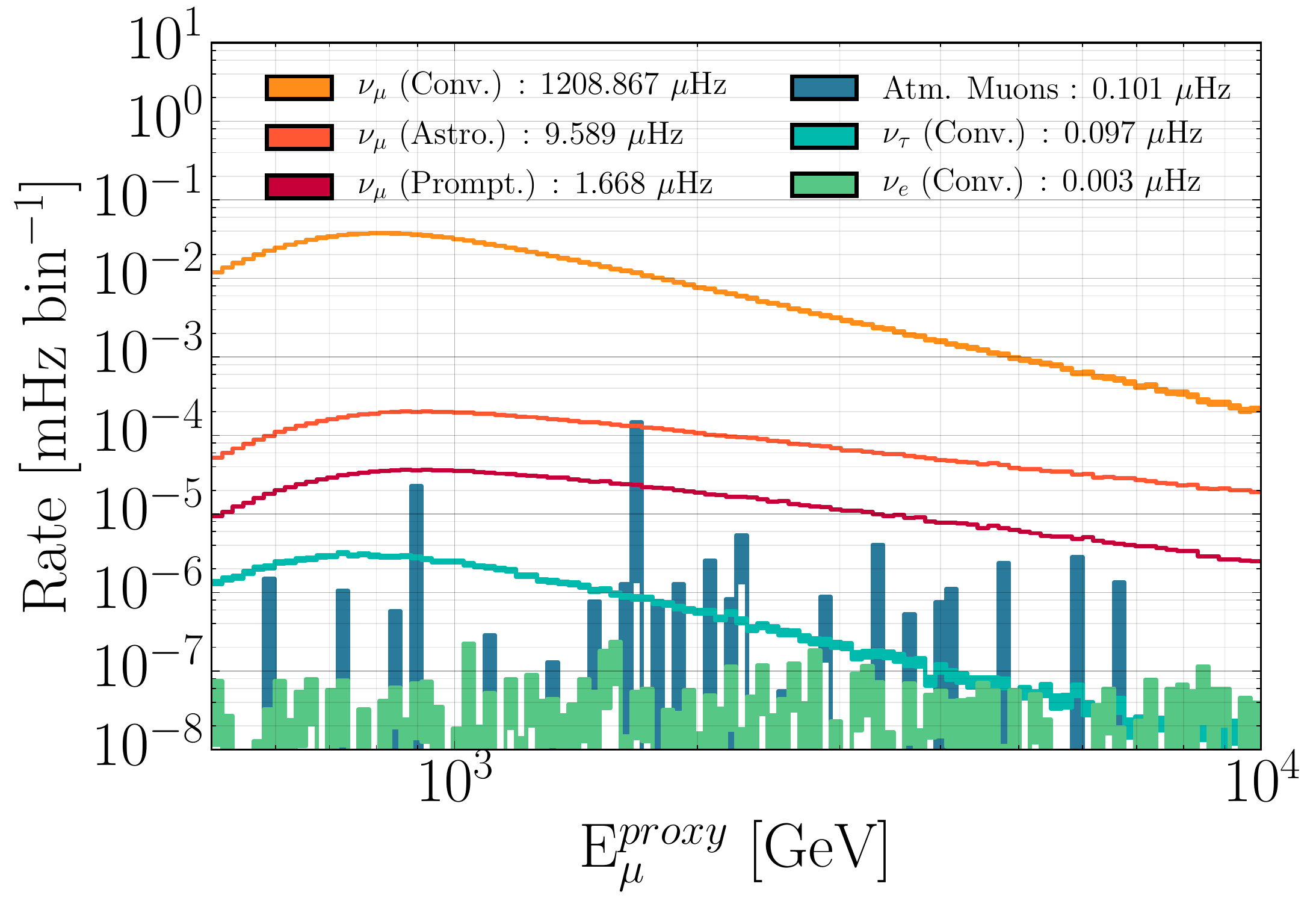}
    \caption{}
    \end{subfigure}
    \caption{The expected event rates from different neutrino components and backgrounds that pass the final filter, assuming the no-sterile model. (a) The event rate as a function of reconstructed $\cos \theta$. (b) The rate as a function of reconstructed energy. Figures from Ref.~\cite{IceCube:2020tka}.}
    \label{fig:eventsample}
\end{figure}

Ultimately, the event sample has a purity of 99.9\%, with very few background events. 
The expected event distributions, as a function of reconstructed energy and zenith, are shown in \Cref{fig:eventsample}. 

\section{Systematic Parameters}

The systematic treatment used in this analysis is near identical to the previous MEOWS analysis. 
The 18 systematic parameters can be divided into four categories: conventional flux parameters, detector parameters, astrophysics parameters, and cross section parameters. 
Each of these systematic parameters are implemented as nuisance parameters in our fits, with a prior associated with each one. 
Minor corrections to the systematic treatment come from a related MEOWS+Decay analysis in IceCube \cite{Moulai:2021zey,IceCube:2022alt}.
They are: updated livetime from 7.6 years to 7.634 years; updated Barr parameter corrections (described in \Cref{sec:conventional}) using atmospheric data from the AIRS satellite; corrected Earth composition model that places the bedrock under IceCube at \SI{3}{\km} under the surface, rather than \SI{30}{\km}.

Here, we summarize each of the systematic parameters included in the MEOWS analysis, and which are also used in our updated analysis.
More detailed descriptions of the systematic parameters can be found in Refs.~\cite{IceCube:2020tka,Axani:2019sbk}.
A table of all the systematic parameters with their central values and priors are listed in \Cref{tab:sysparameters}.

\subsection{Conventional Flux Parameters}
\label{sec:conventional}

In \Cref{sec:atmoshpericneutrinos}, we discussed how a nominal conventional atmospheric flux is calculated assuming some cosmic ray model, hadronic interaction model, and atmospheric density model. 
A total of nine systematic parameters are used to parameterize the uncertainties in these models.
The prompt component is not accounted for in the systematic parameters, as it is a subleading component in the energy range of interest and its uncertainty can be absorbed in the conventional component's systematic uncertainties.

\begin{description}
    \item[Conventional Normalization] \hfill \\
        A 40\% normalization uncertainty is applied to the nominal conventional flux derived by \texttt{MuEx}.

    \item[Cosmic Ray Spectral Slope] \hfill \\
        The cosmic ray model used approximately follows a falling power-law spectrum.
        Uncertainties exist in the spectral shape of this cosmic ray flux, which translates into a spectral shape uncertainty in the produced neutrino flux. 
        This is accounted for by the inclusion of a spectral shape correction term $\Delta \gamma$ in the conventional flux 
        \begin{equation}
            \Phi(E_\nu; \Delta \gamma) = \Phi(E_\nu) \left( \frac{E}{\SI{2.2}{\TeV}} \right)^{-\Delta \gamma},
        \end{equation}
        where $\Phi(E_\nu)$ is the nominal conventional neutrino flux.
        $\Delta \gamma$ is centered at $0$ and given a prior width of $0.03$.

    \item[Barr Gradients (WP, WM, YP, YM, ZP, ZM)] \hfill \\
        The uncertainties in hadronic production are parameterized using the Barr parametrization \cite{Barr:2006it}.
        In the Barr scheme, uncertainties in $\pi^{\pm}$ and $K^{\pm}$  production are grouped into different regions of incident particle energy $E_\textrm{i}$ and $x_{\textrm{lab}}=E_{\textrm{s}}/E_{\textrm{i}}$, where $E_{\textrm{s}}$ is the secondary total energy.
        \Cref{fig:Barr} shows how these regions are divided for $\pi^{\pm}$ and $K^{\pm}$ separately.
        Because the conventional flux is primarily composed of $K^{\pm}$ and the energy range of interest begins in the hundreds of \GeV, only the W, Y, and Z regions are considered.
        For the kaons, the regions for $K^{+}$ and $K^{-}$ are treated separately, so a total of six Barr parameters are constructed: WP, WM, YP, YM, ZP, and ZM, where the ``P'' and ``M'' indicate if it's for the positively charged or negatively charged kaons, respectively.
        These Barr parameters represent a modification to the $K$ production rate at their respective $E_\textrm{i}$ and $x_{\textrm{lab}}$.
        The uncertainties are: 40\% for WP and WM, 30\% for YP and YM, and $12.2\% \times \log_{10} ( E_{\textrm{i}}/\SI{500}{\GeV})$ for ZP and ZM.

        \begin{figure}
            \centering
            \includegraphics[width=0.70\textwidth]{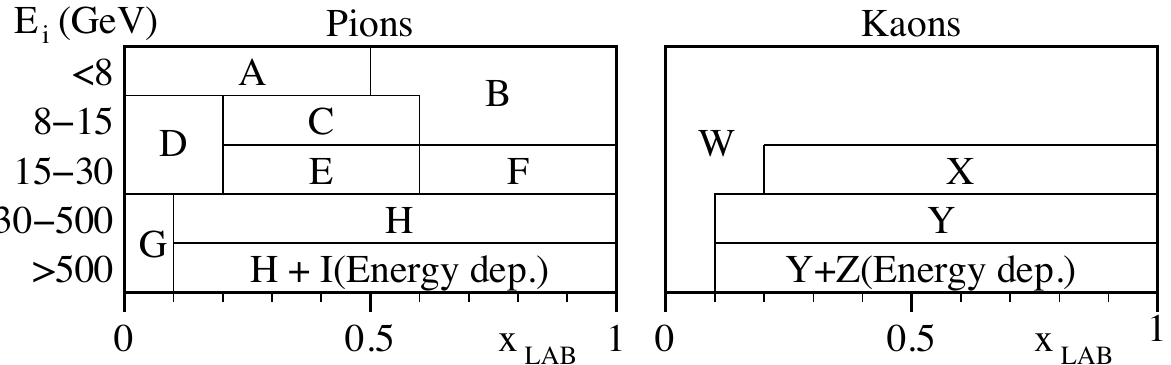}
            \caption{Different uncertainty regions for the meson production rate.
            The boundaries are chosen so that different regions correspond roughly to different physical effects. 
        Figure taken from Ref.~\cite{Barr:2006it}.}
            \label{fig:Barr}
        \end{figure}

    \item[Atmospheric Density] \hfill \\
        Hadronic production and evolution will depend on the atmospheric density profile.
        A nominal profile is taken from the AIRS satellite \cite{AIRS}.
        Uncertainties provided by the AIRS satellite is used to perturb the density profile and extract the effect these uncertainties have on the atmospheric flux.

\end{description}

While only a single cosmic ray and hadronic interaction model were chosen to derive a nominal neutrino flux, several discrete models exist.
It has been found that the combination of the systematic parameters above cover the different possible neutrino fluxes derived by different combinations of these models \cite{Moulai:2021zey}.

\subsection{Detector Parameters}

\begin{description}

    \item[DOM Efficiency] \hfill \\
        The DOM efficiency parameter accounts for the correlated photon efficiency of the DOM's photocathodes globally, as well as any other external properties that may effectively reduce the overall photons detected by the DOMs in IceCube.

    \item[Ice Gradients] \hfill \\
        The ice in which IceCube is embedded has been deposited over the course of thousands of years under varying conditions.
        Therefore, the ice is not uniform and one must expect varying photon propagation properties as you move deeper into the ice.
        The uncertainty on this was parameterized with the ``SnowStorm'' method \cite{IceCube:2019lxi}.
        SnowStorm decomposes the depth dependent ice properties into Fourier modes, where lower order modes correspond to variations that span slowly across the height of the detector, and higher order modes correspond to variations that change rapidly. 
        These variations are able to be decomposed into two parameters, which are simply called ``ice gradient 0'' and ``ice gradient 1.''

    \item[Hole Ice] \hfill \\
        The hole ice refers to the column of ice around the detector strings that were melted to install the DOMs and later refroze.
        This process changed the property of the ice immediately surrounding the DOMs, introducing an angular acceptance effect for photons approaching the DOMs.
        Notably, towards the central axis of the refrozen columns, the ice is more opaque due to bubbles or impurities, affecting photons that are traveling straight up towards the DOMs.
        The effect of these upward-going photons is parameterized with the ``forward hole ice'' systematic.

\end{description}

\subsection{Astrophysics Parameters}

Towards higher energies, astrophysical neutrinos start to become a significant fraction of the neutrino flux, as can be seen in \Cref{fig:eventrate}.
This astrophysical flux is assumed to be isotropic and equal for all three neutrino and antineutrino flavors.
This flux is included and is given by 
\begin{equation}
    \frac{dN_\nu}{dE} = \Phi_{\textrm{astro}} \times \left( \frac{E_\nu}{\SI{100}{\TeV}} \right)^{-\gamma_{\textrm{astro}}}.
\end{equation}
The two relevant astrophysical systematic parameters are the astrophysical normalization $\Phi_{\textrm{astro}}$ and the spectral correction $\Delta \gamma_{\textrm{astro}}$, which modifies the spectral index like $\gamma_{\textrm{astro}} = 2.5 + \Delta \gamma_{\textrm{astro}}$. 
The normalization is centered at 
\begin{equation}
    \Phi_{\textrm{astro}} = \SI{0.787e-18}{\per\GeV\per\steradian\per\second\per\square\cm}
\end{equation}
and the spectral correction is centered at $\Delta \gamma_{\textrm{astro}} = 0$. 
The prior widths for the two parameters are correlated, but in one dimension they are both set to $0.36$.

\subsection{Cross Section Parameters}

\begin{description}

    \item[Neutrino Cross Section] \hfill \\
        As the neutrinos propagate through the Earth, their flux is affected by their interactions with the nucleons in the Earth.
        Uncertainties in the neutrino-nucleon cross section would therefore affect the expected neutrino flux at IceCube.
        These uncertainties are incorporated into the fit, with a $3\%$ uncertainty for the neutrino cross section and a $7.5\%$ uncertainty for the antineutrino cross section.

    \item[Kaon Cross Section] \hfill \\
        Mesons produced by cosmic rays travel for some distance before decaying.
        These interactions affect the final energy of these mesons before decaying, in turn affecting the daughter neutrino energies.
        The interaction cross section between kaons and nuclei have not been measured at these high energies.
        Therefore, a $\pm7.5\%$ uncertainty is placed on the theoretical predictions of this cross section.

\end{description}

\begin{table}
\centering
\begin{tabular}{ l c }
\hline
\hline
\multicolumn{2}{c}{\textbf{Conventional Flux Parameters}}\\
\hline
\hline
Normalization ($\Phi_{\mathrm{conv.}}$)   &  1.0 $\pm$ 0.4\\
\hline
Spectral shift ($\Delta\gamma_{\mathrm{conv.}}$)   &  0.00 $\pm$ 0.03\\
\hline
Atm. Density      &  0.0 $\pm$ 1.0\\
\hline
Barr WP           &  0.0 $\pm$ 0.4\\
\hline
Barr WM           &  0.0 $\pm$ 0.4\\
\hline
Barr YP           &  0.0 $\pm$ 0.3\\
\hline
Barr YM           &  0.0 $\pm$ 0.3\\
\hline
Barr ZP           &  0.0 $\pm$ 0.12\\
\hline
Barr ZM           &  0.0 $\pm$ 0.12\\
\hline
\hline
\multicolumn{2}{c}{\textbf{Detector Parameters}}\\
\hline
\hline
DOM Efficiency    &  0.97 $\pm$ 0.10\\
\hline
Hole Ice (p$_2$)   &  -1.0 $\pm$ 10.0\\
\hline
Ice Gradient 0    &  0.0 $\pm$ 1.0*\\
\hline
Ice Gradient 1    &  0.0 $\pm$ 1.0*\\
\hline
\hline
\multicolumn{2}{c}{\textbf{Astrophysics Parameters}}\\
\hline
\hline
Normalization ($\Phi_{\mathrm{astro.}}$)     &  0.787 $\pm$ 0.36*\\
\hline
Spectral shift ($\Delta\gamma_{\mathrm{astro.}}$)   &  0.0 $\pm$ 0.36*\\
\hline
\hline
\multicolumn{2}{c}{\textbf{Cross Section Parameters}}\\
\hline
\hline
Cross Section $\sigma_{\nu_\mu}$   &  1.00 $\pm$ 0.03\\
\hline
Cross Section $\sigma_{\overline{\nu}_\mu}$    &  1.000 $\pm$ 0.075\\
\hline
Kaon Energy Loss $\sigma_{KA}$   &  0.0 $\pm$ 1.0\\
\hline
\hline
\end{tabular}
\caption{A table of the systematic parameters included in the previous MEOWS analysis and the current MEOWS$+\theta_{34}$ analysis.
The star indicates that the uncertainty is correlated with the adjacent starred parameter.}
\label{tab:sysparameters}
\end{table}

    \chapter{Analysis Procedure and Results}
\label{ch:ICresults}

\section{Physics Parameters}

In our analysis, we fit over both the physics and systematic parameters. 
For the physics parameters, the scan is done over a discrete 3-dimensional grid in\\ $(\Dmqfo, \Umufsq, \Utaufsq)$-space. 
The physics point are sampled as follows:
$\Dmqfo \in [0.1,50]\ \eVq$, in steps of 0.1 in $\log_{10}(\Dmqfo)$ starting at $\Dmqfo=0.1\ \eVq$ with 50 \eVq appended at the end; 
$\Umufsq \in [0.001, 0.5]$, in steps of 0.1 in $ \log_{10}(\Umufsq)$ starting at $\Umufsq=0.001$ with 0.5 appended at the end; and 
$\Utaufsq \in [0.001, 0.5]$, in steps of 0.2 in $ \log_{10}(\Utaufsq)$ starting at $\Utaufsq=0.001$ with 0.5 appended at the end.

While the previous MEOWS analysis presented their results in terms of \sinsqtthtf, we choose to present our results in terms of \Umufsq and \Utaufsq.
The connection between the mixing matrix parameters $|U_{\alpha i}|^2$ and mixing angles $\theta_{i j}$ are given in \Crefrange{eq:usandthetas1}{eq:usandthetas3}.

Additionally, we choose to set $\delta_{24} = \pi$.
We make this choice for two reasons.
First, adding a fourth physics parameter to fit over would be too computationally expensive to be completed in a reasonable amount of time. 
Second, we found the choice of $\delta_{24}$ to have a small effect on the sensitivity when compared to the improvement obtained by the inclusion of $\theta_{34}$.
Increasing $\theta_{34}$ weakens the \numu disappearance while strengthening the \numubar disappearance, resulting in a partial cancellation of the effect.
Further, while the cancellation isn't exact, the fact that the conventional neutrino flux has more \numu than \numubar, and the \numu cross section is greater than the \numubar cross section, means that increasing $\delta_{24}$ weakens the resonance effect.
Therefore, setting $\delta_{24}$ to its maximal value, $\pi$, is the conservative choice. 

\section{Binning \& Likelihood}

The data is binned in two dimensions: reconstructed zenith $\cos\theta$ and reconstructed energy $E$.
The zenith angle $\cos\theta$ ranges between $-1$ and 0, with 20 bins of width 0.05.
Here, $\cos\theta=-1$ corresponds to events coming from directly below the detector, and $\cos\theta=0$ corresponds to events coming from the horizon.
The reconstructed energy ranges from 500 to 9976 \GeV, in 13 bins of width $\log_{10} (E/[\GeV]) = 0.1$.

For a single bin, the Poisson likelihood
\begin{equation}
    \mathcal{L}(\theta, \theta_{\eta}| k) =  \frac{\lambda (\theta, \theta_{\eta})^k e^{- \lambda(\theta, \theta_{\eta})}}{k!}
\end{equation}
gives the likelihood of observing $k$ events, given an expectation of $\lambda (\theta, \theta_{\eta})$, where $\theta$ is some set of physics parameters and $\theta_{\eta}$ is some set of systematic parameters.

When the expectation $\lambda (\theta, \theta_{\eta})$ is determined through MC, finite simulations lead to an uncertainty in $\lambda (\theta, \theta_{\eta})$. This is accounted for with an effective likelihood 
\begin{equation}
    \mathcal{L_{\rm{Eff}}}(\theta, \theta_{\eta}| k) = \left( \frac{\mu}{\sigma^2} \right)^{\frac{\mu^2}{\sigma^2}+1} \Gamma \bigg( k +  \frac{\mu^2}{\sigma^2}+1 \bigg) \bigg[ k! \left(1+\frac{\mu}{\sigma^2} \right)^{k + \frac{\mu^2}{\sigma^2}+1}  \Gamma \left( \frac{\mu^2}{\sigma^2}+1 \right) \bigg]^{-1},
\end{equation}
derived in Ref.~\cite{Arguelles:2019izp}.
$\mu$ and $\sigma^2$ are determined by the weights $w$ of each MC event,
\begin{equation}
    \mu = \sum_i w_i \quad  \sigma^2 = \sum_i w_i^2.
\end{equation}

In addition to the statistical likelihood, the prior likelihood of a set of systematic parameters $\theta_\eta$ is given by 
\begin{equation}
    \Pi(\theta_{\eta})
    =  
    \prod_\eta \frac{1}{\sqrt{2 \pi \sigma_\eta^2}} e^{\frac{-(\theta_\eta - \Theta_\eta)^2}{2\sigma_\eta^2}},
\end{equation}
where $\Theta_\eta$ and $\sigma_\eta$ are the prior central values and widths, respectively. 

The final likelihood is given by the product of the effective likelihood and the prior likelihood 
\begin{equation}
    \mathcal{L}(\theta, \theta_{\eta}| k) =  \mathcal{L}_{\rm{Eff}}(\theta, \theta_{\eta}| k) \Pi(\theta_{\eta}).
    \label{eq:likelihood}
\end{equation}
For simplicity, we drop the $k$ label.

\section{Frequentist Analysis} 
\label{sec:freqanalysis}

In the frequentist analysis, the likelihood ratio
\begin{equation}
    \Lambda = \frac{
        \sup_{\theta_{\eta}} 
        \mathcal{L}(\theta_{0}, \theta_{\eta})
    }{
        \sup_{\theta, \theta_{\eta}} 
        \mathcal{L}(\theta, \theta_{\eta})
    }
    \label{eq:lambda}
\end{equation} 
is used to construct the test statistic. In \Cref{eq:lambda}, the ``null'' hypothesis $\theta_0$ is compared to an alternative hypothesis where the physics parameters $\theta$ are free and $\theta_0 \in \theta$.
In both models, the systematic parameters $\theta_{\eta}$ are set to maximize the likelihood.
In this context, the ``null'' hypothesis does not necessarily have to be the null physics parameters ($\Dmqfo=0, \Umufsq=0, \Utaufsq=0$).

At each sampled physics point $\theta$, the negative likelihood is minimized over the systematic parameters $\theta_{\eta}$. 
The minimization is done with the internal IceCube software \texttt{GolemFit} using the L-BFGS-B algorithm \cite{10.1145/279232.279236}. 

We define the test statistic 
\begin{equation}
    \textrm{TS}
    \equiv
    -2 \log \Lambda
    = -2 (\ell (\theta_{0}) - \ell (\hat{\theta})
    ,
    \label{eq:TS}
\end{equation}
where 
\begin{equation}
    \ell (\theta_{0}) = \log
    [\sup_{\theta_{\eta}} 
    \mathcal{L}(\theta_{0}, \theta_{\eta})], \quad
    \ell (\hat{\theta}) = \log 
    [\sup_{\theta, \theta_{\eta}} 
    \mathcal{L}(\theta, \theta_{\eta})].
\end{equation}

In this thesis, we assume  Wilks' theorem, so that our test statistic TS follows a $\chi^2$-distribution with degrees of freedom equal to the difference in the number of parameters between the null and alternative hypothesis. 
In our case, this is 3. 

The validity of Wilks' theorem is not guaranteed, and simulated pseudo-experiments have to be run to obtain the proper coverage. 
This is computationally expensive, and is not done in this thesis, but is planned for future publications. 
Despite possibly not following a $\chi^2$-distribution, we will refer to the TS as $\chi^2$, for simplicity.

\section{Bayesian Analysis} 
\label{sec:bayesanalysis}

In addition to the frequentist analysis described above, a Bayesian analysis is done using the Bayes factor
\begin{equation}
    K = \frac{\mathcal{E}_i}{\mathcal{E}_j},
    \label{eq:bayesfactor}
\end{equation}
where $\mathcal{E}$ is the ``evidence'' for some model. 
The evidence is given by 
\begin{equation}
    \mathcal{E}=\int d\theta_{\eta} \mathcal{L}(\theta, \theta_{\eta}),
\end{equation}
where $\mathcal{L}(\theta, \theta_{\eta})$, as defined in \Cref{eq:likelihood}, includes the priors of the systematic parameters. 

The Bayes factor quantifies the support of one model over another. 
Using the Bayes factor, we will compare the evidence of each physics parameter $\theta$ with the no-sterile neutrino hypothesis. 
In \Cref{eq:bayesfactor}, we will set the numerator $\mathcal{E}_i$ as the no-sterile hypothesis.
Therefore, a positive value of $\log_{10} K$ means a preference for the null model, while a negative value means a preference for the alternative model.
A qualitative measure, called Jeffreys scale,  of the strength of the evidence against the null model is given in \Cref{tab:jeffreysscale}, following Ref.~\cite{Jeffreys:1939xee}.

\begin{table}
\centering
\begin{tabular}{c|c}
$\log_{10}K$ & Evidence against null \\
 \hline\hline
 >0 & Null supported \\
 \hline
 -1/2 -- 0 & Not worth more than a bare mention \\
 \hline
 -1 -- -1/2 & Substantial \\
 \hline
    -3/2 -- -1 & Strong \\
 \hline
 -2 -- -3/2 & Very strong \\
 \hline
    <-2 & Decisive
\end{tabular}
\caption{Jeffreys scale, from Ref.~\cite{Jeffreys:1939xee}.}
\label{tab:jeffreysscale}
\end{table}

\section{Sensitivity}

The sensitivity to the sterile parameters is calculated in two ways. The first is an approximate sensitivity referred to as the ``Asimov'' sensitivity \cite{Cowan:2010js}. 
In this method, the median sensitivity is estimated by the use of a single representative data set, as opposed to an ensemble of simulated experiments. 
The benefit is, of course, avoiding the computational limitations of simulating a large number of experimental trials.
We used this method to conduct tests before we unblinded the data.

\begin{figure}
    \centering
    \includegraphics[width=.99\textwidth]{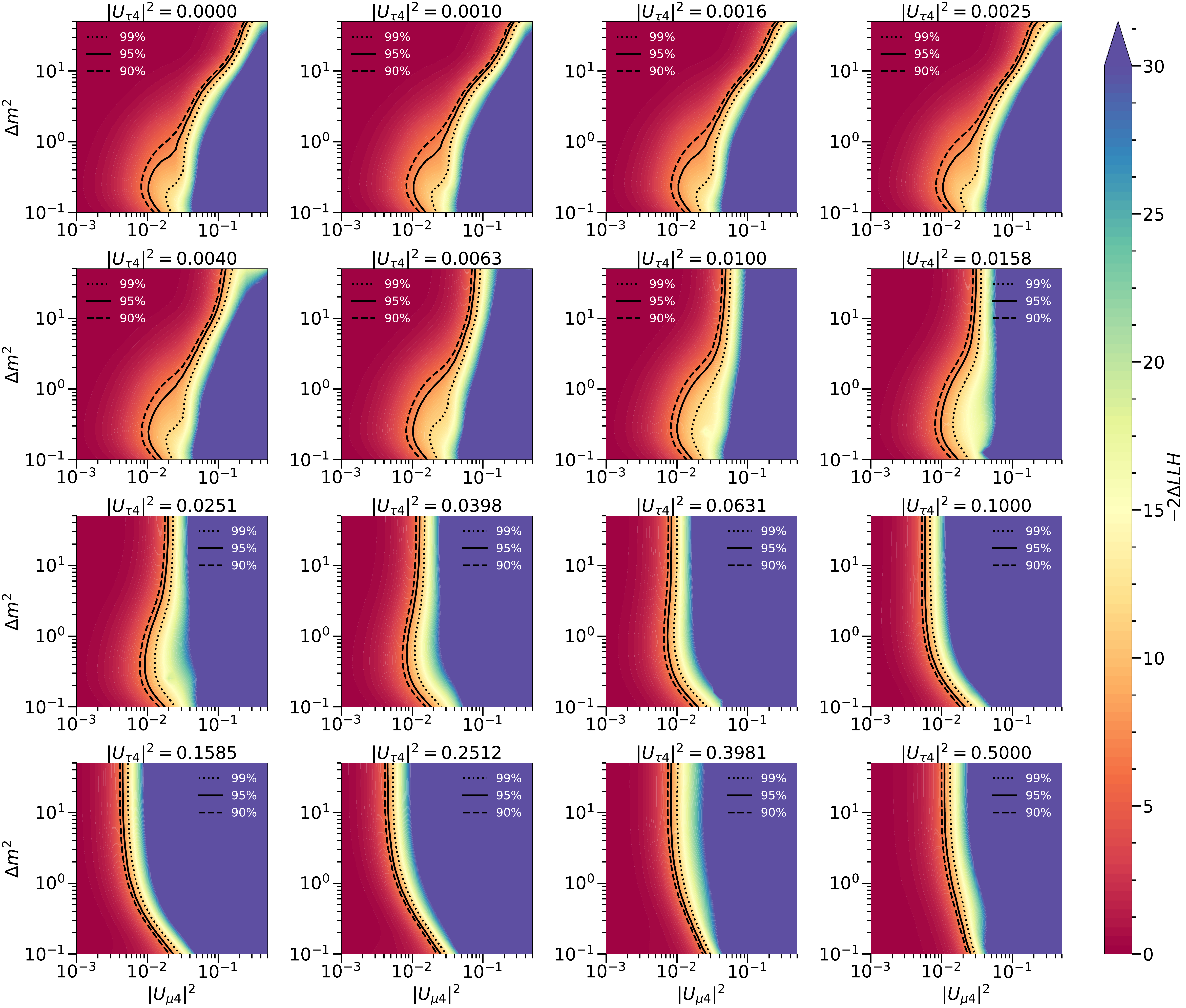}
    \caption{The expected sensitivities for the sterile parameters sampled. Each frame corresponds to a slice of $\Utaufsq$ sampled. The lines shown correspond to the 99\%, 95\%, and 90\% confidence levels assuming Wilks' Theorem.}
    \label{fig:AsimovSensitivitySliced}
\end{figure}

\begin{figure}
    \centering
    \includegraphics[width=0.6\textwidth]{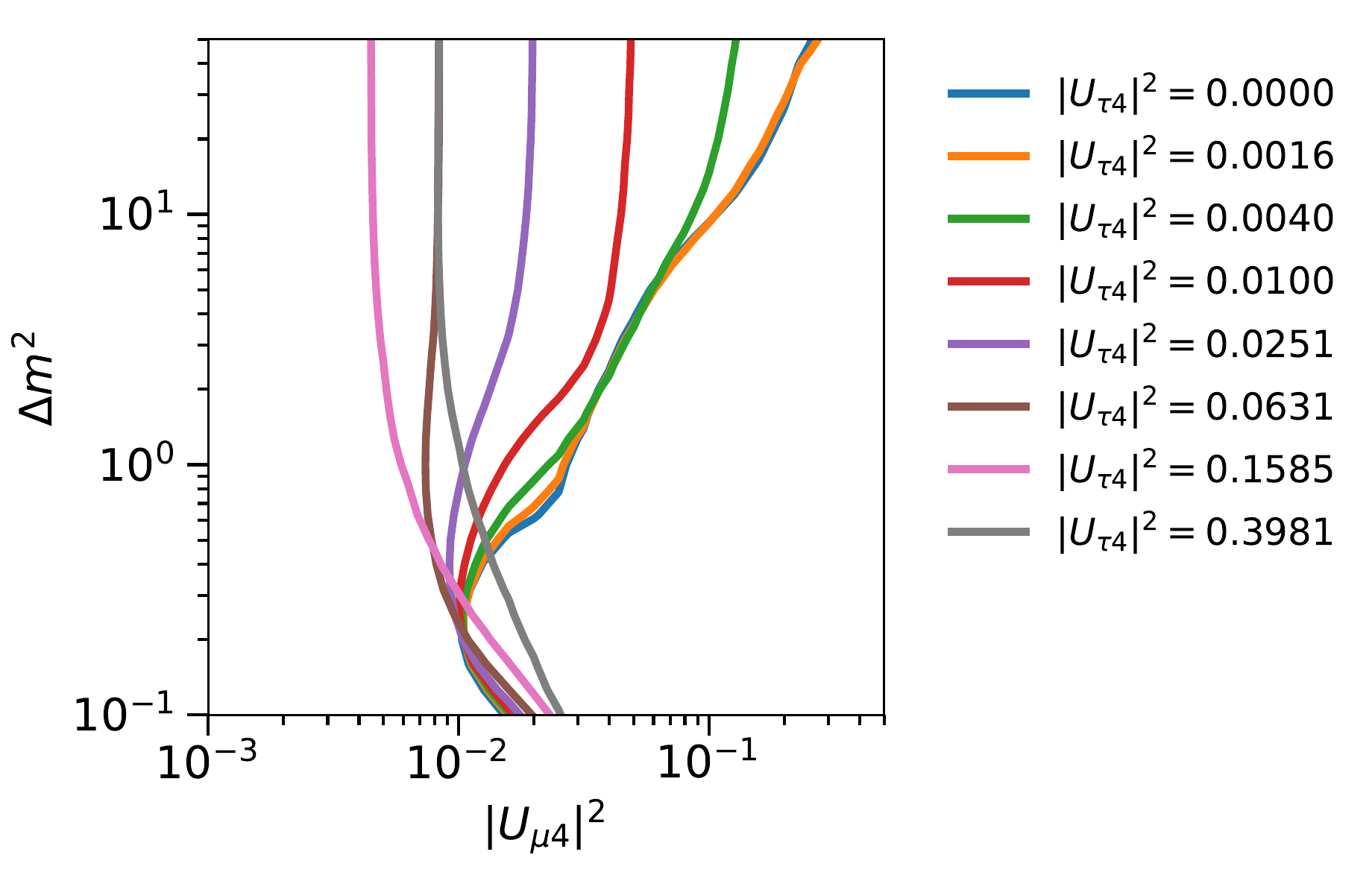}
    \caption{The expected sensitivities for the sterile parameters sampled. The 95\% confidence level is shown for various values of $\Utaufsq$ sampled. As $\Utaufsq$ increases, the sensitivity to the sterile hypothesis improves dramatically.}
    \label{fig:AsimovSensitivity}
\end{figure}

The Asimov data set is chosen to be the expected distribution of events assuming the null hypothesis and central value systematic parameters, with no statistical fluctuation. 
The median Asimov sensitivity is shown at the 99\%, 95\%, and 90\% confidence levels in \Cref{fig:AsimovSensitivitySliced}.
\Cref{fig:AsimovSensitivity} also shows the 95\% confidence level for various values of $\Utaufsq$ overlaid on a single plot. 
\Cref{fig:AsimovSensitivitySliced} and \Cref{fig:AsimovSensitivity} shows how the sensitivity of IceCube improves dramatically as the value of $\Utaufsq$ increases. 

The second method of deriving the sensitivity is to run simulated experiments and finding the average exclusion sensitivities for the ensemble. 
The Asimov sensitivity described above is an approximation of this method, but is substantially faster to compute. 
Therefore, the more accurate method of deriving the sensitivities were not derived until after the results were unblinded. 

\begin{figure}
    \centering
    \includegraphics[width=.99\textwidth]{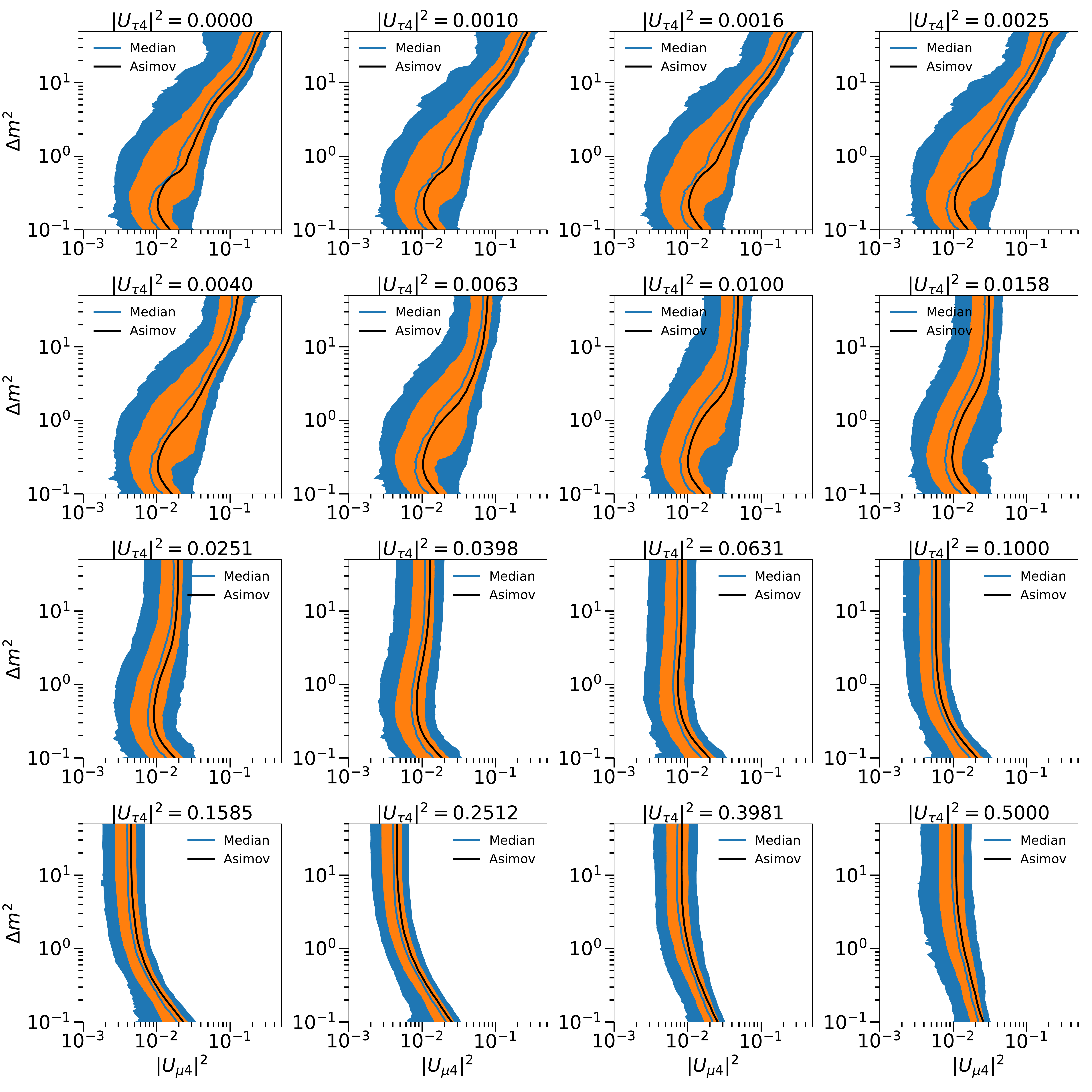}
    \caption{The sensitivity bands derived by generating 400  pseudoexperiments. The blue line is the median sensitivity at the 95\% confidence level, while the orange and blue bands contain 95.45\% ($2\sigma$) and 68.72\% ($1\sigma$) of the sensitivity boundary, respectively. The black lines are the derived Asimov sensitivities.}
    \label{fig:BrazilPlot}
\end{figure}

We ran 400 trials and found the sensitivity distribution as shown in \Cref{fig:BrazilPlot}.
Plotted are the median sensitivity at the 95\%, as well as the band that contains 95.45\% ($2\sigma$) and 68.72\% ($1\sigma$) of the sensitivity curves. 
We also include the Asimov sensitivity as shown previously in \Cref{fig:AsimovSensitivitySliced}.
We find that the Asimov sensitivity did provide a near approximation of the sampled median sensitivity.

\section{Pre-Unblinding Checks}

Before fully unblinding the data, we ran a number of ``blind'' checks on the data to test for significant deviations from the model, without looking at parameters of interest. 
The steps taken, and the results, are:
\begin{enumerate}
    \item We fit over the entire physics and systematic parameter space, keeping the results blind.
    At the best fit physics and systematic point, we run 10,000 realizations. 
    For each realization, we fit for the systematic parameters at the injected physics point. 
    We then compare the observed TS to the distribution of TS obtained from the ensemble test. 
    We chose a p-value of p=0.05 to be a stopping condition. 
    In \Cref{fig:preunblindingstep1}, we show the obtained TS, compared to the observed TS. 
    We find a p-value of 0.82: Step 1 passed.
    
    \begin{figure}
        \centering
        \includegraphics[width=.6\textwidth]{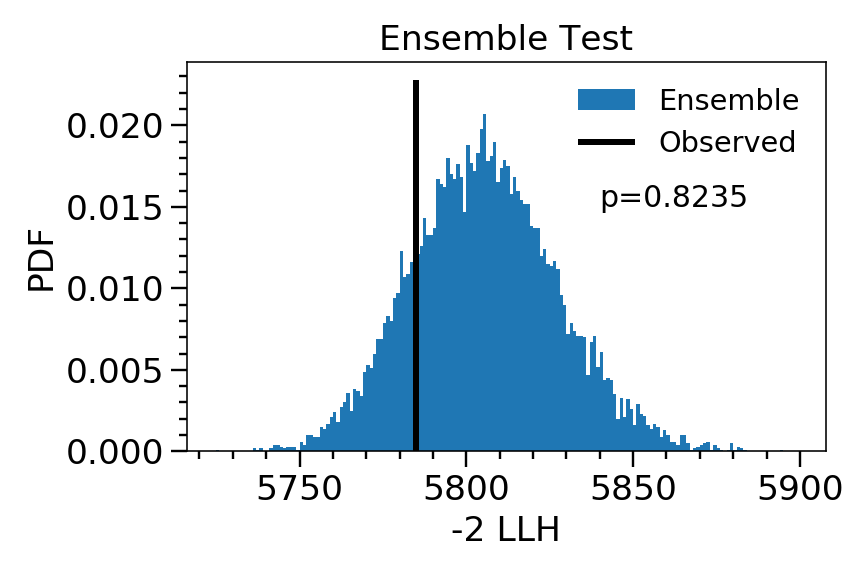}
        \caption{Step 1 of the pre-unblinding check. At the best fit physics and systematic point, 10,000 pseudoexperiments were injected. These pseudoexperiments then had the systematic parameters fitted at the best fit physics point. The plot shows the recovered distribution of the TS, and the black line shows the observed best fit TS.}
        \label{fig:preunblindingstep1}
    \end{figure}
    
    \item For each of the 260 analysis bins, we compare the observed likelihood to the distributions of likelihoods from the ensemble test of Step 1.
    Here, the likelihood used is a Poisson likelihood with the expected event counts taken to be the expectation from the best fit physics and systematic parameters.
    We chose that if six or more bins have a p-value corresponding to greater than $3\sigma$, we would stop. 
    At no single bin was a deviation of $3\sigma$ observed: Step 2 passed. 

    \item If any systematic pulls more than $3\sigma$ or hits a bound, we stop.
    \Cref{fig:preunblindingstep3} shows the best fit systematic parameters. 
    No systematic pulls greater than $3\sigma$ or hit a bound: Step 3 passed.
    
    \begin{figure}
        \centering
        \includegraphics[width=0.99\textwidth]{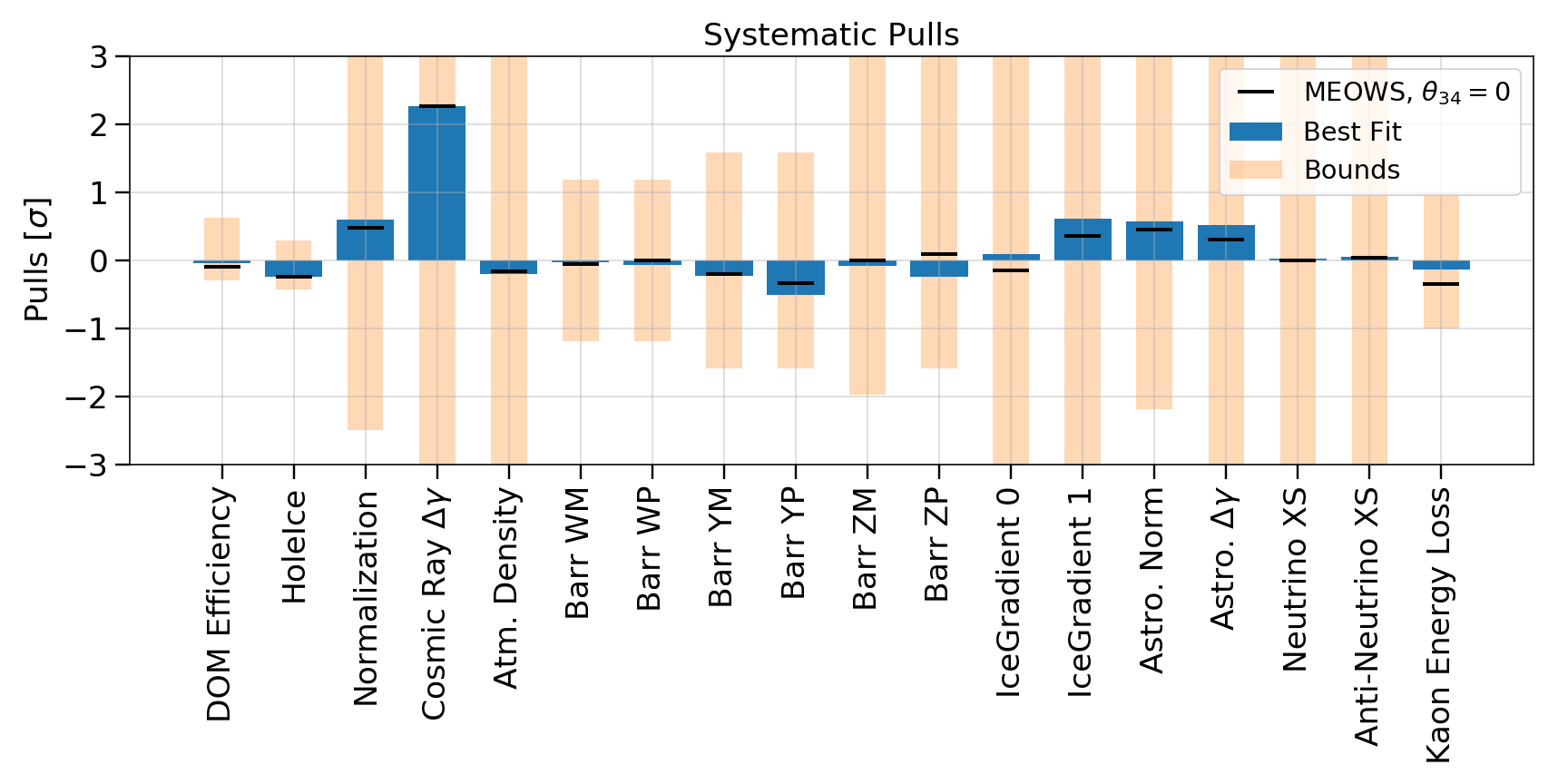}
        \caption{Step 3 of the pre-unblinding check. The plot shows the recovered best fit systematic parameters in blue, in terms of their pull. Not all parameters had bounds that extended beyond $3\sigma$, so their ranges are shown by the orange bars. The black bars show the best fit systematic for the analysis in Refs.~\cite{IceCube:2020phf,IceCube:2020tka}.}
        \label{fig:preunblindingstep3}
    \end{figure}
    
    \item We collapse the observation into two 1D distributions of reconstructed energy and zenith. 
    If any bin pulls greater than $3\sigma$, then we stop. 
    No bins pulled greater than $3\sigma$: step 4 passed. 
    \item We plot the 1D distributions from Step 4, and calculate a $\chi^2$ comparing the data to the best fit expectation. 
    We compare this $\chi^2$ to the distribution of $\chi^2$ obtained from the 10,000 simulated samples. 
    \Cref{fig:preunblindingstep5} shows these distributions.
    We found a p-value of p=0.30 and p=0.39 for the energy and zenith distribution, respectively: step 5 passed.
    \begin{figure}
    \begin{subfigure}{.49\textwidth}
        \centering
        \includegraphics[width=\textwidth]{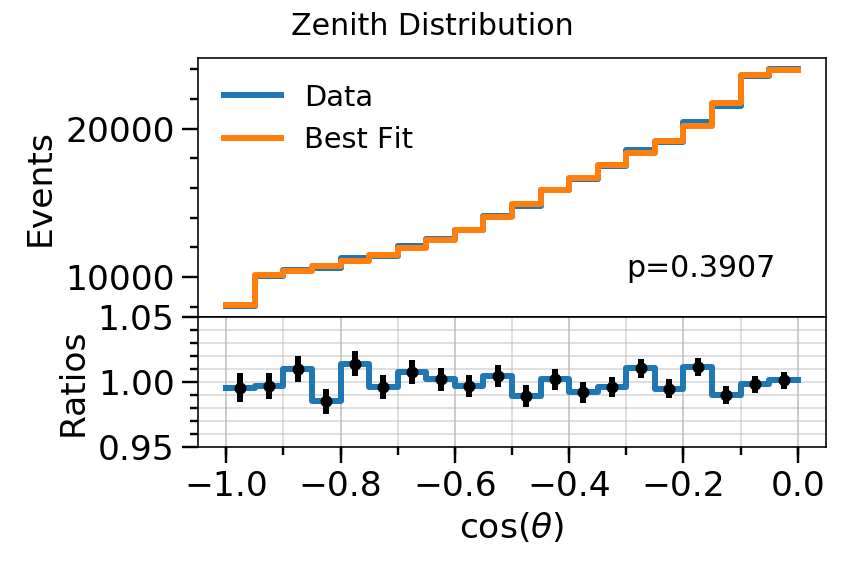}
        \caption{}
    \end{subfigure}
    \begin{subfigure}{.49\textwidth}
        \centering
        \includegraphics[width=\textwidth]{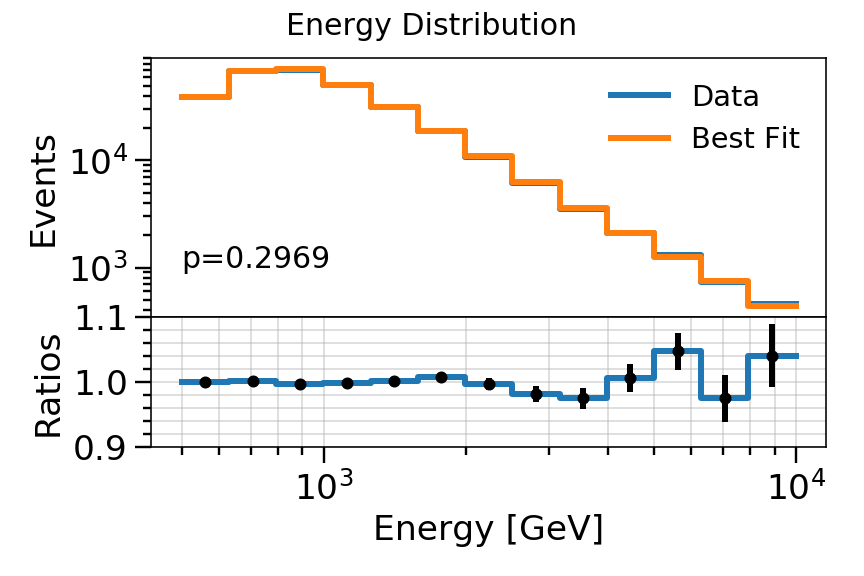}
        \caption{}
    \end{subfigure}
    \caption{Results of step 5 of the pre-unblinding checks. (a) and (b) shows the 1D distributions of zenith and energy, respectively. The distribution of the data is shown along with the expectation at the best fit, and their ratio. The p-value is found by comparing the $\chi^2$ to the obtained $\chi^2$ distribution from the 10,000 pseudoexperiments.}
    \label{fig:preunblindingstep5}
    \end{figure}
\end{enumerate}

With the pre-unblinding tests completed and passing, we then fully unblinded the results. 

\section{Results}

This section presents the results of this thesis analysis.
The author and the IceCube collaboration intend to publish this result, which will lead to updates with higher-statistics tests and simulations..
However, the basic conclusions of the analysis presented here will not change.

\subsection{Frequentist}

\begin{figure}
    \centering
    \includegraphics[width=.99\textwidth]{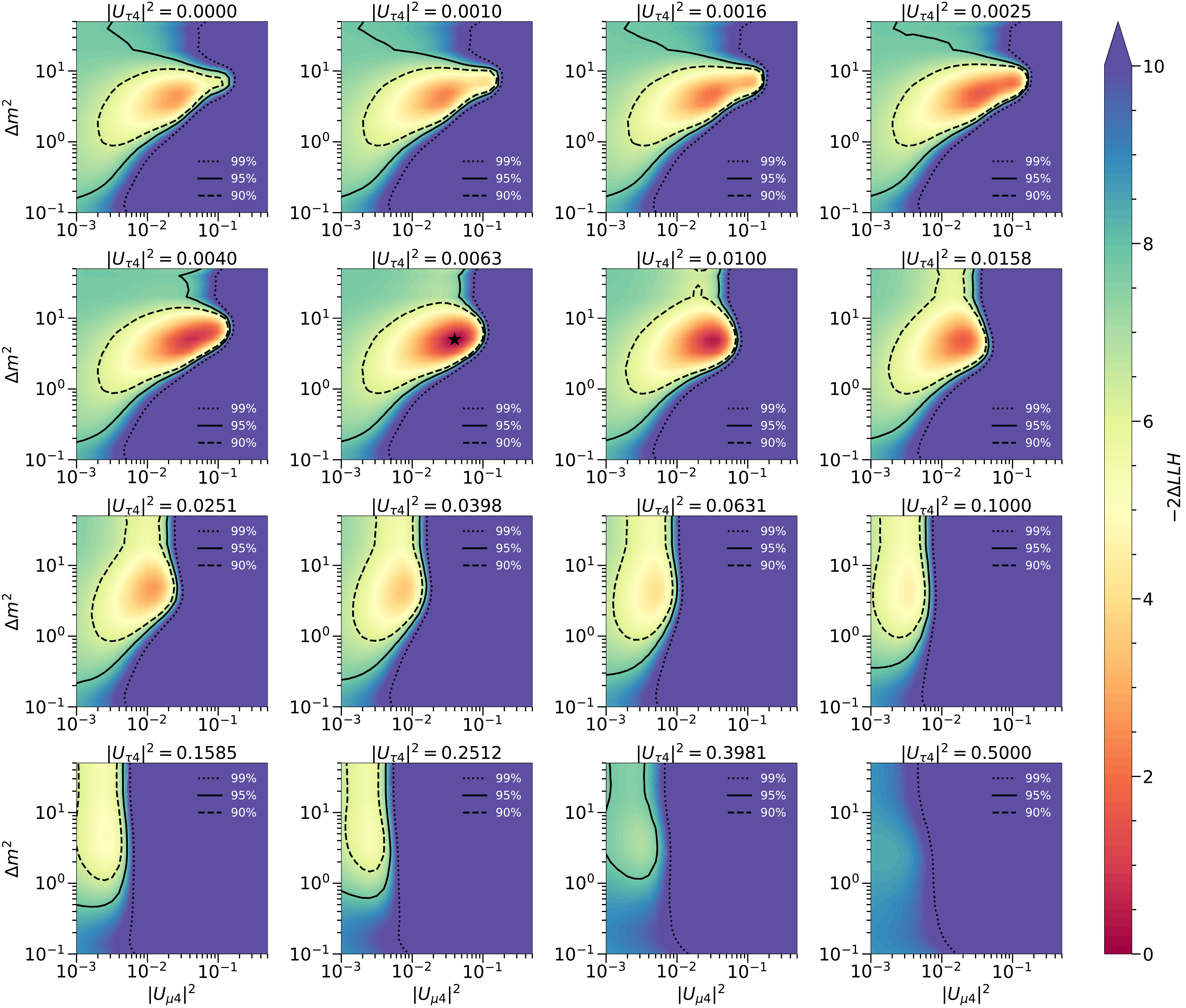}
    \caption{The observed confidence regions at 90\%, 95\%, and 99\% confidence levels. The best fit point is labeled by a star. Here, the three dimensional parameter space is sliced into the sampled values of \Utaufsq.}
    \label{fig:frequentistresults}
\end{figure}

\begin{figure}
    \centering
    \includegraphics[width=.85\textwidth]{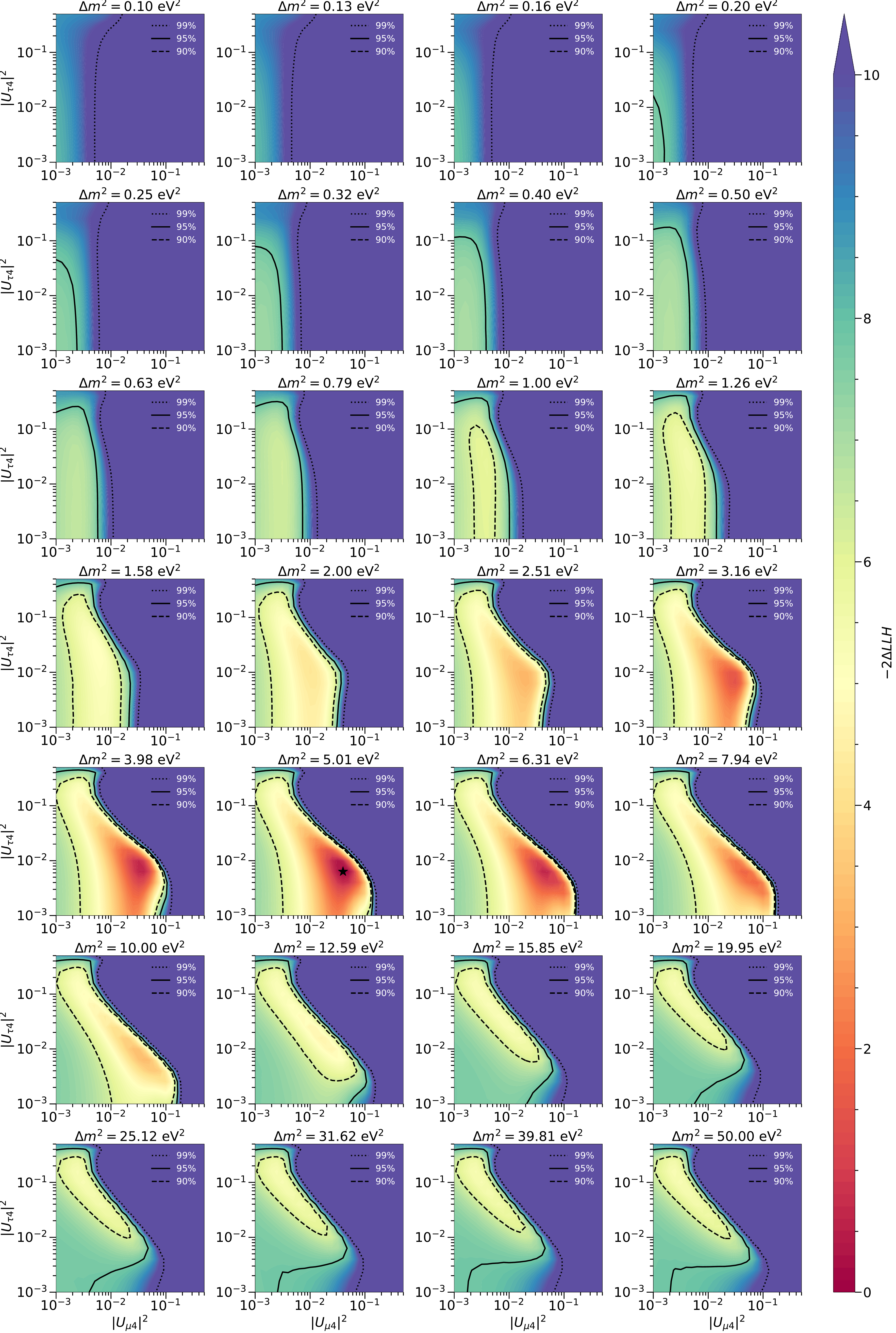}
    \caption{The observed confidence regions at 90\%, 95\%, and 99\% confidence levels. The best fit point is labeled by a star. Here, the three dimensional parameter space is sliced into the sampled values of \Dmqfo.}
    \label{fig:frequentistresultsdm2slices}
\end{figure}

\begin{figure}
    \centering
    \includegraphics[width=.85\textwidth]{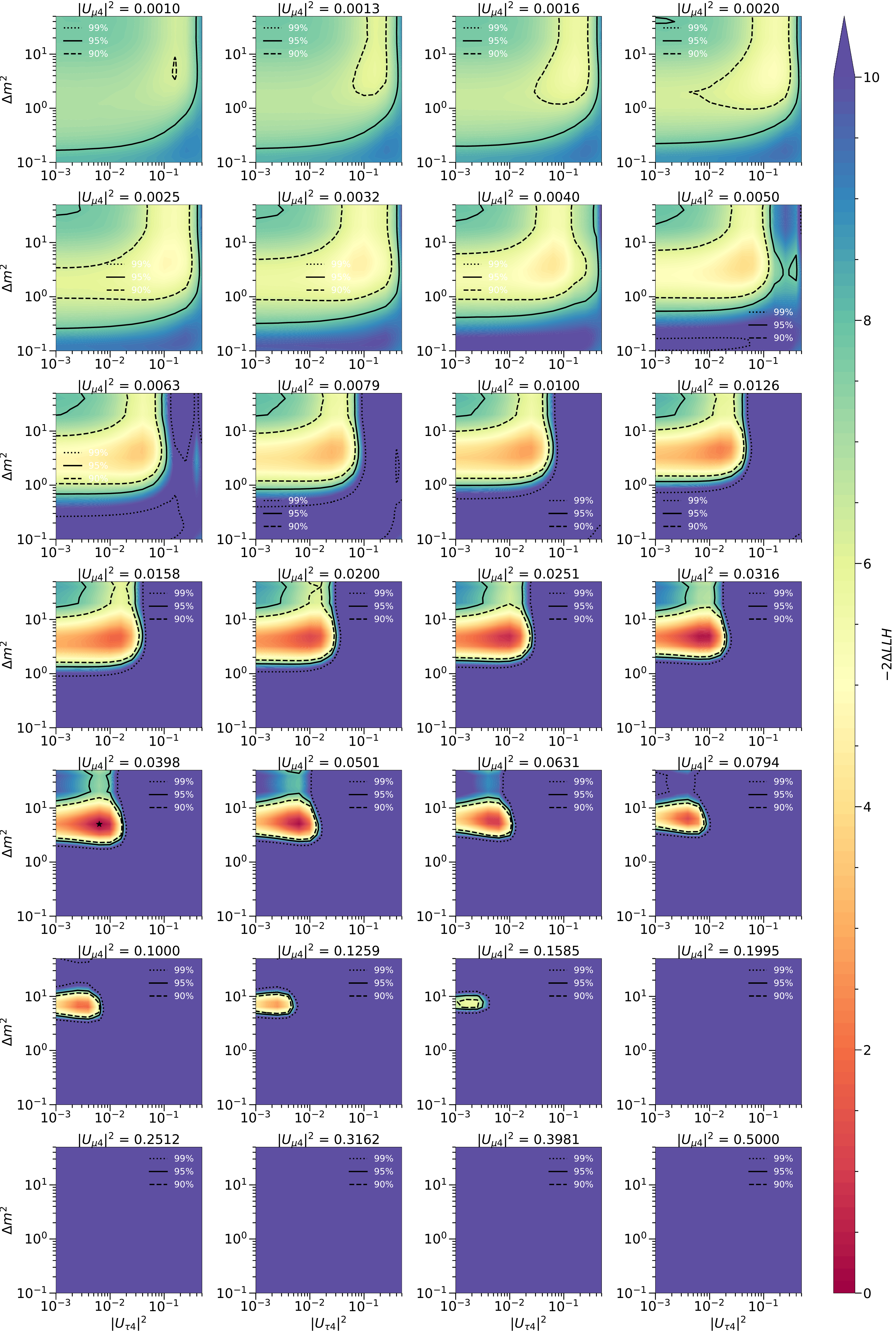}
    \caption{The observed confidence regions at 90\%, 95\%, and 99\% confidence levels. The best fit point is labeled by a star. Here, the three dimensional parameter space is sliced into the sampled values of \Umufsq.}
    \label{fig:frequentistresultssin22thslices}
\end{figure}

For the frequentist scans, the results are shown in \Cref{fig:frequentistresults}. 
The confidence regions are drawn assuming Wilks' Theorem with 3 degrees of freedom, with the 90\%, 95\%, and 99\% confidence levels shown. In \Cref{fig:frequentistresults}, the three dimensional parameter space is sliced into different frames of \Utaufsq. 
In \Cref{fig:frequentistresultsdm2slices} and \Cref{fig:frequentistresultssin22thslices}, we also show the confidence regions sliced in \Dmq and \Umufsq, respectively.
The best fit point is found at $\Dmqfo = 5.0\ \eVq$, $\Umufsq = 0.04$, and $\Utaufsq=0.006$. 
At this point, we obtain a $\Delta \chi^2 = 7.7$.
Assuming Wilks' Theorem with 3 degrees of freedom, this corresponds to a p-value of 5.2\% ($1.94\sigma$). 
Using the pseudoexperiments generated to obtain the sensitivities in \Cref{fig:BrazilPlot}, we can obtain a more accurate p-value of 2.7\% ($2.2\sigma$).

\begin{figure}
    \centering
    \includegraphics[width=0.7\textwidth]{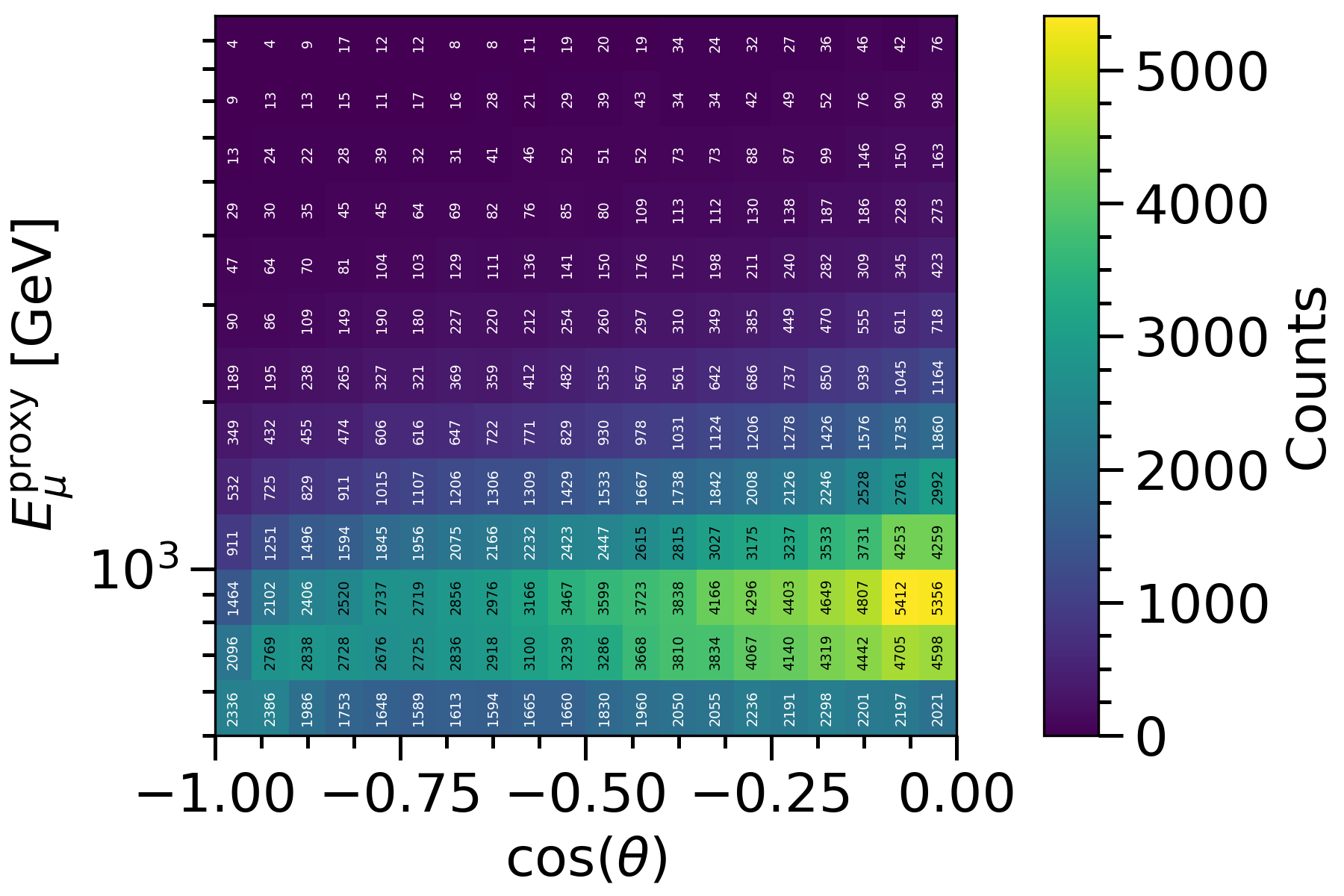}
    \caption{The distribution of events observed in the 7.634 year MEOWS sample.}
    \label{fig:Data}
\end{figure}

\begin{figure}
    \centering
    \includegraphics[width=0.7\textwidth]{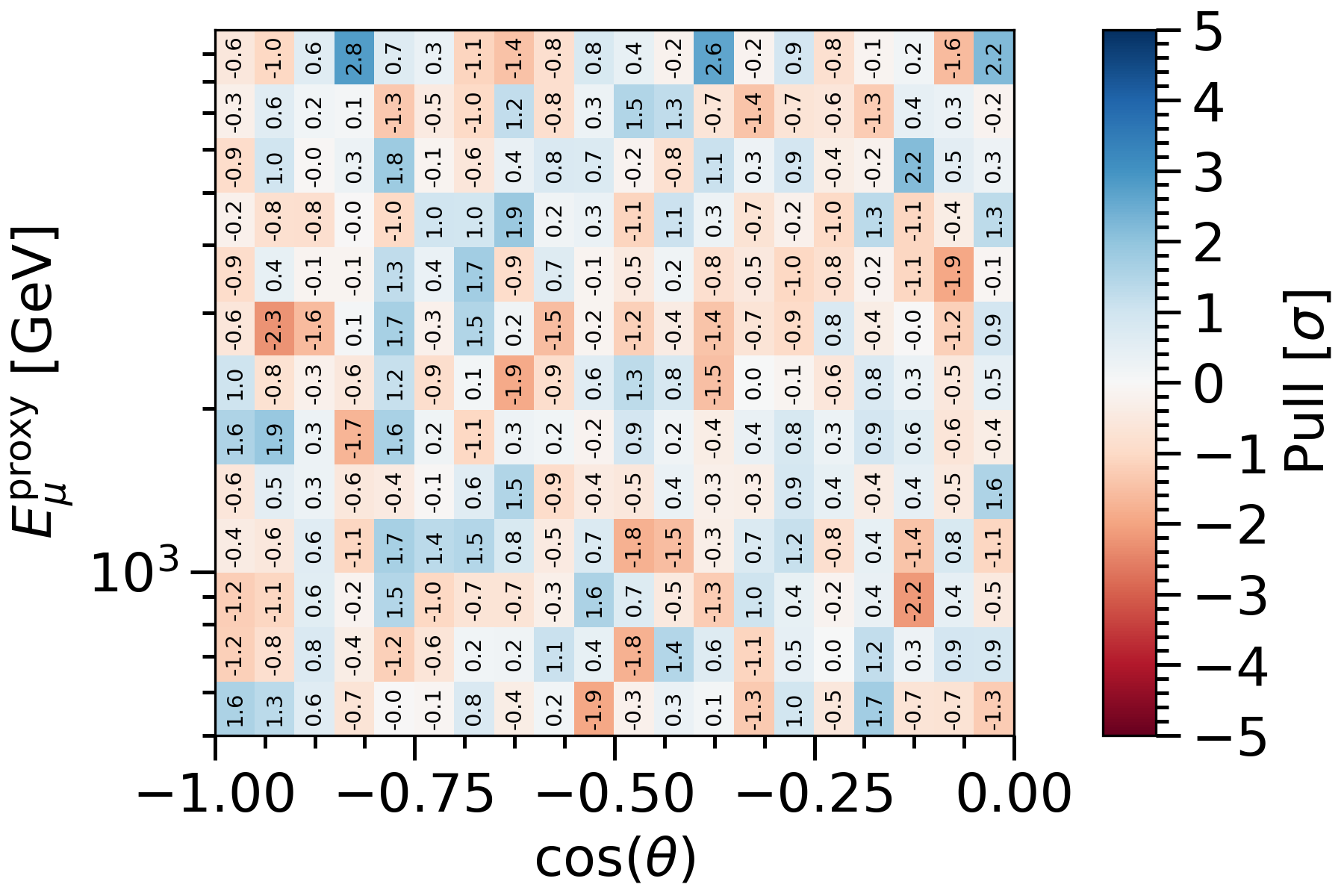}
    \caption{The statistical pulls of the data relative to the best fit physics and systematic parameters.}
    \label{fig:DatavsBFPulls}
\end{figure}

In \Cref{fig:Data}, we show the data distribution. 
The pulls of the data, relative to the best fit physics and systematic parameters, are shown in \Cref{fig:DatavsBFPulls}.



\begin{figure}
    \centering
    \includegraphics[width=.99\textwidth]{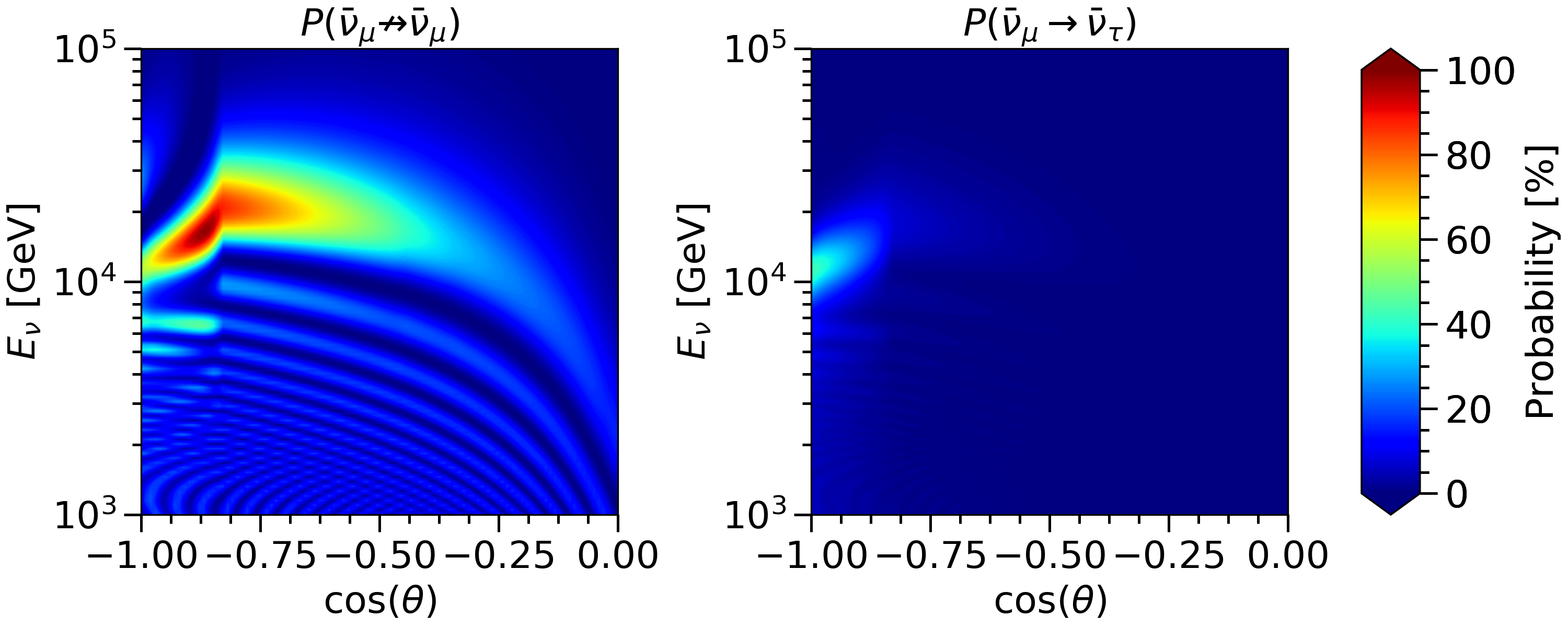}
    \caption{The oscillograms for $\numubar \to \numubar$ and $\numubar \to \nutaubar$ oscillations at the best fit point. The plot on the left shows the ratio of $\numubar$ flux at IceCube between the best fit model and the null model. The plot on the right shows the ratio of the expected $\nutaubar$ flux at the best fit model over the expected $\numubar$ flux with the null model.}
    \label{fig:BFoscillogram}
\end{figure}

In \Cref{fig:BFoscillogram}, we show the oscillogram at this best fit point, compared to the null. 

\begin{figure}
    \centering
    \begin{subfigure}{.49\textwidth}
        \includegraphics[width=\textwidth]{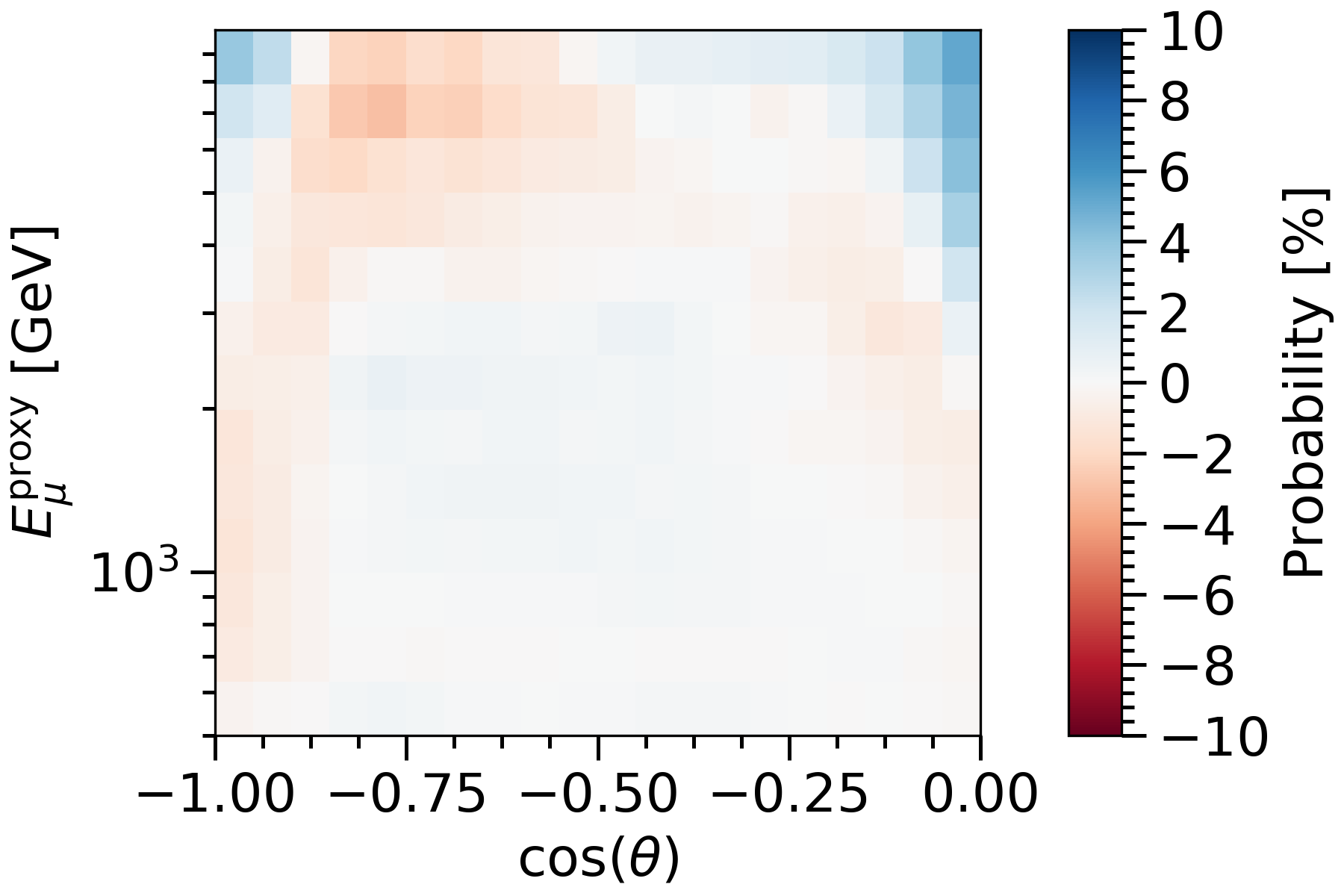}
        \caption{}
        \label{fig:BFvsNullExpectation}
    \end{subfigure}
    \begin{subfigure}{.49\textwidth}
        \includegraphics[width=\textwidth]{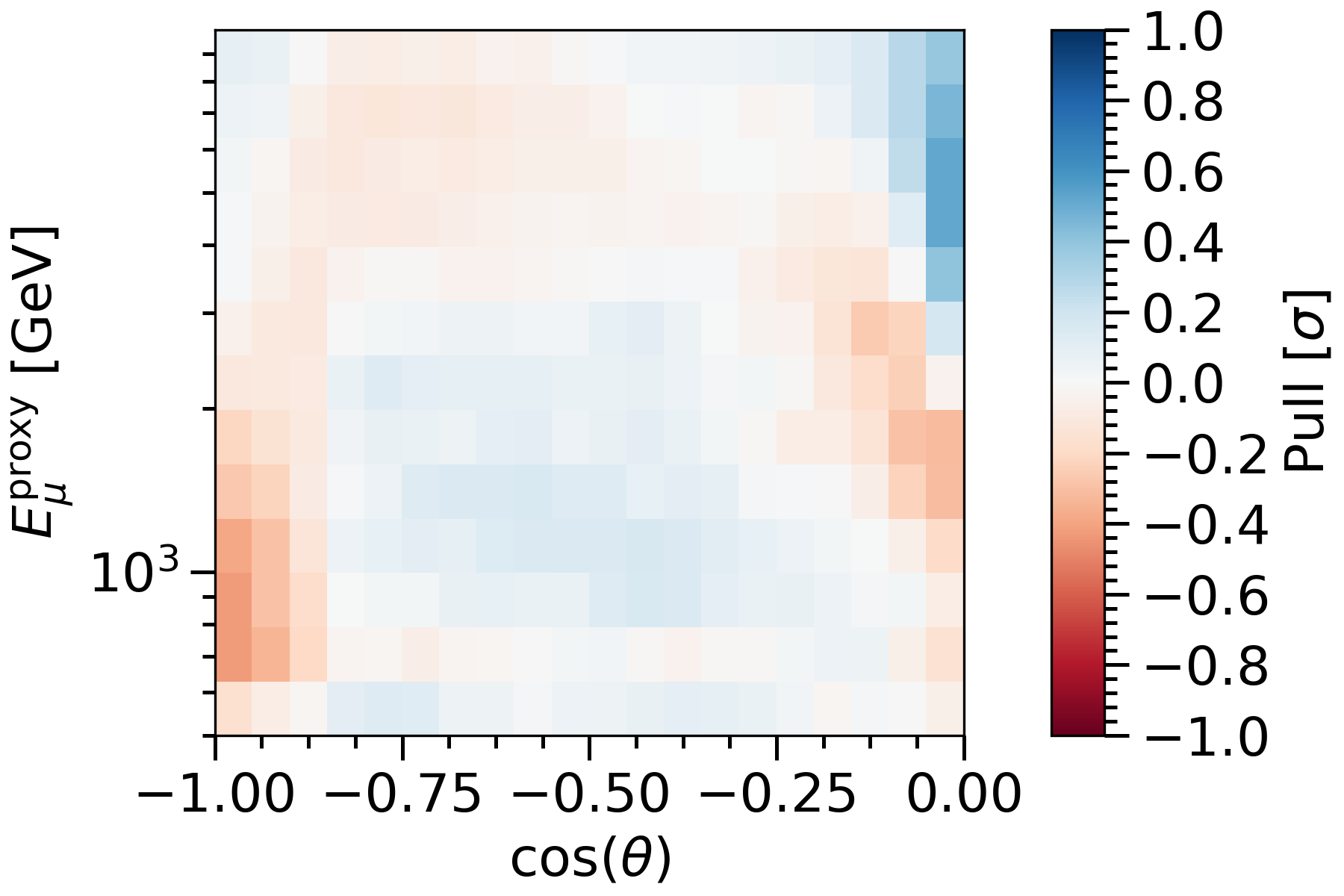}
        \caption{}
        \label{fig:BFvsNullPulls}
    \end{subfigure}

    \bigskip
    \begin{subfigure}{.49\textwidth}
        \includegraphics[width=\textwidth]{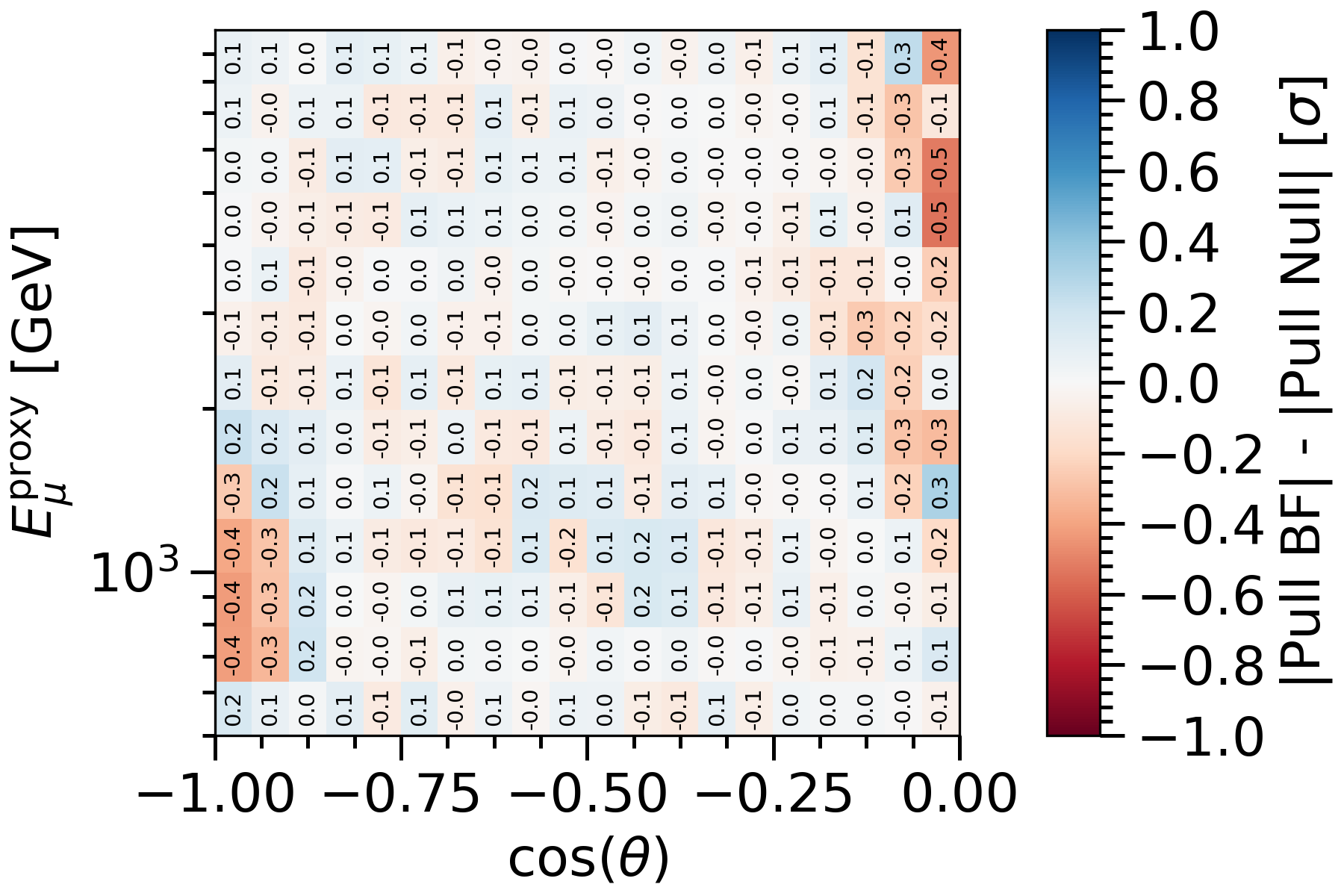}
        \caption{}
        \label{fig:DatavsBFPullsvsNullPulls}
    \end{subfigure}
    \caption{(a) The percent difference of expected events between the best fit sterile parameters versus the null. (b) The statistical pull of the expected event rate at the best fit sterile parameters relative to the null expected event rate. (c) The difference in the absolute values of the data pulls between the sterile and null model.}
\end{figure}

We show in \Cref{fig:BFvsNullExpectation} the percent difference between the expected event rate at the best fit versus the no-sterile hypothesis. 
In both cases, we use the best fit systematic parameters for each physics hypothesis. 
We can see here that, while the disappearance occurs mainly at true neutrino energies $E_\nu > 10\ \TeV$ (see \Cref{fig:BFoscillogram}), the signal appears below $10\ \TeV$ in reconstructed $E_\mu$.
In \Cref{fig:BFvsNullPulls}, we show the statistical pull that the best fit expectation has against the null best fit. 
Finally, in \Cref{fig:DatavsBFPullsvsNullPulls}, we show the difference between the absolute value of the data pulls relative to the null fit versus the absolute value of data pulls relative to the best fit. 
This last figure reveals the bins that are pulling the best fit to the sterile parameters. 

\begin{figure}
	\centering
        \begin{subfigure}{.6\textwidth}
            \centering
            \includegraphics[width=\textwidth]{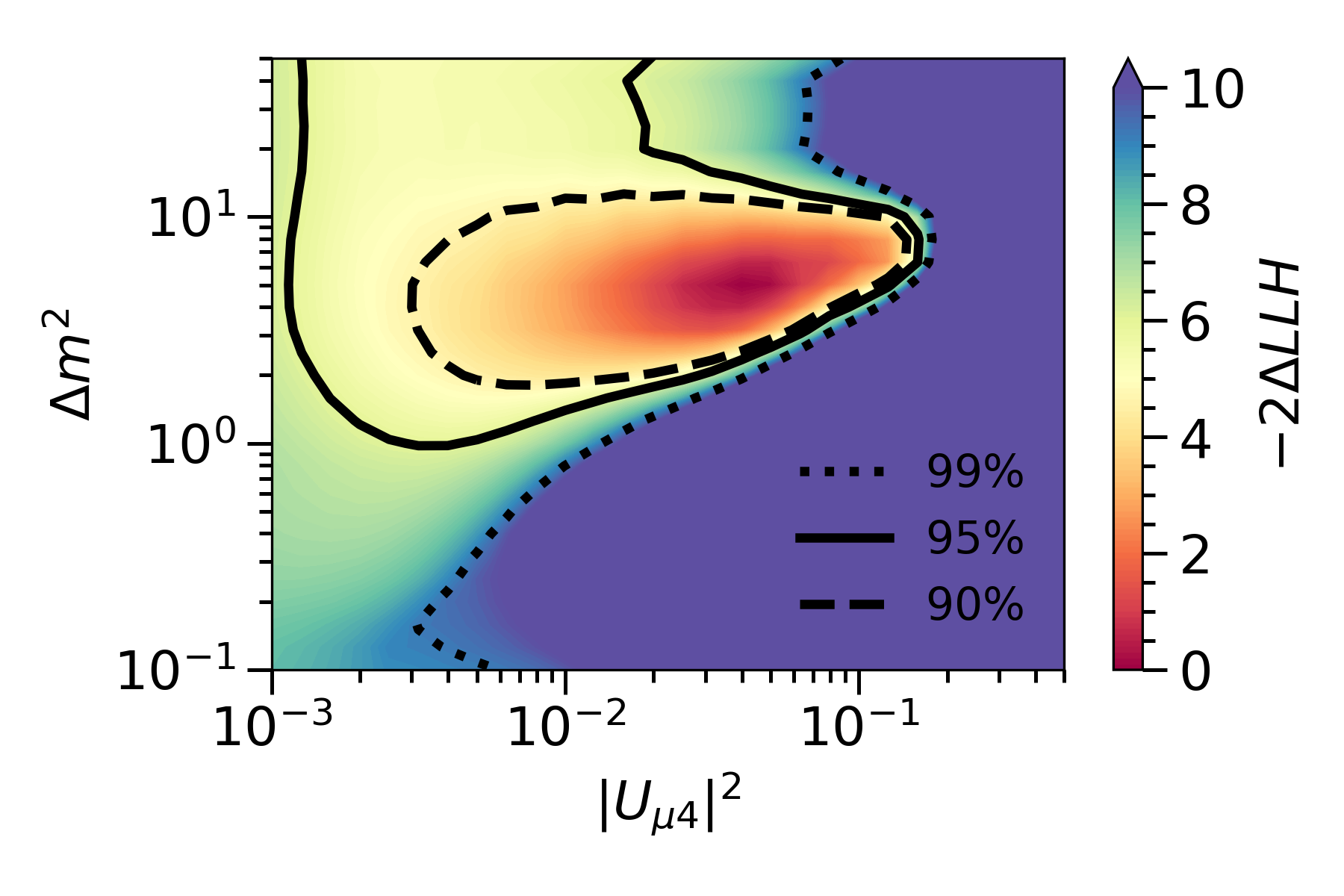}
            \caption{Confidence regions with \Utaufsq profiled.}
            \label{fig:profiledut4sq}
        \end{subfigure}

        \begin{subfigure}{.6\textwidth}
            \centering
            \includegraphics[width=\textwidth]{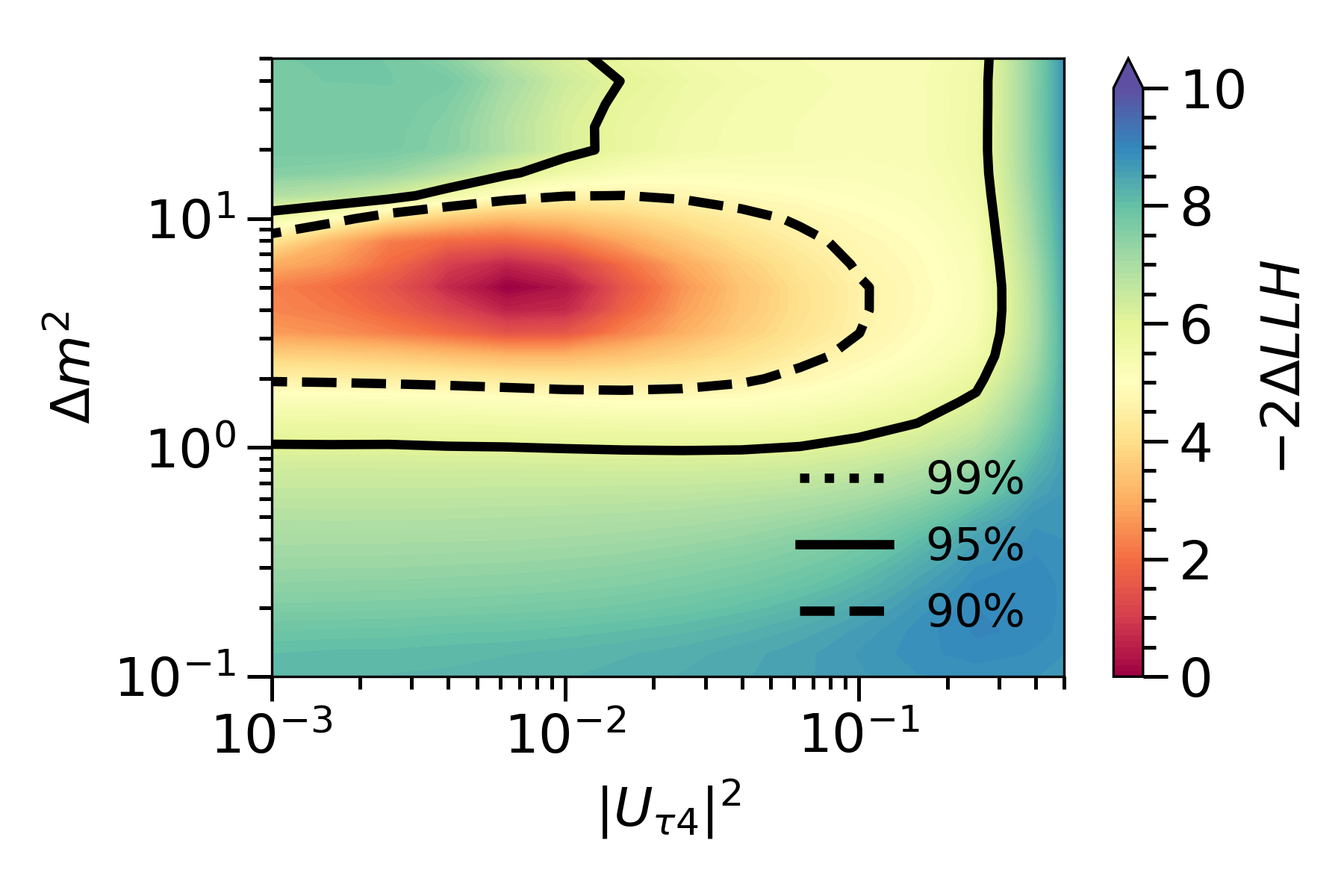}
            \caption{Confidence regions with \Umufsq profiled.}
            \label{fig:profiledum4sq}
        \end{subfigure}

        \begin{subfigure}{.6\textwidth}
            \centering
            \includegraphics[width=\textwidth]{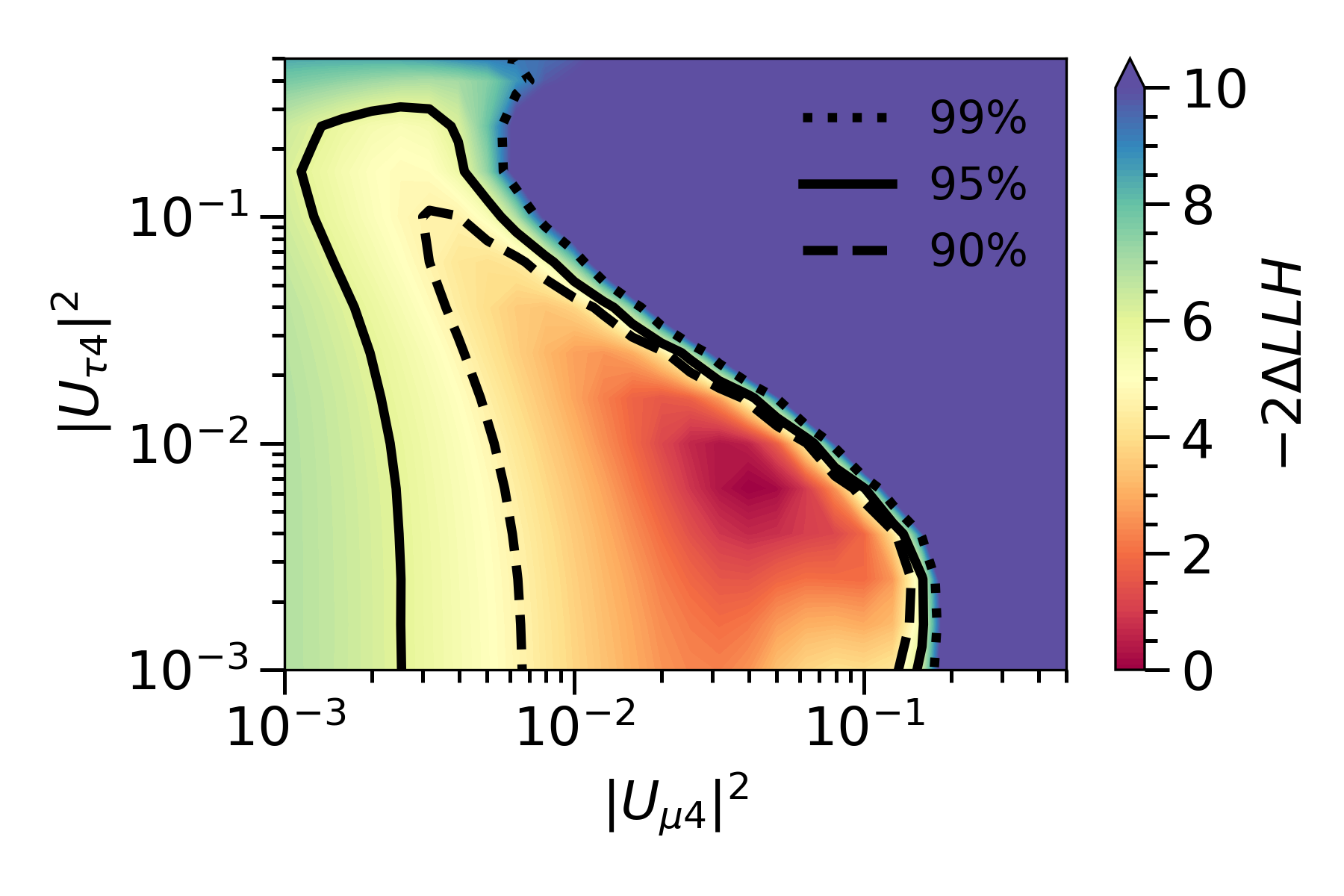}
            \caption{Confidence regions with  \Dmqfo  profiled.}
            \label{fig:profileddm2}
        \end{subfigure}

        \caption{The MEOWS+$\theta_{34}$ results, profiled over each of the fitted physics parameters.}
        \label{fig:profiled}
    \end{figure}

While a small number of experiments have conducted a multidimenstional fit including $\theta_{34}$, the author of this work is not aware of any that release the three (or more) dimensional confidence regions like in \Cref{fig:frequentistresults,fig:frequentistresultsdm2slices,fig:frequentistresultssin22thslices}. 
We therefore include two-dimensional confidence regions, profiled over the third dimension, to allow comparisons. 
In \Cref{fig:profiled}, we show the confidence regions after profiling over each of the physics parameters, with the 90\%, 95\%, and 99\% confidence levels drawn assuming Wilks' Theorem and two degrees of freedom.

\subsection{Bayesian}

\begin{figure}
    \centering
    \includegraphics[width=.99\textwidth]{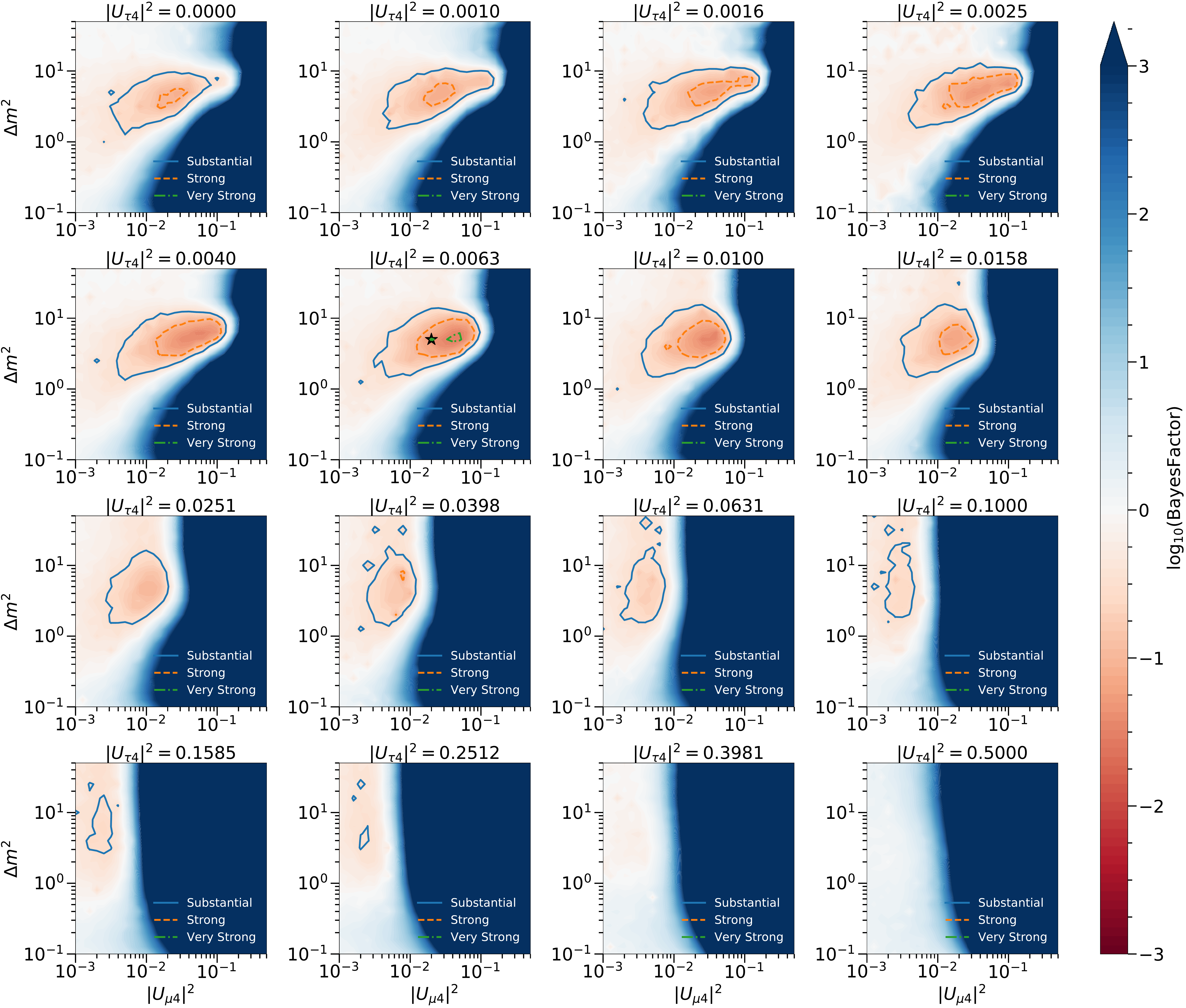}
    \caption{The result of the Bayesian fit, in slices of \Utaufsq.  Here, the Bayes factor is calculated relative to the null hypothesis. A negative value corresponds to a preference to that hypothesis compared to the null. A positive value corresponds to a preference for the null.}
    \label{fig:bayesianresults}
\end{figure}

\begin{figure}
    \centering
    \includegraphics[width=.83\textwidth]{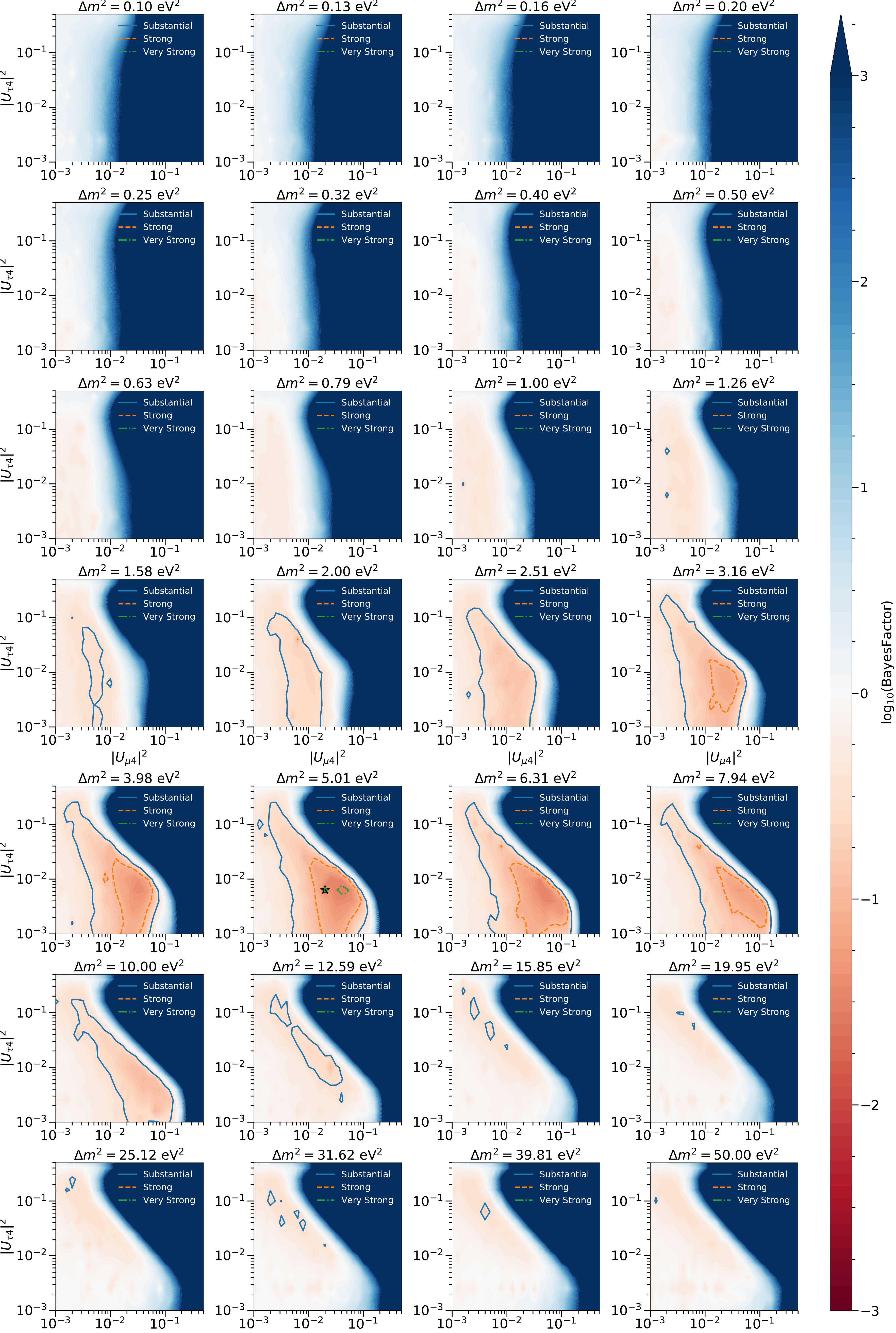}
    \caption{The result of the Bayesian fit, in slices of \Dmqfo.  Here, the Bayes factor is calculated relative to the null hypothesis. A negative value corresponds to a preference to that hypothesis compared to the null. A positive value corresponds to a preference for the null.}
    \label{fig:bayesianresultsdm2}
\end{figure}

\begin{figure}
    \centering
    \includegraphics[width=.83\textwidth]{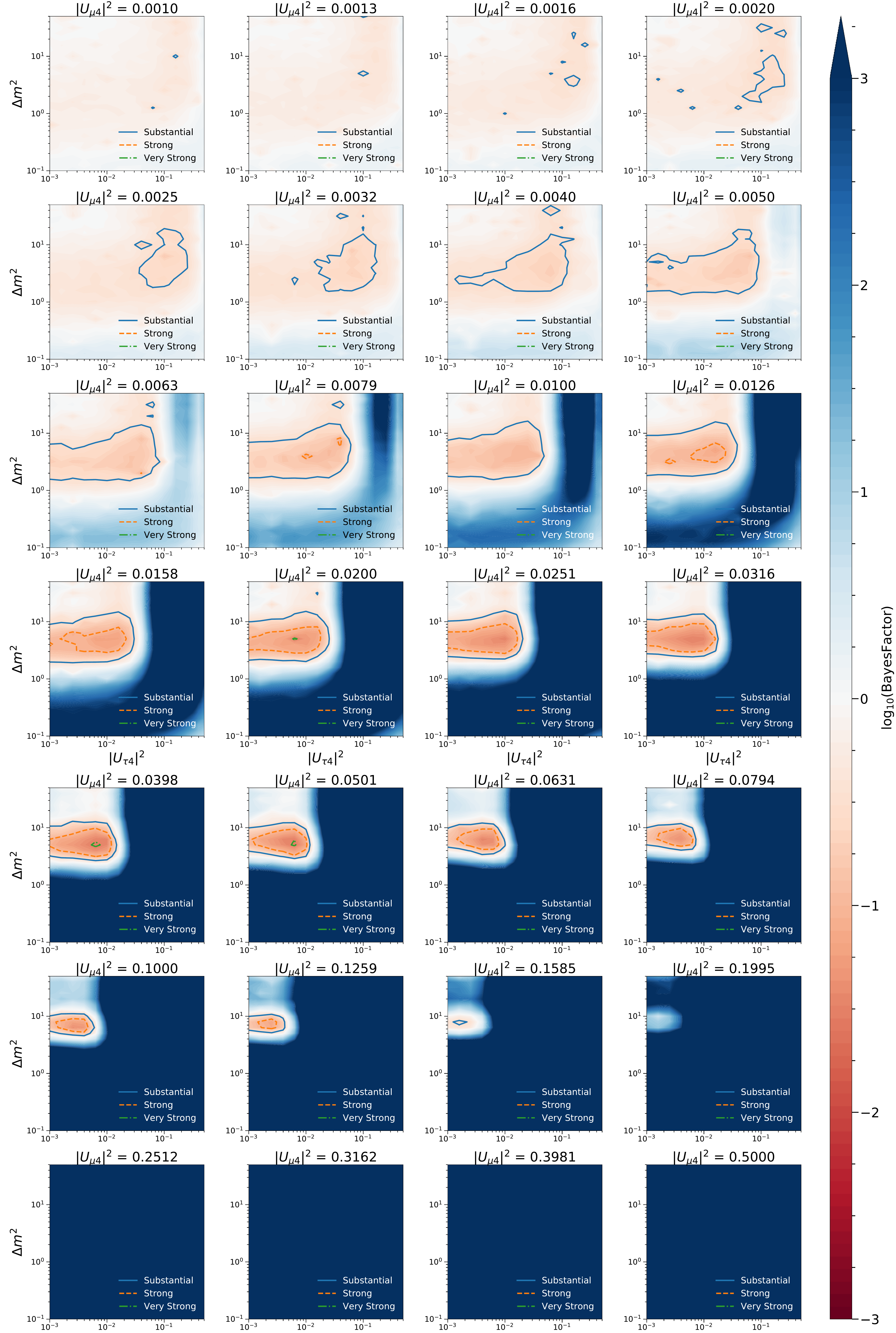}
    \caption{The result of the Bayesian fit, in slices of \Umufsq.  Here, the Bayes factor is calculated relative to the null hypothesis. A negative value corresponds to a preference to that hypothesis compared to the null. A positive value corresponds to a preference for the null.}
    \label{fig:bayeseanresultsum4sq}
\end{figure}

For the Bayesian analysis, the result of the evidence calculations is shown sliced in \Utaufsq in \Cref{fig:bayesianresults}. 
We also show the same results sliced by \Dmqfo in \Cref{fig:bayesianresultsdm2} and \Umufsq in \Cref{fig:bayeseanresultsum4sq}.

\begin{figure}
    \centering
    \includegraphics[width=0.95\textwidth]{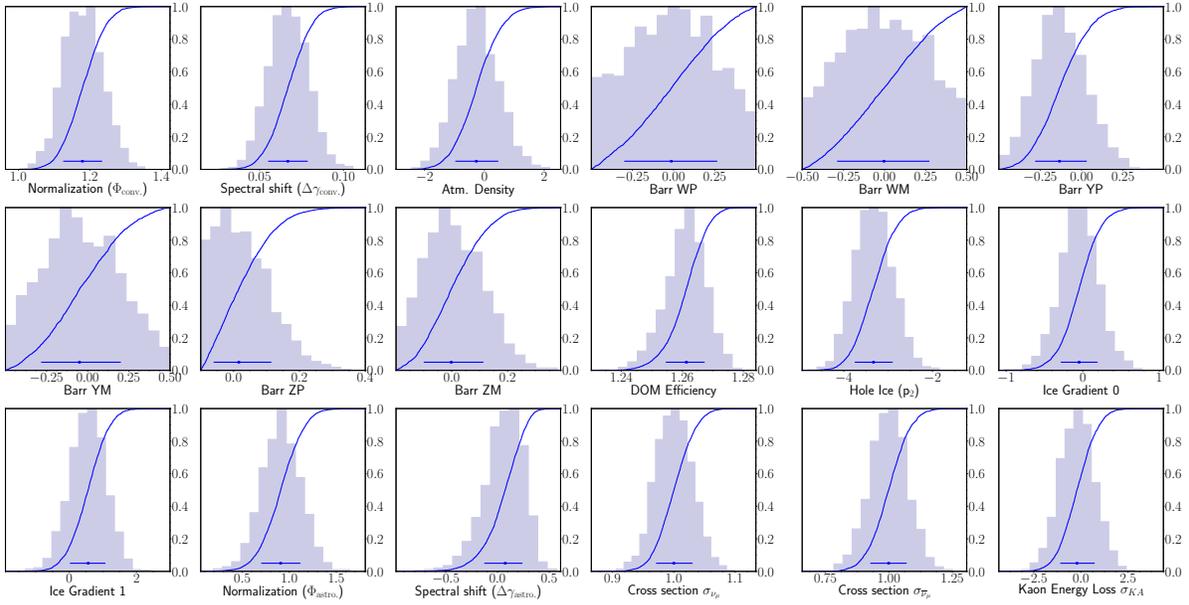}
    \caption{The posterior distributions for each nuisance parameter found at the best fit point. Included also is the cumulative distribution and the $1\sigma$ bounds of each parameter.}
    \label{fig:posteriors}
\end{figure}

\begin{figure}
    \centering
    \includegraphics[width=0.95\textwidth]{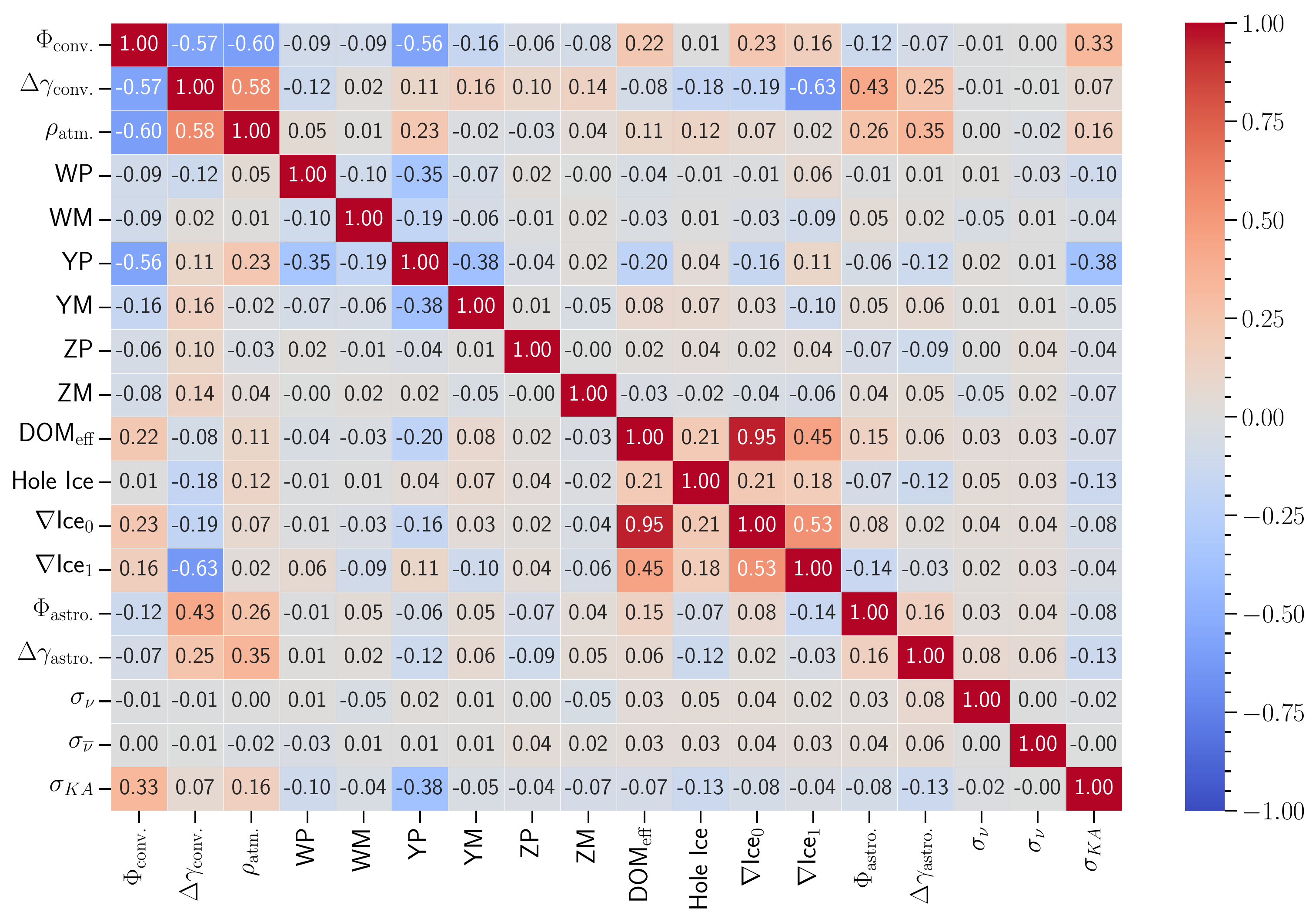}
    \caption{The correlations between the nuisance parameters in the posterior distribution for the best fit point.}
    \label{fig:correlations}
\end{figure}

The point found with the largest evidence is $\Dmqfo = 5.0\ \eVq$, $\Umufsq = 0.02$, and $\Utaufsq=0.006$.
The observed Bayes factor, relative to the null model, was found to be $\log_{10}K=-1.56$.
Following Jeffreys' scale in \Cref{tab:jeffreysscale}, this corresponds to a ``Very Strong'' preference for the sterile model with respect to the non-sterile model. 
We note that the only other point to have a ``Very Strong'' preference for the sterile model, with $\log_{10}K=-1.55$, is found at the same point as the frequentist best fit: $\Dmqfo = 5.0\ \eVq$, $\Umufsq = 0.04$, and $\Utaufsq=0.006$.
A table of the best fit parameters from this analysis is provided in \Cref{tab:bestfitparameters}.
The uncertainties in the nuisance parameters are obtained from the posterior distributions found at the best fit point. 
The posterior distributions for each nuisance parameter is shown in \Cref{fig:posteriors}, and the correlations between these parameters are shown in \Cref{fig:correlations}.

\begin{table}
\centering
\begin{tabular}{ l c }
\hline
\hline
\multicolumn{2}{c}{\textbf{Physics Parameters}}\\
\hline
\hline 
$\Delta m_{41}^2$  & $5.0\ \eVq$\\
\hline
$\Umufsq$ & $0.02$\\
\hline
$\Utaufsq$ & $0.006$\\
\hline
\hline
\multicolumn{2}{c}{\textbf{Conventional Flux Parameters}}\\
\hline
\hline
Normalization ($\Phi_{\mathrm{conv.}}$)   &  1.179 $\pm$ 0.054\\
\hline
Spectral shift ($\Delta\gamma_{\mathrm{conv.}}$)   &  0.067 $\pm$ 0.012\\
\hline
Atm. Density      &  -0.27 $\pm$ 0.72\\
\hline
Barr WP           &  -0.01 $\pm$ 0.28\\
\hline
Barr WM           &  -0.00 $\pm$ 0.28\\
\hline
Barr YP           &  -0.13 $\pm$ 0.16\\
\hline
Barr YM           &  -0.05 $\pm$ 0.24\\
\hline
Barr ZP           &  0.016 $\pm$ 0.088\\
\hline
Barr ZM           &  -0.00 $\pm$ 0.11\\
\hline
\hline
\multicolumn{2}{c}{\textbf{Detector Parameters}}\\
\hline
\hline
DOM Efficiency    &  0.9634 $\pm$ 0.0049\\
\hline
Hole Ice (p$_2$)   &  -3.33 $\pm$ 0.43\\
\hline
Ice Gradient 0    &  -0.05 $\pm$ 0.24\\
\hline
Ice Gradient 1    &  0.56 $\pm$ 0.53\\
\hline
\hline
\multicolumn{2}{c}{\textbf{Astrophysics Parameters}}\\
\hline
\hline
Normalization ($\Phi_{\mathrm{astro.}}$)     &  0.91 $\pm$ 0.21\\
\hline
Spectral shift ($\Delta\gamma_{\mathrm{astro.}}$)   &  0.07 $\pm$ 0.18\\
\hline
\hline
\multicolumn{2}{c}{\textbf{Cross Section Parameters}}\\
\hline
\hline
Cross Section $\sigma_{\nu_\mu}$   &  1.000 $\pm$ 0.030\\
\hline
Cross Section $\sigma_{\overline{\nu}_\mu}$    &  0.999 $\pm$ 0.071\\
\hline
Kaon Energy Loss $\sigma_{KA}$   &  -0.21 $\pm$ 0.91\\
\hline
\hline
\end{tabular}
\caption{Best fit parameters found at the point with the largest evidence.
The uncertainties for the nuisance parameters are obtained from the posterior distribution obtained at the best fit physics point.}
\label{tab:bestfitparameters}
\end{table}

\section{Discussion}

In this chapter we presented the results of an expanded sterile neutrino search in IceCube.

In the frequentist analysis, we find, under the assumption of Wilks' Theorem, a p-value of 5.2\%.
While not reaching the usual standard of $2\sigma$ which the community frequently uses as the standard for a ``signal,'' it's substantial enough to look at more closely.
Interestingly, throwing pseudoexperiments has returned a p-value of only 2.7\%, albeit with not enough realizations thrown to be confident in this value.
As this thesis is being written, presented, and submitted, more pseudoexperiments are being simulated and fitted in order to achieve a more accurate p-value and possibly demonstrate a new experiment with a signal under the 3+1 sterile neutrino model.
We will also conduct tests throughout the sampled parameter space to test if Wilks' Theorem is valid, to more accurately draw the confidence regions.

The Bayesian analysis, on the other hand, seems to show a clearer preference for the sterile model versus the null model, with a Bayes factor that indicates a ``Very Strong'' preference for the sterile model.
With few other sterile neutrino experiments computing a Bayes factor, it's difficult to compare our results with the observation from other experiments.
As this analysis moves forward towards publication, we will see if this observation holds up.
If so, it will likely lead to further study and, possibly, further motivation to search for and study sterile neutrinos.

    \chapter{Conclusion}

In this work, we presented two analyses. 

The first is an update to the sterile neutrino global fits.
Fitting to the global data reveals a strong preference for a 3+1 model versus the null of $6.6\sigma$.
The best fit is at $\Dmqfo = \SI{13.1}{\eV\squared}$, $\Uef=0.30$, and $\Umuf=0.065$.
However, internal tensions make this model unviable.
We explore two other models, 3+2 and 3+1+Decay.
In the 3+2 model, we find no additional improvement to the fit compared to the 3+1 mode. 
In the 3+1+Decay, when the decay width $\Gamma$ is left unconstrained, we find the best fit point at $\Dmqfo = \SI{1.4}{\eV\squared}$, $\Uef=0.3$, $\Umuf=0.09$, and $\tau = \SI{2.7}{\per\eV}$.
This model reduces the tension to $3.6\sigma$, a substantial improvement over the 3+1 tension, but still too high.

In the second analysis, we expanded the previous MEOWS analysis to fit over the mixing parameter $\theta_{34}$ (or \Utaufsq). 
We do two different fits.
In the frequentist fit, we find a best fit point at $\Dmqfo = \SI{5.0}{\eV\squared}$, $\Umufsq = 0.04$, and $\Utaufsq =     0.006$, with a p-value of 5.2\% assuming Wilks' Theorem with 3 degrees of freedom.
A more accurate p-value calculated with pseudoexperiments is currently indicating a lower p-value, but more trials need to be run confirm that value. 
The Bayesian fit, on the other hand, is seeing a stronger preference for the sterile model versus the null.
The extracted Bayes factor of $\log_{10}<-1.56$ indicates a ``Very Strong'' preference for the sterile model, following Jeffreys' scale.
The point with the largest evidence was found at $\Dmqfo = \SI{5.0}{\eV\squared}$, $\Umufsq = 0.02$, and $\Utaufsq = 0.006$, near the frequentist best fit point.

    \appendix
    \chapter{Specific Contributions}

For the MiniBooNE analysis, my contributions included 

\begin{itemize}
    \item Running the data processing chain from start to finish. 
    \item Analyzing reconstructed $\pi^{0}$ invariant mass to find energy shift in the PMTs since the previous data run.
\end{itemize}

With regards to the global fits, my contributions included
\begin{itemize}
    \item Deriving the oscillation formulas for a neutrino decay width of $\Gamma$, and implementing them into the fitting software.
    \item Incorporating the various SBL experiments that released data since I started at MIT.
    \item Updated code to more exact oscillation formulae.
    \item Improved implementation of some already implemented experiments.
\end{itemize}

With respect to the IceCube analysis presented here, my contributions include 
\begin{itemize}
    \item Creating a \nutau simulation set for the analysis. 
    \item Implementing a modification in \texttt{PROPOSAL} to decay $\tau$ leptons as if they were polarized.
    \item Performing the sensitivity studies and analysis of the MEOWS+$\theta_{34}$ search. 
\end{itemize}

This work has been supported by the National Science Foundation.

    \chapter{Neutrino Oscillations Derivation}
\label{sec:NeutrinoOscillationsDerivation}

In this section, we will derive the oscillation formula for $N$ neutrinos. We follow the derivation provided in Ref.~\cite{Thomson:2013zua}, where we also generalize for an arbitrary number of neutrinos. 

We write out our $N \times N$ mixing expression as 
\begin{equation}
    \begin{pmatrix}
        \nu_\alpha \\
        \nu_\beta \\
        \vdots
    \end{pmatrix}
    = 
    \begin{pmatrix}
        U_{\alpha 1} & U_{\alpha 2} & \dots \\
        U_{\beta 1} & U_{\beta 2} & \\
        \vdots & & \ddots
    \end{pmatrix}
    \begin{pmatrix}
        \nu_1 \\
        \nu_2 \\
        \vdots
    \end{pmatrix}.
\end{equation}

Assuming a unitary mixing matrix, we can also write our expression as 
\begin{equation}
    \begin{pmatrix}
        \nu_1 \\
        \nu_2 \\
        \vdots
    \end{pmatrix}
    = 
    \begin{pmatrix}
        U_{\alpha 1}^{*} & U_{\beta 1}^{*} & \dots \\
        U_{\alpha2 }^{*} & U_{\beta 2}^{*} & \\
        \vdots & & \ddots
    \end{pmatrix}
    \begin{pmatrix}
        \nu_\alpha \\
        \nu_\beta \\
        \vdots
    \end{pmatrix}.
\end{equation}

By defining 
\begin{equation}
    U
    \equiv 
    \begin{pmatrix}
        U_{\alpha 1} & U_{\alpha 2} & \dots \\
        U_{\beta 1} & U_{\beta 2} & \\
        \vdots & & \ddots
    \end{pmatrix},
\end{equation}
we can write the unitarity condition \(U^{\dagger} U = U U^{\dagger} = I\) as 
\begin{equation}
    (UU^{\dagger})_{ij} = \sum_{k} U_{ik} U^{\dagger}_{kj} = \sum_{k} U_{ik} U_{jk} ^{*} = \delta_{ij}.
    \label{eq:13.19}
\end{equation}




\begin{figure}
    \vspace{1cm}
    \centering
    \parbox{100pt}
    {
        \begin{fmfgraph*}(100,100)
            \fmfstraight
            \fmfleft{ph1,i1}
            \fmfright{o1,o2,o3}
            \fmf{fermion}{i1,v1,o3}
            \fmf{fermion}{o1,v2,o2}
            \fmf{phantom}{ph1,v2}
            \fmf{boson}{v1,v2}
            \fmflabel{u}{i1}
            \fmflabel{d}{o3}
            \fmflabel{$\alpha^+$}{o1}
            \fmflabel{$\nu_\alpha$}{o2}
            \fmfv{l.a=-115,label=\(\frac{g_{W}}{\sqrt{2}} \)}{v2}
        \end{fmfgraph*}
        }
        \quad\quad \(\mathlarger{\mathlarger{\mathlarger{ = \sum_i }}}\) \quad\quad
        \parbox{100pt}
        {
            \begin{fmfgraph*}(100,100)
                \fmfstraight
                \fmfleft{ph1,i1}
                \fmfright{o1,o2,o3}
                \fmf{fermion}{i1,v1,o3}
                \fmf{fermion}{o1,v2,o2}
                \fmf{phantom}{ph1,v2}
                \fmf{boson}{v1,v2}
                \fmflabel{u}{i1}
                \fmflabel{d}{o3}
                \fmflabel{$\alpha^+$}{o1}
                \fmflabel{$\nu_i$}{o2}
                \fmfv{l.a=-115,label=\(\frac{g_{W}}{\sqrt{2}} U_{\alpha i}^{*}\)}{v2}
            \end{fmfgraph*}
            }
            \vspace*{5mm}
            \caption{The left side shows a \(\beta\)-decay to a neutrino of flavor \(\alpha\). The right side shows this same decay as a sum of contributions from the different mass eigenstates.}
            \label{fig:betadecay}
\end{figure}
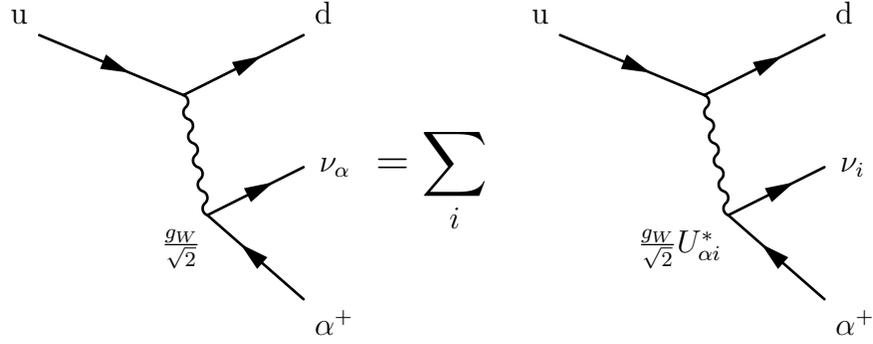

Let's assume that at $t=0$, a neutrino of flavor $\alpha$ is produced, as shown in \Cref{fig:betadecay}. Written in terms of its mass eigenstates, we have 
\begin{equation}
    \label{eq:after13.19}
    |\psi(0) \rangle = |\nu_\alpha \rangle \equiv \sum_{i} U_{\alpha i}^{*} |\nu _i \rangle.
\end{equation}
Note that, since the produced neutrino in \Cref{fig:betadecay} appears as an adjoint spinor, the complex conjugates of the matrix elements are used in \Cref{eq:after13.19}. 

Propagating these mass eigenstates as plane waves gives us the wavefunction  
\begin{equation}
    |\psi(\mathbf{x}, t) \rangle = \sum_{i} U_{\alpha i}^{*} |\nu _i \rangle e^{-i\phi_i},
\end{equation}
where 
\begin{equation}
    \phi_i = p_i \cdot t = E_i t - \mathbf{p_i \cdot x}.
    \label{eq:under13.7}
\end{equation}


\begin{figure}
    \vspace*{3mm}
    \centering
    \begin{fmfgraph*}(100,100)
        \fmfleft{i1,i2}
        \fmfright{o1,o2}
        \fmflabel{$\nu_\beta$}{i2}
        \fmflabel{d}{i1}
        \fmflabel{$\beta^-$}{o2}
        \fmflabel{u}{o1}

        \fmf{fermion}{i1,v1,o1}
        \fmf{fermion}{i2,v2,o2}
        \fmf{photon}{v1,v2}
    \end{fmfgraph*}
    \vspace*{5mm}
    \caption{Inverse \(\beta\)-decay interaction}
    \label{fig:inversebetadecay}
\end{figure}
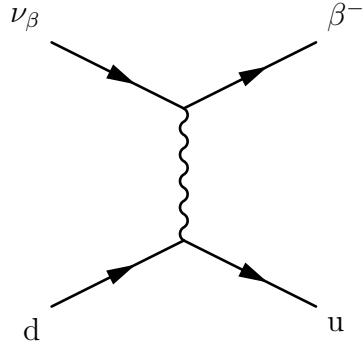

When this neutrino wavefunction later undergoes a charged current interaction at time $t$, as in \Cref{fig:inversebetadecay}, the interaction occurs in the flavor state, so we must expand the mass eigenstate terms to the corresponding weak eigenstates,
\begin{align}
    |\psi(\mathbf{x}, t) \rangle &= \sum_{i} U_{\alpha i}^{*} \left( \sum_\gamma U_{\gamma i}|\nu_\gamma \rangle \right) e^{-i\phi_i} \nonumber \\
    &= \sum_\gamma \left( \sum_{i}  U_{\alpha i}^{*} U_{\gamma i}e^{-i\phi_i} \right) |\nu_\gamma \rangle \nonumber\\
    &= \sum_\gamma c_\gamma |\nu_\gamma \rangle \label{eq:propagated},
\end{align}
where we have grouped together terms corresponding to the weak eigenstates \( |\nu_\gamma \rangle \), and we define 
\begin{equation}
    c_\gamma \equiv \sum_{i}  U_{\alpha i}^{*} U_{\gamma i}e^{-i\phi_i}.
\end{equation}
Note, again, that since the neutrino in \Cref{fig:inversebetadecay} comes in as a spinor, the PMNS matrix elements are doubly-complex conjugated (i.e. the complex conjugation cancels out). 

By writing out our propagated state as in \Cref{eq:propagated}, we can easily obtain the probability of a produced $\nu_\alpha$ oscillating and being detected as a $\nu_\beta$.

\begin{align}
    P(\nu_\alpha \to \nu_\beta ) &= |\langle \nu_\beta |\psi(\mathbf{x}, t) \rangle |^2  \\
    &= c_\beta c_\beta^{*} \\
    &= \left| \sum_{i}  U_{\alpha i}^{*} U_{\beta i}e^{-i\phi_i} \right|^2 \label{eq:13.22}
\end{align}

We can expand \Cref{eq:13.22} by using the following identity: 
\begin{align}
    \left|\sum_i z_i\right|^2 &= \sum_{ij} z_i z_j^{*} \\
    &= \sum_{i} |z_i|^2 +  \sum_{i\neq j} z_i z_j^{*} \\
    &= \sum_{i} |z_i|^2 +  \sum_{i < j} (z_i z_j^{*} + z_i ^{*}z_j) \\
    &= \sum_{i} |z_i|^ 2 + 2 \sum_{i < j} \Re(z_i z_j^{*}). \label{eq:13.23}
\end{align}

With \Cref{eq:13.23}, we can write \Cref{eq:13.22} as 

\begin{equation}
    P(\nu_\alpha \to \nu_\beta ) = \sum_i |U_{\alpha i}^{*} U_{\beta i}|^{2} + 2 \sum_{i<j} \Re(U_{\alpha i}^{*} U_{\beta i}U_{\alpha j} U_{\beta j}^{*}e^{-i(\phi_i-\phi_j)})
    \label{13.24}
\end{equation}

We can also use \Cref{eq:13.23} on \Cref{eq:13.19} to write 
\begin{equation}
    \left|\sum_{i} U_{\alpha i}^{*} U_{\beta i}\right|^{2} = \sum_{i}| U_{\alpha i}^{*} U_{\beta i}|^{2} + 2 \sum_{i<j} \Re( U_{\alpha i}^{*} U_{\beta i} U_{\alpha j} U_{\beta j}^{*}) = \delta_{\alpha \beta},
    \label{after13.24}
\end{equation}
which now lets us write \Cref{13.24} as 
\begin{equation}
    P(\nu_\alpha \to \nu_\beta ) =
    \delta_{\alpha \beta} 
    + 2 \sum_{i<j} \Re(U_{\alpha i}^{*} U_{\beta i}U_{\alpha j} U_{\beta j}^{*} ( e^{i(\phi_j-\phi_i)} - 1 ) ), 
\end{equation}
which we further expand into
\begin{equation}
    \begin{split}
        P(\nu_\alpha \to \nu_\beta )= \delta_{\alpha \beta} 
        &+ 2 \sum_{i<j} \Re(U_{\alpha i}^{*} U_{\beta i}U_{\alpha j} U_{\beta j}^{*}) (\cos(\phi_j-\phi_i) - 1 ) \\
        &-2 \sum_{i<j} \Im(U_{\alpha i}^{*} U_{\beta i}U_{\alpha j} U_{\beta j}^{*}) \sin(\phi_j-\phi_i).
        \label{expanded13.25}
    \end{split}
\end{equation}

By defining \(\Delta_{ji} = \frac{\phi_j - \phi_i}{2}\) and using trigonometric identities, we can rewrite 
\begin{align}
    \cos(2\Delta_{ji}) -1 &= \cos^{2}(\Delta_{ji}) - \sin^{2}(\Delta_{ji}) - 1 \\
    &= -\sin^{2}(\Delta_{ji}) - ( 1 - \cos^{2}(\Delta_{ji}) ) \\
    &= -2 \sin^{2}(\Delta_{ji}). 
\end{align}
Using this in \Cref{expanded13.25},
\begin{equation}
    \begin{split}
        P(\nu_\alpha \to \nu_\beta )= \delta_{\alpha \beta} 
        &- 4 \sum_{i<j} \Re(U_{\alpha i}^{*} U_{\beta i}U_{\alpha j} U_{\beta j}^{*}) \sin^{2}(\Delta_{ji}) \\
        &-2 \sum_{i<j} \Im(U_{\alpha i}^{*} U_{\beta i}U_{\alpha j} U_{\beta j}^{*}) \sin(2\Delta_{ji}).
    \end{split}
    \label{eq:oscequation}
\end{equation}

The final step is to rewrite \(\Delta_{ji}\) with physical quantities. From \Cref{eq:under13.7} we can write 
\begin{equation}
    \Delta_{ji} = \frac{\Delta \phi_{ji}}{2} = (E_2 - E_1) T - (p_2 - p_1) L.
\end{equation}
Let's assume that the momenta of the two neutrinos states $\nu_1$ and $\nu_2$ are equal\footnote{While this simplifies the derivation, the resulting oscillation equations remain the same without this assumption \cite{Giunti:2007ry}.}, $p \equiv p_1 = p_2$. In this case
\begin{equation}
    \begin{split}
        \Delta \phi_{21} &= (E_2 - E_1)T \\
        &= \left( (p^2 + m_2^2)^{\frac{1}{2}} - (p^2 + m_1^2)^{\frac{1}{2}} \right) T \\
        &= \left( p \left(1 + \frac{m_2^2}{p^2} \right)^\frac{1}{2} - p \left(1 + \frac{m_1^2}{p^2} \right)^\frac{1}{2} \right) T.
    \end{split}
    \label{eq:13.11}
\end{equation}
Because we assume that $m \ll p$, we are justified in doing the Taylor expansion
\begin{equation}
    \left(1 + \frac{m^2}{p^2} \right)^\frac{1}{2} \approx 1 +\frac{m^2}{2 p^2}.
\end{equation}
When this approximation is inputted into \Cref{eq:13.11}, we find
\begin{equation}
    \begin{split}
        \Delta \phi_{21} &\approx \left( p \left(1 + \frac{m_2^2}{2p^2} \right) - p \left(1 + \frac{m_1^2}{2p^2} \right) \right) T \\
        &= \frac{m_2^2 - m_1^2}{2 p} T \\
        &\approx \frac{m_2^2 - m_1^2}{2 p} L.
    \end{split}
\end{equation}
Where in the final step we use the approximation that $T = L$ because the neutrinos are traveling near the speed of light. 
With the final approximation $p=E$, we can write 
\begin{equation}
    \Delta \phi_{ji} = \frac{\Delta m_{ji}^2 L}{2 E}.
    \label{eq:13.12}
\end{equation}

Using \Cref{eq:13.12}, we can write \Cref{eq:oscequation} as
\begin{equation}
    \begin{split}
        P(\nu_\alpha \to \nu_\beta )= \delta_{\alpha \beta} 
        &- 4 \sum_{i<j} \Re(U_{\alpha i}^{*} U_{\beta i}U_{\alpha j} U_{\beta j}^{*}) \sin^{2}\left(\frac{\Delta m_{ji}^2 L}{4 E} \right) \\
        &-2 \sum_{i<j} \Im(U_{\alpha i}^{*} U_{\beta i}U_{\alpha j} U_{\beta j}^{*}) \sin \left(\frac{\Delta m_{ji}^2 L}{2 E} \right).
    \end{split}
    \label{eq:finaloscequationnatural}
\end{equation}

\Cref{eq:finaloscequationnatural} is the final neutrino oscillation formula, written in natural units. In the field of neutrino physics, the standard is to give the mass-squared splitting $\Delta m_{ji}^2$ in units of \si{\eV}, the energy $E$ in \si{\GeV}, and the distance $L$ in kilometers.
For the rest of this text we will use these units, unless otherwise stated.
With these unit conversions in mind, \Cref{eq:finaloscequationnatural} can be rewritten as 
\begin{equation}
    \begin{split}
        P(\nu_\alpha \to \nu_\beta )= \delta_{\alpha \beta} 
        &- 4 \sum_{i<j} \Re(U_{\alpha i}^{*} U_{\beta i}U_{\alpha j} U_{\beta j}^{*}) \sin^{2}\left( 1.27 \Delta m_{ji}^2 [\si{\eV}] \frac{L [\si{\kilo\meter}]}{E [\si{\GeV}]} \right) \\
        &-2 \sum_{i<j} \Im(U_{\alpha i}^{*} U_{\beta i}U_{\alpha j} U_{\beta j}^{*}) \sin \left(2.54 \Delta m_{ji}^2 [\si{\eV}] \frac{L [\si{\km}]}{E [\si{\GeV}]} \right).
    \end{split}
    \label{eq:appendixfinaloscequation}
\end{equation}

While \Cref{eq:appendixfinaloscequation} has a term for each combination of $\Delta m^{2}_{ij}$, they are not all independent. 
The relation between the different mass-squared splittings is given by 
\begin{equation}
    \Delta m^{2}_{ij} = \sum_{k=i}^{j-1} \Delta m^{2}_{k~k+1},
\end{equation}
so that there are only N-1 independent mass-squared splittings. 

    \chapter{MiniBooNE Supplementary Material}
\label{ch:miniboonesupp}

We include here the Supplementary Material for the MiniBooNE publication presented in \Cref{chapter:miniboone}.
\includepdf[pages={-}]{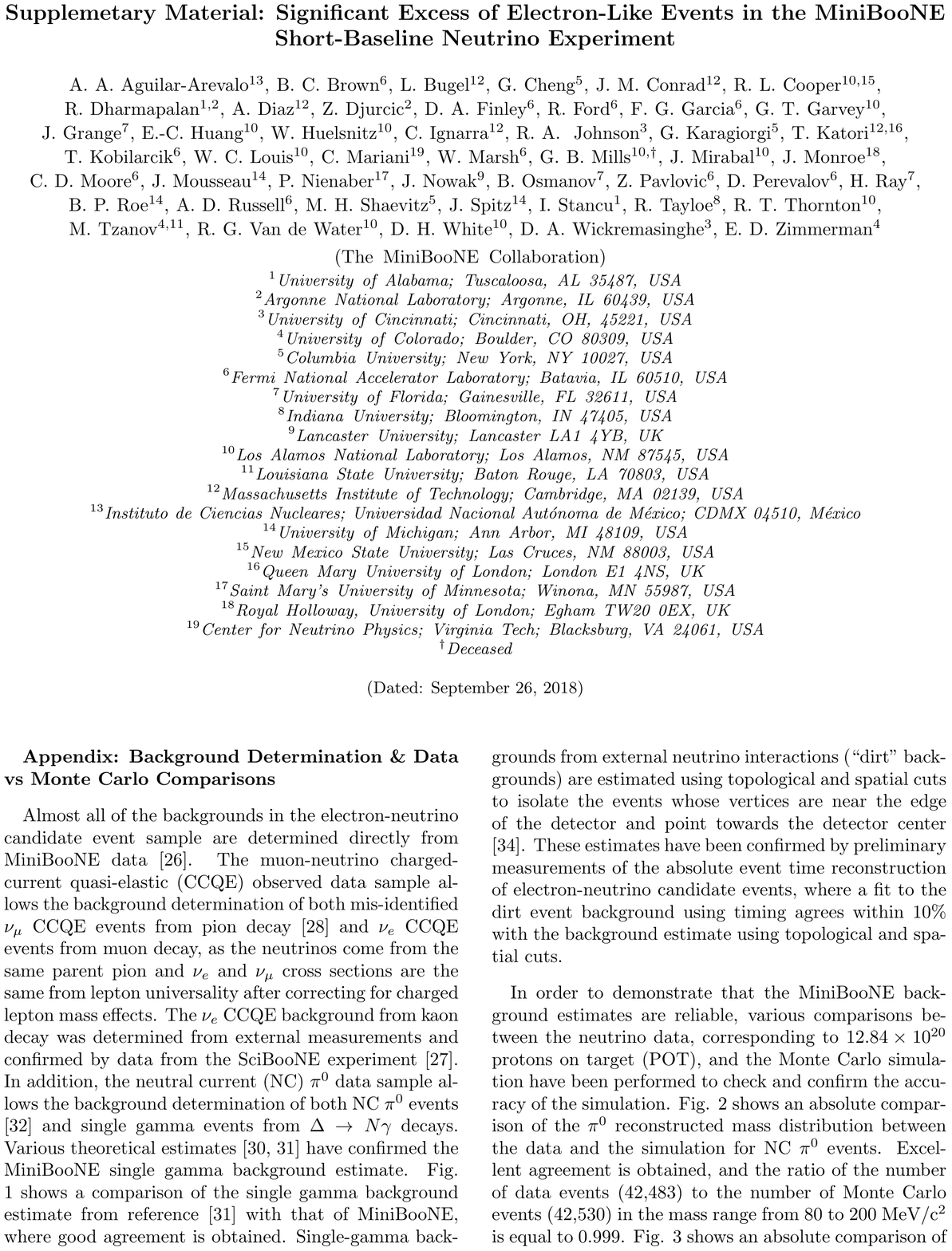}

\end{fmffile}
\begin{singlespace}
\printbibliography
\end{singlespace}

\end{document}